\documentclass[12pt]{iopart}
\usepackage{graphicx}
\usepackage{epsfig}
\usepackage{dcolumn}
\usepackage{bm}
\usepackage{amsfonts}
\usepackage{amssymb}
\usepackage{amsmath}
\usepackage{color}
\usepackage{imakeidx}
\usepackage{hyperref}
\usepackage{etoolbox}
\usepackage{cite}

\usepackage[export]{adjustbox}
\definecolor{steelblue}{RGB}{25,25,112}
\definecolor{DarkGreen}{rgb}{0.0,0.5,0.0}
\hypersetup{colorlinks,linkcolor={blue},citecolor={blue},urlcolor={blue}}  

\makeatletter
\newcommand{\mainmatter}{%
  \setcounter{footnote}{0}%
  \patchcmd{\@makefntext}{\fnsymbol}{\arabic}{}{}%
  \patchcmd{\@thefnmark}{\fnsymbol}{\arabic}{}{}%
  \def\@makefnmark{\textsuperscript{\arabic{footnote}}}%
}
\makeatother

\makeatletter
\def\@mkboth#1#2{}
\newlength\appendixwidth
\preto\appendix{\addtocontents{toc}{\protect\patchl@section}}
\newcommand{\patchl@section}{%
  \settowidth{\appendixwidth}{\textbf{Appendix }}%
  \addtolength{\appendixwidth}{1.5em}%
  \patchcmd{\l@section}{1.5em}{\appendixwidth}{}{\ddt}%
}
\makeatother

\makeatletter
\newcommand*{\shifttext}[2]{%
	\settowidth{\@tempdima}{#2}%
	\makebox[\@tempdima]{\hspace*{#1}#2}%
}
\makeatother

\newcommand{\metalambda}{%
	\mathop{%
		\rlap{$\Lambda$}%
		\mkern2mu
		\shifttext{2.5pt}{$\Lambda$}%
	}%
}

\makeindex

\begin{document}

\title[In the Realm of the Hubble tension $-$ a Review of Solutions]{In the Realm of the Hubble tension $-$ \\ a Review of Solutions~\footnote{In honor of the seminal work by E. Hubble~\cite{1936rene.book.....H}.}}

\author{Eleonora Di Valentino$^1$$^*$, Olga Mena$^2$, Supriya Pan$^3$, Luca Visinelli$^{4}$, Weiqiang Yang$^{5}$, Alessandro Melchiorri$^6$, David F. Mota$^7$, Adam G. Riess$^{8,9}$, Joseph Silk$^{8,10,11}$}

\ead{$^*$eleonora.di-valentino@durham.ac.uk}
\address{$^1$ Institute for Particle Physics Phenomenology, Department of Physics, Durham University, Durham DH1 3LE, UK.}
\address{$^2$ IFIC, Universidad de Valencia-CSIC, 46071, Valencia, Spain}
\address{$^3$ Department of Mathematics, Presidency University, 86/1 College Street, Kolkata 700073, India}
\address{$^{4}$ INFN, Laboratori Nazionali di Frascati, C.P. 13, I-100044 Frascati, Italy}
\address{$^{5}$ Department of Physics, Liaoning Normal University, Dalian, 116029, P.\ R.\ China}
\address{$^6$ Physics Department and INFN, Universit\`a di Roma ``La Sapienza'', Ple Aldo Moro 2, 00185, Rome, Italy}
\address{$^7$ Institute of Theoretical Astrophysics, University of Oslo, 0315 Oslo, Norway}
\address{$^8$ Department of Physics and Astronomy, Johns Hopkins University, Baltimore, MD 21218, USA}
\address{$^9$ Space Telescope Science Institute, 3700 San Martin Drive, Baltimore, MD 21218, USA}
\address{$^{10}$ Institut d'Astrophysique de Paris (UMR7095: CNRS \& UPMC- Sorbonne Universities), F-75014, Paris, France}
\address{$^{11}$ BIPAC, Department of Physics, University of Oxford, Keble Road, Oxford OX1~3RH, UK}

\maketitle
\normalsize

\newpage 
\begin{abstract}
The simplest $\Lambda$CDM model provides a good fit to a large span of cosmological data but harbors large areas of phenomenology and ignorance. With the improvement of the number and the accuracy of observations, discrepancies among key cosmological parameters of the model have emerged. The most statistically significant tension is the 4$\sigma$ to 6$\sigma$ disagreement between predictions of the Hubble constant, $H_0$, made by the early time probes in concert with the ``vanilla'' $\Lambda$CDM Cosmological model, and a number of late time, model-independent determinations of $H_0$ from local measurements of distances and redshifts. The high precision and consistency of the data at both ends present strong challenges to the possible solution space and demands a hypothesis with enough rigor to explain multiple observations -- whether these invoke new physics, unexpected large-scale structures or multiple, unrelated errors. A thorough review of the problem including a discussion of recent Hubble constant estimates and a summary of the proposed theoretical solutions is presented here. We include more than 1000 references, indicating that the interest in this area has grown considerably just during the last few years. We classify the many proposals to resolve the tension in these categories: Early Dark Energy, Late Dark Energy, Dark energy models with 6 degrees of freedom and their extensions, Models with extra relativistic degrees of freedom, Models with Extra Interactions, Unified cosmologies, Modified gravity, Inflationary models, Modified recombination history, Physics of the critical Phenomena, and Alternative proposals. Some are formally successful, improving the fit to the data in light of their additional degrees of freedom, restoring agreement within $1-2\sigma$ between \textit{Planck} 2018, using the Cosmic Microwave Background power spectra data, Baryon Acoustic Oscillations, Pantheon SN data, and R20, the latest SH0ES Team~\cite{Riess:2020fzl} measurement of the Hubble constant ($H_0 = 73.2 \pm 1.3{\rm\,km\,s^{-1}\,Mpc^{-1}}$ at 68\% confidence level). However, there are many more unsuccessful models which leave the discrepancy well above the $3\sigma$ disagreement level. In many cases, reduced tension comes not simply from a change in the value of $H_0$ but also due to an increase in its uncertainty due to degeneracy with additional physics, complicating the picture and pointing to the need for additional probes. While no specific proposal makes a strong case for being highly likely or far better than all others, solutions involving early or dynamical dark energy, neutrino interactions, interacting cosmologies, primordial magnetic fields, and modified gravity provide the best options until a better alternative comes along. 
\end{abstract}
\submitto{\CQG}

\newpage 
\tableofcontents
\clearpage
\mainmatter


\section{Introduction}

Although the standard cosmological scenario, the so-called $\Lambda$-Cold Dark Matter ($\Lambda$CDM) model, provides a remarkable fit to the bulk of available cosmological data, we should not forget that there is little understanding of the nature of its largest components. The aphorism, ``All models are wrong but some are useful'' (see e.g.\ Ref.~\cite{doi:10.1080/01621459.1976.10480949}) may be especially appropriate for $\Lambda$CDM which lacks the deep underpinnings a model requires to approach fundamental physics laws. Specifically, there are three ingredients, i.e.\ Inflation~\cite{Brout:1977ix, Guth:1980zm, Sato:1980yn}, Dark Matter (DM)~\cite{Rubin:1970zza, Trimble:1987ee} and Dark Energy (DE)~\cite{Riess:1998cb, Perlmutter:1998np}, for which the physical evidence comes from cosmological and astrophysical observations only. In addition, in the standard $\Lambda$CDM model we assume, these ingredients take on their simplest (i.e.\ ``Vanilla'') form (until there is strong evidence to the contrary), adopting an effective theory perspective for an underlying physical theory (yet to be discovered). With the increase of experimental sensitivity, deviations from the standard scenario therefore may be expected and could provide the means to reach a deeper understanding of the theory. In this predicament, we must be careful not to cling to the model too tightly or to risk missing the appearance of departures from the paradigm.

In this context, several tensions present between the different cosmological probes become interesting because, if not due to systematic errors (and as we shall later show, their explanation would appear to require multiple, unrelated errors), they could indicate a failure of the canonical $\Lambda$CDM model. Currently, the most notable anomalies worth consideration are those arising when the {\it Planck} satellite measurements~\cite{Aghanim:2018eyx} of the Cosmic Microwave Background (CMB) anisotropies are compared to low redshift probes, or compared within the {\it Planck} data itself. The {\it Planck} experiment has measured the CMB power spectra with an exquisite precision, but the constraints for the cosmological parameters are always model-dependent.\footnote{To date, very few conclusions about the kinematics and/or dynamics of the Universe have been made without model assumptions in cosmology, typically in the form of a $\Lambda$CDM model or in the form of a Friedmann-Lema\^{i}tre-Robertson-Walker (FLRW) metric. The claimed $\sim$1\% precision in cosmology is achieved at the expense of strong model assumptions. Additionally, the data reduction in the large cosmological surveys (employed before the cosmological model fit) is often achieved within the context of a $\Lambda$CDM fiducial model.} This means that, if there is no evidence for systematic errors in the data, a better model may be found which, if used for analysing the measured power spectra, would make tensions and anomalies disappear. In particular, extensively discussed in the literature, are the tensions present between the {\it Planck} data in a $\Lambda$CDM context~\cite{Aghanim:2018eyx} and local determinations of the Hubble constant, e.g.\ Ref.~\cite{Riess:2020fzl} (here R20), and the weak lensing experiments~\cite{Joudaki:2016kym, Hildebrandt:2016iqg, Hildebrandt:2018yau, Asgari:2020wuj, Abbott:2017wau} for the $S_8$ parameter. In addition, there are the {\it Planck} internal lensing anomalies related to the excess of lensing in the temperature power spectrum, producing a tension between the cosmological parameters extracted in the high-$\ell$ and low-$\ell$ multipole ranges: $A_{\rm lens}>1$ at about $2.8\sigma$~\cite{Calabrese:2008rt, Aghanim:2018eyx} and a closed Universe (i.e.\ a Universe with $\Omega_k<0$) is preferred at more than $3.4\sigma$ without the inclusion of additional constraints~\cite{Aghanim:2018eyx, DiValentino:2019qzk, Handley:2019tkm}.

In this review, we shall focus on the Hubble constant $H_0$ tension between the late time and early time measurements of the Universe because this is the most statistically significant, long-lasting and widely persisting tension, with 4$\sigma$ to 6$\sigma$ disagreement depending on the datasets considered. Indeed, this tension has existed since the first release of results from {\it Planck} in 2013~\cite{Ade:2013zuv} and has grown in significance with the improvement of the data. We consider a broad range of investigations performed over the last few years by the scientific community, and discuss how the Hubble constant value can be either resolved or reconciled in various cosmological models. 

After a presentation of the most recent experimental measurements of the Hubble constant in Section~\ref{sec:exp}, we revise the possibility of a local solution and the sound horizon problem in Section~\ref{sec:local}. At this point, we classify many proposals to resolve the Hubble puzzle in different categories: we discuss the Early Dark Energy models in Section~\ref{earlyDE}, the Late Dark Energy proposals in Section~\ref{lateDE}, the dark energy models with 6 degrees of freedom and their extensions in Section~\ref{DE6Dof}, models predicting extra relativistic degrees of freedom that can be parameterized by the effective number of neutrino species $N_{\rm eff}$ in Section~\ref{DR}, models with Extra Interactions between the different components of the Universe in Section~\ref{InteractSolut}, Unified cosmologies in Section~\ref{Unif}, modified gravity scenarios in Section~\ref{MG}, inflationary models in Section~\ref{inflat}, models of modified recombination history in Section~\ref{RecombH}, models based on the physics of the critical Phenomena in Section~\ref{CritPhen}, and finally in Section~\ref{others} we present other alternative proposals.

At the beginning of each section, we shall present an illustrative figure showing the estimated values of the present matter energy density parameter $\Omega_mh^2$, the Hubble constant $H_0$, and the sound horizon $r_dh$ for the several models described in the corresponding section. In these figures, we shall also depict a cyan horizontal band corresponding to the $H_0$ value measured in R20~\cite{Riess:2020fzl}, a yellow vertical band to the $\Omega_mh^2$ value estimated by {\it Planck} 2018~\cite{Aghanim:2018eyx} in a $\Lambda$CDM scenario, and a light green horizontal band associated with the $r_d h$ value measured by the Baryonic Acoustic Oscillation (BAO) data. The points sharing the same symbol refer to the very same model in the same paper, and the different colors refer to different dataset combinations. These plots are useful to have a clear visualization of the overall agreement of the proposed model with the current cosmological probes.
In addition, we shall also present a figure with a whisker plot illustrating the 68\% marginalized Hubble constant values obtained in the several cases reported in the section. 
We present our conclusions in Section~\ref{concl}.

Finally, in the Appendix we show Table~\ref{tabnotation} with the notation convention used in this review, two additional tables (i.e.\ Table~\ref{listPlanck} and Table~\ref{listPlanck+}) where we classify the several theoretical or phenomenological proposals depending on the agreement among their predictions of the Hubble constant and the value of $H_0$ reported in Ref.~\cite{Riess:2020fzl} and a useful plot in Figure~\ref{fig:all} for the readers. In particular, in Table~\ref{listPlanck} we report the results from those analyses that account for {\it Planck} data only, and in Table~\ref{listPlanck+} those that consider a combination of {\it Planck} plus additional observational probes. In Figure~\ref{fig:all} we show the combined effort made by the entire scientific community to solve or alleviate the Hubble constant tension until today.\footnote{This figure has been made by combining all the similar Figures in the review, i.e.\ Figures~\ref{fig:chapter4_H0Om},~\ref{fig:chapter5_H0Om},~\ref{fig:chapter6_H0Om},~\ref{fig:chapter7_H0Om},~\ref{fig:chapter8a_H0Om},~\ref{fig:chapter8b_H0Om},~\ref{fig:chapter9_H0Om},~\ref{fig:chapter11-14_H0Om} that are shown in the next sections.} A sample code for producing the whisker plots associated with this work is made publicly available online at \href{https://github.com/lucavisinelli/H0TensionRealm}{github.com/lucavisinelli/H0TensionRealm}.

\begin{figure*}
\includegraphics[width=0.9\textwidth]{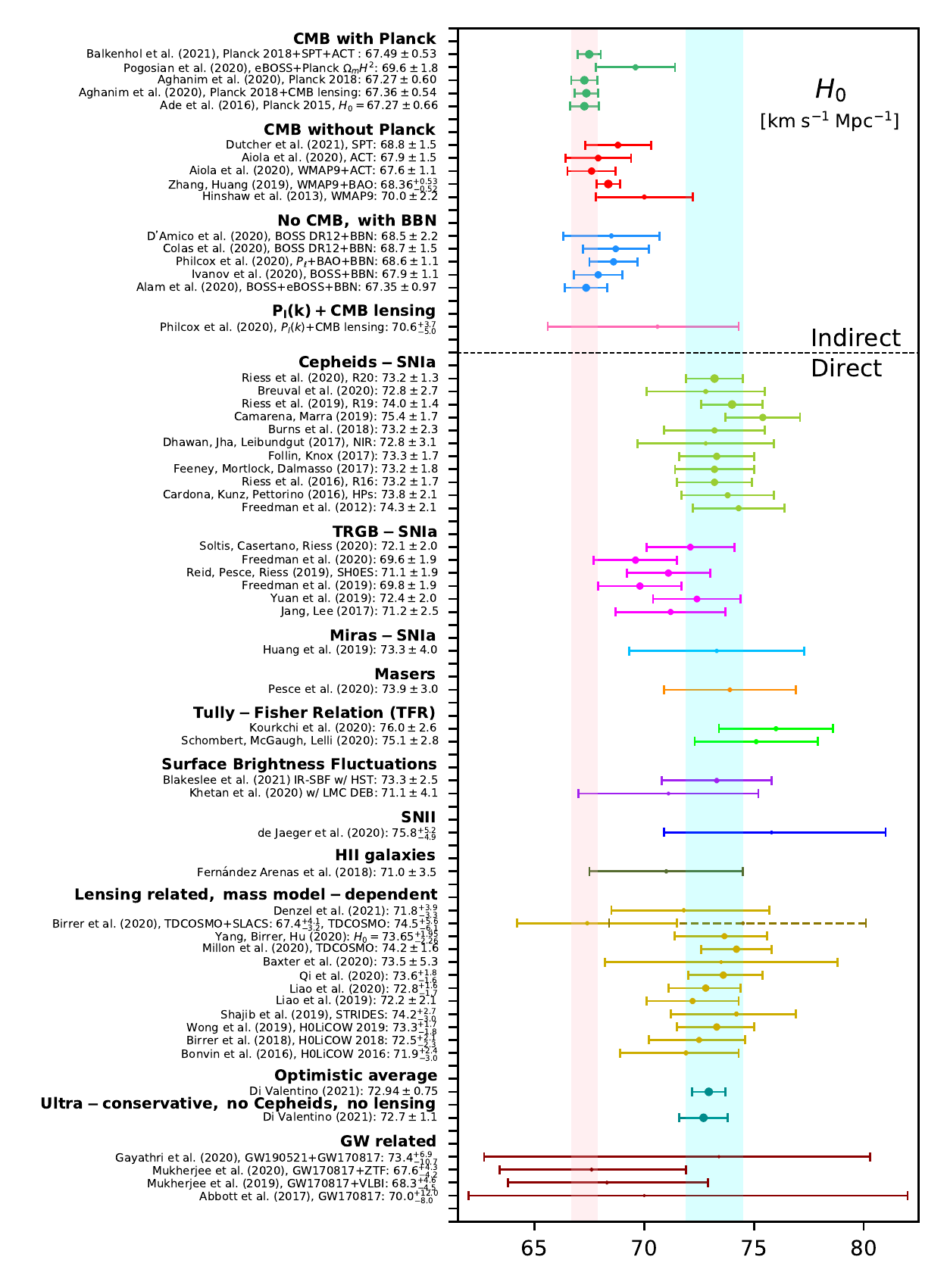}
\caption{Whisker plot with 68\% CL constraints of the Hubble constant $H_0$ through direct and indirect measurements by different astronomical missions and groups performed over the years. The cyan vertical band corresponds to the $H_0$ value from SH0ES Team~\cite{Riess:2020fzl} (R20, $H_0 = 73.2 \pm 1.3{\rm\,km\,s^{-1}\,Mpc^{-1}}$ at 68\% CL) and the light pink vertical band corresponds to the $H_0$ value as reported by {\it Planck} 2018 team~\cite{Aghanim:2018eyx} within a $\Lambda$CDM scenario. A sample code for producing similar figures with {\it any choice of the data} is made publicly available online at \href{https://github.com/lucavisinelli/H0TensionRealm}{github.com/lucavisinelli/H0TensionRealm}.}
\label{fig:H0-CMB-Local}
\end{figure*}

\section{Experimental measurements of \texorpdfstring{$H_0$}{H0}}
\label{sec:exp}

\begin{figure*}
\includegraphics[width=\textwidth]{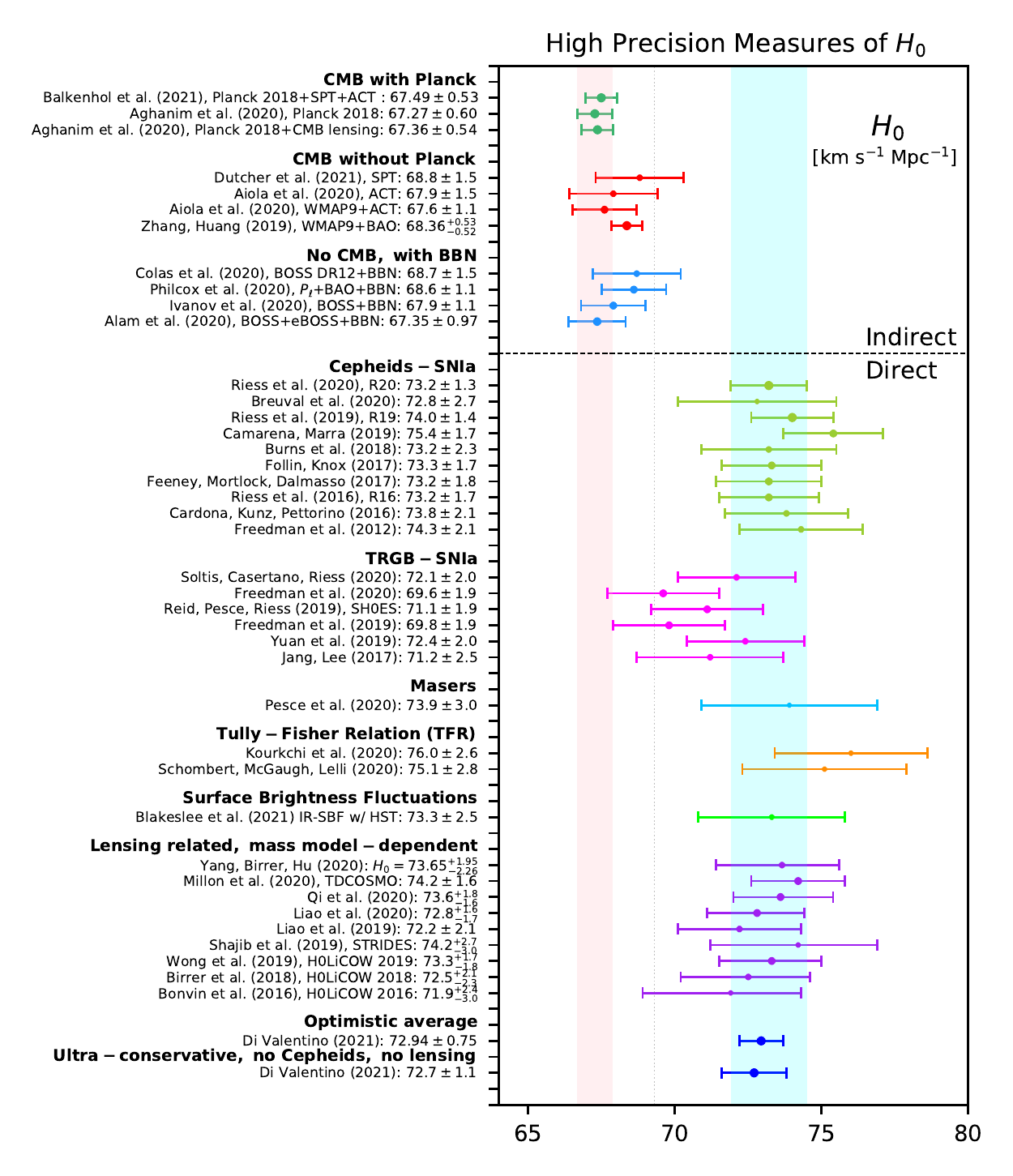}
\caption{Filtered version of Fig.~\ref{fig:H0-CMB-Local} showing the 68\% CL constraints of the Hubble constant $H_0$ with error bars less than $3 {\rm \,km\,s^{-1}\,Mpc^{-1}}$ for the direct measurements and less than $1.5 {\rm \,km\,s^{-1}\,Mpc^{-1}}$ for the indirect estimates. Similar to Fig.~\ref{fig:H0-CMB-Local}, the cyan vertical band corresponds to the $H_0$ value from SH0ES Team~\cite{Riess:2020fzl} (R20, $H_0 = 73.2 \pm 1.3{\rm\,km\,s^{-1}\,Mpc^{-1}}$ at 68\% CL) and the light pink vertical band corresponds to the $H_0$ value as reported by {\it Planck} 2018 team~\cite{Aghanim:2018eyx} within a $\Lambda$CDM scenario. A dotted vertical line for $H_0= 69.3 {\rm\,km\,s^{-1}\,Mpc^{-1}} $ has been added for a quick visualization of the division for the $H_0$ values obtained in the different measurements. }
\label{fig:H0-CMB-Local_2}
\end{figure*}

Within the class of cosmological models described by the Friedmann-Lema\^{i}tre-Robertson-Walker (FLRW) metric, the physical scale of the Universe is a time-dependent quantity whose knowledge allows us to convert all relative quantities to absolute ones. At a given time there should be only one correct distance scale of the background Universe. In principle, scales measured at different times should appear consistent when interpreted in the context of an accurate, time-dependent cosmological model. The Hubble constant (or Hubble-Lema\^itre constant) is the name given to the present expansion rate which sets the distance scale, defined as $H_0 \equiv a^{-1}\mathrm{d}a/\mathrm{d}t$ when the scale factor of the expanding Universe, $a=1$ (or $z=0)$. Figure~\ref{fig:H0-CMB-Local} (and~\ref{fig:H0-CMB-Local_2} for the filtered version) provide a useful reference for the following discussion of the Hubble constant landscape.

Because the Hubble constant tension appears to manifest as a difference between its value predicted via the use of measurements in concert with Early Universe physics (described by $\Lambda$CDM) and the value measured in the Late Universe (with or without the use of the late-time behavior of $\Lambda$CDM) we shall briefly review these two sets of inferences. To be explicit in our phenomenological definition, Early and Late do not refer to the redshift when the measurement is made but rather to the epoch of the $\Lambda$CDM model that is invoked. For example, a useful test is to consider whether a specific measurement has any dependence on the number of neutrinos included in $\Lambda$CDM (in this dichotomy Early does and Late does not).

\subsection{Early}

We consider here as ``Early'' predictions for $H_0$ those relying, in principle or in practice, on the accuracy of a number of assumptions of the $\Lambda$CDM model used to describe the Universe at $z>1000$, including a number of ansatzes about the properties of neutrinos (e.g.\ there are 3 active species known with a minimal total mass of $0.06{\rm \,eV}$ assuming normal hierarchy~\cite{Zyla:2020zbs}), particle interactions, the absence of primordial magnetic fields, a null running of the scalar spectral index, no additional relativistic particles or degrees of freedom, etc. Certain types and scales of breakdowns in these assumptions may be apparent within the CMB power spectra (and are not seen) though others may not. Many of these same ansatzes are used to relate local measurements of ``primordial'' abundances to the baryon density~\cite{Cooke:2017cwo}. The $\Lambda$CDM model is further used to describe the evolution of the Universe at $0<z<1000$ to predict the expansion rate, $H(z)$ and its present value, $H_0$, from the parameters derived from the CMB data and the early model. The late Universe form of $\Lambda$CDM makes use of different ansatzes than at early times including descriptions of dark matter (no interactions, stable, cold) and dark energy (as a cosmological constant). Again, some of these are tested but not to the precision with which they are relied upon in the model. For this reason the Hubble constant tension can identify a failure of the standard $\Lambda$CDM scenario at early or late epochs. 

First we review the status of $H_0$ predictions from a variety of CMB experiments beginning with {\it Planck} which is the de-facto ``Gold standard'' experiment. The most widely cited prediction from {\it Planck} in a flat $\Lambda$CDM model for the Hubble constant is $H_0=67.27 \pm 0.60{\rm \,km\,s^{-1}\,Mpc^{-1}}$ at 68\% confidence level (CL) for {\it Planck} 2018~\cite{Aghanim:2018eyx}, while it is $H_0=67.36 \pm 0.54{\rm \,km\,s^{-1}\,Mpc^{-1}}$ at 68\% CL for {\it Planck} 2018 + CMB lensing~\cite{Aghanim:2018eyx}, i.e.\ with the inclusion of the four-point correlation function or trispectrum data.\footnote{We will use {\it Planck} 2015 to indicate the full {\it Planck} 2015 TT, TE, EE + lowTEB dataset combination, and {\it Planck} 2015 TT for {\it Planck} 2015 TT + lowTEB. {\rm Here, TT is the temperature power spectrum, EE is the E-mode polarization auto-power spectra, and TE is the temperature-E-mode cross-power spectra.} Similarly, we will indicate with {\it Planck} 2018 the full {\it Planck} 2018 TT, TE, EE + low-$\ell$ + lowE combination and we will use {\it Planck} 2018 TT for the {\it Planck} 2018 TT + low-$\ell$ + lowE combination.}
The previous CMB satellite experiment Wilkinson Microwave Anisotropy Probe (WMAP)~\cite{Hinshaw:2012aka}, in its nine-year data release, assuming the same $\Lambda$CDM model, preferred a value for the Hubble constant $H_0=70.0 \pm 2.2 {\rm \,km\,s^{-1}\,Mpc^{-1}}$ at 68\% CL, a value that can be in agreement with both {\it Planck} and R20 because of its very large error bars. This conclusion used to apply
to another CMB experiment from the ground, South Pole Telescope (SPTPol)~\cite{Henning:2017nuy}, that reports a value of $H_0=71.3 \pm 2.1{\rm \,km\,s^{-1}\,Mpc^{-1}}$ at 68\% CL, considering the full datasets in TE and EE. However, the result from SPT-3G~\cite{Dutcher:2021vtw} improves from those in Ref.~\cite{Henning:2017nuy} and leads to a value of $H_0 = 68.8 \pm 1.5{\rm \,km\,s^{-1}\, Mpc^{-1}}$ at 68\% CL.
The recent SPTPol result is competitive with those from other ground-based experiments such as
the combination of the Atacama Cosmology Telescope (ACT), a ground based telescope, and WMAP. Indeed, the combination of ACT (from $\ell=600$ in TT and $\ell=350$ in TE/EE) and WMAP data, with a Gaussian prior on $\tau$ instead of the low-$\ell$ polarization likelihood, results in $H_0=67.6 \pm 1.1{\rm \,km\,s^{-1}\,Mpc^{-1}}$ at 68\% CL~\cite{Aiola:2020azj}, always assuming a $\Lambda$CDM model, or $H_0=67.9 \pm 1.5{\rm \,km\,s^{-1}\,Mpc^{-1}}$ at 68\% CL for ACT alone.
Finally, a combination of ground based CMB experiments SPT, SPTPol, and the Atacama Cosmology Telescope Polarimeter (ACTPol) gives $H_0=69.72 \pm 1.63{\rm \,km\,s^{-1}\,Mpc^{-1}}$ at 68\% CL~\cite{Wang:2019isw}, while SPTPol + ACTPol, when combined with the {\it Planck} dataset, gives $H_0 = 67.49 \pm 0.53{\rm \,km\,s^{-1}\, Mpc^{-1}}$ at 68\% CL~\cite{Balkenhol:2021eds}.

We may also consider less precise constraints that arise exclusively from measurements of the polarization of the CMB, i.e.\ from the EE CMB power spectra~\cite{Addison:2021amj}, always assuming a $\Lambda$CDM model: {\it Planck} EE gives $H_0=70.0 \pm 2.7{\rm \,km\,s^{-1}\,Mpc^{-1}}$ at 68\% CL, ACTPol $H_0=72.4^{+3.9}_{-4.8}{\rm \,km\,s^{-1}\,Mpc^{-1}}$ at 68\% CL, and SPTPol $H_0=73.1^{+3.3}_{-3.9}{\rm \,km\,s^{-1}\,Mpc^{-1}}$ at 68\% CL, but their combination finds $H_0=68.7\pm1.3{\rm \,km\,s^{-1}\,Mpc^{-1}}$ at 68\% CL for the different directions of correlations~\cite{Addison:2021amj}.

Measurements of Baryon Acoustic Oscillations (BAO) (or other features in galaxy power spectra) at any redshift are ``scale-free'', primarily constraining the product of the sound horizon and the $H_0$ value, but neither without a prior on the other. When the prior comes from the CMB, or baryon abundance estimates, the determination of $H_0$ depends on the above ansatz at $z>1000$ and we will consider the result as belonging to the Early or indirect class. As such, there are $H_0$ estimates from a reanalysis of the Baryon Oscillation Spectroscopic Survey (BOSS) Data Release 12 (DR12) on anisotropic galaxy clustering in Fourier space~\cite{Ivanov:2019pdj}, that provide $H_0=67.9 \pm 1.1{\rm \,km\,s^{-1}\,Mpc^{-1}}$ at 68\% CL using a prior on the physical baryon density $\omega_b$, derived from measurements of primordial deuterium abundance~\cite{Cooke:2017cwo} ($D/H = (2.527\pm0.030)\times 10^{-5}$) assuming the standard Big Bang Nucleosynthesis (BBN) picture, and a $\Lambda$CDM model with a total neutrino mass free to vary in a small CMB-motivated range and a fixed primordial power spectrum tilt $n_s$ to the {\it Planck} best-fit.
The same lower $H_0$ is confirmed also from a reanalysis of the BOSS DR12 data using the Effective Field Theory of Large-Scale Structure (EFTofLSS) formalism~\cite{DAmico:2019fhj}, predicting the clustering of Cosmological Large-Scale Structure in the mildly non-linear regime, that results in $H_0=68.5 \pm 2.2{\rm \,km\,s^{-1}\,Mpc^{-1}}$ at 68\% CL, always assuming BBN, and fixing the values of the baryon/dark-matter ratio, $\Omega_b/\Omega_c$, and $n_s$ to the {\it Planck} 2018 best-fit. A companion paper~\cite{Colas:2019ret} gives instead $H_0=68.7 \pm 1.5{\rm \,km\,s^{-1}\,Mpc^{-1}}$ at 68\% CL, assuming a BBN prior on $\Omega_b h^2$ instead of $\Omega_b/\Omega_c$.
In addition, the combination of BAO from Main Galaxy Sample (MGS)~\cite{Strauss:2002dj}, BOSS galaxy and extended BOSS (eBOSS), with the BBN prior independent from the CMB anisotropies, provides $H_0=67.35 \pm 0.97{\rm \,km\,s^{-1}\,Mpc^{-1}}$ at 68\% CL in a $\Lambda$CDM scenario~\cite{Alam:2020sor}. 
Moreover, a lower Hubble constant $H_0=68.19 \pm 0.36{\rm \,km\,s^{-1}\,Mpc^{-1}}$ at 68\% CL~\cite{Alam:2020sor} is also obtained within the $\Lambda$CDM scheme when combining together {\it Planck} 2018, the Pantheon sample~\cite{Scolnic:2017caz} of 1048 Type Ia supernovae (SNIa), Sloan Digital Sky Survey (SDSS) BAO + Redshift Space Distortions (RSD), and the Dark Energy Survey (DES) $3\times2$pt data~\cite{Troxel:2017xyo, Abbott:2017wau, Krause:2017ekm}.
We have to note here that SNIa data is similar to BAO in that it is scale-free and cannot directly measure $H_0$ nor is Early or Late until its luminosity is calibrated at one end or the other.
These lower Hubble constant values are in agreement with previous estimates, when other BAO data~\cite{Beutler:2011hx,Ross:2014qpa,Alam:2016hwk} were included in the dataset combinations (see also Refs.~\cite{Abbott:2017smn,Wang:2017yfu,Zhang:2018jfu,Zhang:2019cww,Schoneberg:2019wmt,Cuceu:2019for}).
For a flat $\Lambda$CDM model, the combination of WMAP + BAO (6dF Galaxy Survey, MGS, the BOSS DR12 galaxies and the eBOSS DR14 quasars) also gives a lower value $H_0=68.36^{+0.53}_{-0.52}{\rm \,km\,s^{-1}\,Mpc^{-1}}$ at 68\% CL~\cite{Zhang:2018air}. Lastly, a combination of galaxy cluster sparsity, cluster gas mass fraction and BAO gives $H_0=69.6\pm1.7{\rm \,km\,s^{-1}\,Mpc^{-1}}$ at 68\% CL~\cite{Corasaniti:2021ihg}.

By combining the unreconstructed BOSS DR12 galaxy power spectra $P_\ell(k)$, modeled using the EFTofLSS, assuming a weak Gaussian prior on the amplitude of the scalar primordial power spectrum $A_s$ centered on the {\it Planck} best-fit, and a $\Omega_m$ prior from Pantheon, Ref.~\cite{Philcox:2020xbv} finds $H_0 = 65.1 ^{+3.0}_{-5.4}{\rm \,km\,s^{-1}\,Mpc^{-1}}$ at 68\% CL.
In addition, the same analysis is performed with a $\Omega_m$ prior from uncalibrated BAO (6dFGS, MGS, and eBOSS DR14 Lyman-$\alpha$ measurements) giving $H_0 = 65.6 ^{+3.4}_{-5.5}{\rm \,km\,s^{-1}\,Mpc^{-1}}$ at 68\% CL~\cite{Philcox:2020xbv}. Finally, considering the combination of $P_\ell(k)$ with the {\it Planck} 2018 CMB-marginalized lensing likelihood~\cite{Aghanim:2019ame}, and a prior on $A_s$ twice tighter than before, Ref.~\cite{Philcox:2020xbv} obtains $H_0 = 70.6 ^{+3.7}_{-5.0}{\rm \,km\,s^{-1}\,Mpc^{-1}}$ at 68\% CL.
This result is shifted and slightly stronger (for the addition of galaxy information) with respect to another sound horizon independent measurement as obtained in Ref.~\cite{Baxter:2020qlr}, that, analysing the same CMB lensing data from {\it Planck}, using conservative external priors on $\Omega_m$ from Pantheon and $A_s$ from {\it Planck} 2018, and varying the total neutrino mass, finds $H_0 = 73.5 \pm 5.3{\rm \,km\,s^{-1}\,Mpc^{-1}}$ at 68\% CL.
Finally, for the combination $P_\ell(k)$ + BAO + BBN, Ref.~\cite{Philcox:2020vvt} finds $H_0 = 68.6 \pm 1.1{\rm \,km\,s^{-1}\,Mpc^{-1}}$ at 68\% CL within a $\Lambda$CDM model plus a total neutrino mass free to vary, using a prior on the physical baryon density $\omega_b$ but neglecting any knowledge on the power spectrum tilt $n_s$.

Using the latest BAO data, including the eBOSS DR16 measurements~\cite{Alam:2020sor}, and a prior on $\Omega_mh^2$ based on the {\it Planck} 2018 best fit in a $\Lambda$CDM model, Ref.~\cite{Pogosian:2020ded} finds $H_0 = 69.6 \pm 1.8{\rm \,km\,s^{-1}\,Mpc^{-1}}$ at 68\% CL. Considering Pantheon SNIa apparent magnitude + DES-3yr binned SNIa apparent magnitude + $H(z)$ + BAO in Ref.~\cite{Cao:2021ldv} the authors find $H_0=68.8\pm1.8{\rm \,km\,s^{-1}\,Mpc^{-1}}$ at 68\% CL. In Ref.~\cite{Lemos:2018smw} the authors apply the inverse distance ladder to fit a parametric form of $H(z)$ to BAO and SNIa data, using priors on the sound horizon at the drag epoch $r_d$ from {\it Planck}, obtaining $H_0=68.42\pm0.88{\rm \,km\,s^{-1}\,Mpc^{-1}}$ at 68\% CL, and from WMAP, obtaining $H_0=67.9\pm1.0{\rm \,km\,s^{-1}\,Mpc^{-1}}$ at 68\% CL.

It may be worth noting that Early inferences of $H_0$ tend to increase (rather than decrease) from the baseline value derived from the {\it Planck} 2018 temperature anisotropy data with the inclusion of polarization data, BAO data, or additional freedom in $\Lambda$CDM (see Figure~\ref{fig:H0-CMB-Local}).

\subsubsection{CMB $-$ systematics in Planck?}

The {\it Planck} CMB angular spectra provide the most precise constraints on the cosmological parameters. However, as with any experimental measurement, it is not free from systematic errors. Let us therefore briefly discuss here what are these errors and whether they may have a significant impact in the determination of $H_0$ under the $\Lambda$CDM assumption.

First of all, the {\it Planck} collaboration~\cite{Aghanim:2019ame} presented the results using two different likelihood pipelines for the data at multipoles $\ell>30$: {\tt Plik} and {\tt CamSpec} (now updated in Ref.~\cite{Efstathiou:2019mdh}). It is important to stress here that, while both likelihood codes in principle should use the same measurements, in reality they consider different sky masks and chunks of data. Moreover, they treat foregrounds in a significant different way, especially for what concerns polarization. In the case of {\tt Plik}, for example, foregrounds and calibration efficiencies are treated by varying $21$ additional parameters, while in {\tt CamSpec} only $9$ parameters are varied. This is because in {\tt CamSpec}, the foregrounds in polarization are subtracted in the map domain, and it does not include the $100 \times 100\,$GHz TT spectrum.
The cosmological constraints on $\Lambda$CDM parameters from {\tt Plik} and {\tt CamSpec} differ at most
by $0.5 \sigma$ in case of the baryon density and just by $0.1 \sigma$ for the Hubble constant~\cite{Aghanim:2019ame}. 
While the choice between {\tt Plik} or {\tt CamSpec} seems to have little effect in reducing the Hubble tension, it is important to stress that just a different likelihood assumption could in principle shift by $0.5 \sigma$ any constraint coming from the CMB.

A more worrying systematic could, on the contrary, be responsible for the so-called $A_{\rm lens}$ anomaly. Introduced in Ref.~\cite{Calabrese:2008rt}, the $A_{\rm lens}$ parameter is an ``unphysical'' parameter that simply rescales by hand the effects of gravitational lensing on the CMB angular power spectra, and can be measured by the smoothing of the peaks in the damping tail. For $A_{\rm lens}=0$ one has no lensing effect, while for $A_{\rm lens}=1$ one simply recovers the value expected in the cosmological model of choice. Interestingly, the {\it Planck} CMB power spectra show a preference for $A_{\rm lens}>1$ at more than two standard deviations using both {\tt Plik} and {\tt CamSpec}. Perhaps, even more interesting is that the inclusion of BAO data provides evidence for $A_{\rm lens}>1$ at more than $99 \%$ CL (about 99\% for the {\tt CamSpec} likelihood pipeline).
Having $A_{\rm lens}>1$ can not be easily explained theoretically since it would require either a closed Universe (that would challenge several other datasets and the simplest inflationary models~\cite{DiValentino:2019qzk}) or even more exotic solutions such as the modifications to GR~\cite{Ade:2015rim,DiValentino:2015bja,Aghanim:2018eyx,Moshafi:2020rkq}. Moreover, this lensing anomaly is not seen in the {\it Planck} trispectrum data (CMB lensing) that offer a complementary and independent measurement. If not due to new physics, the $A_{\rm lens}$ anomaly is probably due to a small but still undetected systematic error in the {\it Planck} data. Can this systematic help in reducing the Hubble tension? The answer is affirmative. When $A_{\rm lens}$ is included in the analysis, the {\it Planck} and {\it Planck} + BAO constraints on $H_0$ are indeed slightly shifted towards higher values to $H_0=68.3\pm0.7{\rm \,km\,s^{-1}\,Mpc^{-1}}$ and $H_0=68.22\pm0.49{\rm \,km\,s^{-1}\,Mpc^{-1}}$ at 68\% CL, respectively, using either {\tt Plik} or {\tt CamSpec}. Assuming the {\it Planck} constraints, the introduction of $A_{\rm lens}$ would therefore reduce from $4.2$ $\sigma$ to $3.3 \sigma$ the current tension with the R20 constraint of $H_0=73.2 \pm 1.3{\rm \,km\,s^{-1}\,Mpc^{-1}}$ at 68\% CL~\cite{Riess:2020fzl}.

However, a proper physical interpretation of $A_{\rm lens}$ is still unavailable. If, indeed, $A_{\rm lens}$ demands for new physics, then one may actually derive a {\it smaller} value of $H_0$ from the {\it Planck} satellite. In a physical model based on General Relativity (GR), more lensing is now inevitably connected to an increase in the cold dark matter density and this changes the previous constraints. Just as an example, if a closed Universe is the explanation for $A_{\rm lens}>1$, then the Hubble constant from {\it Planck} could be as low as $\sim 55 {\rm \,km\,s^{-1}\,Mpc^{-1}}$~\cite{Aghanim:2018eyx, DiValentino:2019qzk, Handley:2019tkm}. Nonetheless, as we discuss in this review, (exotic) modified gravity models have been proposed that could explain at the very same time the {\it Planck} lensing anomaly and the Hubble tension.
On the other hand, if $A_{\rm lens}$ is due to systematics, then there is still the question, if the same systematic is fully described by $A_{\rm lens}$, or if further extensions are needed and how they could impact the final constraints on $H_0$.

In a few words, one can conclude that systematics in the {\it Planck} data (as in any other experimental measurement) could certainly be present and are actually suggested by the $A_{\rm lens}$ anomaly. However, at the moment, there is no indication for a systematic that could increase the mean value of the Hubble constant from {\it Planck} by significantly more than $1 {\rm \,km\,s^{-1}\,Mpc^{-1}}$ under the $\Lambda$CDM assumption. The Hubble tension, even if weakened in statistical significance, would probably remain.

\subsection{Late}
\label{sec:late}

The best-established and only strictly empirical method to measure $H_0$ locally comes from measuring the distance-redshift relation, usually undertaken by building a ``distance ladder''. The most often utilized approach is to use geometry (e.g.\ parallax) to calibrate the luminosities of specific star types (e.g.\ pulsating Cepheid variables and exploding Type Ia supernovae or SNIa) which can be seen at great distances where their redshifts measure cosmic expansion. Cepheids are most often used to reach distances of $10-40\,$Mpc because they are the brightest objects in the optical with luminosities reaching in excess of 100,000 solar luminosities and offer the highest precision per object of about 3\% in distance at a given pulsation period.\footnote{For a discussion about the cosmological model insensitivity of the local measure of the Hubble constant $H_0$ from the Cepheid distance ladder see Ref.~\cite{Dhawan:2020xmp}.} SNIa exceed a billion solar luminosities and are nearly as precise per object but they are rare in any volume, such as the local one, thus often serve as the last rung on the distance ladder. These methods treat stars as empirical, standardized candles, i.e.\ the premise that once empirically standardized, the same type has the same luminosity, without reference to stellar modeling or astrophysics theory. One may consider the failure of this premise to be anti-Copernican and harder to imagine than a failure of $\Lambda$CDM!

The Hubble Space Telescope (HST) provided the first capability to measure Cepheids beyond a few Mpc to reach the nearest SNIa hosts (and the hosts of other long-range distance indicators) and the final result of the Hubble Space Telescope Key Project was $(72 \pm 8){\rm \,km\,s^{-1}\,Mpc^{-1}}$~\cite{Freedman:2000cf}, a result later recalibrated to use improved geometric distance calibration to the Large Magellanic Cloud (LMC) to yield $(74.3 \pm 2.2){\rm \,km\,s^{-1}\,Mpc^{-1}}$~\cite{Freedman_2012}, see also Ref.~\cite{Sandage:2006cv}. However, these efforts were severely limited by the reach of the first generation of Hubble instruments to observing Cepheids in the hosts of just a few well-observed, well-standardizable SNIa.

The SH0ES Project started in 2005 and advanced this approach by 
\begin{enumerate}
\item increasing the sample of high quality calibrations of SNIa by Cepheids from a few to 19 (R16)~\cite{Riess:2016jrr}, 
\item increasing the number of independent geometric calibrations of Cepheids to five (R18)~\cite{Riess:2018uxu} including by extending the range of parallax measurements to Cepheids using spatial scanning of HST, 
\item measuring the fluxes of Cepheids with geometric distance measurements and those in supernova hosts with the same instrument to negate calibration errors (R19)~\cite{Riess:2019cxk}, 
\item measuring Cepheids in the near-infrared to reduce systematics related to dust and reddening laws. 
\end{enumerate}
Improved geometric distance estimates to the LMC using detached eclipsing binaries~\cite{Pietrzy_ski_2019}, to NGC 4258 using water masers~\cite{Reid:2019tiq} and to Milky Way Cepheids from European Space Agency (ESA) Gaia parallaxes~\cite{lindegren2020gaia} have greatly advanced this work in recent years. 
The values of $H_0$ by this route have ranged between 73 $-$ 74 ${\rm \,km\,s^{-1}\,Mpc^{-1}}$, with the present status based on the improved ESA Gaia mission Early Data Release 3 (EDR3) of parallax measurements using 75 Milky Way Cepheids with Hubble Space Telescope photometry and EDR3 parallaxes~\cite{gaiacollaboration2020gaia}, that gives $H_0=73.2 \pm 1.3{\rm \,km\,s^{-1}\,Mpc^{-1}}$ at 68\% CL~\cite{Riess:2020fzl}, in tension at $4.2\sigma$ with the {\it Planck} value in a $\Lambda$CDM scenario. We will refer to this new measurement as R20 and this will be a reference throughout the review. This value is also close to the conservative average (excludes R20) and optimistic average (includes R20) we present later in this section so this is a reasonable overall benchmark.

There have been numerous reanalyses of the SH0ES data using different formalisms, statistical methods of inference, or replacement of parts of the dataset, but none has produced a significant indication of a change in $H_0$. The larger value of $H_0$ is seen in the reanalysis of the R16 Cepheid data by using Bayesian hyper-parameters~\cite{Cardona:2016ems} $H_0=73.75 \pm 2.11{\rm \,km\,s^{-1}\,Mpc^{-1}}$ at 68\% CL, and the local determination of the Hubble constant~\cite{Camarena:2019moy} achieved using the cosmographic expansion of the luminosity distance, that gives $H_0=75.35\pm 1.68{\rm \,km\,s^{-1}\,Mpc^{-1}}$ at 68\% CL. There is a measurement obtained replacing the sample of SNIa measured in the optical with that measured in the near-infrared (NIR) where SNIa are better standard candles~\cite{Dhawan:2017ywl}, i.e.\ $H_0=72.8 \pm 1.6 ({\rm stat}) \pm 2.7 ({\rm sys}) {\rm \,m\,s^{-1}\,Mpc^{-1}}$ at 68\% CL. Other measurements based on the Cepheids-SNIa include Ref.~\cite{Burns:2018ggj}, that finds $H_0 = 73.2 \pm 2.3{\rm \,km\,s^{-1}\,Mpc^{-1}}$ at 68\% CL, analysing the final data release of the Carnegie Supernova Project I and a different method for standardizing SNIa light curves. A number of reanalyses including a notable one that leaves the reddening laws in distant galaxies uninformed by the Milky Way is performed in Ref.~\cite{Follin:2017ljs}, that finds $H_0=73.3 \pm 1.7{\rm \,km\,s^{-1}\,Mpc^{-1}}$ at 68\% CL. These are in agreement with R16, showing that systematic bias or uncertainty in the Cepheid calibration step of the distance ladder measurement can not explain the Hubble tension. Reference~\cite{Feeney:2017sgx} produces an estimate of the Hubble constant based on a Bayesian hierarchical model of the local distance ladder, that gives $H_0=73.15\pm1.78{\rm \,km\,s^{-1}\,Mpc^{-1}}$ at 68\% CL, allowing outliers to be modeled. These measurements generally made use of the Cepheid photometry presented by the SH0ES Team. However, the previously cited result for $H_0$ of $74.3 \pm 2.2{\rm \,km\,s^{-1}\,Mpc^{-1}}$ from Ref.~\cite{Freedman_2012} used an independent set of Cepheid data from that of the SH0ES Team, obtained with different instruments on HST, and with photometry measured with different algorithms (and by different investigators) which removes the dependence of the tension on any one set of Cepheid measurements. Similarly, Ref.~\cite{Javanmardi:2021viq} has undertaken a complete reanalysis of SH0ES Cepheid measurements starting at the pixel level from the Hubble Space Telescope data and using different methods for measuring Cepheid photometry, correcting for bias, developing new Cepheid light curve templates, etc, and the result agreed with the prior SH0ES analysis in R16 to 0.5 $\sigma$ or 0.02 mag (1\% in distance) indicating that the measurements are robust.

Using the Gaia Data Release 2 parallaxes~\cite{Brown:2018dum} of Cepheid companions (in binaries or host clusters rather than of the Cepheids themselves) to obtain a Galactic calibration of the Leavitt law in the V, J, H, $K_S$, and Wesenheit $W_H$ bands, it is possible to derive a Hubble constant measurement anchored to Milky Way Cepheids. When all Cepheid companions are considered, the authors in Ref.~\cite{Breuval:2020trd} obtain $H_0=72.8 \pm 1.9 ({\rm stat + sys}) \pm 1.9 ({\rm parallax\ zero\!-\!point}){\rm \,km\,s^{-1}\,Mpc^{-1}}$ at 68\% CL.

There have been alternative distance ladders which substitute another type of star for Cepheids. There are such measurements obtained using the Tip of the Red Giant Branch (TRGB) in lieu of Cepheids, performed by different teams, and these are in the range of $\sim$ 70-72 ${\rm \,km\,s^{-1}\,Mpc^{-1}}$. We have the 2017 measurement of the Hubble constant based on the calibration of the SNIa using the TRGB obtained by Ref.~\cite{Jang:2017dxn}, that is $H_0=71.17 \pm 1.66 ({\rm random}) \pm 1.87 ({\rm sys}) {\rm \,km\,s^{-1}\,Mpc^{-1}}$ at 68\% CL. There is the 2019 determination made by Ref.~\cite{Freedman:2019jwv} which measures TRGB in a nine SNIa hosts, adds 5 from ~\cite{Jang:2017dxn}, and calibrates TRGB in the LMC which yields $H_0=69.8 \pm 0.8 ({\rm stat}) \pm 1.7 ({\rm sys}){\rm \,km\,s^{-1}\,Mpc^{-1}}$ at 68\% CL and Ref.~\cite{Freedman:2020dne} (F20), for which $H_0=69.6 \pm 0.8 ({\rm stat}) \pm 1.7 ({\rm sys}) {\rm \,km\,s^{-1}\,Mpc^{-1}}$ at 68\% CL, or the same but with a different accounting of the LMC extinction of the TRGB using reddening maps derived from Red Clump stars by~\cite{Yuan:2019npk} gives $H_0=72.4\pm2.0 {\rm \,km\,s^{-1}\,Mpc^{-1}}$ at 68\% CL. A value of $H_0\sim72 {\rm \,km\,s^{-1}\,Mpc^{-1}}$ also results from the revised OGLE Team LMC reddening maps~\cite{Skowron_2021, Soltis:2020gpl}. The addition of two new TRGB measurements in NGC 1404 and NGC 5643, host to 4 SNIa~\cite{hoyt2021carnegie} appears to raise the F20 value of $H_0$ by $\sim$1\% to $\sim$ 70 ${\rm \,km\,s^{-1}\,Mpc^{-1}}$ but the revised value is not tabulated. Even the lower mean value from F20 from the higher LMC extinction gives $H_0$ measurements in agreement with both {\it Planck} and R20 estimates within 95\% CL, and therefore can not discriminate between the two. Furthermore, if the luminosity of SNIa is calibrated with the TRGB luminosity, that is, calibrated with the Gaia EDR3 trigonometric parallax of Omega Centauri, in Ref.~\cite{Soltis:2020gpl} is obtained the Hubble constant $H_0 = 72.1 \pm 2.0{\rm \,km\,s^{-1}\,Mpc^{-1}}$ at 68\% CL.
Another determination of $H_0$ using velocities and TRGB distances to 33 galaxies located between the Local Group and the Virgo cluster is given by Ref.~\cite{Kim:2020gai} and it is equal to $H_0=65.9 \pm 3.5 ({\rm stat}) \pm 2.4 ({\rm sys}) {\rm \,km\,s^{-1}\,Mpc^{-1}}$ at 68\% CL, i.e.\ in agreement with both {\it Planck} and R20 within $2\sigma$.

An alternative to either Cepheids or TRGB is MIRAS (variable red giant stars)~\cite{Huang:2019yhh}. These stars come from older stellar populations than Cepheid variables and have been calibrated directly in the maser host, NGC 4258 and used to calibrate SNIa in the host NGC 1559, to yield $H_0=73.3\pm4.0 {\rm \,km\,s^{-1}\,Mpc^{-1}}$ at 68\% CL.

There has been some discussion of whether the SNIa used at either ends of the distance ladder are consistent because of the possibility of differences in the SNIa environments and related impact on their luminosity (Refs.~\cite{Rigault:2014kaa,Rigault:2018ffm,Nicolas:2020lql}). Such differences will depend on the specific samples used to measure $H_0$. In Ref~\cite{Jones:2018vbn} the authors analysed the residual, host dependencies on the sample used by the SH0ES Team and found expectable deviations in $H_0$ at the level of 0.3\% and thus which do not appear to encompass a large fraction of the difference. 

There are also distance ladders which substitute SNIa for another long range indicator calibrated by Cepheids and TRGB such as the use of the Surface Brightness Fluctuations (SBF) method, which gives $H_0=70.50 \pm 2.37 ({\rm stat}) \pm 3.38 ({\rm sys}) {\rm \,km\,s^{-1}\,Mpc^{-1}}$ at 68\% CL~\cite{Khetan:2020hmh} from legacy SBF data and $H_0=73.3 \pm 0.7 \pm 2.4 {\rm \,km\,s^{-1}\,Mpc^{-1}}$ at 68\% CL~\cite{Blakeslee:2021rqi} from a new sample of NIR data from HST. Moreover, in Ref.~\cite{Blakeslee:2021rqi} a reanalysis of the result obtained by Ref.~\cite{Khetan:2020hmh} is performed, improving the LMC distance, and finding $H_0=71.1 \pm 2.4 ({\rm stat}) \pm 3.4 ({\rm sys}) {\rm \,km\,s^{-1}\,Mpc^{-1}}$ at 68\% CL. Likewise is the use of Tully-Fisher Relation, i.e.\ on the correlation between the rotation rate of spiral galaxies and their absolute luminosity, used to measure the distances after calibration from TRGB and Cepheids. Considering the optical and the infrared bands, Ref.~\cite{Kourkchi:2020iyz} finds $H_0 = 76.0 \pm 1.1 ({\rm stat}) \pm 2.3 ({\rm sys}) {\rm \,km\,s^{-1}\,Mpc^{-1}}$ at 68\% CL, while using the Baryonic Tully-Fisher relation, Ref.~\cite{Schombert:2020pxm} finds $H_0 = 75.1 \pm 2.3 ({\rm stat}) \pm 1.5 ({\rm sys}){\rm \,km\,s^{-1}\,Mpc^{-1}}$ at 68\% CL.
Lastly, the authors of Ref.~\cite{deJaeger:2020zpb} have presented another measurement of $H_0$ independent of SNIa using Type II supernovae (SN II) as standardisable candles, providing the result $H_0 = 75.8^{+5.2}_{-4.9}{\rm \,km\,s^{-1}\,Mpc^{-1}}$ at 68\% CL. A further Hubble constant determination is given in Ref.~\cite{Fernandez-Arenas:2017isq}, that uses as a standard candle the relation between the integrated H$\beta$ line luminosity and the velocity dispersion of the ionized gas of HII galaxies and giant HII regions, finding $H_0 = 71.0 \pm 2.8 ({\rm random}) \pm 2.1 ({\rm sys}){\rm \,km\,s^{-1}\,Mpc^{-1}}$ at 68\% CL. 

Finally, the Megamaser Cosmology Project (MCP)~\cite{Pesce:2020xfe} measures the Hubble constant using geometric distance measurements to six Megamaser-hosting galaxies. This approach avoids any distance ladder (i.e.\ multiple objects) by providing geometric distance {\it directly} into the Hubble flow and finds $H_0 = 73.9 \pm 3.0{\rm \,km\,s^{-1}\,Mpc^{-1}}$ at 68\% CL for maser host redshifts in the CMB rest frame, and a value of a few higher or lower for different methods of mapping peculiar velocities. The use of the 2M++ peculiar velocity maps in particular gives a value that is lower than this by $\sim$ 2-3 ${\rm \,km\,s^{-1}\,Mpc^{-1}}$~\cite{Boruah:2020fhl}.

The above methods have been fully or largely empirical and we may view these as being largely independent of astrophysical modeling other than the assumptions of a FLRW metric for computing distances. Although the systematic uncertainty of the distance ladder measurement has also been debated, recent surveys including various $H_0$ measurements robustly conclude that the discrepancy in the value of $H_0$ between early- and late-Universe observations ranges between $4\sigma$ and $6\sigma$~\cite{Verde:2019ivm,Riess:2020sih,DiValentino:2020vnx}. The distance ladder method also seems to be insensitive to the choice of the cosmology underlying Cepheids calibration~\cite{Follin:2017ljs}. Now we consider Late Universe approaches to measuring $H_0$ with some dependence on astrophysical modeling problems, though the models are not the same as $\Lambda$CDM.

\subsubsection{(Astrophysical) Model-Dependent:}

Methods that make use of significant astrophysical input (rather than strict empirical fitting) present additional challenges to the quantification of systematic uncertainties. In these cases one must measure the allowed theory space
using a wide range of plausible, if not preferable assumptions. This is not common to such analyses which often use ``one that works''. However, there have been great recent strides in quantifying the systematic uncertainty due to astrophysical inputs.

The time delays seen for strongly lensed images and their different path lengths can be modeled to measure the Hubble constant, though model-dependence results from imperfect knowledge of the foreground and lens mass distributions, i.e.\ how and where the dark matter is distributed between the observed and the image plane. The mass distribution problem is not settled and has a significant role in the inference of $H_0$ in this approach. Assuming lens models where the lens mass follows either a power-law or a Navarro-Frenk-White (NFW)~\cite{Navarro:1996gj} profile plus stars distribution, the most conventional assumption, the H0LiCOW ($H_0$ Lenses in COSMOGRAIL's Wellspring) experiment~\cite{Suyu:2016qxx} uses the time-delay in strong lensing to perform a cosmographic analysis of multiply-imaged quasars, improving the Hubble constant measurement from
$H_0=71.9_{-3.0}^{+2.4}{\rm \,km\,s^{-1}\,Mpc^{-1}}$ at 68\% CL in 2016~\cite{Bonvin:2016crt}, to
$H_0=72.5_{-2.3}^{+2.1}{\rm \,km\,s^{-1}\,Mpc^{-1}}$ at 68\% CL in 2018~\cite{Birrer:2018vtm}, and to $H_0=73.3_{-1.8}^{+1.7}{\rm \,km\,s^{-1}\,Mpc^{-1}}$ at 68\% CL in 2019~\cite{Wong:2019kwg}. 
A reanalysis of H0LiCOW's four lenses, which have both measurements of time-delay distance and distance inferred from stellar kinematics, has been performed in Ref.~\cite{Yang:2020eoh}, that finds $H_0=73.65_{-2.26}^{+1.95}{\rm \,km\,s^{-1}\,Mpc^{-1}}$ at 68\% CL.
A blind time-delay cosmographic analysis for the strong lens system DES $J0408-5354$ (STRIDES) is instead presented in Ref.~\cite{Shajib:2019toy} and, assuming a flat $\Lambda$CDM cosmology, gives $H_0=74.2_{-3.0}^{+2.7}{\rm \,km\,s^{-1}\,Mpc^{-1}}$ at 68\% CL.
Compressing the cumulative distribution function of time-delays using Principal Component Analysis, fitting a Gaussian Processes Regressor, and assuming a flat Universe, the fit of 27 doubly-imaged quasars results in $H_0=71_{-3}^{+2} {\rm \,km\,s^{-1}\,Mpc^{-1}}$ at 68\% CL~\cite{Harvey:2020lwf}. 
The combination of 6 lenses from H0LiCOW and 1 from STRIDES (called TDCOSMO) and a power-law model measures $H_0=74.2\pm1.6 {\rm \,km\,s^{-1}\,Mpc^{-1}}$ at 68\% CL~\cite{Millon:2019slk}.
However, without the use of conventional, locally determined priors on the lens mass distribution, the constraints become weaker and relatively undiscriminating such as those from TDCOSMO giving $H_0=74.5^{+5.6}_{-6.1} {\rm \,km\,s^{-1}\,Mpc^{-1}}$ at 68\% CL~\cite{Birrer:2020tax}, or TDCOSMO + SLACS analysis, where knowledge of the mass distribution in galaxies is discarded and replaced with that inferred from a specific set of galaxies, the SLACS sample of 33 strong gravitational lenses. This route places only weak constraints on the lens mass profiles and finds $H_0=67.4 ^{+4.1}_{-3.2}{\rm \,km\,s^{-1}\,Mpc^{-1}}$ at 68\% CL~\cite{Birrer:2020tax}. Its mean value is more similar to the one of {\it Planck} 2018, but in agreement with R20 at $1.3\sigma$, i.e.\ unable to discriminate between the two measurements now, but it is expected to be able to resolve the Hubble tension at $3-5\sigma$ in the future~\cite{Birrer:2020jyr} with the use of kinematic information to constrain the mass profiles. 
Another time-delay strong lensing measurement of the Hubble constant has been obtained analysing 8 strong lensing systems in~\cite{Denzel:2020zuq}, and is equal to $H_0=71.8 ^{+3.9}_{-3.3}{\rm \,km\,s^{-1}\,Mpc^{-1}}$ at 68\% CL.
An alternative use of lensing is to observe time delays of SN images behind a combination of a cluster lens and galaxy lens. Unfortunately, only one such object has been seen, SN Refsdal~\cite{Kelly:2014mwa}, and the uncertainty per object in $H_0$ is large, 7\% to 10\% and most sensitive to the model of the mass distribution in the cluster and nearest galaxy and ``blind'' predictions of new images of Refsdal by different models did not statistically agree to within their errors~\cite{Rodney:2015uyq}.

A determination of $H_0$ which is independent of late-time behavior of $\Lambda$CDM has been obtained in~\cite{Liao:2019qoc} from strongly lensed quasar systems from the H0LiCOW program and Pantheon SNIa compilation using Gaussian process regression, estimating $H_0=72.2\pm2.1{\rm \,km\,s^{-1}\,Mpc^{-1}}$ at 68\% CL. An updated result using the H0LiCOW dataset consisting of six lenses~\cite{Liao:2020zko} gives instead $H_0=72.8^{+1.6}_{-1.7}{\rm \,km\,s^{-1}\,Mpc^{-1}}$ at 68\% CL.

There are also estimates of the Hubble constant based on determining the change in age of the oldest elliptical galaxies as a function of redshift, so-called ``Cosmic Chronometers'' (CC).
Such galaxies are demonstrated to be largely ``passively'' evolving~\cite{Moresco:2018xdr} (i.e.\ stars form in one episode and then simply age) so that the oldest age at given redshifts may be directly equated with the change in the age of the Universe between those redshifts.
Spectra of these galaxies are used to measure the 4000 angstrom break whose size has been modeled to depend on age but also depends on metallicity, and star formation history, but it is weakly dependent on the initial mass function.
The break occurs due to the superposition of the spectral energy distribution of older stars where absorption features just blueward of the break produce the appearance of a jump. Stars of different masses and with different metallicities produce different depths of absorption and hence contributions to the break. The relation between the size of the break and age, metallicity and star formation history (i.e.\ how many stars of what range of mass form how often) is given by a stellar population synthesis model (summing stellar spectra in proportion to an estimated interstellar mass function, i.e.\ the initial ratios of small to large stars). Assuming the correct mean metallicity and functional form of the star formation history (and negligible residual star formation), the aging, $\mathrm{d}t$, is estimated across the change in redshift $\mathrm{d}z$ where $H(z)$ is proportional to ${\mathrm{d}z / \mathrm{d}t}$ and the value at $z=0$ may be estimated. In principle there is a great deal of astrophysics involved in this estimate including the time scale of star formation (exponential decline rate, truncation, new potential episodes due to refueling from mergers, etc), the estimation of metallicity with redshift, the spectral energy distribution of stars at a given metallicity and their initial mass function, both as a function of redshift, and questions related to alterations in the passive model due to merging and downsizing of galaxies. However, it has been shown in Ref.~\cite{Moresco:2020fbm} that the spectra have enough information to largely constrain both the metallicity and the star formation history (especially in super red galaxies as shown in Ref.~\cite{Moresco:2016mzx}), while the initial mass function has still to be assumed.
This method is ultimately challenging to independently test (e.g.\ with null tests to see if they can recover known aging as can be done for distance indicators comparing them to each other)\footnote{See along this line Ref.~\cite{Simon:2004tf}, where CC measurements of $H(z)$ were provided before the BAO ones and are in very good agreement, except for the overall normalisation if BAO are calibrated using the sound horizon of Planck's $\Lambda$CDM, while CC bounds are cosmology independent~\cite{Jimenez:2019cll}.} but new ideas may help.

Because this idea is new, there has not yet been enough independent effort to produce such measurements of $H(z)$, as all are sourced from the same compilation, to adequately sample the variance of the model space. 
This situation appears to be improving as an initial effort to quantify these systematics has been done by Ref.~\cite{Moresco:2020fbm} demonstrating systematic uncertainties most limited by stellar libraries and metallicity ranging from 5\% to 15\% in $H(z)$. However, many earlier measurements were based on a single model of stellar population synthesis~\cite{Bruzual:2003tq} and did not consider all of the modeling uncertainties. A recent analysis~\cite{Moresco:2020fbm} that incorporates the systematic uncertainty shows that the uncertainty in $H_0$ is $\sim$ 6\% if one incorporates the systematic errors (on diagonal) and 8\% (optimistic scenario that excludes worst model) after including the covariance of these uncertainties across redshift. The uncertainty from transforming these measures from $H(z)$ to $H_0$ is an additional $\sim$ 4\% for a total uncertainty in $H_0$ with present data of 9\%.

An additional concern is sample selection bias. Because the value of $H_0$ in early studies appeared to have some dependence on the mass range of the galaxies~\cite{Moresco:2010wh} seen at low redshift in SDSS data, it is important to correct surveys for mass incompleteness bias when harvesting passive galaxies from higher redshift surveys which will be more severely magnitude limited (easier to find more massive galaxies at a given redshift and a noisy measurement of mass is more likely higher of higher mass at higher redshift where the volume is greater). These measurements with the same data compilation, often in conjunction with other probes and different redshift space interpolation generally finds $H_0=66-73 {\rm \,km\,s^{-1}\,Mpc^{-1}}$ and an uncertainty of 6 km s$^{-1}$ Mpc$^{-1}$ following the inclusions of systematic uncertainties ~\cite{Moresco:2010wh,Yu:2017iju,Gomez-Valent:2018hwc,Gomez-Valent:2019lny,Haridasu:2018gqm,Dutta:2019pio,Krishnan:2020obg,Li:2019nux,Renzi:2020fnx,Nunes:2020uex,Zhang:2020uan}.\footnote{{See also Ref.~\cite{Colgain:2021ngq} for a discussion about the model independent determinations of $H_0$.}} It is probably safe to say at present this technique does not weigh heavily on the Hubble tension.

There is an estimate of $H_0$ based on modeling the extragalactic background light and its role in attenuating $\gamma$-rays that yields~\cite{Dominguez:2013mfa}, i.e.\ $H_0=71.8_{-5.6}^{+4.6}({\rm stat})_{-13.8}^{+7.2}({\rm sys}){\rm \,km\,s^{-1}\,Mpc^{-1}}$ at 68\% CL, and the updated value~\cite{Dominguez:2019jqc}, i.e.\ $H_0=67.4_{-6.2}^{+6.0}{\rm \,km\,s^{-1}\,Mpc^{-1}}$ at 68\% CL. However, the extragalactic background light is challenging to model and plays a dominant role in this approach.
Finally, Ref.~\cite{Qi:2020rmm}, combining the observations of ultra-compact structure in radio quasars and strong gravitational lensing with quasars acting as background sources, finds in a flat Universe $H_0 = 73.6^{+1.8}_{-1.6}{\rm \,km\,s^{-1}\,Mpc^{-1}}$ at 68\% CL.

In Ref.~\cite{Wan:2021umh}, using X-ray and Sunyaev-Zel'dovich (SZ) effect signals measured with Chandra, {\it Planck} and Bolocam for a sample of 14 massive, dynamically relaxed galaxy clusters, $H_0 = 67.3^{+21.3}_{-13.3}{\rm \,km\,s^{-1}\,Mpc^{-1}}$ at 68\% CL is obtained including the temperature calibration uncertainty, while $H_0 = 72.3\pm7.6{\rm \,km\,s^{-1}\,Mpc^{-1}}$ at 68\% CL only statistically.

In Ref.~\cite{Riess:2020sih} it has been pointed out that if some of the late Universe measurements are averaged together, by not considering each time a different method or geometric calibration or team, the Hubble constant tension between these averaged values and {\it Planck} will range between 4.5$\sigma$ and 6.3$\sigma$. In particular, in Ref.~\cite{Verde:2019ivm} an optimistic average of the late time Universe measurements gives $H_0 = 73.3 \pm 0.8{\rm \,km\,s^{-1}\,Mpc^{-1}}$ at 68\% CL, and in Ref.~\cite{DiValentino:2020vnx} $H_0 = 72.94 \pm 0.75{\rm \,km\,s^{-1}\,Mpc^{-1}}$ at 68\% CL, showing a $5.9\sigma$ level of disagreement with the standard $\Lambda$CDM model. A conservative estimate may be made by leaving out the most precise and most model-dependent results, i.e.\ excluding the measurements based on Cepheids-SNIa and Time-Delay Lensing, and gives $H_0 = 72.7 \pm 1.1{\rm \,km\,s^{-1}\,Mpc^{-1}}$ at 68\% CL~\cite{DiValentino:2020vnx}. In fact, even if multiple and/or unrelated systematic errors in the different experiments could be present (see for example the discussion in Refs.~\cite{Lin:2019htv,Efstathiou:2020wxn,breuval2021influence}), it seems unlikely these can resolve the Hubble tension, lowering all the late time measurements to agree with the early ones.

\subsubsection{Standard Sirens:}

An approach that does not require any form of cosmic distance ladder (see Ref.~\cite{Freedman:2000cf}) is the combination of the distance to the source inferred purely from the gravitational-wave signal, with the recession velocity inferred from measurements of the redshift using electromagnetic data. Gravitational-waves (GW) can therefore be used as standard sirens to estimate the luminosity distance out to cosmological scales directly, without the use of intermediate astronomical distance measurements. 
Unfortunately, there has only been one high-confidence event to date, GW170817, and it is too nearby ($z < 0.01$) to yield a good constraint on the Hubble expansion, though it has been attempted many times yielding results that sit between the Early and Late and with large uncertainties that encompass both. The authors of Ref.~\cite{Abbott:2017xzu} have used the detection of the GW170817 event in both gravitational waves and electromagnetic signals to determine $H_0 = 70.0 ^{+12.0}_{-8.0}{\rm \,km\,s^{-1}\,Mpc^{-1}}$ at 68\% CL. In Ref.~\cite{Mukherjee:2019qmm} the authors showed that, introducing a peculiar velocity correction for GW sources, the GW170817 event, combined with the Very Large Baseline Interferometry observation, gives $H_0 = 68.3^{+4.6}_{-4.5}{\rm \,km\,s^{-1}\,Mpc^{-1}}$ at 68\% CL.
Other constraints on the Hubble constant are those presented in Refs.~\cite{Gayathri:2020fbl,Mukherjee:2020kki,Gayathri:2021isv}. These bounds assume that the event ``ZTF19abanrhr'', reported by the Zwicky Transient Facility, is identified as the electromagnetic counterpart of the observed black hole merger GW190521, but such an association is still controversial~\cite{Ashton:2020kyr}. Another interesting observables are the so-called ``dark sirens'', i.e.\ compact binaries coalescences without electromagnetic counterpart, from LIGO/Virgo, that give $H_0 = 75^{+25}_{-22}{\rm \,km\,s^{-1}\,Mpc^{-1}}$ at 68\% CL alone~\cite{Finke:2021aom}, or $H_0 = 70^{+11}_{-7}{\rm \,km\,s^{-1}\,Mpc^{-1}}$ at 68\% CL~\cite{Finke:2021aom} in combination with GW170817.

\subsubsection{Systematics:}

It is hard to conceive of a single type of systematic error that would apply to the measurements of the disparate phenomena reviewed above as to effectively resolve the Hubble constant tension. We stress that the high quality of the measurements of the last decade demand a specific hypothesis for the nature of such a systematic that can be tested against the data rather than a non-specific statement of ``unknown unknowns'' which makes no testable predictions. We may consider greatly underestimated experimental errors in this same category as measurement error is as integral to the experiments as the measured value. Because the tension remains with the removal of the measurements of any single type of object, mode or calibration (e.g.\ SNIa, Cepheids, CMB, the distance to the LMC, etc) it is challenging to devise a single error that would suffice and we are not aware of a specific proposal that is not ruled out by the data. Of course multiple, unrelated systematic errors have a great deal more flexibility to resolve the tension but become less likely by their inherent independence. It is beyond the scope here to consider and review all such possible combinations. Such a resolution might argue for a true value of $H_0$ ``in the middle'', e.g.\ $\sim$ 70 ${\rm \,km\,s^{-1}\,Mpc^{-1}}$, as the easiest to accommodate, as was the resolution of the 1980's debate between 50 and 100. However, the analogy with the present situation breaks down because in the past case the tension was within the same types of measurements and at the same redshifts and thus pointed directly to systematics and away from the possibility of cosmological discovery of new physics. Nevertheless it is important to continue to broaden the measurements as a hedge against such a multiple-error scenario.

In summary, we conclude the case for an observational difference between the Early and Late Universe appears strong, is hard to dismiss, and merits an explanation. Even adopting a conservative view of the present situation, the agreement between early and late determinations of $H_0$, to high $\sim$ 1\% precision, is a critical test of $\Lambda$CDM, which none have suggested has been passed. Thus it is important to explore what may or may not be discovered if this fundamental test is ever passed.

\section{The Local Solution and the Sound Horizon Problem}
\label{sec:local}

The different $H_0$ measurements have motivated the scientific community to look for alternative cosmological scenarios that could reconcile or alleviate the $H_0$ tension.\footnote{This is a tension which historically has been called the Hubble Tension for ease of comparison but it could have easily been referred to as a sound horizon tension, among other possibilities. The only danger through naming is to neglect correlations and covariances between different measurements.}

\subsection{Inhomogeneous and Anisotropic Solutions}
\label{sec:InhomogeneousAndAnisotropicSolutions}

An underdense local Universe, corresponding to the simplest possibility for solving the Hubble constant tension for a sample-variance effect, has been definitely ruled out, because empirical and theoretical estimates of such fluctuations are a factor of $\sim 20$ too small. Such a void would need to extend to $z>0.5$ or higher to not be apparent in the Hubble diagram of SNIa or BAO measurements.
Considering a large-volume cosmological N-body simulation\footnote{For cosmological N-body simulations, the readers might be interested to Refs.~\cite{Llinares:2013jza,Barrera-Hinojosa:2019mzo,Barrera-Hinojosa:2020arz}.} to model the local measurements and to quantify the variance due to local density fluctuations and inhomogeneous selection of SNIa, in Ref.~\cite{Wu:2017fpr} it has been found that the extreme underdensity required for such a void is very unlikely to exist in the LSS fluctuations of a $\Lambda$CDM Universe aside from the conflict with the observations.
In Ref.~\cite{Kenworthy:2019qwq} the evidence in the Hubble diagram of large scale outflows caused by local voids has been studied, finding that the SNIa luminosity distance-redshift relation is in disagreement at $4 - 5\sigma$ with large local underdensities that can explain the Hubble tension. These findings agree with Ref.~\cite{Lukovic:2019ryg}, that concludes that a large local void alone is a very unlikely explanation, and with Ref.~\cite{Cai:2020tpy}, where the void matter distribution is described by an inhomogeneous but isotropic Lema\^{i}tre-Tolman-Bondi (LTB) metric.

Previous work has questioned the isotropy of the expansion of the Universe by estimating the anisotropy in the Hubble constant from SNIa data~\cite{Kalus:2012zu, Bengaly:2015dza, Bengaly:2015nwa} and from large samples of galaxies and clusters~\cite{Bolejko:2015gmk, Migkas:2020fza}, coming to diverse conclusions regarding the level of anisotropy. When the Pantheon dataset is analysed, a non-zero anisotropy is found which is mostly due to the non-uniform angular distribution of SNIa in the sample~\cite{Andrade:2018eta}.

In Refs.~\cite{Clarkson:2011uk, Heinesen:2020bej}, a consistent analysis that does not take into account an underlying FLRW metric has computed the luminosity distance cosmography for a general spacetime under a minimal set of assumptions. This is achieved by a series  expansion of the luminosity distance for a general spacetime with no assumptions on the metric tensor and allows to relax the assumptions of an isotropic expansion rate. In this metric-free analysis, the effective deceleration parameter can be negative without the need for a cosmological constant. A direct testing of the geometric assumptions for the FLRW metric using this method has yet to be carried out. This framework has been recently tested against cosmological numerical simulations~\cite{Clarkson:2011br, Adamek:2018rru, Macpherson:2021gbh}.

A different type of inhomogeneity relates to the non-linear time evolution in general relativity. Local inhomogeneities could drive a portion of volume away from an initial ``background'' FLRW model, which would serve as an approximation to the actual spacetime metric. How well the FLRW metric approximates the actual lumpy spacetime metric, the ``fitting problem'', was first discussed in Refs.~\cite{Ellis:1984bqf, Ellis:1987zz}. Inhomogeneities back-react on the large scale metric to produce an effective stress-energy tensor that adds up to the large scale stress-energy tensor. Different studies that attempt to assess the magnitude of such a backreaction of local structure on large scale cosmological dynamics reach conflicting results~\cite{Green:2014aga, Buchert:2015iva}, with the discrepancy being partly due to the differences in the quantification of backreaction in the different schemes~\cite{Clifton:2019cep}. Various frameworks for investigating the fitting problem have been proposed, see e.g.\ Refs.~\cite{Zalaletdinov:1992cg, Zalaletdinov:1996aj, Gasperini:2009wp, Gasperini:2009mu, Korzynski:2009db, Gasperini:2011us}, including the Buchert's scheme~\cite{Buchert:1999er, Buchert:2001sa, Buchert:2011sx, Buchert:2019mvq} which is treated in relation to the Hubble tension in Section~\ref{sec:scaling_solutions}.

\subsection{The Sound Horizon Problem}

In the following sections we will briefly review some of the most discussed models in the literature. Before going through all the possibilities, a word of caution is mandatory here: the solution to the Hubble constant tension can introduce a further disagreement with the BAO data, or the so-called ``sound horizon problem''.

The Hubble constant value is estimated from the CMB data, assuming a model, in three passages:
\begin{enumerate}
    \item from the measurements of the baryon density and the matter density, derivation of the sound horizon at the CMB last-scattering $r_{\rm s}^*$ at redshift $z_*$,
    \item from the position of the CMB acoustic peaks, derivation of the comoving angular diameter distance to last scattering $D_A^*=r_{\rm s}^*/\theta_s^*$,
    \item from $D_A^*=\int_0^{z_*} \mathrm{d}z/H(z)$, a derivation of $H(z)$ is available for all the redshifts $z$.
\end{enumerate}

\noindent BAO data can also provide a measurement of the Hubble constant, since these measurements constrain the product $Hr_d$.\footnote{For a study of the Hubble constant tension between CMB lensing and BAO measurements, see Ref.~\cite{Wu:2020nxz}.} This implies that in order to be in agreement with the CMB, which requires a low value of the Hubble constant value, the BAO constraints on the sound horizon at the baryon drag epoch lie on the high allowed region, i.e.\ around $147\,$Mpc. Contrarily, to be in agreement with R20, BAO data prefer a lower value for the sound horizon, i.e.\ around $137\,$Mpc.
Therefore, to reach an agreement among all the datasets, both a larger $H_0$ value and a lower sound horizon are needed from the CMB assuming a specific model, see Ref.~\cite{Bernal:2016gxb}. 

In Ref.~\cite{Mortsell:2018mfj} it has been argued that late time dark energy modifications of the expansion history are slightly disfavoured. Instead, in a pre-CMB decoupling scenario, an extra dark energy component can better solve the $H_0$ tension. The same thing happens if Modified Gravity modifications are accounted for, see e.g.\ Ref.~\cite{Lin:2018nxe}.

Following this direction, guidance to model building can instead be found in Ref.~\cite{Knox:2019rjx}. If different solutions are divided into post-recombination and pre-recombination solutions of the Hubble tension, the post-recombination modifications of the expansion history, such as the $w$CDM model where the DE equation of state is free to vary (see Section~\ref{sec:wCDM}), do not change the sound horizon, therefore they are unlikely to be a possible direction for fitting all the datasets. 
More promising are instead the pre-recombination solutions, as extra radiation at recombination as parameterized by $N_{\rm eff}$ or an Early Dark Energy component, since these non-standard cosmologies can increase $H_0$ while reducing $r_{\rm s}$. Unfortunately, these solutions are unable to solve completely the $H_0$ tension with R20~\cite{Arendse:2019hev}.

Many modifications to the $\Lambda$CDM model have been proposed in order to solve the Hubble constant tension, focusing on the scenarios that can reduce the sound horizon $r_{\rm s}$ at recombination. Nevertheless, it has been pointed out in a recent article~\cite{Jedamzik:2020zmd}
that models which only reduce $r_{\rm s}$ can never fully resolve the Hubble constant tension, if they are expected to be in agreement at the same time with the other cosmological datasets, such as BAO or weak lensing observations. For this very same reason different proposed models in the literature are often classified as either early or late time modifications of the expansion history, in order to take into account the sound horizon problem appearing when BAO data are considered~\cite{Knox:2019rjx,Arendse:2019hev}.\footnote{\, The BAO data are extracted under the assumption of a $\Lambda$CDM scenario, and their reliability has been tested for Early Time solutions~\cite{Bernal:2020vbb} and dark energy models that can be parameterized by $w_0-w_a$~\cite{Alam:2020sor}. Therefore, we should be careful in excluding all the Late Time solutions only using this argument (see also Ref.~\cite{Heinesen:2019phg}).}

``Late time solutions'' of the Hubble constant tension refer to the modifications of the expansion history after recombination, that increase the $H_0$ value leaving the sound horizon unaltered. These late solutions are well-known for solving successfully the Hubble constant tension, but being in disagreement with the BAO + Pantheon data~\cite{Knox:2019rjx,Arendse:2019hev}. In the following sections we shall present some of the most studied models in the literature belonging to this class of solutions.

We offer a brief comment that some local determinations of $H_0$ and constraints on $H(z)$ that use SNIa (e.g.\ from SH0ES and Pantheon SNIa) have covariance, sharing SNIa and light curve parameters 
which define the Hubble expansion at $0.02 < z < 0.15$ and that it is not strictly valid to use both constraints simultaneously and independently without proper account of their interdependence ~\cite{Dhawan:2020xmp, Efstathiou:2021ocp}.
This is likely to have consequences  particularly for late-time solutions that allow for a sudden or rapid change in $H(z)$ at $z<0.1$ which would impact both constraints. There are two approaches that may be used in principle to account for the covariance. One may use an inverse distance ladder starting in the Early Universe to calibrate SNIa in the Hubble flow (in the context of any cosmological model to predict $H(z)$) and thus predict the absolute peak magnitude $M_B$ of SNIa needed to match its empirical calibration from the local distance ladder. However, we caution that the value of $M_B$ derived is specific to a SNIa light curve fitting formalism and therefore it is crucial to measure $M_B$ consistently and to account for the covariance of SNIa data in both the local and Hubble flow samples. Alternatively one may use the SNIa distance ladder to directly calibrate Hubble flow SNe so that their constraining power and covariance are fully contained in the SNIa sample, i.e. a single set of distances, redshifts and their covariance which may then be used to constrain a cosmological or cosmographic model as done in ~\cite{Dhawan:2020xmp}. This approach will be formally included in a future SH0ES + Pantheon data release. A good {\it approximation} to this latter approach (neglecting only the SNIa-SNIa data covariance) is to i) subtract from Pantheon distance moduli the quantity $5 \log_{10}(H_0/70.0)$ in magnitudes, where e.g.\ $H_0=73.2{\rm \,km\,s^{-1}\,Mpc^{-1}}$~\cite{Riess:2020fzl} as $70.0$ was the Pantheon reference; ii) include covariance between every SN, namely a coherent 1.7\% (the uncertainty on $H_0$ from the calibration procedure only), which corresponds to a magnitude of 0.037. This later step adds a fixed quantity $(0.037)^2$ to the covariance matrix of errors which is already provided by the Pantheon collaboration. Here we note that the benchmark local $H_0$ determination from R21 uses a value of $q_0=-0.55$ derived from Pantheon, so this approximation is not strictly combining independent information, but any non-pathological alternative expansion of $H(z)$ consistent with either BAO, SNIa or CMB+$\Lambda$CDM would affect $H_0$ at the $\leq$ 1\% level.

``Early time solutions'', instead, modify the expansion history before the recombination period, changing both $H_0$ and $r_{\rm s}$ in the appropriate direction to solve the Hubble tension and the sound horizon problem simultaneously. Namely, a lower value of the sound horizon $r_{\rm s}$ is needed to allow $H_0$ to be in agreement with R20 and BAO + Pantheon at the same time. This can be achieved by increasing the expansion rate $H(z)$ before decoupling by, for instance, allowing an energy injection around the recombination epoch~\cite{Evslin:2017qdn,Aylor:2018drw}. This class of early time solutions is known to be able to alleviate, but not to solve, the $H_0$ tension below the $3\sigma$ significance~\cite{Arendse:2019hev,Lin:2021sfs}.

Finally, while in Ref.~\cite{Beenakker:2021vff} the authors explored a set of 7 assumptions that a model needs to break in order to alleviate the Hubble tension, in Ref.~\cite{Bernal:2021yli} the authors propose the use of new cosmic triangle plots to simultaneously represent independent constraints on key quantities related to the Hubble parameter ($t_U$, $r_{\rm s}$, and $\Omega_m$) useful to find its solution.

\section{Early Dark Energy}
\label{earlyDE}

The presence of a dark energy component during the early evolution of the Universe would affect the clustering of both dark matter and the baryon-photon fluid, suppressing the clustering power on small length-scales~\cite{Caldwell:2003vp,Bartelmann:2005fc,Doran:2005sn}.
These Early Dark Energy (EDE) models are able to solve the Hubble tension, reducing at the same time the sound horizon~\cite{Karwal:2016vyq}.

Since the EDE component must arise dynamically around the epoch of matter-radiation equality, these cosmologies could suffer from a ``cosmic-coincidence'' problem (see e.g.\ Ref.~\cite{Pettorino:2013ia}). A possibility proposed for solving this fine-tuning is to have EDE generated by a scalar field that conformally couples to neutrinos~\cite{Sakstein:2019fmf}. Indeed, in this scenario there will be a large injection of energy when neutrinos become non-relativistic, that could be around the time of matter-radiation equality for neutrinos with masses $m_{\nu} \sim 0.2{\rm \,eV}$. The model proposed, therefore, exploits a possible natural coincidence. A similar solution to the fine-tuning problem is provided by the early neutrino dark energy model proposed in Ref.~\cite{Das:2020wfe} (see also previous work of Refs.~\cite{Fardon:2003eh, Kaplan:2004dq, Peccei:2004sz}), where the dark energy density is controlled by the value of neutrino mass. Another possibility is instead proposed by Ref.~\cite{Tian:2021omz}, where the onset and ending of EDE are triggered by the radiation-matter transition, solving the fine-tuning.
Finally, in Ref.~\cite{Sabla:2021nfy} the coincidence problem is solved with an assisted quintessence, showing that this scaling possibility, that naturally explains the EDE, restores the Hubble constant tension.

\begin{figure*}
\centering
\includegraphics[width=0.85\textwidth, right]{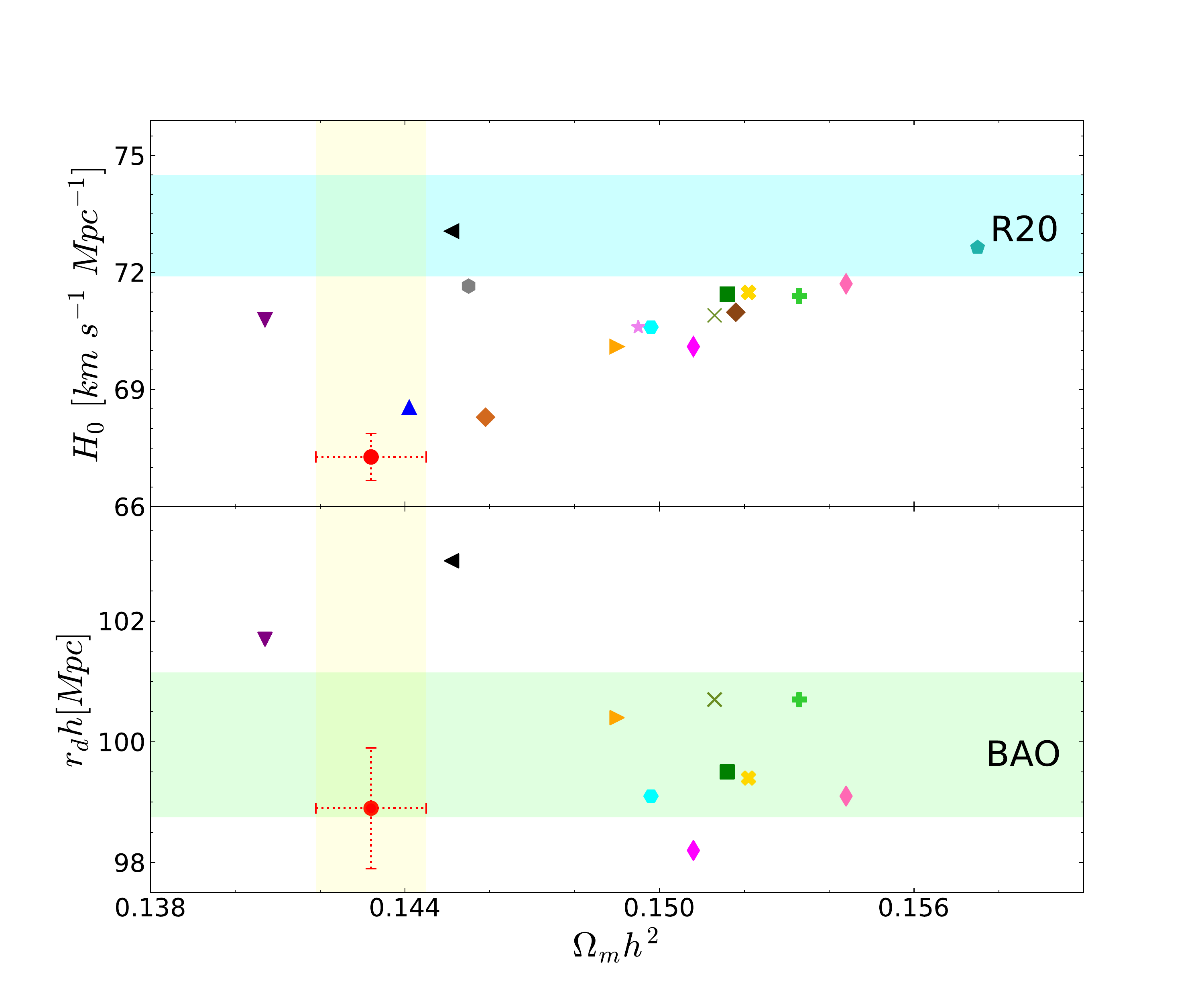}
\includegraphics[width=0.85\textwidth, right]{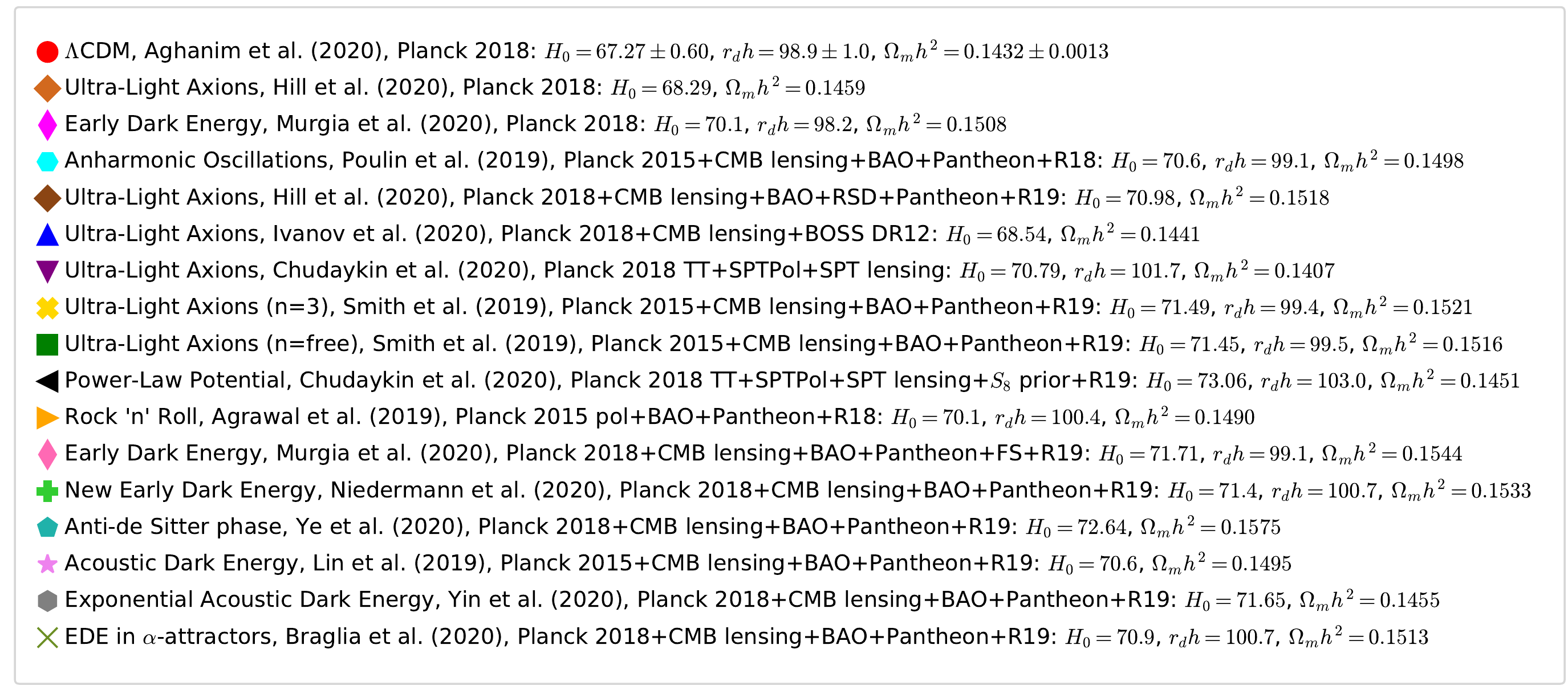}
\caption{Estimated values of the current matter energy density $\Omega_mh^2$, Hubble constant $H_0$ and sound horizon $r_dh$ in terms of various data points for different models discussed throughout Section~\ref{earlyDE}. The cyan horizontal band corresponds to the $H_0$ value measured by R20~\cite{Riess:2020fzl}, the yellow vertical band to the $\Omega_mh^2$ value estimated by {\it Planck} 2018~\cite{Aghanim:2018eyx} in a $\Lambda$CDM scenario, and the light green horizontal band to the $r_d h$ value measured by BAO data. The points sharing the same symbol refer to the same model in the same paper, and the different colors indicate a different dataset combination.}
\label{fig:chapter4_H0Om}
\end{figure*}

\begin{figure*}
\includegraphics[width=\textwidth]{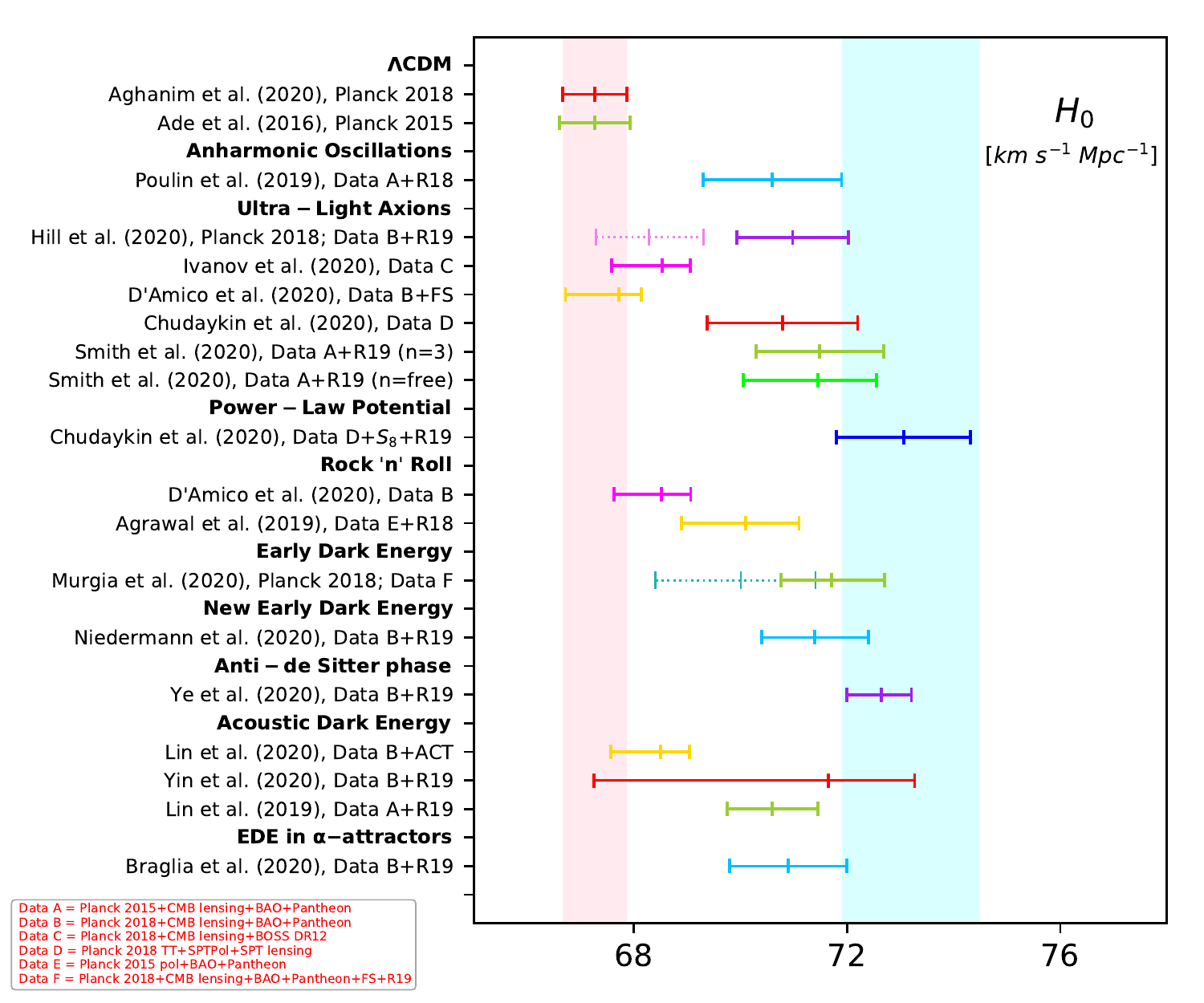}
\caption{Whisker plot with the 68\% marginalized Hubble constant constraints for the models of Section~\ref{earlyDE}. The cyan vertical band corresponds to the $H_0$ value measured by R20~\cite{Riess:2020fzl} and the light pink vertical band corresponds to the $H_0$ value estimated by {\it Planck} 2018~\cite{Aghanim:2018eyx} in a $\Lambda$CDM scenario. For each line, when more than one error bar is shown, the dotted one corresponds to the {\it Planck} only constraint on the Hubble constant, while the solid one to the different dataset combinations reported in the red legend, in order to appreciate the shift due to the additional datasets. }
\label{fig:chapter4_whisker}
\end{figure*}

In Figures~\ref{fig:chapter4_H0Om} and~\ref{fig:chapter4_whisker} we provide a very useful assessment of the models discussed in this Section~\ref{earlyDE} in light of the Hubble constant tension, as explained in the Introduction. 


\subsection{Anharmonic Oscillations}
\label{sec:AnharmonicOscillations}

An injection of energy at early times (approximately at $z \gtrsim 3000$), where the dark energy component behaves like a cosmological constant and then dilutes away as radiation, has been shown to be an effective possibility for reducing the $H_0$ tension. For example, the authors in Ref.~\cite{Poulin:2018cxd} proposed a physical EDE model based on a scalar field $\phi$ with a potential having an oscillating feature of the form~\cite{Kamionkowski:2014zda}:
\begin{equation}
    \label{EDE-Model1}
    V (\phi) \propto \Bigl[1- \cos \left(\phi/f \right) \Bigr]^n\,,
\end{equation}
where $f$ is an unknown energy scale and $n>0$. At early times, the scalar field is frozen and behaves like a cosmological constant until it starts to oscillate at a critical redshift $z_c$, after which it behaves as a fluid with an equation of state $w_n = (n-1)/(n+1)$~\cite{Turner:1983he}. The energy density parameter and the equation of state of the scalar field as a function of the scale factor $a_c \equiv (1+z_c)^{-1}$ at which the transition occurs are, respectively~\cite{Poulin:2018dzj}:
\begin{eqnarray}
    \Omega_{\phi}(a) &=& \frac{2 \Omega _{\phi}(a_c)}{\left(a/a_c\right)^{3( 1 + w_n)}+1}\,, \label{EDE-density-parameter}\\
    w_\phi(a) &=& -1 + \frac{1+w_n}{1+(a_c/a)^{3(1+w_n)}}\,.\label{EDE-eos}
\end{eqnarray}
At early times $a \rightarrow 0$, the scalar field behaves as a cosmological constant with the equation of state $w_{\phi}(a) \rightarrow -1$, while for $a \gg a_c$ we have $w_{\phi}(a) \rightarrow w_n$. Hence, the energy density is constant at early times, and decays as $a^{-3(1+w_n)}$ when the scalar field becomes dynamical~\cite{Turner:2001mx}. The EDE component dilutes like matter ($w_n =0$) for $n =1$ , like radiation ($w_n = 1/3$) for $n=2$, and faster than radiation for $n \geq 3$; for $n \rightarrow \infty$, the scalar field behaves like a stiff fluid with the equation of state $w_n \rightarrow 1$, and corresponds to a scalar ``kination'' field~\cite{Barrow:1982ei} whose energy density is dominated by its kinetic term and dilutes as $a^{-6}$.

The authors of Ref.~\cite{Poulin:2018cxd} showed that $n=3$ is the solution preferred by the data, and {\it Planck} 2015 + CMB lensing + BAO + Pantheon + R18 gives $H_0 = 70.6 \pm 1.3{\rm \,km\,s^{-1}\,Mpc^{-1}}$ at 68\% CL, solving the Hubble tension within $2 \sigma$. We should stress here that this result includes the R18 prior on the Hubble constant.

\subsection{Ultra-Light Axions}

Extremely light pseudoscalar particles known as ``axions'' can arise from various mechanisms such as the breaking of ``accidental'' symmetries~\cite{Choi:2006qj, Choi:2009jt} or from manifold compactification within string theory~\cite{Witten:1984dg, Svrcek:2006yi, Douglas:2006es, Arvanitaki:2009fg, Marsh:2015xka}. We discuss the QCD axion in Section~\ref{sec:QCDaxion}, while for now we consider an axion-like field $\phi$ of mass $m$ that does not necessarily relate to QCD. Axion-like particles can explain the dark matter observed~\cite{Arias:2012az, Visinelli:2017imh} and, at a different mass scale, they are a candidate for dark energy~\cite{Visinelli:2018utg}.

Ref.~\cite{Smith:2019ihp} attempts to alleviate the Hubble tension by considering sub-dominant oscillating scalar field moving under a potential inspired by the one that generically arises in string theory for an axion-like field:
\begin{equation}
    \label{eq:axionlikeV}
    V (\phi) = m^2f^2 \Bigl[1- \cos \left(\phi/f \right) \Bigr]^n\,,
\end{equation}
where $f$ is an energy scale. The axion-like potential is recovered for the case $n=1$. A fit to the {\it Planck} 2015 + CMB lensing + BAO + Pantheon + R19 datasets gives $H_0 = 71.49 \pm 1.20{\rm \,km\,s^{-1}\,Mpc^{-1}}$ at 68\% CL for $n=3$, and $H_0 = 71.45 ^{+1.10}_{-1.40}{\rm \,km\,s^{-1}\,Mpc^{-1}}$ at 68\% CL for free $n$~\cite{Smith:2019ihp}, apparently reducing the Hubble tension at one standard deviations. Indeed, as in the previous case, we should stress that the R19 prior is included in the analysis, possibly biasing the final result towards higher $H_0$ values.

Although the expressions in Eqs.~\eqref{EDE-Model1}-\eqref{eq:axionlikeV} share a similar dependence on the field $\phi$, the results presented in Ref.~\cite{Smith:2019ihp} differ from those in Ref.~\cite{Poulin:2018cxd} because in the latter an approximate form of the scalar field evolution equations was used, while the authors in Ref.~\cite{Smith:2019ihp} investigate the scenario by directly solving the linearized scalar field equations without relying on approximations.

An update of these results that considers more recent data is performed in Ref.~\cite{Hill:2020osr}. In this case, while the fit of a full combination {\it Planck} 2018 + CMB lensing + BAO + RSD + Pantheon + R19 gives $H_0 = 70.98\pm1.05{\rm \,km\,s^{-1}\,Mpc^{-1}}$ at 68\% CL, in tension at $1.3\sigma$ with R20, also including a prior on the Hubble constant, {\it Planck} 2018 data alone provides a value of $H_0 = 68.29^{+1.02}_{-1.00}{\rm \,km\,s^{-1}\,Mpc^{-1}}$ at 68\% CL, in disagreement at $2.9\sigma$ with R20. The authors therefore conclude that this EDE model, apart from showing a disagreement with all current cosmological datasets, does not solve the $H_0$ tension. These findings are confirmed by Refs.~\cite{Lucca:2020fgp, Haridasu:2020pms}, where additional dataset combinations and model extensions are considered, and in Ref.~\cite{Ivanov:2020ril}, where {\it Planck} 2018 + CMB lensing + BOSS DR12 gives $H_0 = 68.54 ^{+0.52}_{-0.95}{\rm \,km\,s^{-1}\,Mpc^{-1}}$ at 68\% CL, with a disagreement at $3.3\sigma$ with R20.

A different conclusion is instead reached in Ref.~\cite{Smith:2020rxx}, where the authors revisit the impact of EDE on galaxy clustering using BOSS galaxy power spectra, properly analysed adopting the EFTofLSS, and {\it Planck} 2018. They found that the conclusions can change with the choice of priors on the EDE parameter space, and that EDE and $\Lambda$CDM provide a statistically indistinguishable fits, with almost the same $\chi^2$, for EFTofLSS + {\it Planck} 2018 + SNIa. Unfortunately, a Bayesian model comparison accounting for the numbers of extra parameters in the EDE model is missing. However, in Ref.~\cite{DAmico:2020ods} the authors analyse the same model, finding for {\it Planck} 2018 + CMB lensing + BAO + Pantheon + Full Shape (FS) of BOSS DR12 $H_0 = 67.72 ^{+0.42}_{-1.00}{\rm \,km\,s^{-1}\,Mpc^{-1}}$ at 68\% CL, in disagreement with R20 at $3.9\sigma$.

In Ref.~\cite{Murgia:2020ryi}, moreover, it has been pointed out that the 1-parameter EDE cosmology can solve the tension between {\it Planck} and R20 and be favoured by the full dataset combination. In particular, {\it Planck} 2018 gives $H_0 = 70.10 ^{+1.4}_{-1.6}{\rm \,km\,s^{-1}\,Mpc^{-1}}$ at 68\% CL~\cite{Murgia:2020ryi}, alleviating the tension with R20 at $1.6\sigma$, and {\it Planck} 2018 + CMB lensing + BAO + Pantheon + FS of BOSS DR12 + R19 gives $H_0 = 71.71 ^{+1.0}_{-0.95}{\rm \,km\,s^{-1}\,Mpc^{-1}}$ at 68\% CL~\cite{Murgia:2020ryi}, in full agreement with R20.

A complementary analysis is performed in~\cite{Chudaykin:2020igl}, that for {\it Planck} 2018 TT (up to $\ell = 1000$) + SPTPol (TE and EE) + SPT Lensing gives $H_0 = 70.79 \pm 1.41{\rm \,km\,s^{-1}\,Mpc^{-1}}$ at 68\% CL, solving the tension with R20 within $1.3\sigma$. 

Finally, in Ref.~\cite{Kaloper:2019lpl} it is argued that a mechanism in which an EDE dumps most of its energy content into radiation in the redshift range $z=[3000,5000]$ can solve the Hubble tension, and this might be an observational signal of the Weak Gravity Conjecture.

\subsubsection{Dissipative Axion:}

The authors of Ref.~\cite{Berghaus:2019cls} present a concrete realization of a particle physics model for EDE. In more detail, an axion-like particle acts as a dark energy component which mimics EDE at the background level and behaves as a cosmological constant at early times, before decaying to dark gauge bosons through sphaleron processes mediated by a new non-Abelian gauge group. Although in this ``dissipative axion'' model the Hubble tension can potentially be alleviated, a proper comparison with {\it Planck} 2018 data is to date missing.

\subsubsection{Axion Interacting With a Dilaton:}

Another possible realization of the EDE scenario is an axion interacting with a dilaton, as proposed in Ref.~\cite{Alexander:2019rsc}. Starting from string theory, the authors showed that the dynamics of an interacting dilaton-axion scenario naturally realizes the EDE potential. Despite its promising potential, a comparison with {\it Planck} 2018 data is absent.

\subsection{Power-Law Potential}

In Ref.~\cite{Chudaykin:2020acu}, the authors consider an alternative EDE scenario with a potential of the form:
\begin{equation}
V_n (\phi) = V_0 \frac{\phi^{2n}}{2^n}\,,
\end{equation}
where $V_0$ is the amplitude of the potential and $n$ is a power-law index. This potential approximates the anharmonic potential in Eq.~\eqref{EDE-Model1} in the limit $\phi/f \ll 1$. A fit to the {\it Planck} 2018 TT (up to $\ell=1000$) + SPTPol (TE and EE) + SPTLensing + $S_8$ prior (from KiDS, VIKING-450 and DES of~\cite{Joudaki:2019pmv}) + R19 datasets with $n=3$ gives $H_0=73.06\pm1.26{\rm \,km\,s^{-1}\,Mpc^{-1}}$ at 68\% CL~\cite{Chudaykin:2020acu}, solving the Hubble tension within $1\sigma$. However, since the derived $H_0$ value is obtained assuming the R19 prior, it is difficult to properly assess the ability of the model to solve the tension.

\subsection{Rock `n' Roll}

Reference~\cite{Agrawal:2019lmo} considers a scenario in which a scalar field evolves under a potential of the form $V\propto \phi^{2n}$. Depending on the value of the index $n$, the scalar field asymptotically evolves to either an oscillatory (rocking) behavior or to a rolling solution with a nearly constant equation of state. The presence of the scalar field injects energy close to recombination, effectively reducing the sound horizon and increasing the Hubble constant value. The potential of the model is parameterized as:
\begin{equation}
    \label{eq:potential}
    V(\phi) = V_0 \left(\frac{\phi}{M_{\rm Pl}}\right)^{2n} + V_{\Lambda},
\end{equation}
with a constant value of $V_0$ and $V_{\Lambda}$, and where $M_{\rm Pl}=1/\sqrt{8 \pi G_N}$ is the reduced Planck mass. Within this model, and for $n=2$, the {\it Planck} data and the R20 measurement are in better agreement than in the canonical $\Lambda$CDM framework, provided a modest tuning to justify the absence of lower orders in the potential~\cite{Agrawal:2019lmo}.

Indeed, for this scenario, {\it Planck} 2015 + BAO + Pantheon + R18 data provides the constraint $H_0=70.1^{+1.0}_{-1.2}{\rm \,km\,s^{-1}\,Mpc^{-1}}$ at 68\% CL~\cite{Agrawal:2019lmo}, apparently reducing the Hubble tension at $1.9\sigma$. However, again, we note the presence of the R18 prior in the analysis. An updated analysis is performed in Ref.~\cite{DAmico:2020ods}, where {\it Planck} 2018 + CMB lensing + BAO + Pantheon gives $H_0 = 68.52 ^{+0.55}_{-0.89}{\rm \,km\,s^{-1}\,Mpc^{-1}}$ at 68\% CL, in disagreement with R20 at $3.3\sigma$.

\subsection{New Early Dark Energy}

In Ref.~\cite{Niedermann:2019olb} the authors propose a model in which a first-order phase transition occurs in a dark sector before recombination and avoid to imprint unobserved large-scale anisotropies in the CMB. Such a transition would produces a short phase of New EDE (NEDE) which could address the Hubble tension. Similarly to previously considered mechanisms for ending inflation~\cite{Linde:1990gz, Adams:1990ds, Copeland:1994vg}, the potential considered involves two scalar fields and it is of the form:
\begin{equation}
    V(\psi,\phi) =\frac{\lambda}{4}\psi^4+\frac{1}{2}\beta M^2 \psi^2-\frac{1}{3}\alpha M \psi^3 + \frac{1}{2}m^2\phi^2 +\frac{1}{2}\tilde\lambda \phi^2\psi^2 \,,
\end{equation}
where $\psi$ is the field responsible for the tunneling and $\phi$ is the trigger field required to modulate the tunneling. The parameters of the potential are subject to the restrictions $\alpha^2 > 4 \,\beta \lambda$ and $\beta>0$. The background field changes from the cosmological constant equation of state $w_\Lambda = -1$ to a constant $w_{\rm NEDE}^*$ around the time $t_{\rm tr}$. Such a sudden transition can be modeled through the equation of state:
\begin{equation}
w_{\rm NEDE}(t) =
\begin{cases}
-1\,, & \quad \hbox{for $t \leq t_{\rm tr}$} \,;\\
w_{\rm NEDE}^*\,, & \quad \hbox{for $t>t_{\rm tr}$} \,,
\end{cases}
\end{equation}
where we expect that the NEDE energy density redshifts faster than radiation, as $1/3 \leq w_{\rm NEDE}^* \leq 1$. In the approximation of a sudden transition, the background energy density evolves as:
\begin{equation} \label{eq:rho_EDE_bg}
\bar{\rho}_{\rm NEDE}(t) = \bar{\rho}^*_{\rm NEDE} \left( \frac{a(t_*)}{a(t)}\right)^{3[1+w_{\rm NEDE}(t)]}\,,
\end{equation}
with a constant parameter $\rho_{\rm NEDE}^*$. 

A fit to {\it Planck} 2018 + CMB lensing + BAO + Pantheon data gives $H_0=69.6^{+1.0}_{-1.3}{\rm \,km\,s^{-1}\,Mpc^{-1}}$ at 68\% CL, reducing the Hubble tension within $2.3\sigma$~\cite{Niedermann:2020dwg}. Including R19, the Hubble constant becomes $H_0=71.4\pm1.0{\rm \,km\,s^{-1}\,Mpc^{-1}}$ at 68\% CL, with an evidence at $4\sigma$ for NEDE, and reducing the tension with R20 at $1.1\sigma$.

\subsection{Chain Early Dark Energy}

Chain EDE proposes an alternative mechanism in which a scalar field tunnels rapidly via a series of first order phase transitions through many $(N \gg 1)$ successive metastable minima of ever lower energy. This kind of model was previously employed as a mechanism for inflation~\cite{Freese:2004vs}. Building on this, an alternative model of EDE called Chain EDE has been proposed in Ref.~\cite{Freese:2021rjq} as a solution to the Hubble constant tension. In the model, the Hubble tension could be resolved without inducing large anisotropies in the CMB by invoking $N \gtrsim 10^4$ such phase transitions~\cite{Freese:2021rjq}. However, a full data analysis for this model is currently missing.

\subsection{Anti-de Sitter phase}

In Ref.~\cite{Ye:2020btb} the authors propose a phenomenological EDE model with an Anti-de Sitter (AdS) phase around the recombination period as a solution to the Hubble tension. AdS vacua are theoretically important because they naturally emerge within the string theory framework (for late-time AdS see Refs.~\cite{Visinelli:2019qqu,Calderon:2020hoc, Akarsu:2019hmw}). 

This EDE model with an AdS phase will make the energy injection more efficient without spoiling the fit to CMB data. We have $w_{\rm DE} > -1$ when the EDE field rolls down to $V<0$. Therefore, $\rho_\phi \sim a^{-3(1+w)}$ redshifts very rapidly. In~\cite{Ye:2020btb} the potential is modeled as:
\begin{equation}
V(\phi) = \begin{cases}
V_0\left(\frac{\phi}{M_{\rm Pl}}\right)^4-V_{\rm AdS}\,,& \quad \hbox{for $\frac{\phi}{M_{\rm Pl}}\leq \left(\frac{V_{\rm AdS}}{V_0}\right)^{1/4}$}\,,\\
0\,, & \quad \hbox{for $\frac{\phi}{M_{\rm Pl}}>\left(\frac{V_{\rm AdS}}{V_0}\right)^{1/4}$}\,,
\end{cases}
\end{equation}
where $V_{\rm AdS}$ is the depth of the AdS well. 

While a constraint on $H_0$ from {\it Planck} data alone is missing, {\it Planck} 2018 + CMB lensing + BAO + Pantheon + R19 gives $H_0=72.64^{+0.57}_{-0.64}{\rm \,km\,s^{-1}\,Mpc^{-1}}$ at 68\% CL~\cite{Ye:2020btb} solving the tension within one standard deviation. However, the presence of the R19 Gaussian prior in the analysis makes difficult to assess the consistency between the measurements. An extended model considering the temperature of the CMB $T_0$ free to vary has been studied in Ref.~\cite{Ye:2020oix}, finding consistent results.

\subsection{Graduated Dark Energy}

In Ref.~\cite{Akarsu:2019hmw} the graduated Dark Energy model (gDE) is introduced, inspired by Ref.~\cite{Barrow:1990vx}. A limiting case of the gDE is a sign-switching cosmological constant, that can be appealing from the string theory perspective. Using the {\it Planck} information as a BAO data point at redshift $z = 1090$, and using also SNIa JLA~\cite{Betoule:2014frx} + BAO + CC measurements, the authors in Ref.~\cite{Akarsu:2019hmw} argue that this model is in agreement with the local $H_0$ measurements. However, a complete and robust data analysis considering the perturbations and the full {\it Planck} 2018 data is missing to date.

\subsection{Acoustic Dark Energy}

Acoustic Dark Energy (ADE) has been proposed in Ref.~\cite{Lin:2019qug} to alleviate the Hubble tension. The authors consider a general phenomenological model of perturbations in a dark fluid which becomes important around matter-radiation equality. The presence of ADE impacts on the CMB through the gravitational effects on the acoustic oscillations. More concretely, ADE consists of a perfect dark fluid specified by its background equation of state $w_{\rm ADE}(a)$ and its rest frame sound speed $c_s^2$. The ADE equation of state changes around the scale factor $a=a_c$, ranging from $w_{\rm ADE}=-1$ to $w_f$ as:
\begin{equation}
\label{eqn:eos}
w_{\rm ADE}(a) = -1 + \frac{1+w_f }{[1+(a_c/a)^{3(1+w_f )/p}]^{p}} \,,
\end{equation}
where the index $p$ controls the rapidity of the transition, such that small values lead to sharper transitions. For $p=1$, the model described in Section~\ref{sec:AnharmonicOscillations} is obtained. In Ref.~\cite{Lin:2019qug}, the ADE model with $p=1/2$ is analysed by fitting against {\it Planck} 2015 + CMB lensing + BAO + Pantheon + R19 data, obtaining $H_0=70.60\pm0.85{\rm \,km\,s^{-1}\,Mpc^{-1}}$ at 68\% CL~\cite{Lin:2019qug} and reducing the Hubble tension to $1.6\sigma$. An updated analysis without the Gaussian prior on the Hubble constant is presented in Ref.~\cite{Lin:2020jcb}, where the combination of {\it Planck} 2018 + CMB lensing + ACT + Pantheon + BAO gives $H_0=68.50^{+0.55}_{-0.93}{\rm \,km\,s^{-1}\,Mpc^{-1}}$ at 68\% CL, restoring the tension with R20 at the $3.6\sigma$ level.

\subsubsection{Exponential Acoustic Dark Energy:} Acoustic Dark Energy in which the equation of state has an exponential dependence on the scale factor (eADE) has been explored in Ref.~\cite{Yin:2020dwl}:
\begin{equation}
\label{eqn:eoseADE}
w_{\rm eADE}(a) = -1 + 2 ^{1-\frac{a_c}{2a}} \,,
\end{equation}
where $a_c$ corresponds to the critical scale factor at which the eADE fluid becomes dominant. The equation of state evolves from the value $w=-1$ before the transition to $w\approx 1$ at present time. The fractional energy density evolves as
\begin{equation}
    \Omega_{\rm eADE}(a) = 2f_c\,\frac{(c_s^2+1)^2-(w_{\rm eADE}(a)+1)^2}{(c_s^2+1)^2}\,,
\end{equation}
where $f_c$ is the fractional contribution of eADE at $a_c$, and $c_s$ is the sound speed. For this model, a fit to {\it Planck} 2018 + CMB lensing + BAO + Pantheon + R19 datasets provides the constraint $H_0=71.65^{+1.62}_{-4.40}{\rm \,km\,s^{-1}\,Mpc^{-1}}$ at 68\% CL~\cite{Yin:2020dwl}, solving within $1\sigma$ the tension with R20. However, the analysis already includes a Gaussian prior on the Hubble constant.

\subsection{EDE in \texorpdfstring{$\alpha$}{alpha}-attractors}

In the framework of inflation (see Section~\ref{inflat}), it is possible to introduce a class of models that possess an attractor point predicting the value of the scalar spectral tilt $n_s$ and the tensor-to-scalar ratio $r$, independently from the specific functional form of the inflaton potential $V(\phi)$~\cite{Kallosh:2013hoa, Kallosh:2013yoa, Galante:2014ifa}. An EDE model can also be extended to include $\alpha$-attractors, with a potential for the EDE scalar field of the form~\cite{Braglia:2020bym}:
\begin{equation}
\label{eq:potential2}
V(\phi)=\Lambda+V_0\frac{(1+\beta)^{2n} \tanh\left(\phi/\sqrt{6\alpha}M_\textup{\rm Pl}\right)^{2 p}}
{\left[1+ \beta \tanh\left(\phi/\sqrt{6\alpha}M_\textup{\rm Pl}\right)\right]^{2n}} \,,
\end{equation}
where $V_0,\,p,\,n$, $\alpha$ and $\beta$ are constants. The shape of the potential, away from the plateau and around its minimum, regulates the shape of the energy injection and it is thus crucial to successfully alleviate the Hubble tension. For the choice $p=2$ and $n=4$, the scalar field oscillates at the bottom of the potential, making this case more similar to the original EDE proposal~\cite{Poulin:2018cxd}. For these values of the model parameters, the analysis of {\it Planck} 2018 + CMB lensing + BAO + Pantheon + R19 data gives $H_0=70.9\pm1.1{\rm \,km\,s^{-1}\,Mpc^{-1}}$ at 68\% CL, softening the tension with R20 down to the $1.4\sigma$ level~\cite{Braglia:2020bym}. Note, however, that this result already incorporates a Gaussian prior in the Hubble constant.

\begin{figure*}
\centering
\includegraphics[width=0.85\textwidth, right]{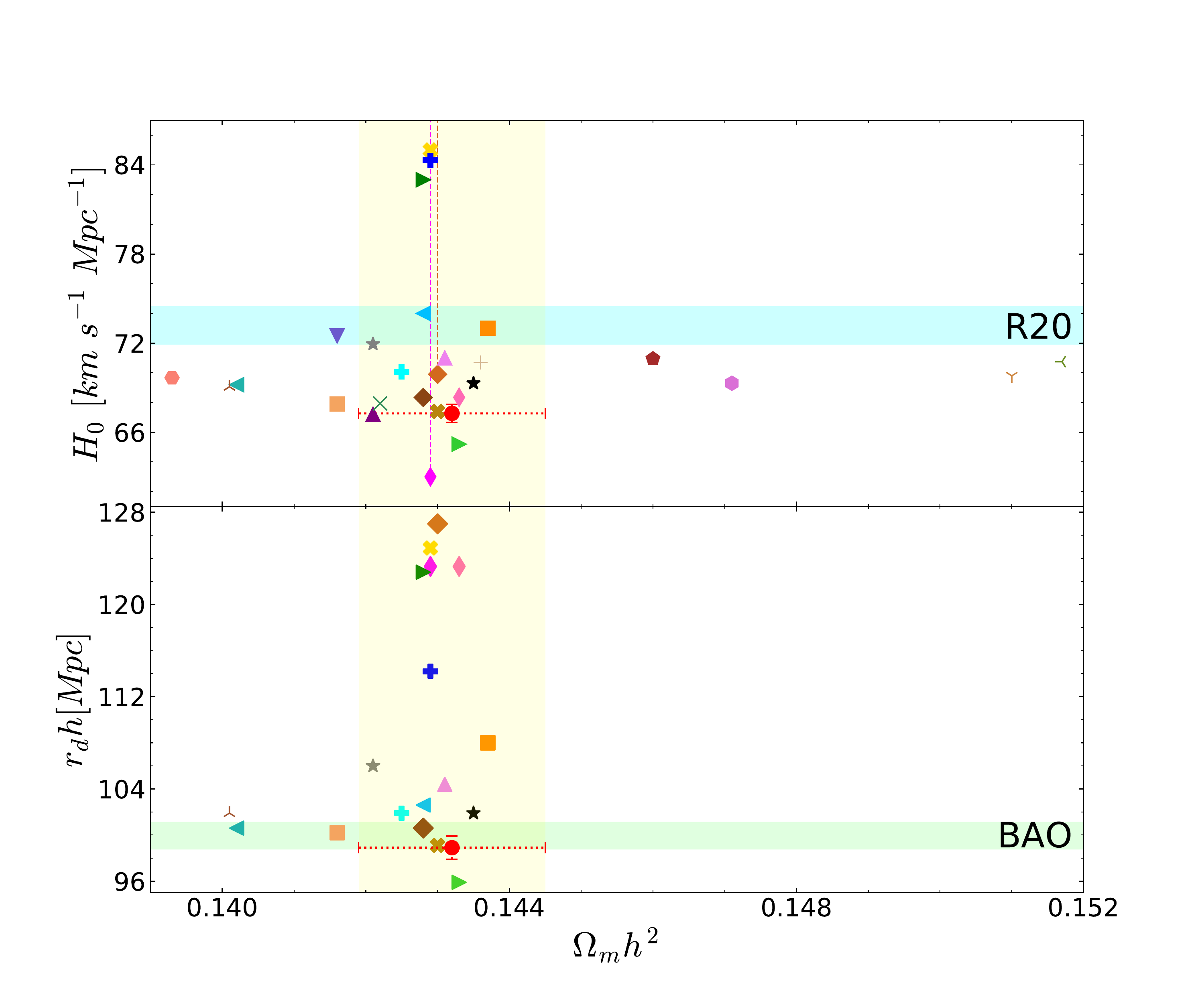}
\includegraphics[width=0.85\textwidth, right]{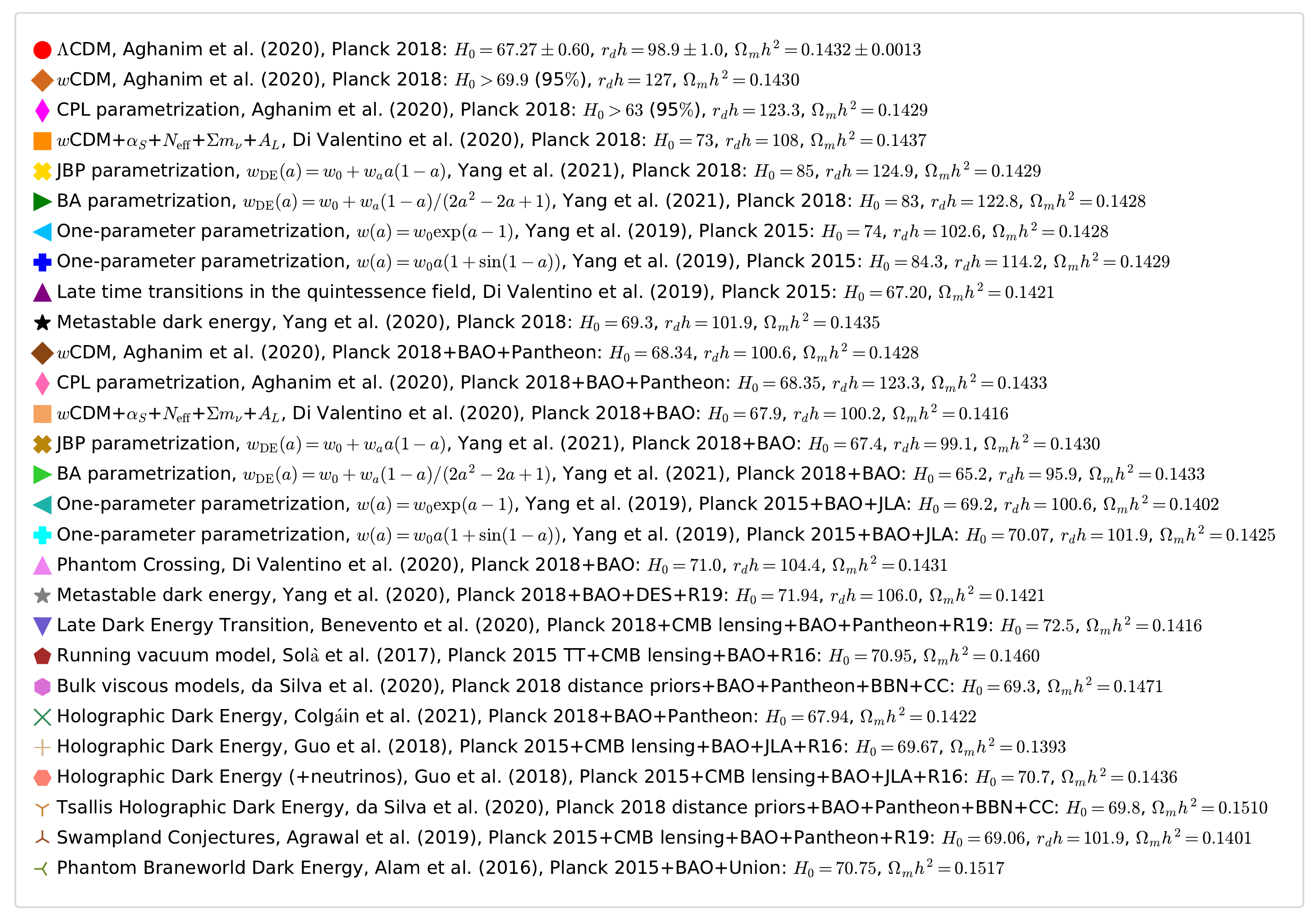}
\caption{Estimated values of the current matter energy density $\Omega_mh^2$, Hubble constant $H_0$ and sound horizon $r_dh$ in terms of various data points for different models discussed throughout Section~\ref{lateDE}. The cyan horizontal band corresponds to the $H_0$ value measured by R20~\cite{Riess:2020fzl}, the yellow vertical band to the $\Omega_mh^2$ value estimated by {\it Planck} 2018~\cite{Aghanim:2018eyx} in a $\Lambda$CDM scenario, and the light green horizontal band to the $r_dh$ value measured by BAO data. The points sharing the same symbol refer to the same model in the same paper, and the different colors indicate a different dataset combination.}
\label{fig:chapter5_H0Om}
\end{figure*}

\begin{figure*}
\includegraphics[width=\textwidth]{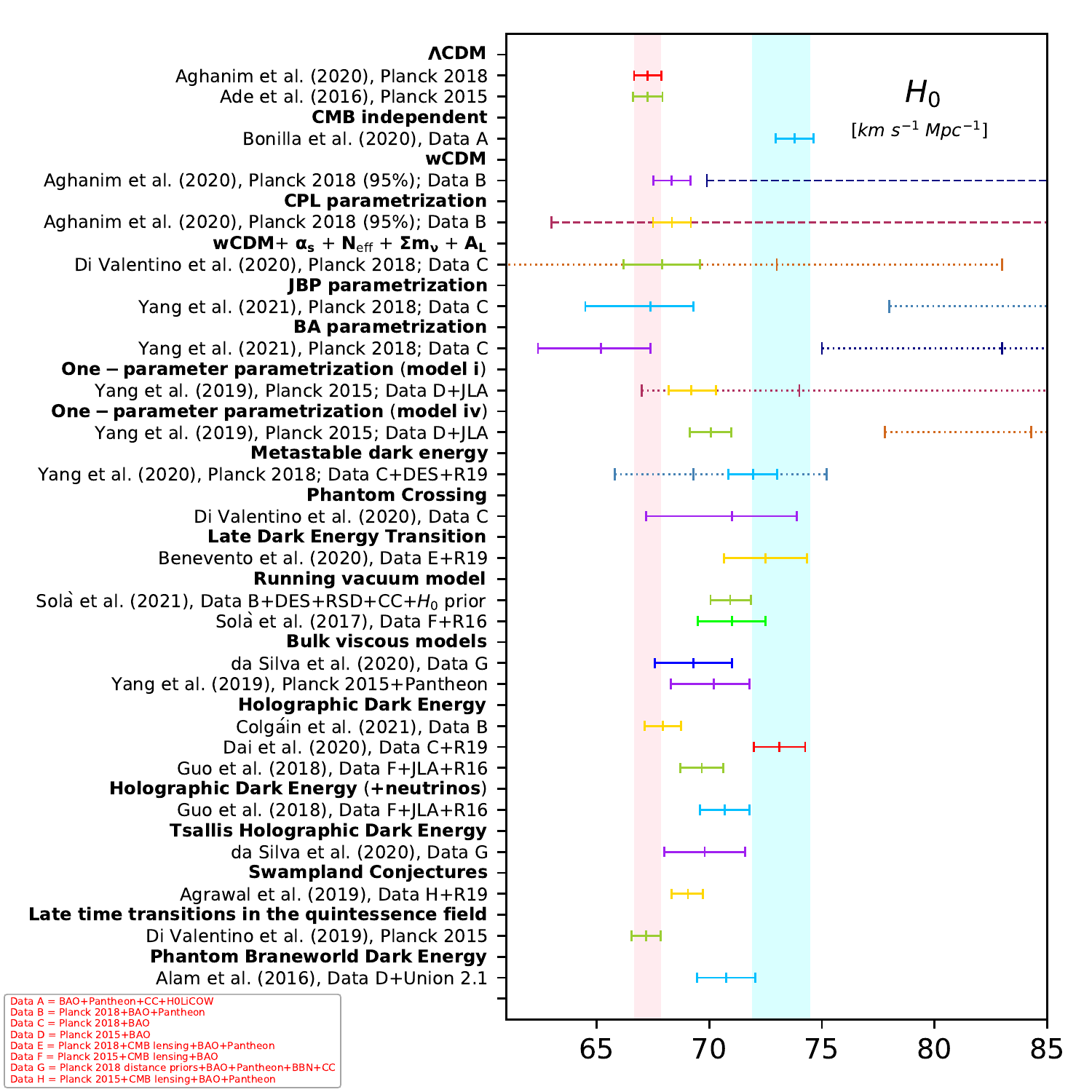}
\caption{Whisker plot with the 68\% (95\% if dashed) marginalized Hubble constant constraints for the models of Section~\ref{lateDE}. The cyan vertical band shows the $H_0$ value measured by R20~\cite{Riess:2020fzl} and the light pink vertical band corresponds to the $H_0$ value estimated by {\it Planck} 2018~\cite{Aghanim:2018eyx} in a $\Lambda$CDM scenario. For each line, when more than one error bar is shown, the dotted one corresponds to the {\it Planck} only constraint on the Hubble constant, while the solid one to the different dataset combinations reported in the red legend, in order to appreciate the shift due to the additional datasets.}
\label{fig:chapter5_whisker}
\end{figure*}

\section{Late Dark Energy}
\label{lateDE}

A dark energy component with a time-varying equation of state $w_{\rm DE}(z) \equiv p_{\rm DE}/\rho_{\rm DE}$ modifies the Hubble rate through the first Friedmann equation:
\begin{equation}
H^2(z) = H_0^2\,\left[ \Omega_{r}(1+z)^4 + \Omega_{m}(1+z)^3 + \Omega_{\rm DE} f(z) + \Omega_{k}(1+z)^2 \right]\,,
\end{equation}
where
\begin{equation}
    \label{eq:definef}
    f(z)=\exp\left[3\int_0^{\ln(1+z)} \mathrm{d}\ln(1+z') (1+w_{\rm DE}(z'))\right],
\end{equation}
and $\Omega_{r}$, $\Omega_m$, $\Omega_{\rm DE}$ and $\Omega_{k}$ are the density parameters, evaluated at present time, for radiation, matter (cold dark matter+baryons), dark energy and curvature, respectively, satisfying $\Omega_{r} + \Omega_m + \Omega_{\rm DE} + \Omega_{k} = 1$. We have also defined the Hubble rate $H(z)$ at redshift $z$ as:
\begin{equation}
    \label{def:Hubble_rate}
    H(z) \equiv \frac{1}{a(z)}\frac{\mathrm{d}a}{\mathrm{d}t}\,,
\end{equation}
so that at present time $H(z=0) \equiv H_0$.

In Ref.~\cite{Zhao:2017cud}, it has been argued that the Hubble tension can be interpreted as an evidence for a non-constant dynamical dark energy at $3.5\sigma$ (see also Ref.~\cite{DiValentino:2017gzb}). A different approach aimed at reconstructing the dark energy properties using Gaussian processes constrains the Hubble constant as $H_0=73.78\pm 0.84{\rm \,km\,s^{-1}\,Mpc^{-1}}$ at 68\% CL, using a joint analysis of the geometrical cosmological probes such as SNIa, CC, BAO, and the H0LiCOW lenses sample~\cite{Bonilla:2020wbn}. Finally, a reconstruction of the dynamical Dark Energy using the latest measurements has been studied also in Ref.~\cite{Wang:2018fng}.

In Figures~\ref{fig:chapter5_H0Om} and~\ref{fig:chapter5_whisker} we provide a very comprehensive status of the models discussed in this Section~\ref{lateDE} in light of the Hubble constant tension, as explained in the Introduction.

\subsection{\texorpdfstring{$w$}{w}CDM model}
\label{sec:wCDM}

We first consider a model in which the equation of state for the dark energy component is independent of redshift and generally differs from the cosmological constant value, $w_{\rm DE}(z) \equiv w_0 \neq -1$. This simple extension of $\Lambda$CDM is referred to as the $w$CDM model, where ``$w$'' stands for the equation of state $w_0$. Here, Eq.~\eqref{eq:definef} gives:

\begin{equation}
    f(z) = (1+z)^{3(1+w_0)}\,.
\end{equation}
For the case $w_0 = -1$, the function $f(z)$ is also independent of redshift and the dark energy component acts as a cosmological constant of density parameter $\Omega_{\rm DE}$.

A likelihood analysis with {\it Planck} 2018 data for this model assumes a constant equation of state for DE, $w_0 = -1.58^{+0.16}_{-0.35}$ at 68\% CL and $H_0 > 69.9{\rm \,km\,s^{-1}\,Mpc^{-1}}$ at 95\% CL~\cite{Aghanim:2018eyx}. Such a $w$CDM scenario would therefore solve the $H_0$ tension within two standard deviations. The Hubble constant is in fact almost unconstrained in the $w$CDM scenario, due to the geometrical degeneracy between $w_{\rm DE}$ and $H_0$. Therefore, this scenario can perfectly accommodate a Hubble constant in agreement with R20, at the price of a phantom-like dark energy equation of state, i.e.\ $w_0 < -1$. Such a result implies that the energy density of dark energy is increasing over time, so that the scale factor of the Universe would reach infinity in a finite time and the Universe would end in a ``big rip''~\cite{Caldwell:2003vq}. In addition, the Hamiltonian of the theory could have vacuum instabilities due to negative kinetic terms. Nevertheless, the reader should keep in mind that despite of these many theoretical problems, there exist models with an effective energy density with a phantom-like equation of state which avoid the aforementioned difficulties, see e.g.\ Refs.~\cite{Parker:2003as, Parker:2000pr, Caldwell:2005xb, Csaki:2005vq, Kaloper:2009nc}. This model is however in tension with additional datasets, and considering {\it Planck} 2018 + Pantheon + BAO, the Hubble constant will be $H_0 = 68.34\pm0.82{\rm \,km\,s^{-1}\,Mpc^{-1}}$ at 68\% CL~\cite{Aghanim:2018eyx}, in $3.2\sigma$ tension with R20. Other approaches in the literature that explored the ability of a phantom dark energy to solve the Hubble tension considered a redshift-binned dark energy model~\cite{Qing-Guo:2016ykt}, a $w$CDM model in which $w_0$ is fixed to some specific values~\cite{Vagnozzi:2019ezj}, taking into account previously unconsidered systematic effects affecting the SNIa measurements~\cite{Martinelli:2019krf}, exploiting the $H_0-w_0$ degeneracy~\cite{Alestas:2020mvb}, {reanalysing the BOSS DR12 data using the EFTofLSS formalism~\cite{DAmico:2020kxu}}, considering an extreme combination of Hubble measurements~\cite{DiValentino:2020vnx}, and exploring the epoch that possibly sourced the $H_0$ tension~\cite{Haridasu:2020pms}.

\subsection{$w_0w_a$CDM or CPL parameterization}

We now discuss some models in which the equation of state for DE depends on the redshift. Among such models, we first consider the Chevallier - Polarski - Linder parameterization (CPL)~\cite{Chevallier:2000qy,Linder:2002et}:
\begin{equation}
    \label{eq:CPL}
    w_{\rm DE} (a) = w_0+(1-a)w_a \,,
\end{equation}
where $a$ is the cosmological scale factor normalized to unity today, $w_0$ is the DE equation of state today, and $w_a$ describes its evolution with time. We refer to this scenario as the $w_0w_a$CDM model. For example, if $w_a < 0$ ($w_a > 0$), $w_{\rm DE}(a)$ becomes more negative (positive) as we look backwards in time. Within the CPL parameterization, {\it Planck} 2018 provides the constraints $w_0=-1.21^{+0.33}_{-0.60}$ and $w_a<-0.85$ at 68\% CL~\cite{Yang:2021flj},\footnote{\, Such constraints for the DE equation of state parameters are also in agreement with the bounds obtained in Ref.~\cite{Menci:2020ybl} using the abundance of massive Galaxies at high redshifts.} and $H_0>63{\rm \,km\,s^{-1}\,Mpc^{-1}}$ at 95\% CL, in agreement with R20 within $2\sigma$. However, when additional datasets are considered, {\it Planck} 2018 + Pantheon + BAO gives $H_0 = 68.35\pm0.84{\rm \,km\,s^{-1}\,Mpc^{-1}}$ at 68\% CL~\cite{Aghanim:2018eyx}, in $3.2\sigma$ tension with R20.

Other studies that account how the CPL parameterization of the DE equation of state addresses the Hubble tension explore the $w_0w_a$CDM model by changing the pivot redshift~\cite{Yang:2018prh}, taking into account unconsidered systematic effects affecting the SNIa~\cite{Martinelli:2019krf}, exploiting the degeneracy between $H_0$ and $w_{\rm DE}$~\cite{Alestas:2020mvb}, considering an extreme combination of Hubble measurements~\cite{DiValentino:2020vnx}, demanding a higher power of polarizations with respect to $\Lambda$CDM to be in agreement with R19~\cite{Kitazawa:2020qdx}, or showing how this solution worsens the $\Omega_{m}-\sigma_8$ growth tension~\cite{Alestas:2021xes}.

\subsection{Dark energy in extended parameter spaces}
\label{sec:Dark-energy-in-extended-parameters-space}

In order to identify the optimal extension of the minimal $\Lambda$CDM model to alleviate the $H_0$ tension, leading to a better fit to observations, one can allow to vary more than one well-motivated cosmological parameters simultaneously. In other words, one should try a combination of parameters that can ameliorate the Hubble tension without considering only one specific mechanism. Indeed, many assumptions and simplifications made in the six parameter description of the $\Lambda$CDM model may not be fully justified, and perhaps could hide some physical aspects essential in the evolution of the Universe. In a multi-parameter space, the biases introduced by the choice of the model are easily avoided~\cite{DiValentino:2015ola, DiValentino:2016hlg, DiValentino:2017zyq, DiValentino:2019dzu, DiValentino:2020hov}.

To begin with, the authors consider an 11-parameter space model in which the $\Lambda$CDM model is augmented by the running of the scalar spectral index $\alpha_s$, the total neutrino mass $\Sigma m_\nu$, the effective number of relativistic degrees of freedom $N_{\rm eff}$ (see Section~\ref{DR} for details), a constant dark energy equation of state $w_0$, and the $A_{\rm lens}$ parameter~\cite{Calabrese:2008rt}. In this scenario, a fit of the 11-parameter space model to the {\it Planck} 2018 data results in $H_0=73^{+10}_{-20}{\rm \,km\,s^{-1}\,Mpc^{-1}}$ at 68\% CL, in agreement with R20 within $1\sigma$. When additional data are considered, {\it Planck} 2018 + BAO gives $H_0 = 67.9 \pm 1.7{\rm \,km\,s^{-1}\,Mpc^{-1}}$ at 68\% CL~\cite{DiValentino:2019dzu}, in $2.5\sigma$ tension with R20, and {\it Planck} 2018 + Pantheon gives $H_0 = 66.9 \pm 2.0{\rm \,km\,s^{-1}\,Mpc^{-1}}$ at 68\% CL~\cite{DiValentino:2019dzu}, in $2.6\sigma$ tension. 

The $\Lambda$CDM model can be further extended by considering, instead, a dynamical dark energy equation of state $w_{\rm DE}(z)$, parameterized by the CPL relation in Eq.~\eqref{eq:CPL}. This is the same as considering the 11-parameter space model but with a DE equation of state modeled with the CPL relation. In this 12-parameter space, a fit to the {\it Planck} 2018 data gives $H_0=72 \pm 20{\rm \, km\,s^{-1}\,Mpc^{-1}}$ at 68\% CL~\cite{DiValentino:2019dzu}, in agreement with R20 within $1\sigma$. Again, the results prefer a phantom-like DE at more than three standard deviations. When additional data are considered, {\it Planck} 2018 + BAO gives $H_0 = 64.8 
^{+2.5}_{-2.9}{\rm \,km\,s^{-1}\,Mpc^{-1}}$ at 68\% CL~\cite{DiValentino:2019dzu}, in $3\sigma$ tension with R20, and {\it Planck} 2018 + Pantheon gives $H_0 = 66.8 \pm 2.1{\rm \,km\,s^{-1}\,Mpc^{-1}}$ at 68\% CL~\cite{DiValentino:2019dzu}, in $2.6\sigma$ tension.

\subsection{Dynamical dark energy parameterisations with two free parameters}

Dynamical dark energy parameterizations with two free parameters have been extensively studied in the literature, see for instance~\cite{Chevallier:2000qy, Linder:2002et, Cooray:1999da, Efstathiou:1999tm, Astier:2000as, Weller:2001gf, Jassal:2005qc,Nesseris:2005ur, Feng:2004ff,Barboza:2008rh, Feng:2011zzo, Li:2012vn, Feng:2012gf, Yang:2017amu,Yang:2017alx,Pan:2017zoh,Rezaei:2017yyj,Vagnozzi:2018jhn,Du:2018tia}. Apart from the most well known dynamical dark energy prescribed by the CPL parameterization with two free parameters~\cite{Chevallier:2000qy, Linder:2002et, Cooray:1999da}, some other two-parameter parameterizations have recently been confronted with the latest {\it Planck} 2018 data in Ref.~\cite{Yang:2021flj}, namely:

\begin{itemize}
\item The JBP parameterization of the dark energy equation of state proposed by Jassal-Bagla-Padmanabhan~\cite{Jassal:2005qc}:
\begin{equation}
    \label{jbp}
    w_{\rm DE}(a) = w_0 + w_a \,a\, (1-a)\,,
\end{equation}
that, when analysed in light of {\it Planck} 2018 measurements, provides $H_0=85^{+13}_{-7}{\rm \, km\,s^{-1}\,Mpc^{-1}}$ at 68\% CL, while {\it Planck} 2018 + BAO gives $H_0 = 67.4 
^{+1.9}_{-2.9}{\rm \,km\,s^{-1}\,Mpc^{-1}}$ at 68\% CL.

\item The Logarithmic dark energy equation of state parameterization proposed by Efstathiou~\cite{Efstathiou:1999tm}:
\begin{equation}
    \label{log}
    w_{\rm DE} (a) = w_0 - w_a \ln a\,,
\end{equation}
for which the {\it Planck} 2018 data analysis results in a value of the Hubble constant $H_0=83^{+15}_{-8}{\rm \, km\,s^{-1}\,Mpc^{-1}}$ at 68\% CL, while {\it Planck} 2018 + BAO gives $H_0 = 64.8 \pm 2.1 {\rm \,km\,s^{-1}\,Mpc^{-1}}$ at 68\% CL.

\item The BA parameterization proposed by Barboza and Alcaniz~\cite{Barboza:2008rh}:
\begin{equation}
    \label{ba}
    w_{\rm DE} (a) = w_0 + w_a \left(\frac{1-a}{2a^2-2a+1}. \right)\,,
\end{equation}
for which the {\it Planck} 2018 data analysis results in a value of the Hubble constant $H_0=83^{+15}_{-8}{\rm \, km\,s^{-1}\,Mpc^{-1}}$ at 68\% CL, while {\it Planck} 2018 + BAO gives $H_0 = 65.2^{+2.2}_{-2.8}{\rm \,km\,s^{-1}\,Mpc^{-1}}$ at 68\% CL.
\end{itemize}
\noindent All of the parameterizations above are in agreement with R20 within $2\sigma$ for {\it Planck} 2018 only, while for {\it Planck} 2018 + BAO are in tension at $2.5\sigma$, $3.4\sigma$ and $3.1\sigma$, respectively.

\subsection{Dynamical dark energy parameterizations with one free parameter}

Compared to the dynamical dark energy parameterizations with two free parameters, there are only a few dynamical dark energy parameterizations with one free parameter~\cite{Gong:2005de,Yang:2018qmz}. However, some recent investigations clearly demonstrate that dynamical dark energy parameterizations with a single parameter are very effective in alleviating the Hubble tension (see Ref.~\cite{Yang:2018qmz}). 

The models considered in~\cite{Yang:2018qmz} are reported in Table~\ref{tabParams}, together with their results on the value of $H_0$ are all at 68\% CL. Note, that $w_0$ is the present value of the dark energy equation of state, that means $w_0 = w_{\rm DE} (a = 1)$.
\begin{table}[!ht]
\begin{center}
    \def\arraystretch{1.5}
	\begin{tabular}{llll}
    \hline\hline
		Model & Equation of state & Hubble constant $H_0$ &\\
	&  & {\it Planck} 2015 & + BAO + JLA \\
		\hline
		i) & $w_{\rm DE} (a)=w_0\exp(a-1)$ & $74^{+11}_{-7}{\rm \, km\,s^{-1}\,Mpc^{-1}}$ & $69.2^{+1.1}_{-1.0}{\rm \, km\,s^{-1}\,Mpc^{-1}}$ \\
		ii) & $w_{\rm DE} (a)=w_0a[1-\log(a)]$ & $81^{+12}_{-9}{\rm \, km\,s^{-1}\,Mpc^{-1}}$& $69.0^{+1.0}_{-1.1}{\rm \, km\,s^{-1}\,Mpc^{-1}}$ \\
		iii) & $w_{\rm DE} (a)=w_0a\exp(1-a)$, & $84^{+10}_{-8}{\rm \, km\,s^{-1}\,Mpc^{-1}}$& $69.4\pm1.0{\rm \, km\,s^{-1}\,Mpc^{-1}}$ \\
		iv) & $w_{\rm DE} (a)=w_0a[1+\sin(1-a)]$ & $84.3^{+9.9}_{-6.5}{\rm \, km\,s^{-1}\,Mpc^{-1}}$ & $70.07^{+0.91}_{-0.94}{\rm \, km\,s^{-1}\,Mpc^{-1}}$\\
		v) & $w_{\rm DE} (a)=w_0a[1+\arcsin(1-a)]$ & $83^{+12}_{-7}{\rm \, km\,s^{-1}\,Mpc^{-1}}$& $69.6^{+1.0}_{-1.2}{\rm \, km\,s^{-1}\,Mpc^{-1}}$\\
		\hline
		\hline
	\end{tabular}
	\caption{The models considered in Ref.~\cite{Yang:2018qmz} and the Hubble constant obtained by analysing the {\it Planck} 2015 data and its combination with BAO and JLA.}
	\label{tabParams}
\end{center}
\end{table}

All the models considered in Ref.~\cite{Yang:2018qmz} are in agreement with R20 within $1\sigma$ for {\it Planck} 2015 only, always at the price of a phantom dark energy equation of state today, and within $2.6\sigma$ for {\it Planck} 2015 + BAO + JLA. However, a re-analysis with the most recent {\it Planck} 2018 dataset is still missing in the literature.

\subsection{Metastable dark energy}
\label{metastable}

Another possibility to solve the Hubble constant problem relies on metastable DE models, where the DE energy density can decay or increase depending only on its intrinsic nature and not on the external parameters~\cite{Shafieloo:2016bpk,Li:2019san,Szydlowski:2017wlv,Szydlowski:2018kbk,Yang:2020zuk}. In the simplest model of metastable DE, the DE energy density evolves as~\cite{Shafieloo:2016bpk, Li:2019san}:
\begin{equation}
    \label{model-metastable}
    \frac{\mathrm{d}\rho_{\rm DE}}{\mathrm{d}t} = - \Gamma \rho_{\rm DE}\,,
\end{equation}
where $\Gamma$ is a constant decay rate and $t$ denotes cosmic time. The equation of state obtained from Eq.~\eqref{model-metastable} is:
\begin{equation}
    \label{meta-eos}
    w_{\rm DE} = -1 -\frac{1}{3 H}\frac{\mathrm{d}\ln \rho_{\rm DE}}{\mathrm{d} t} = -1 + \frac{\Gamma}{3H}\,.
\end{equation}
The fit against Pantheon + BAO data provides $H_0=75.01^{+4.71}_{-5.80}{\rm \,km\,s^{-1}\,Mpc^{-1}}$ at 68\% CL.
However, when CMB distance priors from {\it Planck} 2018 are included, the Hubble tension is restored at more than $3\sigma$~\cite{Li:2019san}.

Reference~\cite{Yang:2020zuk} performs an analysis of this metastable DE model against {\it Planck} 2018 ({\it Planck} 2018 + BAO + DES + R19) data which leads to a value of the Hubble constant $H_0=69.3^{+5.9}_{-3.5}{\rm \,km\,s^{-1}\,Mpc^{-1}}$ ($H_0=71.94\pm1.08{\rm \,km\,s^{-1}\,Mpc^{-1}}$) at 68\% CL, solving the tension with R20 within $1\sigma$. Notice that the alleviation of the tension for {\it Planck} 2018 alone is mainly due to the large error bars in $H_0$.

\subsection{Phantom Crossing}

If a Phantom Crossing model is accounted for, the Hubble tension can be solved within one standard deviation, without spoiling the agreement with the BAO data~\cite{DiValentino:2020naf}. If the dark energy density is Taylor-expanded around an extremum at scale factor $a=a_m$ as:
\begin{equation}
\rho_{\rm DE} (a)=\rho_0+\rho_2(a-a_m)^2+\rho_3(a-a_m)^3 =\rho_0[1+\alpha (a-a_m)^2 +\beta (a-a_m)^3]\,,
\end{equation}
where $\rho_0$, $\rho_2$, $\rho_3$ are constants and $\alpha \equiv \rho_2/\rho_0$, $\beta \equiv \rho_3/\rho_0$, the DE equation of state results in:
\begin{equation}
w_{\rm DE} (a) = -1 - \frac{a\,[2\alpha(a-a_m)+3\beta(a-a_m)^2]}{3[1+\alpha(a-a_m)^2+\beta(a-a_m)^3]}\,.
\end{equation}
For this particular parameterization, an analysis to {\it Planck} 2018 + BAO measurements results in a Hubble constant value of $H_0=71.0^{+2.9}_{-3.8}{\rm \,km\,s^{-1}\,Mpc^{-1}}$ at 68\% CL~\cite{DiValentino:2020naf}. A full dataset combination of {\it Planck} 2018 + CMB lensing + BAO + R19 + Pantheon gives instead $H_0=70.25\pm0.78{\rm \,km\,s^{-1}\,Mpc^{-1}}$ at 68\% CL~\cite{DiValentino:2020naf}, in agreement with R20 at $2\sigma$.

\subsection{Late Dark Energy Transition}

Another possibility for solving the Hubble tension is to consider a Late Dark Energy Transition, in which the equation of state for dark energy sharply changes from the cosmological constant value $w_{\rm DE} =-1$ to a phantom-like value $w_{\rm DE} < -1$ at redshift $z \sim \mathcal{O}(0.1)$~\cite{Mortonson:2009qq, Benevento:2020fev, Alestas:2020zol, Dhawan:2020xmp}. Such a transition is referred to as a ``hockey stick'' because of the shape of the equation of state $w_{\rm DE} (z)$.

Starting from the prediction for the Hubble constant $\tilde H_0$ in $\Lambda$CDM, a late DE transition leads to the actual Hubble constant $H_0 = (1+\delta)\tilde H_0$ where $\delta$ is the fractional change in the Hubble constant. To model this, one considers a DE energy density content $\rho_{\rm DE} (z)$ that transitions from the cosmological constant value $\rho_{\Lambda} = \Omega_{\Lambda}\rho_{\rm crit,0}$ to a phantom-like fluid at redshift $z_t$. The transition is modulated by a smooth step function $f(z)$ as:
\begin{eqnarray}
\rho_{\rm DE} (z) &=& \rho_{\Lambda}\,\left[1+f(z)\right]\,;\\
f(z) &=& \frac{2 \delta}{\Omega_{\Lambda}}\frac{S(z)}{S(0)}\,;\\
S(z) &=& \frac{1}{2} \left[ 1- \tanh \left(\frac{z-z_{t}}{\Delta z} \right) \right]\,,
\end{eqnarray}
where $\Delta z$ is the duration of the transition. For $z\gg z_t$, the expansion history is indistinguishable from the $\Lambda$CDM scenario. In Ref.~\cite{Benevento:2020fev} it has been shown that the combination {\it Planck} 2018 + CMB lensing + BAO + Pantheon + R19 provides $H_0=72.5\pm 1.85{\rm \,km\,s^{-1}\,Mpc^{-1}}$ at 68\% CL, solving the Hubble tension within one standard deviation. However, notice that this result incorporates already a Gaussian prior on the Hubble constant corresponding to R19.

In Ref.~\cite{Alestas:2020zol}, a sudden change in the dark energy equation of state by a quantity $\Delta w$ is considered as a possible solution to the Hubble tension. The equation of state is modeled as:
\begin{equation}
    w_{\rm DE} (z) = -1 + \Delta w \,\Theta(z_t-z)\,,
\end{equation}
where $\Theta$ is the Heaviside step function. 

The possibility that a Late Phantom Transition ever occurred has been recently challenged in Ref.~\cite{Camarena:2021jlr}, where it has been shown that a ``hockey stick'' dark energy can not solve the Hubble crisis because the SNIa absolute magnitude $M_B$ considered to obtain R19 is inconsistent with the $M_B$ necessary to fit SNIa, BAO and CMB data.

However, if a corresponding transition for the SNIa absolute magnitude $M$ is accounted for, as in Ref.~\cite{Alestas:2020zol}:
\begin{equation}
M (z) = M_C + \Delta M \,\Theta(z_t-z)\,,
\end{equation}
then the Late Phantom Transition approach is again a viable possibility to address the Hubble tension.
However, a full CMB data analysis is currently missing.

\subsection{Running vacuum model}
\label{sec:RVM}

The running vacuum model was proposed in Refs.~\cite{Shapiro:2000dz, Shapiro:1999zt} to solve the ``coincidence problem'' by using a Quantum Field Theory approach in cosmology, where the vacuum energy density can be derived from a general renormalization group equation whose beta-function takes the form of an adiabatic expansion in powers of the Hubble rate and its time derivatives (see also the explanations in Refs.~\cite{Shapiro:2004ch,Sola:2007sv,Shapiro:2009dh} and the analysis in Refs.~\cite{Sola:2011qr,Basilakos:2012ra,Mimoso:2013zhp}). Therefore, in this model the cosmological constant is assumed to be an affine power-law function of the Hubble rate, $\Lambda=\Lambda(H)$. The story of the running vacuum model and related ideas can be found in the reviews~\cite{Sola:2013gha,Sola:2015rra}, while the extensions for a curved spacetime and for a string Universe are carried out in Refs.~\cite{Moreno-Pulido:2020anb} and \cite{Mavromatos:2020kzj}, respectively. Another extension of the $\Lambda$CDM model that accounts for this parameterizations are the dynamical quasi-vacuum models ($w$DVMs), in which the Hubble tension is reduced because of the phantom-like behavior of DE~\cite{Sola:2017znb}. The analysis of {\it Planck} 2015 + CMB lensing 2015 + BAO + R16 provides, indeed, $H_0=70.95 \pm 1.46{\rm \,km\,s^{-1}\,Mpc^{-1}}$ at 68\% CL~\cite{Sola:2017znb}, solving the Hubble tension at $1.1\sigma$. However, in this result the R16 Gaussian prior on the Hubble constant and the BAO data are both considered.

Another extension named as RRVM of type-II, where the vacuum dynamics is not caused by an interaction between the vacuum and matter sectors, but by the running of the gravitational coupling G, has been studied in Ref.~\cite{Sola:2021txs}, where {\it Planck} 2018 + Pantheon + DES + BAO + RSD + CC + a prior on $H_0$ from~\cite{Reid:2019tiq} gives $H_0=70.93 ^{+0.93}_{-0.87}{\rm \,km\,s^{-1}\,Mpc^{-1}}$ at 68\% CL, in agreement at $1.4\sigma$ with R20, but already including a prior on the Hubble constant.

\subsection{Transitional Dark Energy model}

When a parametric model where a transition in the dark energy equation of state is accounted for, in order to be consistent with $H_0\sim 73{\rm \,km\,s^{-1}\,Mpc^{-1}}$, the DE component is not yet present until redshifts around $z = 2$, but its energy density has instead a rapid change between $z= 0.5$ and $z=2$~\cite{Keeley:2019esp}. This result has been obtained with a model-independent Gaussian regression analysis process using {\it Planck} 2015, BAO, Pantheon and R16 data, but a complete analysis with perturbations and the full {\it Planck} 2018 data is absent.

\subsection{Negative Dark Energy}
\label{sec:negativeDE}

In Ref.~\cite{Dutta:2018vmq} the authors assume that the Universe follows a $\Lambda$CDM cosmology at higher redshifts ($z\ge 4$), in agreement with the {\it Planck} measurements, and reanalyse the low redshift cosmological data in order to reconstruct a Hubble rate $H(z)$ which is in full agreement with R16. Once the energy density for the DE component is computed as a function of redshift without assuming a specific model, they find a local minimum of the DE energy density with a negative value. While this scenario could be ascribed to a negative cosmological constant plus an evolving dark energy component, the model considered deserves further investigations since these findings compromise its stability. A model which comprises a negative cosmological constant plus a time-evolving quintessence field has been considered in Refs.~\cite{Visinelli:2019qqu,Calderon:2020hoc} and tested against BAO surveys and the Pantheon SNIa data, however, a test against the full {\it Planck} dataset is still missing.

\subsection{Bulk viscous models}

Bulk viscous models have been proposed to alleviate the $H_0$ tension. 
A bulk viscous fluid is characterized by its energy density $\rho$ and a pressure term $p$ which comprises two components, the first being the conventional pressure term $p_{\rm con} = w_0 \rho$, where $w_0$ is a constant equation of state, and the second one being a viscosity component $p_{\rm vis} = -\xi (t) {u^{\mu}}_{; \mu}$ that depends on the coefficient of bulk viscosity $\xi(t)>0$ and on the four-velocity of the fluid $u^\mu$~\cite{Brevik:2005bj}. Therefore, the effective pressure term $p$ takes the form $p = w_0 \rho - \xi (\rho) {u^{\mu}}_{; \mu}$.

The bulk viscosity can play an effective role in describing the evolution of the Universe in its early and late phases~\cite{Haro:2015ljc} (see the review in Ref.~\cite{Brevik:2017msy} for more details). For any bulk viscous fluid
as described above, its evolution in the FLRW Universe is given by:
\begin{equation}
    \frac{\mathrm{d}\rho}{\mathrm{d}t} + 3 H(1+w_0)\rho = 9H^2\xi (t)\,,
\end{equation}
where $H$ has been defined in Eq.~\eqref{def:Hubble_rate}. In general, two different kinds of bulk viscous models are considered: one where dark energy has a viscous nature~\cite{Wang:2017klo, daSilva:2020mvk} but matter has an independent evolution, or alternatively, a unified bulk viscous model in which dark matter and dark energy can not be distinguished~\cite{Yang:2019qza}. Both scenarios can alleviate the $H_0$ tension. 

Considering that dark energy has a viscous nature, where the viscosity coefficient is proportional to the Hubble parameter $\xi (t)=\eta_0\,H$, as introduced in Ref.~\cite{Wang:2017klo}, the authors of Ref.~\cite{daSilva:2020mvk} have found that for this model, the combination of {\it Planck} 2018 CMB distance priors + Pantheon + BAO + BBN + CC results in $H_0=69.3\pm1.7{\rm\,km\,s^{-1}\,Mpc^{-1}}$ at 68\% CL, reducing the Hubble tension to the $1.9\sigma$ level.

A model in which the bulk viscosity is proportional to the energy density and inversely proportional to the Hubble parameter, $\xi (t) = \eta_0\,\sqrt{\rho_{\rm DE}}/H$, has been introduced in Refs.~\cite{Mostaghel:2016lcd, Mostaghel:2018pia} and considered in light of the Hubble tension in Ref.~\cite{daSilva:2020mvk}. An analysis that fits the {\it Planck} 2018 CMB distance priors + Pantheon + BAO + BBN + CC data provides $H_0=69.3\pm1.7{\rm\,km\,s^{-1}\,Mpc^{-1}}$ at 68\% CL, thus reducing the tension with R20 down to $1.9\sigma$~\cite{daSilva:2020mvk}.

The bulk viscosity can be a thermodynamic function $\xi (t)=\eta_0\,\rho_{\rm DE}^\nu$, as introduced in Ref.~\cite{Velten:2012uv}. A fit to the combination of {\it Planck} 2018 CMB distance priors + Pantheon + BAO + BBN + CC data on this model gives $H_0=69.2\pm1.7{\rm\,km\,s^{-1}\,Mpc^{-1}}$ at 68\% CL~\cite{daSilva:2020mvk}, alleviating the Hubble tension down to $1.9\sigma$ as in the two previous models. Note that a CMB-only analysis for all these models is currently missing.

In Ref.~\cite{Yang:2019qza} the authors have investigated a unified cosmic scenario endowed with a bulk viscosity in which the bulk viscosity coefficient follows a general law $\xi (t) = \alpha \rho^m$ (see also Ref.~\cite{Velten:2012uv}). We see that in the scenario where $w_0 = 0$ and $m$ is a free parameter, {\it Planck} 2015 + Pantheon leads to $H_ 0 = 68.0 \pm 1.1{\rm\,km\,s^{-1}\,Mpc^{-1}}$ at 68\% CL~\cite{Yang:2019qza} and is hence in disagreement with R20 at $3.1\sigma$. For the case of a free $w_0$ with $m =0$, instead, {\it Planck} 2015 + Pantheon leads to $H_0 = 70.2^{+1.6}_{-1.9}{\rm\,km\,s^{-1}\,Mpc^{-1}}$ at 68\% CL~\cite{Yang:2019qza}, solving the tension at $1.4 \sigma$. Lastly, for the case in which both $w_0$ and $m$ are free parameters, {\it Planck} 2015 + Pantheon gives $H_0 = 68.0^{+2.7}_{-2.4}{\rm\,km\,s^{-1}\,Mpc^{-1}}$ at 68\% CL~\cite{Yang:2019qza}, alleviating the tension with R20 at $ 1.7\sigma$.

Models in which a viscous inhomogeneous fluid describes the content of the late Universe are adopted in Ref.~\cite{Elizalde:2020mfs}. In the models studied, the pressure of the single fluid considered is a function of both the Hubble rate and density, with parameters that are fixed through a Bayesian Learning method over measured values of $H(z)$ for $z \lesssim 2.5$. For the models considered, this method yields $H_0 = 73.4 \pm 0.1{\rm\,km\,s^{-1}\,Mpc^{-1}}$ and $H_0 = 73.52 \pm 0.15{\rm\,km\,s^{-1}\,Mpc^{-1}}$ at 68\% CL~\cite{Elizalde:2020mfs}, respectively. Note however that an analysis that uses {\it Planck} 2018 data is still missing.

\subsection{Holographic Dark Energy}

An interesting DE candidate that was proposed following the holographic principle is the holographic dark energy (HDE)~\cite{Li:2004rb,Huang:2004mx,Zhang:2014ija}. The model was extensively studied for its ability to explain the late-time cosmic acceleration (see the review of Ref.~\cite{Wang:2016och}). In this model, the dark energy equation of state is given by:
\begin{equation}
    w_{\rm DE}(z) = -\frac{1}{3}-\frac{2}{3c}\sqrt{\Omega_{\rm DE}(z)}\,,
\end{equation}
where $c$ is a dimensionless parameter.

In Ref.~\cite{Guo:2018ans}, the authors argue that the HDE model can alleviate the tension with the local measurements of $H_0$. A fit to the {\it Planck} 2015 + CMB lensing + BAO + JLA + R16 data for HDE returns $H_0 = 69.67^{+0.95}_{-0.94}{\rm\,km\,s^{-1}\,Mpc^{-1}}$ at 68\% CL~\cite{Guo:2018ans}, which alleviates the tension with R20 down to the $2.2\sigma$ level. The fit within the extended model HDE$+N_{\rm eff}+m^{\rm eff}_{\nu, \rm sterile}$ (in which a massive sterile neutrino is included) to the same dataset gives $H_0 = 70.70 \pm 1.10{\rm\,km\,s^{-1}\,Mpc^{-1}}$ at 68\% CL~\cite{Guo:2018ans}, alleviating the tension with R20 at $1.5\sigma$. 

An updated analysis for HDE using {\it Planck} 2018 data can be found in Ref.~\cite{Dai:2020rfo}, where {\it Planck} 2018 + BAO + R19 gives $H_0 = 73.12 \pm 1.14{\rm\,km\,s^{-1}\,Mpc^{-1}}$ at 68\% CL, in agreement with R20. Ref.~\cite{Dai:2020rfo} also considers the effect of including Pantheon data, finding that this inclusion shifts the value of $H_0$ to a lower value. In fact, Ref.~\cite{Dai:2020rfo} further uncovered that both the subsets of 
the Pantheon data with $z > 0.2$ and $z < 0.2$ prefer a higher value of $H_0$, but they have a large negative correlation in between which has not yet been fully understood.
Ref.~\cite{Dai:2020rfo} also considered an analysis that includes Pantheon data, finding that this inclusion shifts the value of $H_0$ to a lower value. In fact, the analysis in Ref.~\cite{Colgain:2021beg} considers {\it Planck} 2018 + BAO + Pantheon and obtains
$H_0 = 67.94\pm0.80{\rm\,km\,s^{-1}\,Mpc^{-1}}$ at 68\% CL~\cite{Colgain:2021beg}, at $3.5\sigma$ tension with R20 once the Pantheon data are included.

In this context one may be interested in the stability of the de Sitter state in the dark energy models following a Holographic approach~\cite{10.1093/mnrasl/slz158}. As argued in Ref.~\cite{10.1093/mnrasl/slz158}, unlike in the $\Lambda$CDM model where the de Sitter state is assumed to be stable in the distance future, the dark energy model following a holographic approach could alleviate the Hubble constant tension leading to an unstable de Sitter state in the distance future. This instability is actually responsible for a turning point~\cite{vanPutten:2017bqf} which seems crucial in capturing the $H_0$ tension quantitatively and providing a common ground with the Swampland conjectures.

\subsubsection{Tsallis Holographic Dark Energy:}

An extension of the previous holographic model following Tsallis statistics~\cite{Tsallis:2012js}, dubbed as Tsallis Holographic dark energy, has been found to alleviate the $H_0$ tension~\cite{daSilva:2020bdc}. For this model, {\it Planck} 2018 CMB distance priors + BAO + BBN + CC + Pantheon gives $H_0 = 69.8 \pm 1.8{\rm\,km\,s^{-1}\,Mpc^{-1}}$ at 68\% CL which alleviates the tension with R20 at $1.5\sigma$. A full {\it Planck} data analysis is however missing.

\subsection{Swampland Conjectures}

String theory is a potential candidate for a UV-complete theory. A large number of string vacua are expected, therefore providing a consistent low-energy effective field theory (EFT) limit~\cite{Susskind:2003kw}. These well-behaved solutions that populate the ``landscape'' are conjectured to be surrounded by a ``swampland'' of semi-classical EFTs for which a consistent theory of quantum gravity does not exist~\cite{Vafa:2005ui}. Various recipes have been conjectures in the attempt to understand the conditions under which a given EFT does not lie in the swampland, such as the weak-gravity conjecture~\cite{ArkaniHamed:2006dz} and a set of swampland conjectures~\cite{Ooguri:2006in, Klaewer:2016kiy, Ooguri:2016pdq, Freivogel:2016qwc}. In particular, two of these swampland conjectures constrain the excursion range $\Delta \phi$ of a scalar field $\phi$ in field space as well as the logarithmic gradient of the scalar field potential $V(\phi)$. The first ``distance'' conjecture avoids that a tower of light states emerges when a scalar field moves by a distance $\Delta \phi \geq O(1)$ (in Planck units)~\cite{Grimm:2018ohb, Heidenreich:2018kpg, Blumenhagen:2018hsh}. The second of these swampland criteria establishes that a scalar field potential $V$ arising from a consistent quantum theory of gravity should satisfy $|V_\phi| \geq c V$, where $c \sim \mathcal{O} (1)$ (in Planck units) is a positive constant and $V_\phi = \mathrm{d}V(\phi)/\mathrm{d}\phi$ (see Refs.~\cite{Kinney:2018nny, Danielsson:2018qpa} for the implications of the swampland conjecture in cosmology).

Scalar field models obeying the swampland conjectures have recently gained considerable attention in relation with the proposed solutions to the Hubble tension. In fact, one could construct physically viable scalar field models that could explain the dark energy effects at late time and satisfy the swampland criteria~\cite{Agrawal:2018own, Agrawal:2019dlm, Anchordoqui:2019amx, Anchordoqui:2020sqo}.

In Ref.~\cite{Agrawal:2019dlm}, it is found that a scalar field model, satisfying the swampland criteria with a fixed value of $c$, can alleviate the Hubble tension. For example, an analysis using {\it Planck} 2015 + CMB lensing + BAO + Pantheon + R19 fixing $c=0.1$ gives $H_0 = 69.06^{+0.66}_{-0.73}{\rm\,km\,s^{-1}\,Mpc^{-1}}$ at 68\% CL, reducing the tension down to $2.8\sigma$ level (notice that R19 is already included in the analysis). 

The authors in Ref.~\cite{Colgain:2018wgk} use low-redshift measurements of $H_0$ to fit a polynomial expansion of the Hubble rate $H(z)$ and test the different features against $\Lambda$CDM to alleviate the Hubble tension (see also Ref.~\cite{Colgain:2019joh}). In particular, the functions for $H(z)$ in Ref.~\cite{Colgain:2018wgk} possess a turning point at a critical redshift $z_c=z_c(\Omega_m)$, where $\Omega_m$ is the fractional matter density today. Unfortunately, a full data analysis for this model is missing to date.

A consequence of the swampland criteria within string theory is to consider a quintessence field instead of a cosmological constant. In Ref.~\cite{Banerjee:2020xcn}, this possibility is investigated for solving the Hubble tension, concluding that quintessence models always prefer a lower Hubble constant value than that obtained within the standard $\Lambda$CDM. The addition of an exponential coupling to the dark matter sector does not change this result.

\subsection{Late time transitions in the quintessence field}

In Ref.~\cite{DiValentino:2019exe}, a quintessence field which transits from a matter-like to a cosmological constant-like behavior between recombination and the present time has been proposed to alleviate the Hubble tension. The authors model a transition with the effective DE equation of state:
\begin{equation}
    w_{\rm DE}(a) = \frac{w_{\phi 0}}{1 + (\frac{a}{a_{\rm tr}})^{-\frac{2}{\Delta}}}\,,
\end{equation}
where $a_{\rm tr}$ is the scale factor of the transition and $\Delta$ defines its duration.
They conclude that {\it Planck} 2015 data exclude this model as a possible solution of the Hubble tension, since the best fit value of the Hubble constant is $H_0=67.20\pm 0.64{\rm\,km\,s^{-1}\,Mpc^{-1}}$ at 68\% CL~\cite{DiValentino:2019exe}, at $4.3\sigma$ tension with R20.

A similar observation was found in the context of a minimally coupled slowly or moderately rolling quintessence field with a smooth potential~\cite{Miao:2018zpw}. The authors of Ref.~\cite{Miao:2018zpw} considered the curvature parameter in the analysis and found that the $H_0$ tension in such models remains at more than $3\sigma$.

\subsection{Phantom Braneworld Dark Energy}

In Ref.~\cite{Alam:2016wpf} a braneworld scenario, introduced in~\cite{Sahni:2002dx}, has been proposed to increase the Hubble constant estimate. In this model the observable Universe is situated in a four-dimensional brane embedded in a fifth dimension, the `bulk'. The braneworld dark energy has an equation of state phantom-like, and the accelerated expansion of the Universe is therefore a consequence of this modification of gravity.
Using a combination of {\it Planck} 2015 CMB distance priors, Union 2.1 SNIa and BAO in Ref.~\cite{Alam:2016wpf} the authors find for this scenario $H_0=70.75\pm1.30{\rm\,km\,s^{-1}\,Mpc^{-1}}$ at 68\% CL, solving the Hubble tension within $1.4\sigma$. A full 2018 CMB data analysis is however missing.

\subsection{Frame dependent dark energy}

In Ref.~\cite{Adler:2019fnp}, a frame dependent, although scale invariant, dark energy theory was proposed to alleviate the Hubble tension. In this late time model, the Hubble constant can take extremely large values comparable with the local measurements of $H_0$, offering at the same time an excellent fit to the CMB spectra. The model has some interesting implications, however it needs to be robustly investigated by means of a full data analysis.

\subsection{Chameleon dark energy}

In Ref.~\cite{Cai:2021wgv} a chameleon field (see also Refs.~\cite{Khoury:2003rn, Khoury:2003aq, Brax:2007vm, Banerjee:2008rs, Das:2008iq, Brax:2010kv, Upadhye:2012vh, Wang:2012kj, Khoury:2013yya}) has been proposed to alleviate the Hubble tension. In this paper, the possibility that a matter overdensity, coupled to the chameleon dark energy, can increase the Hubble constant locally, introducing the cosmic inhomogeneity in the Hubble expansion rate at late-time, is taken into consideration. A full data analysis is however missing, but it does not go against the No-Go theorem of general chameleon~\cite{Wang:2012kj}.

\begin{figure*}
\centering

\includegraphics[width=0.85\textwidth, right]{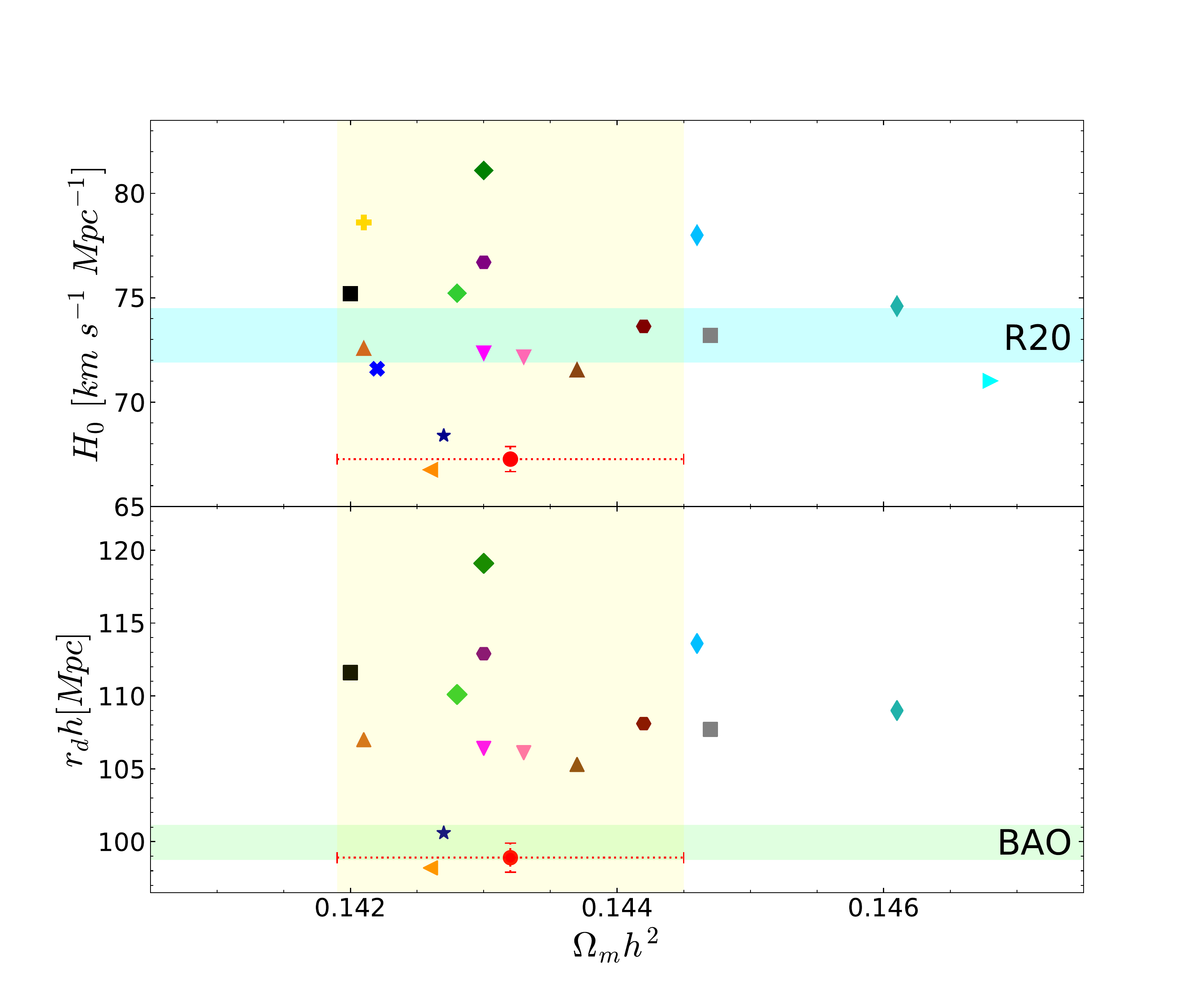}
\includegraphics[width=0.85\textwidth, right]{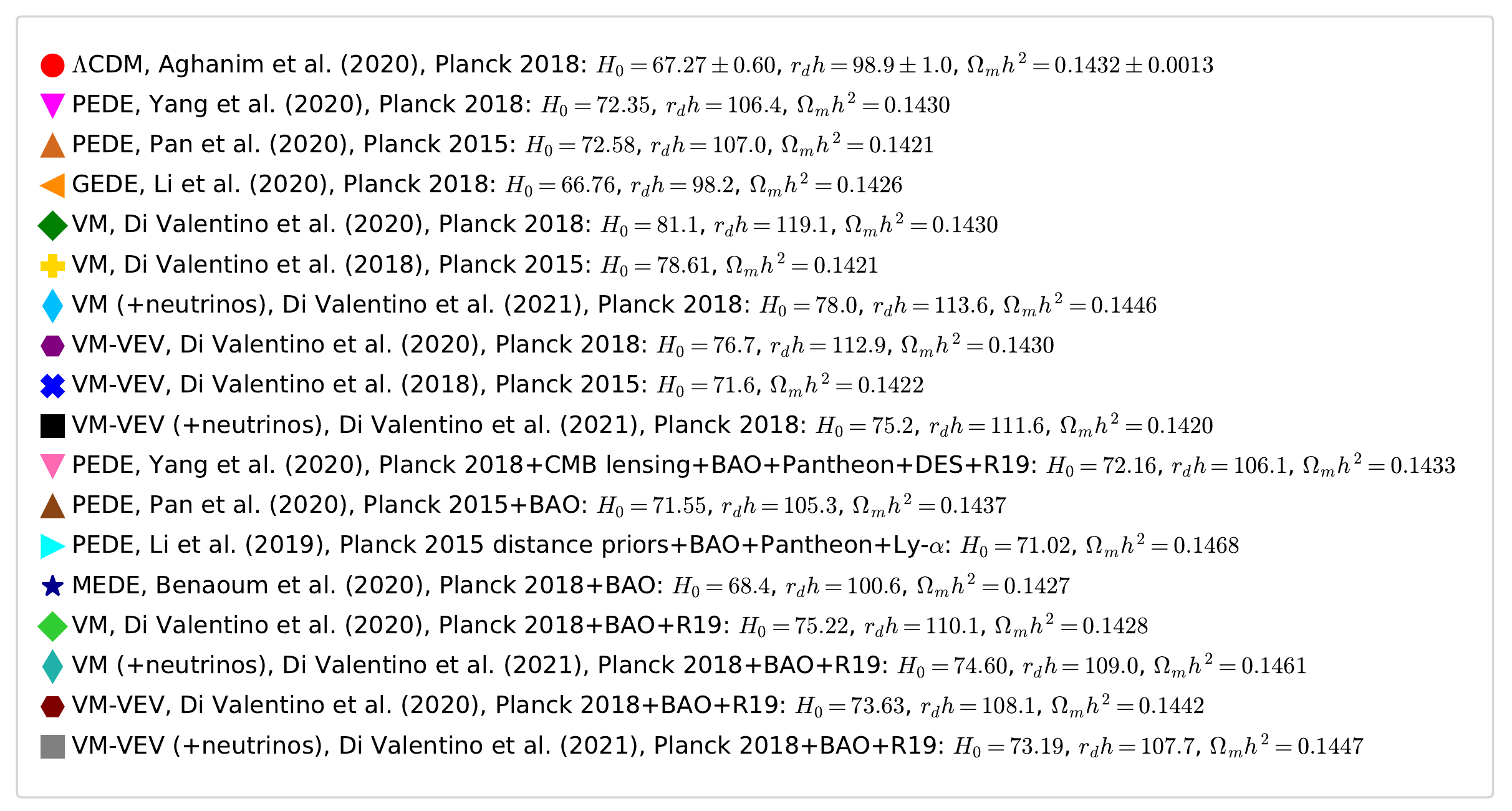}
\caption{Estimated values of the current matter energy density $\Omega_mh^2$, Hubble constant $H_0$ and sound horizon $r_dh$ in terms of various data points for different models discussed throughout the Section~\ref{DE6Dof}. The cyan horizontal band corresponds to the $H_0$ value measured by R20~\cite{Riess:2020fzl}, the yellow vertical band to the $\Omega_mh^2$ value estimated by {\it Planck} 2018~\cite{Aghanim:2018eyx} in a $\Lambda$CDM scenario, and the light green horizontal band to the $r_dh$ value measured by BAO data. The points sharing the same symbol refer to the same model in the same paper, and the different colors indicate a different dataset combination.}
\label{fig:chapter6_H0Om}
\end{figure*}
\begin{figure*}
\includegraphics[width=\textwidth]{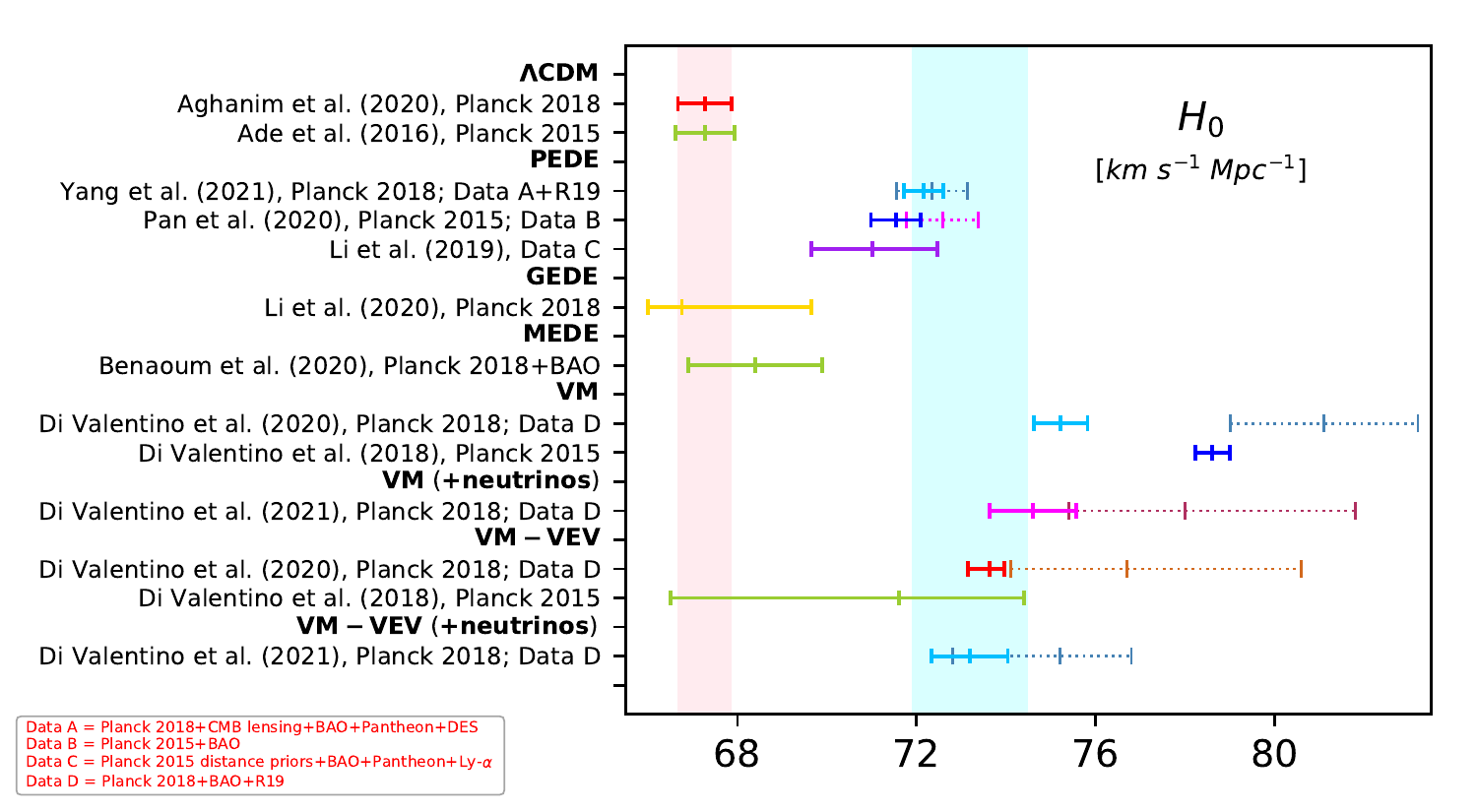}
\caption{Whisker plot with the 68\% marginalized Hubble constant constraints for the models of Section~\ref{DE6Dof}. The cyan vertical band shows the $H_0$ value measured by R20~\cite{Riess:2020fzl} and the light pink vertical band denotes the $H_0$ value estimated by {\it Planck} 2018~\cite{Aghanim:2018eyx} in a $\Lambda$CDM scenario. For each line, when more than one error bar is shown, the dotted one corresponds to the {\it Planck} only constraint on the Hubble constant, while the solid one to the different dataset combinations reported in the red legend, in order to appreciate the shift due to the additional datasets.}
\label{fig:chapter6_whisker}
\end{figure*}

\section{Dark energy models with 6 degrees of freedom and their extensions} \label{DE6Dof}

To alleviate the Hubble constant tension, some dark energy models with no extra degrees of freedom with respect to the $\Lambda$CDM scenario have been proposed. Having the same number of degrees of freedom means that they are not disfavored by a Bayesian model comparison analysis. In Figures~\ref{fig:chapter6_H0Om} and~\ref{fig:chapter6_whisker} we have classified the models according to the values of a number of key parameters, as described in the Introduction.

\subsection{Phenomenologically Emergent Dark Energy}

A famous possibility is a Phenomenologically Emergent Dark Energy (PEDE) model, in which a redshift-dependent dark energy component emerges at late times. In this model, firstly introduced in Ref.~\cite{Li:2019yem}, the fractional DE energy density has the following parameterization:
\begin{equation}
    \label{Omega-DE}
    \tilde{\Omega}_{\rm PEDE} (z) \equiv \frac{\rho_{\rm DE}}{\rho_{{\rm crit},0}} = \Omega_{\rm PEDE,0} \Bigl[1- \tanh (\log_{10} (1+z)) \Bigr]\,,
\end{equation}
where $\Omega_{\rm PEDE,0} = \Omega_{\rm PEDE} (z=0)$ and $\rho_{{\rm crit},0} = 3H_0^2M_{\rm Pl}^2$ is the present critical energy density. The fluid described by Eq.~\eqref{Omega-DE} has a phantom-like equation of state that asymptotically approaches the cosmological constant value $w_\Lambda = -1$ as time proceeds~\cite{Li:2019yem}:
\begin{equation}
    \label{de-eos}
    w_{\rm PEDE} (z) = -1 -\frac{1}{3 \ln 10} \, \Bigl[ 1+ \tanh \bigl(\log_{10} (1+z)\bigr) \Bigr]\,.
\end{equation}
Using the {\it Planck} 2015 CMB distance priors + Pantheon + BAO + Ly-$\alpha$ data, Ref.~\cite{Li:2019yem} finds $H_0=71.02^{+1.45}_{-1.37}{\rm \,km\,s^{-1}\,Mpc^{-1}}$ at 68\% CL, solving the Hubble tension at $1.1\sigma$. Considering a full CMB analysis for this scenario, {\it Planck} 2015 alone gives instead $H_0=72.58^{+0.79}_{-0.80}{\rm \,km\,s^{-1}\,Mpc^{-1}}$ at 68\% CL~\cite{Pan:2019hac}, solving the Hubble tension within $1\sigma$, and {\it Planck} 2015 + BAO gives $H_0=71.55^{+0.55}_{-0.57}{\rm \,km\,s^{-1}\,Mpc^{-1}}$ at 68\% CL, in agreement with R20 at $1.2\sigma$. This result is in agreement with Ref.~\cite{Hernandez-Almada:2020uyr}, where CC measurements are considered. The very same model has been updated in Ref.~\cite{Yang:2021egn}, which finds $H_0=72.35^{+0.78}_{-0.79}{\rm \,km\,s^{-1}\,Mpc^{-1}}$ at 68\% CL for the {\it Planck} 2018 data, and $H_0=72.16\pm0.44{\rm \,km\,s^{-1}\,Mpc^{-1}}$ at 68\% CL for {\it Planck} 2018 + CMB lensing + BAO + Pantheon + DES + R19, confirming the agreement with R20 within one standard deviation. However, in Ref.~\cite{Rezaei:2020mrj} it has been argued that, while at the background level the flat-PEDE model fits the data as well as the $\Lambda$CDM scenario, at the perturbation level the PEDE model can not fit the observational data in cluster scales compared to the $\Lambda$CDM. Extensions of this model considering neutrinos or a non-zero curvature of the Universe can be found in Refs.~\cite{Yang:2021egn, Rezaei:2020mrj, Liu:2020vgn}.

\subsubsection{Generalized Emergent Dark Energy:}

A generalization of the PEDE model, including one more degree of freedom $\Delta$, known as Generalized Emergent Dark Energy (GEDE) can be found in Ref.~\cite{Li:2020ybr}. In the GEDE model, the evolution for the dark energy density is written as~\cite{Li:2020ybr}:
\begin{equation}
    \label{eq:odez}
    {\widetilde{\Omega}_{\rm{GEDE}}(z)} \equiv \frac{\rho_{\rm DE}}{\rho_{{\rm crit},0}} =\, \Omega_{\rm{GEDE,0}}\,\frac{ 1 - {\rm{tanh}}\left(\Delta \, {\rm{log}}_{10}(\frac{1+z}{1+z_t}) \right) }{{1+ {\rm{tanh}}\left(\Delta \, {\rm{log}}_{10}({1+z_t}) \right)}}\,,
\end{equation}
where the redshift $z_t$ marks the transition at which the densities in dark energy and matter equate, ${\widetilde{\Omega}_{\rm{GEDE}}(z_t)}\,=\,\Omega_m(1+z_t)^3$. The redshift $z_t$ is thus not a free parameter. For $\Delta\,=\,0$ this model recovers the $\Lambda$CDM scenario, while for $\Delta\,=\,1$ and $z_t=0$, the PEDE model is recovered. 
The GEDE equation of state is:
\begin{equation}
    w_{\rm GEDE} (z) = -1 -\frac{\Delta}{3 \ln10} \,\left({1+{\rm{tanh}}\left[\Delta \, {\rm{log}}_{10}\left(\frac{1+z}{1+z_t}\right) \right] }\right)\,. 
\end{equation}
The analysis of {\it Planck} 2018 data at the background level for this GEDE scenario provides  $H_0=66.76^{+2.9}_{-0.76}{\rm \,km\,s^{-1}\,Mpc^{-1}}$ at 68\% CL~\cite{Li:2020ybr}, reducing the Hubble tension down to the $2\sigma$ level. However, this result is mostly driven by a volume effect, due to the increased volume of the parameter space. This result is in agreement with the results of Ref.~\cite{Hernandez-Almada:2020uyr}, where CC data are considered. An updated result of this scenario is presented in Ref.~\cite{Yang:2021eud} considering also the effects of the dark energy perturbations where for {\it Planck} 2018, $H_0 = 85^{+12}_{-6}{\rm \,km\,s^{-1}\,Mpc^{-1}}$ at 68\% CL.

\subsubsection{Modified Emergent Dark Energy:}

Another generalization of the PEDE model which includes one additional degree of freedom $\alpha$ is the Modified Emergent Dark Energy (MEDE), in which the dark energy equation of state can be written as~\cite{Benaoum:2020qsi}:
\begin{equation}
    w_{\rm MEDE} (z) = -1 - \frac{\alpha}{3 \ln 10} \,\Bigl(1+ \tanh \bigl[\alpha \log_{10} \left(1+z \right) \bigr] \Bigr)\,.
\end{equation}
If $\alpha\,=\,0$, the model reduces to the $\Lambda$CDM scenario, while for $\alpha\,=\,1$ the PEDE model is recovered. A fit to {\it Planck} 2018 + BAO data within the MEDE model provides $H_0=68.4\pm 1.5{\rm \,km\,s^{-1}\,Mpc^{-1}}$ at 68\% CL and reduces the Hubble tension to $2.4\sigma$~\cite{Benaoum:2020qsi}.

\subsection{Vacuum Metamorphosis}

The Vacuum Metamorphosis (VM) model is a cosmological scenario which is physically motivated by quantum gravitational effects, where a gravitational phase transition occurs at late times~\cite{Parker:2003as, Parker:2000pr, Caldwell:2005xb}. The phase transition is induced when the Ricci scalar curvature $R$ is of the order of the mass squared of the field $m^2$, after which $R$ is frozen. The value of $m^2$ determines the matter density today $\Omega_m$, and therefore the VM model has the same number of free parameters than the flat $\Lambda$CDM scenario.

It has been found that this specific model can be an excellent candidate to solve the $H_0$ tension~\cite{DiValentino:2017rcr}. In more detail, the expansion rate above and below the phase transition reads as\cite{DiValentino:2017rcr}:

\begin{equation}
\frac{H^2}{H_0^2} =
\begin{cases}
\Omega_m (1+z)^3 \!+\! \Omega_r(1+z)^4 \!+\! M\left\{1 \!-\!\left[3\left(\frac{4}{3\Omega_m}\right)^4 M(1-M)^3\right]^{-1}\right\}, &\, \hbox{for $z > z_t^{\rm ph}$}\,;\\
(1-M)(1+z)^4+M, &\, \hbox{for $z\leq z_t^{\rm ph}$}\,,
\end{cases}
\end{equation}
where $M = m^2/(12H_0^2)$ and the phase transition occurs at the redshift
\begin{equation}
    z_t^{\rm ph} = -1 + \frac{3\Omega_m}{4(1-M)}\,.
\end{equation}
The effective DE equation of state is~\cite{DiValentino:2017rcr}: 
\begin{equation}
    w_{\rm VMDE}(z)=-1-\frac{1}{3}\frac{3\Omega_m (1+z)^{3}-4(1-M)(1+z)^{4}}{M+(1-M)(1+z)^{4}-\Omega_m (1+z)^{3}} \,, 
\end{equation}
below the phase transition, while $w_{\rm VMDE}(z)=-1$ above the phase transition.

For the VM scenario, {\it Planck} 2015 gives $H_0=78.61\pm 0.38{\rm \,km\,s^{-1}\,Mpc^{-1}}$ at 68\% CL~\cite{DiValentino:2017rcr}, reducing the tension at $3.9\sigma$. An updated analysis is performed in Ref.~\cite{DiValentino:2020kha}, where a fit to the {\it Planck} 2018 data gives $H_0=81.1\pm 2.1{\rm \,km\,s^{-1}\,Mpc^{-1}}$ at 68\% CL, reducing the tension at $3.1\sigma$, and an analysis to {\it Planck} 2018 + BAO + R19 gives $H_0=75.22\pm 0.60{\rm \,km\,s^{-1}\,Mpc^{-1}}$ at 68\% CL, in agreement at $1.4\sigma$ with R20. An extension of the model considering a non-zero curvature of the Universe can be found in Ref.~\cite{DiValentino:2020kha}. The extension considering the neutrino sector, instead, explored in Ref.~\cite{DiValentino:2021zxy}, provides, for {\it Planck} 2018, $H_0=78.0^{+3.8}_{-2.6} {\rm \,km\,s^{-1}\,Mpc^{-1}}$ at 68\% CL, alleviating the tension at $1.7\sigma$, and, for {\it Planck} 2018 + BAO + R19, $H_0=74.60\pm0.97 {\rm \,km\,s^{-1}\,Mpc^{-1}}$ at 68\% CL, in agreement with R20 within $1\sigma$.

\subsubsection{Elaborated Vacuum Metamorphosis:}

While in the original VM, $M$ is not a free parameter, but fixed by:
\begin{equation}
\Omega_m=\frac{4}{3}\left[3M(1-M)^3\right]^{1/4}~,
\end{equation}
a scenario where the model has one more free parameter $M$, can also be regarded as a possible cosmological scenario. For the VM scenario {\it Planck} 2015 gives $H_0=71.6^{+2.8}_{-5.1}{\rm\,km\,s^{-1}\,Mpc^{-1}}$ at 68\% CL~\cite{DiValentino:2017rcr}, solving the Hubble tension within $1\sigma$. This result is confirmed by the updated analysis performed in Ref.~\cite{DiValentino:2020kha}, where {\it Planck} 2018 gives $H_0=76.7^{+3.9}_{-2.6}{\rm\,km\,s^{-1}\,Mpc^{-1}}$ at 68\% CL, and {\it Planck} 2018 + BAO + R19 $H_0=73.63^{+0.33}_{-0.48}{\rm\,km\,s^{-1}\,Mpc^{-1}}$ at 68\% CL, in agreement within $1\sigma$ with R20. An extension of this model, considering a curvature component, can be found in Ref.~\cite{DiValentino:2020kha}. The extension considering the neutrino sector, instead, explored in Ref.~\cite{DiValentino:2021zxy}, provides, for {\it Planck} 2018, $H_0=75.2^{+1.6}_{-2.4} {\rm \,km\,s^{-1}\,Mpc^{-1}}$ at 68\% CL, and, for {\it Planck} 2018 + BAO + R19, $H_0=73.19\pm0.85 {\rm \,km\,s^{-1}\,Mpc^{-1}}$ at 68\% CL, both in agreement with R20 within $1\sigma$.

\begin{figure*}
\centering
\includegraphics[width=0.85\textwidth, right]{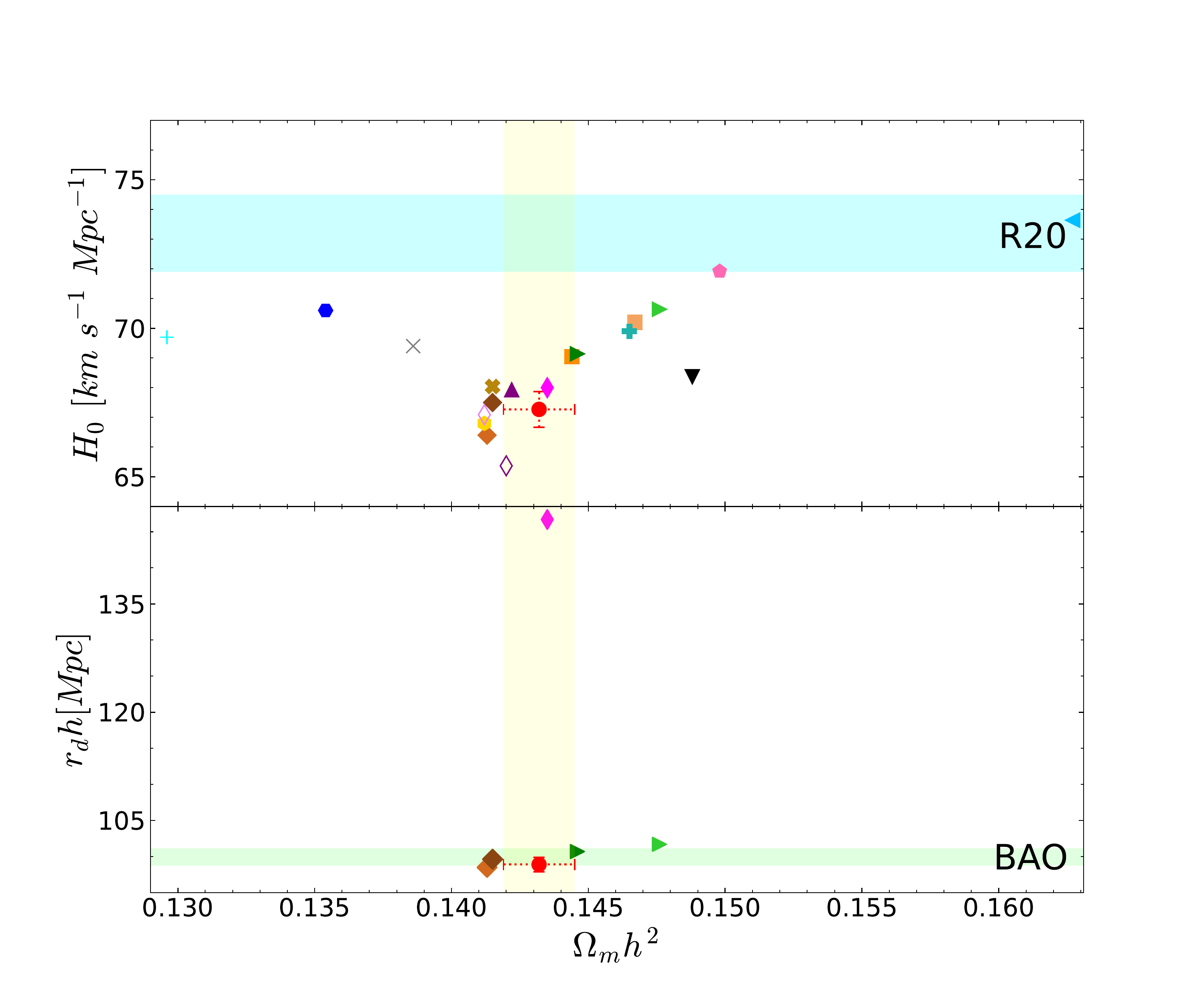}
\includegraphics[width=0.85\textwidth, right]{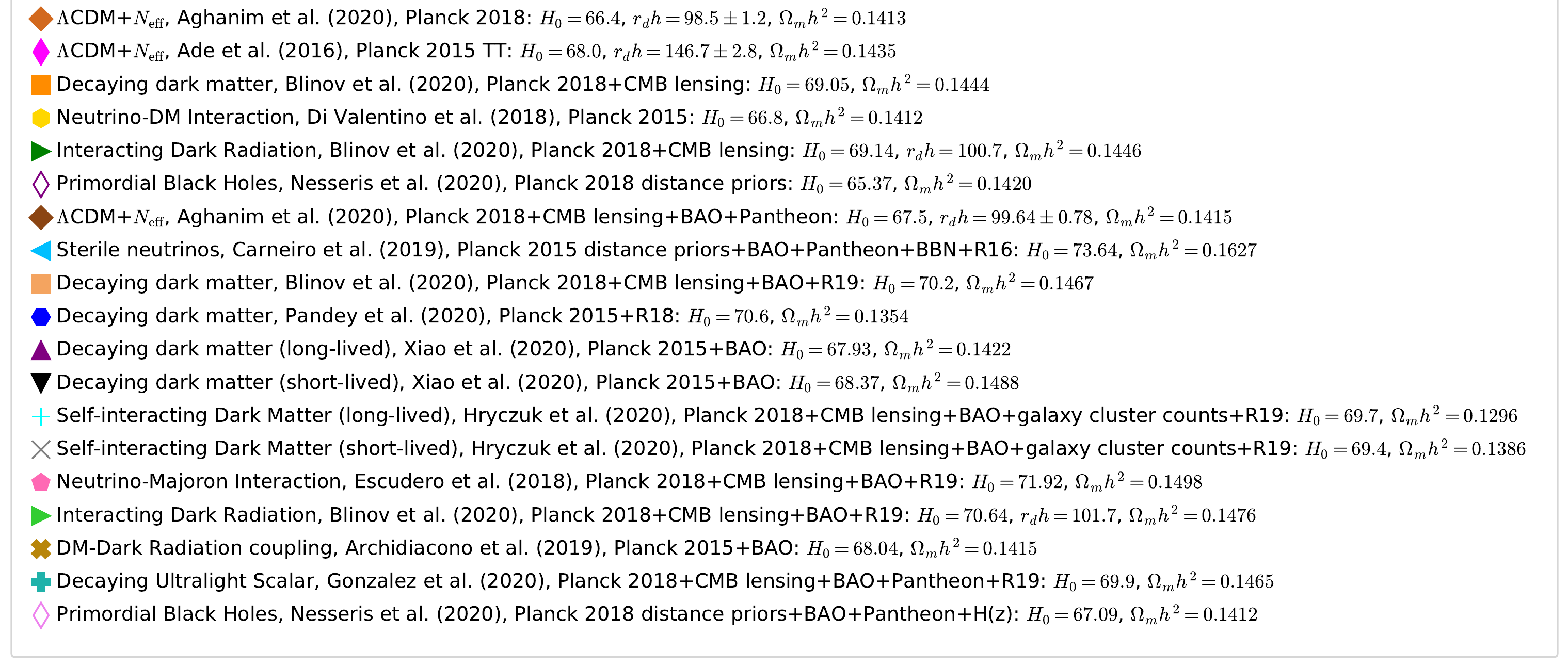}
\caption{Estimated values of the current matter energy density $\Omega_mh^2$, Hubble constant $H_0$ and sound horizon $r_dh$ in terms of various data points for different models discussed throughout the Section~\ref{DR}. The cyan horizontal band corresponds to the $H_0$ value measured by R20~\cite{Riess:2020fzl}, the yellow vertical band to the $\Omega_mh^2$ value estimated by {\it Planck} 2018~\cite{Aghanim:2018eyx} in a $\Lambda$CDM scenario, and the light green horizontal band to the $r_dh$ value measured by BAO data. The points sharing the same symbol refer to the same model in the same paper, and the different colors indicate a different dataset combination.}
\label{fig:chapter7_H0Om}
\end{figure*}
\begin{figure*}
\includegraphics[width=\textwidth]{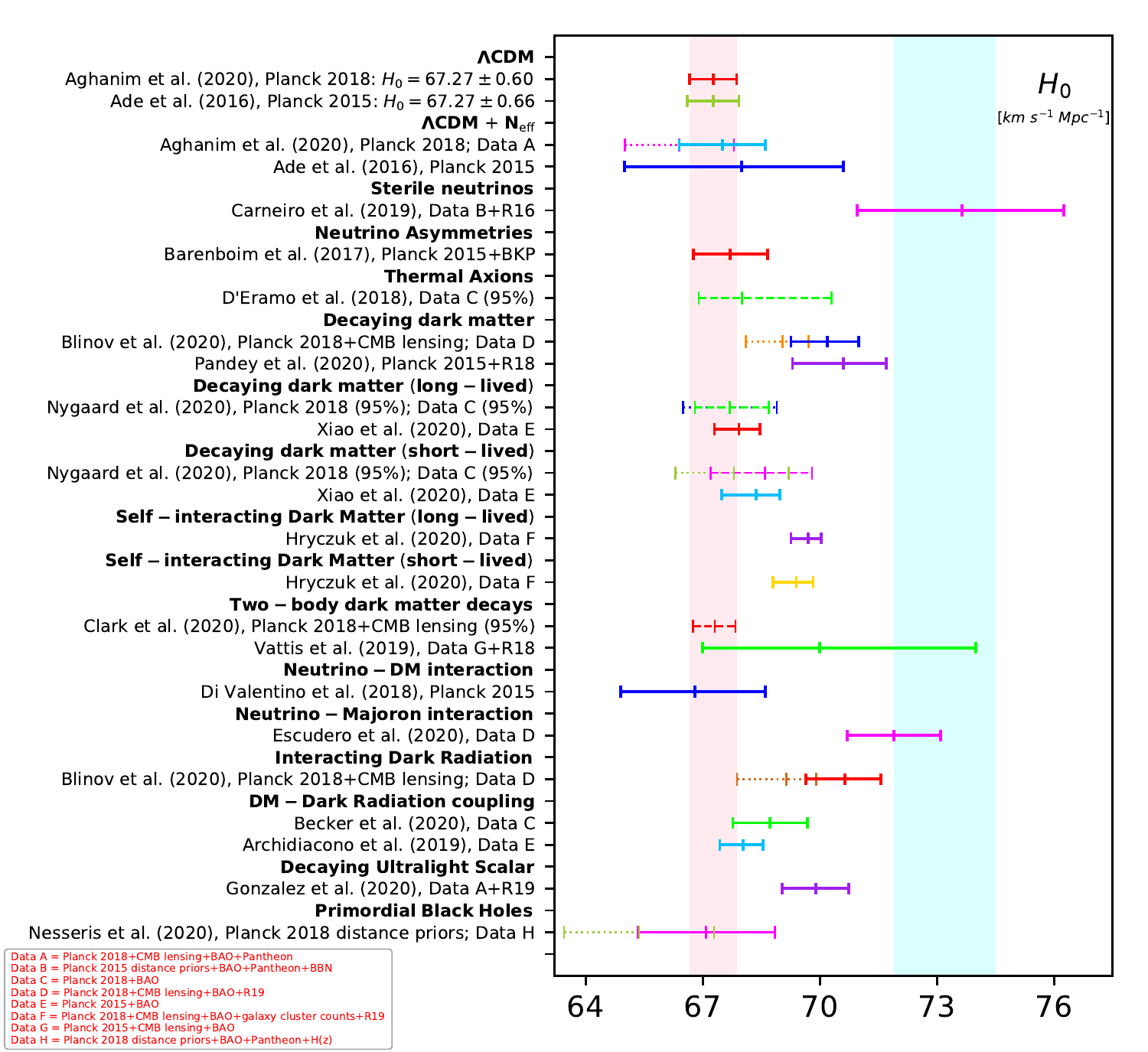}
\caption{Whisker plot with the 68\% (95\% if dashed) marginalized Hubble constant constraints for the models of Section~\ref{DR}. The cyan vertical band corresponds to the $H_0$ value measured by R20~\cite{Riess:2020fzl} and the light pink vertical band corresponds to the $H_0$ value estimated by {\it Planck} 2018~\cite{Aghanim:2018eyx} in a $\Lambda$CDM scenario. For each line, when more than one error bar is shown, the dotted one corresponds to the {\it Planck} only constraint on the Hubble constant, while the solid one to the different dataset combinations reported in the red legend, in order to appreciate the shift due to the additional datasets.}
\label{fig:chapter7_whisker}
\end{figure*}

\section{Models With Extra Relativistic Degrees of Freedom}
\label{DR}

One classical extension of the standard $\Lambda$CDM model considered for the $H_0$ tension resolution, is the possibility of having extra ``dark'' radiation at the recombination period, usually quantified by the number of relativistic degrees of freedom, $N_{\rm eff}$~\cite{Steigman:1977kc}.
The radiation density $\rho_r$ can be written as a function of the photon density $\rho_\gamma$, where we consider the ratio $T_\nu / T_\gamma=(4/11)^{1/3}$ between the background temperatures of neutrinos and photons under the approximation of instantaneous neutrino decoupling:
\begin{equation}
    \rho_r = \rho_\gamma\,\Bigg[1+\frac{7}{8}\Bigg(\frac{4}{11}\Bigg)^{4/3} N_{\rm eff}\Bigg]\,.
\end{equation}
For three active massless neutrino families we usually expect $N_{\rm eff}^{\rm SM} \simeq 3.046$~\cite{Mangano:2005cc,deSalas:2016ztq,Akita:2020szl}, albeit the latest calculations provide $N_{\rm eff}^{\rm SM} =3.0440 \pm 0.0002$~\cite{Froustey:2020mcq,Bennett:2020zkv}, where the uncertainty is due to errors associated to the numerical solution procedure, increased by the errors on the measurement of the solar mixing angle. Note, that additional relativistic degrees of freedom other than the three standard model neutrinos will produce more radiation. Its effect will be the smearing and shifting of the acoustic peaks in the damping tail of the CMB temperature power spectrum, and a delay in the matter to radiation equivalence~\cite{Hou:2011ec, Archidiacono:2013fha, Lattanzi:2017ubx, Vagnozzi:2017ovm}, with a corresponding increase of the early Integrated Sachs-Wolfe (eISW) effect, and therefore of the amplitude of the peak around multipoles $\ell\sim 200$. Because of the strong degeneracy between $N_{\rm eff}$ and the Hubble constant, it would be possible to have a larger value of $H_0$ from the CMB perspective at the price of additional radiation present at recombination (see for example, Ref.~\cite{Green:2019glg} and the sub-sections below). 

While $N_{\rm eff}$ was a possible way of solving the Hubble constant tension at $1.8\sigma$ with the {\it Planck} 2015 TT data ($H_0=68.0^{+2.6}_{-3.0}{\rm\,km\,s^{-1}\,Mpc^{-1}}$ at 68\% CL)~\cite{Ade:2015xua}, with the new {\it Planck} 2018 polarization measurements there is still a disagreement at about $3.6\sigma$ with R20. In particular, {\it Planck} 2018 ({\it Planck} 2018 + CMB lensing + BAO + Pantheon) provides the constraint $N_{\rm eff}=2.92\pm0.19$ at 68\% CL, and $H_0=66.4\pm 1.4{\rm\,km\,s^{-1}\,Mpc^{-1}}$ at 68\% CL~\cite{Aghanim:2018eyx} ($H_0=67.5\pm 1.1{\rm\,km\,s^{-1}\,Mpc^{-1}}$ at 68\% CL), while one would need $N_{\rm eff}\approx 3.95$ to obtain a value of $H_0$ from {\it Planck} 2018 + BAO + Pantheon in perfect agreement with R19~\cite{Vagnozzi:2019ezj}.\footnote{\, For a recent comparison of the abilities of dark radiation models versus dark energy modifications in solving the Hubble constant tension, see Ref.~\cite{Seto:2021xua}.}

To understand the ability of the models of this section in alleviating the Hubble tension, in Figures~\ref{fig:chapter7_H0Om} and~\ref{fig:chapter7_whisker} we have classified them in terms of various key cosmological parameters, as explained in the Introduction.

\subsection{Sterile neutrinos}

A possibility for having extra relativistic degrees of freedom at recombination is the presence of additional sterile neutrinos~\cite{Dodelson:1993je, Abazajian:2012ys, Anchordoqui:2012qu, Jacques:2013xr}, since they are not forbidden by any fundamental symmetry within the Standard Model of elementary particles. They can contribute to $N_{\rm eff}$ in a fractional way, if not fully thermalized, and can still be in agreement with the CMB data, while a thermalized sterile neutrino with $\Delta N_{\rm eff} = N_{\rm eff} - N_{\rm eff}^{\rm SM}=1$ is ruled out at about $6\,\sigma$. 

Light sterile neutrinos are strongly motivated by neutrino short baseline (SBL) oscillation anomalies. In fact, there is a $6.1\,\sigma$ indication for an electron-neutrino appearance, when combining together the MiniBooNE~\cite{Aguilar-Arevalo:2018gpe} and the Liquid Scintillator Neutrino Detector (LSND)~\cite{Aguilar:2001ty} data, even if this result has been challenged by STEREO~\cite{Almazan:2018wln} and PROSPECT~\cite{Ashenfelter:2018iov}. These anomalous datasets can be explained with one sterile neutrino~\cite{Hannestad:2013pha, Battye:2013xqa, DiValentino:2013qma, Bridle:2016isd, Knee:2018rvj} with $\Delta N_{\rm eff} \approx 1$, in strong contradiction with cosmological constraints. A possibility is the presence of some non-standard interactions~\cite{Archidiacono:2016kkh, Chu:2015ipa, Paul:2018njm}, low-temperature reheating~\cite{deSalas:2015glj}, or other special mechanisms. Nevertheless, in Ref.~\cite{Carneiro:2018xwq}, the authors showed that the combination of the {\it Planck} 2015 CMB distance priors + BAO + Pantheon + BBN + R16 gives $N_{\rm eff} \approx 4$ and $H_0=73.64^{+2.61}_{-2.68}{\rm\,km\,s^{-1}\,Mpc^{-1}}$ at 68\% CL, perfectly in agreement with R20. A full {\it Planck} 2018 analysis is however in contradiction with this result. 

In Ref.~\cite{Gelmini:2019deq}, instead, the authors showed that additional radiation produced just before the BBN by an unstable sterile neutrino with a mass of the order of tens of MeV can alleviate the Hubble tension by increasing $N_{\rm eff}$ in the right amount. Also, in Ref.~\cite{Gelmini:2020ekg}, it is shown that, in the presence of a large lepton asymmetry, sterile neutrinos with the masses in the $(150 - 450)\,$MeV range can increase $N_{\rm eff}$ by $0.2-0.4$ and reduce the Hubble tension.

\subsection{Neutrino Asymmetries}

A large lepton number asymmetry $\xi$ in one or more neutrino species contributes to $\Delta N_{\rm eff}$ in the following way, accounting for the thermal distributions of two light mass eigenstates~\cite{Schwarz:2012yw,DiValentino:2011sv,Stuke:2011wz,Schwarz:2009ii}:
\begin{equation}
\Delta N_{\rm eff} 
= \frac{15}{7} \sum_{i=1,2} \left( \frac{\xi_i}{\pi} \right)^2 \left[ 2 + \left( \frac{\xi_i}{\pi} \right)^2 \right]\,.
\end{equation}
Given the role of the extra dark radiation in solving the Hubble tension, in Ref.~\cite{Barenboim:2016lxv} it has been investigated the possibility that a primordial lepton asymmetry can alleviate the tension between the CMB and R16. The combination of {\it Planck} 2015 + BICEP2 \& Keck Array (BKP)~\cite{Ade:2015fwj} results in $H_0=67.71\pm 0.95{\rm \,km\,s^{-1}\,Mpc^{-1}}$ at 68\% CL~\cite{Barenboim:2016lxv}, showing still a disagreement at $3.4\sigma$ with R20.

\subsection{Thermal Axions}
\label{sec:QCDaxion}

The QCD axion~\cite{Weinberg:1977ma, Wilczek:1977pj} is a hypothetical particle that emerges from the Peccei-Quinn solution to the strong CP problem~\cite{Peccei:1977hh, Peccei:1977ur} (see Ref.~\cite{DiLuzio:2020wdo} for a review). These particles may be copiously produced in the early Universe through either non-thermal mechanisms comprising CDM or thermal mechanisms which lead to a population of relativistic axions that would contribute to $N_{\rm eff}$~\cite{Green:2019glg, Melchiorri:2007cd, Hannestad:2007dd, Hannestad:2008js, Hannestad:2010yi, Archidiacono:2013cha, Giusarma:2014zza, DiValentino:2015wba, Ferreira:2018vjj, Giare:2020vzo}.

The production of thermal axions might proceed through various mechanisms. Since the QCD axion couples to the gluon through a model-independent interaction, thermal axions are produced via scatterings off gluons for the decoupling temperature $T_D \gtrsim 1\,$GeV~\cite{Masso:2002np}, and through the scattering off pions and nucleons at lower temperatures, regardless of the QCD axion model (see Ref.~\cite{DiLuzio:2021vjd} for recent updates). Since the coupling of the QCD axion with other SM particles is model-dependent, other production mechanisms via scattering off photons~\cite{Turner:1986tb} and SM leptons~\cite{DEramo:2018vss} or quarks~\cite{Salvio:2013iaa} might also arise. The energy density of radiation in the late Universe leads to a deviation in the effective number of relativistic degrees of freedom given by:
\begin{equation}
    \Delta N_{\rm eff}=\frac{8}{7}\left(\frac{11}{4}\right)^\frac{4}{3}\frac{\rho_a}{\rho_\gamma}\,,
\end{equation}
where the axion energy density is found in terms of the current axion number density $n_a$ and the internal degrees of freedom of the relativistic gas of bosons $g$, and it reads as:
\begin{equation}
    \rho_a = \frac{\pi^2}{30}\left(\frac{\pi^2n_a}{\zeta(3)g}\right)^\frac{4}{3}\,.
\end{equation}

The assessment of a model in which hot axions are produced from the coupling with muons leads to an alleviation of the Hubble tension at $3\sigma$ level~\cite{DEramo:2018vss}. In this $\Lambda$CDM+$N_{\rm eff}$ scenario, the fit to {\it Planck} 2018 + BAO data results in a value for the Hubble constant corresponding to $H_0=68.0^{+2.3}_{-1.1}{\rm \,km\,s^{-1}\,Mpc^{-1}}$ at 95\% CL.

\subsection{Decaying dark matter}

Cosmological scenarios in which it is present a decaying dark matter component~\cite{Ichiki:2004vi} could be alternative proposals to reduce the Hubble constant tension. For instance, a dark matter fluid could decay into invisible massless particles after recombination while still avoiding photon overproduction~\cite{Berezhiani:2015yta}. This model has been explored in Refs.~\cite{Anchordoqui:2015lqa, Chudaykin:2016yfk, Chudaykin:2017ptd} making use of different combinations of the {\it Planck} 2015 data with other cosmological probes. In Ref.~\cite{Anchordoqui:2020djl}, a model of cold dark matter decaying before recombination is inspected and it is shown to not to be a satisfactory solution to the Hubble tension. Nevertheless, the best case scenario to alleviate the Hubble constant occurs when the dark matter particles decay exclusively into dark radiation.

A cosmological model where a fraction of the dark matter density decays into dark radiation increasing $\Delta N_{\rm eff}$, as proposed by Ref.~\cite{Bjaelde:2012wi}, has been considered as a solution for the Hubble tension by many authors. Such a decaying scenario in terms of the background equations reads as:
\begin{eqnarray}
\dot\rho_{\rm DM} + 3 {\cal H} \rho_{\rm DM} &=& -Q\,; \\
\dot\rho_{\rm DR} + 4 {\cal H} \rho_{\rm DR} &=& Q\,,
\end{eqnarray}
where here and in the following, a dot stands for a differentiation with respect to conformal time, and the source term $Q= \Gamma \rho_{\rm DM}$, depends on a constant decay rate $\Gamma$ (see Refs.~\cite{Poulin:2016nat,Enqvist:2015ara,Bringmann:2018jpr} for different functional forms of $\Gamma$). Within this scenario, the authors of Ref.~\cite{Pandey:2019plg} have found from the analysis of {\it Planck} 2015 + R18 a value for the Hubble constant $H_0=70.6^{+1.1}_{-1.3}{\rm \,km\,s^{-1}\,Mpc^{-1}}$ at 68\% CL, reducing the Hubble tension down to the $1.5\sigma$ level. Notice however that this result may be biased due to the fact that it already includes a R18 prior on the Hubble constant.

If we consider two different regimes, i.e.\ one for long-lived decaying cold dark matter (with a lifetime longer than the epoch corresponding to recombination) and another one for short-lived decaying cold dark matter particles, for which the mass-energy density decreases significantly well before recombination, the latter will leave a strong imprint on the CMB. In Ref.~\cite{Xiao:2019ccl}, {\it Planck} 2015 + BAO gives $H_0=67.93^{+0.53}_{-0.63}{\rm \,km\,s^{-1}\,Mpc^{-1}}$ at 68\% CL for long-lived decaying cold dark matter, in disagreement with R20 at more than $3.8\sigma$, while $H_0=68.37^{+0.61}_{-0.89}{\rm \,km\,s^{-1}\,Mpc^{-1}}$ at 68\% CL for short-lived decaying cold dark matter, in disagreement with R20 at $3.5\sigma$. An update is presented in Ref.~\cite{Nygaard:2020sow}, where {\it Planck} 2018 provides $H_0=67.7\pm 1.2{\rm \,km\,s^{-1}\,Mpc^{-1}}$ ($H_0=67.8^{+1.4}_{-1.5}{\rm \,km\,s^{-1}\,Mpc^{-1}}$) at 95\% CL for long (short)-lived decaying cold dark matter particles, in disagreement with R20 at more than $3.7\sigma$ ($3.6\sigma$), and {\it Planck} 2018 + BAO instead lead to $H_0=67.7^{+1.0}_{-0.9}{\rm \,km\,s^{-1}\,Mpc^{-1}}$ ($H_0=68.6^{+1.2}_{-1.4}{\rm \,km\,s^{-1}\,Mpc^{-1}}$) at 95\% CL, in disagreement with R20 at more than $3.9\sigma$ ($3.2\sigma$). If, instead, dark matter is composed of decaying warm dark matter particles, it has been shown in Ref.~\cite{Blinov:2020uvz} that for a dark matter particle mass $m=40\,$eV, {\it Planck} 2018 + CMB lensing ({\it Planck} 2018 + CMB lensing + BAO + R19) provides the constraint $H_0=69.05^{+0.66}_{-0.95}{\rm \,km\,s^{-1}\,Mpc^{-1}}$ ($H_0=70.20^{+0.79}_{-0.94}{\rm \,km\,s^{-1}\,Mpc^{-1}}$) at 68\% CL, reducing the Hubble tension down to the $2.8\sigma$ ($2\sigma$) significance level.

\subsubsection{Self-interacting Dark Matter:}

Another possibility proposed for solving the $H_0$ tension is to consider a self-interacting dark matter sector (SIDM)~\cite{Tulin:2017ara,Spergel:1999mh,Buckley:2009in} with a light force mediator coupled to dark radiation. In this way, there will be a second epoch of hidden dark matter annihilation into dark radiation long after the standard thermal freeze-out, affecting the visible sectors only gravitationally. In Ref.~\cite{Binder:2017lkj}, this scenario is proposed to alleviate the Hubble tension, without performing a data analysis.

Such an analysis has been carried out in Ref.~\cite{Hryczuk:2020jhi}, where a model with self-interacting dark matter particles exchanging a light mediator, produced by the decay of a messenger WIMP-like state, is considered. From the cosmological perspective, this paradigm is very similar to a decaying cold dark matter one and therefore the analysis is applied to the latter. The combination of {\it Planck} 2018 + BAO + R19 + {\it Planck} galaxy cluster counts measurements gives 
$H_0=69.4^{+0.43}_{-0.60}{\rm \,km\,s^{-1}\,Mpc^{-1}}$ ($H_0=69.7^{+0.33}_{-0.44}{\rm \,km\,s^{-1}\,Mpc^{-1}}$) at 68\% CL~\cite{Hryczuk:2020jhi} for the short-(long) lived case, reducing the disagreement with R20 at $2.7\sigma$, even if a Gaussian prior on $H_0$ is included in the data analyses.

\subsubsection{Two-body dark matter decays:}

Different from the two previous cases is the model presented in Ref.~\cite{Blackadder:2014wpa} and analysed in Ref.~\cite{Vattis:2019efj}, where a parent particle decays into two daughter particles, one massless and one massive, with the form $\psi \rightarrow \chi + \gamma$. This decay is well-known within the context of Super Weakly Interacting Massive particles (Super WIMPs)~\cite{Feng:2003xh}. This decaying dark matter model has therefore two free parameters: the fraction $\epsilon$ of the energy of the parent particle which is transferred to the massless particle, and the lifetime $\tau= 1/ \Gamma$, where $\Gamma$ is the decay rate. Assuming there are no decays prior the recombination period, the energy densities of the parent particle $\rho_0$ and of the massless daughter particle $\rho_1$ evolve as:
\begin{eqnarray}
    \frac{\mathrm{d} \rho_0}{\mathrm{d}t} + 3 H \rho_0 &=& - \Gamma \rho_0\,;\\
    \frac{\mathrm{d} \rho_1}{\mathrm{d}t} + 4 H \rho_1 &=& \epsilon \, \Gamma \rho_0\,;\\
    3H^2(a) &=& \kappa^2 \sum_i \rho_i(a)\,,
\end{eqnarray}
where, referring to the massive daughter particle with the subscript $2$, the total energy density is:
\begin{equation} 
    \sum_i \rho_i(a) = \rho_0(a) + \rho_1(a) + \rho_2(a) + \rho_r(a) + \rho_\nu(a) + \rho_b(a) + \rho_\Lambda\,.
\end{equation}

In this scenario, an analysis with {\it Planck} 2015 + CMB lensing + R18 + BAO provides $H_0=70^{+4}_{-3}{\rm \,km\,s^{-1}\,Mpc^{-1}}$ at 68\% CL, in agreement with R20 within $1\sigma$~\cite{Vattis:2019efj}. However, since both the R18 prior on the Hubble constant and the BAO data are present in the joint analysis, so it is difficult to assess how well the model can solve the Hubble tension for the CMB dataset alone.\footnote{\, Another analysis of the model, excluding the CMB measurements is performed in Ref.~\cite{Haridasu:2020xaa}, where Pantheon + H0LiCOW provides $H_0=72.1^{+1.6}_{-1.7}{\rm \,km\,s^{-1}\,Mpc^{-1}}$ at 68\% CL.} An updated CMB result is nevertheless present in Ref.~\cite{Clark:2020miy}, where taking into account the late-Universe decaying dark matter effects like ISW and lensing, an analysis with {\it Planck} 2018 + CMB lensing data results in $H_0=67.31^{+0.53}_{-0.56}{\rm \,km\,s^{-1}\,Mpc^{-1}}$ at 95\% CL, consistent with the $\Lambda$CDM value and in disagreement with R20 at $4.2\sigma$, excluding at a high significance the preferred region of the earlier analysis carried out in Ref.~\cite{Vattis:2019efj}.

\subsubsection{Light Gravitino scenarios:}

In Ref.~\cite{Gu:2020ozv} the authors study the keV gravitino dark matter model arguing that this could reduce the Hubble tension at around the $3\sigma$ level. The bino, the superpartner of the $U(1)$ weak hypercharge gauge field, can have a late decay into a gravitino nearly relativistic in the early Universe, increasing the radiation density by:
\begin{equation}
    \rho^{\rm extra}_R=f \times \rho_{3/2} \times (\gamma_{3/2}-1)\,,
\end{equation}
where $\gamma_{3/2}$ will be the boost factor of the gravitino from the bino decay, and $f$ will be the fraction of the non-thermal gravitino density in the total gravitino production. Therefore, the gravitino contributes to the effective number of relativistic degrees of freedom $\Delta N_{\mathrm{eff}}$ as~\cite{Hooper:2011aj}:
\begin{equation}
\Delta N_{\mathrm{eff}}=f \times\left(\gamma_{3/2}-1\right) / 0.16 \,~.
\end{equation}
A similar contribution to $\Delta N_{\mathrm{eff}}$ is expected from a light dark matter candidate suggested in Ref.~\cite{Alcaniz:2019kah} to solve both the Lithium problem and reconcile the Hubble tension. Another particle physics model to address the Hubble tension is presented in Ref.~\cite{Choi:2019jck}, where the authors consider a gravitino as decaying dark matter and a quintessence dark energy axion, nevertheless a full {\it Planck} 2018 analysis is however still missing.

\subsubsection{Decaying $Z^\prime$:}

The authors of Ref.~\cite{Escudero:2019gzq} studied the cosmological implications of a $L_\mu - L_\tau$ gauge boson.\footnote{\, See also Ref.~\cite{Anchordoqui:2020znj} for the implications in light of the Hubble tension of the interplay between the cosmological determination of $\Delta N_{\rm eff}$ and $Z^\prime$.} They consider the evolution of a light and weakly coupled $Z^\prime$ and its contribution to $\Delta N_{\rm eff}$. There are two qualitatively distinct scenarios:

\begin{itemize}
\item{\it Early Universe Equilibrium:} The $Z^\prime$ population thermalizes at early times and decays into neutrinos, leading to the effective number of neutrino species:
\begin{equation}
      N_{\rm eff} = \frac{8}{7} \left( \frac{11}{4} \right)^{4/3} \frac{ \rho_{\nu} }{ \rho_\gamma}\, ,
\end{equation}
where $\rho_\nu$ is modified by the entropy transferred from $Z^\prime$ decays.

\item{\it Late Equilibration:} The $Z^\prime$ population will be produced through the freeze-in mechanism and eventually thermalizes with neutrinos, increasing $\Delta N_{\rm eff} \simeq 0.21$ through $Z^\prime \to \bar \nu \nu$ decays.
\end{itemize}
Nevertheless, complete {\it Planck} 2018 analyses for these models are missing in the literature and therefore is not possible to quantify their effectiveness in alleviating the Hubble constant tension.

\subsubsection{Dynamical Dark Matter:}

The authors in Ref.~\cite{Dienes:2011ja} discuss a scenario in which the observed DM comprises a vast array of interacting fields, each with different values of their masses, couplings, and abundances. Within such a ``dynamical'' dark matter model, a generalization of the decaying dark matter scenario, it is possible to address the Hubble tension issue, providing a self-sustaining framework to unify short-lived and long-lived decaying dark matter models~\cite{Desai:2019pvs}.

\subsubsection{Degenerate Decaying Fermion Dark Matter:}

A sub-keV decaying fermion as a dark matter candidate has been proposed in Ref.~\cite{Choi:2020tqp}. Such a scenario could address both the Hubble tension issue and the core-cusp problem. Despite that the theoretical framework seems appealing, the strength of the method can only be quantitatively evaluated once a full {\it Planck} 2018 analysis is performed in this cosmological context.

\subsection{Neutrino-Dark Matter Interactions}
\label{sec:dmnu}

Neutrinos interacting with dark matter have been investigated in the literature~\cite{Mangano:2006mp, Serra:2009uu, Diamanti:2012tg, Wilkinson:2014ksa, Campo:2018dfh, Campo:2017nwh, Pandey:2018wvh, Choi:2019ixb, Stadler:2019dii, Mosbech:2020ahp}, because dark matter annihilations into neutrinos can mimic an increase in the value of the dark radiation $N_{\rm eff}$~\cite{Boehm:2012gr, Boehm:2013jpa}, and therefore solve the Hubble tension. In Ref.~\cite{DiValentino:2017oaw}, it has been shown that varying $N_{\rm eff}$, {\it Planck} 2015 gives $H_0=66.8^{+1.8}_{-1.9}{\rm\,km\,s^{-1}\,Mpc^{-1}}$ at 68\% CL, reducing the tension with R20 down to the $2.9\sigma$ level.

\subsubsection{Neutrino-Majoron Interactions:}

The interaction between Majorons and neutrinos has also been proposed to alleviate the Hubble tension.
The massive Majoron is a pseudo-Goldstone boson arising from the spontaneous breaking of global lepton number~\cite{Chikashige:1980ui}, that can thermalize with neutrinos after BBN via inverse neutrino decays~\cite{Chacko:2003dt}, increasing $\Delta N_{\rm eff}$. For this scenario, {\it Planck} 2018 + CMB lensing + BAO + R19 gives $H_0=71.92\pm1.2{\rm\,km\,s^{-1}\,Mpc^{-1}}$ at 68\% CL~\cite{Escudero:2019gvw}, reducing the tension with R20 within $1\sigma$, but this result includes already a prior on the Hubble constant. Unfortunately the result for {\it Planck} 2018 alone is absent in the literature. This very same scenario is analysed also in Refs.~\cite{Arias-Aragon:2020qip,Huang:2021dba,Escudero:2021rfi}.

\subsubsection{FIMPs Decay into Neutrinos:}

Another model proposed to solve the Hubble constant tension has been explored in Ref.~\cite{Boyarsky:2021yoh}, where it is shown that feebly interacting massive particles~\cite{Bernal:2017kxu,Lanfranchi:2020crw,Agrawal:2021dbo} can affect $N_{\rm eff}$. In particular the authors focus on Heavy Neutral Leptons, that in the pure mixing cases can give $\Delta N_{\rm eff}=0.4$ and alleviate the Hubble tension. Unfortunately a data analysis for this model is missing.

\subsection{Interacting Dark Radiation} 

An interacting dark radiation component increasing $\Delta N_{\rm eff}$ has been proposed to alleviate the Hubble tension in Ref.~\cite{Blinov:2020hmc}. In this model, the energy density in relativistic particles is~\cite{Brust:2017nmv}:
\begin{equation}
    \rho_r = \rho_\gamma\,\Bigg[1+\frac{7}{8}\Bigg(\frac{4}{11}\Bigg)^{4/3} \Bigg(N_{\rm eff}+N_{\rm fl}\Bigg)\Bigg]\,,
\end{equation}
where $N_{\rm fl}$ is the interacting counterpart of the dark radiation component. If $N_{\rm eff}=3.046$ and $N_{\rm fl}$ is free to vary, a fit to {\it Planck} 2018 + CMB lensing ({\it Planck} 2018 + CMB lensing + BAO + R19) provides $H_0=69.14^{+0.77}_{-1.26}{\rm\,km\,s^{-1}\,Mpc^{-1}}$ ($H_0=70.64^{+0.93}_{-1.00}{\rm\,km\,s^{-1}\,Mpc^{-1}}$) at 68\% CL~\cite{Blinov:2020hmc}, alleviating the tension with R20 down to the $2.7\sigma$ ($1.6\sigma$) level.

\subsection{Coupled DM - Dark Radiation scenarios}

The possibility of a dark matter interacting with massless relics from the dark sector, i.e.\ dark radiation~\cite{Cyr-Racine:2013fsa, Buen-Abad:2015ova, Lesgourgues:2015wza, Schewtschenko:2015rno, Chacko:2016kgg, Krall:2017xcw, Archidiacono:2017slj, Buen-Abad:2017gxg, Choi:2020pyy, Ko:2016uft, Ko:2016fcd, Tang:2016mot, Ko:2017uyb}, has been proposed to solve the Hubble tension. One example is given by the ETHOS formalism~\cite{Cyr-Racine:2015ihg}, in which it is assumed that a single dark matter species interacts with a relativistic component via 2-to-2 scattering processes of the form DM + DR $\leftrightarrow$ DM + DR, with a comoving interaction rate that depends on temperature as $\Gamma_{\rm DR-DM} \propto T^n$. For the case $n=0$, corresponding to a class of non-Abelian dark matter models, the analysis of {\it Planck} 2015 + BAO datasets gives $H_0=68.04^{+0.50}_{-0.60}{\rm\,km\,s^{-1}\,Mpc^{-1}}$ at 68\% CL~\cite{Archidiacono:2019wdp}, in disagreement with R20 at $3.7\sigma$. An update of this work is presented in Ref.~\cite{Becker:2020hzj}, where {\it Planck} 2018 + BAO results in $H_0=68.73\pm0.96{\rm\,km\,s^{-1}\,Mpc^{-1}}$ at 68\% CL, reducing the Hubble tension with R20 at $2.8\sigma$.

\subsection{Cannibal Dark Matter}

A scattering process that allows three particles to annihilate into two has been proposed as a possible dark matter scenario in Ref.~\cite{Carlson:1992fn}. Dark matter ``cannibalism'' is a generic feature arising in any hidden sector in which a mass gap exists between two species.

This model has been considered for solving the Hubble tension, because it can increase the dark radiation component in the Universe~\cite{Buen-Abad:2018mas}. The thermal history can be divided into three different phases:
\begin{itemize}
    \item The cannibal dark matter paradigm, made of a real scalar field $\phi$, behaves as a radiation fluid, indistinguishable from an extra contribution to $N_{\rm eff}$, while its temperature is above the $\phi$ mass.
    \item A cannibalistic phase happens at a given scale factor, when the $\phi$-fluid cools below the mass of the particles: the interaction $3\to2$ starts processing mass into temperature and the temperature drops logarithmically.
    \item The $3\to2$ interactions decouple and the temperature drops as in the ordinary non-relativistic matter case.
\end{itemize}
Due to its strongly exothermic nature, cannibal dark matter acts as a warm dark matter component for a long period and turns non-relativistic at later times than cold dark matter. In the simplest scenario, this strongly suppresses structure growth, so cannibal dark matter can not be all of the dark matter~\cite{Machacek:1994vg}.
In the analysis of Ref.~\cite{Buen-Abad:2018mas}, only $\sim 1\%$ of the dark matter is considered as cannibalistic. As in many of the previous scenarios, a quantitative assessment of the ability of this model to solve the Hubble constant tension can not be performed, since a full {\it Planck} 2018 analysis for these models is absent in the literature.

\subsection{Decaying Ultralight Scalar}

A class of models that takes advantage of both the $\Delta N_{\rm eff}$ and the EDE models, improving on their downsides, has been proposed in Ref.~\cite{Gonzalez:2020fdy}. The authors study the Decaying Ultralight Scalar (dULS) model, which does not suffer from the EDE fine-tuning. In this model, the dark sector contains an ultralight scalar field $\phi$ of mass $m \gtrsim H_{\rm eq} \sim 10^{-28}\,$eV, where $H_{\rm eq}$ is the Hubble rate at matter-radiation equality, which resonantly decays into an Abelian gauge field $A_\mu$ when the axion field starts to oscillate at $m \sim H$. Mimicking the perturbative preheating stage after inflation, see e.g.\ Refs.~\cite{Greene:1997fu, Freese:2017ace}, an effective description of the model in terms of coupled fluid equations is:
\begin{eqnarray}
    \frac{\mathrm{d}\rho_\phi}{\mathrm{d}t} + 3 H(1+w_\phi)\rho_\phi &=& -\Gamma(t)\rho_\phi\,;\\
    \frac{\mathrm{d}\rho_A}{\mathrm{d}t} + 4 H\rho_A &=& \Gamma(t)\rho_\phi\,,
\end{eqnarray}
where $\Gamma(t)$ is a time-dependent decay rate and the equation of state for the ultralight scalar field transitions from $w_\phi = -1$ to $w_\phi = 0$ at $m \sim H$. A combined analysis to {\it Planck} 2018 + CMB lensing + BAO + Pantheon + R19 data gives $H_0=69.9^{+0.84}_{-0.86}{\rm\,km\,s^{-1}\,Mpc^{-1}}$ at 68\% CL~\cite{Gonzalez:2020fdy}, reducing the $H_0$ tension at $2.2$ standard deviations. However, a Gaussian prior for the Hubble constant coming from the Sh0ES measurement is also included in the analysis.

\subsection{Ultralight dark photon}

In Ref.~\cite{Flambaum:2019cih} the possibility that an extra radiation density could be due to extra vector fields, different from the visible photon field, has been considered for solving the Hubble tension. These vector fields, called dark photon fields, must interact very weakly with the visible matter and should have a small mass (see Refs.~\cite{Holdom:1985ag,Galison:1983pa}), contributing as dark matter. In Ref.~\cite{Anchordoqui:2019yzc} this model has been studied considering the BBN observations concluding that even if the addition of three dark massive vector fields can help to soften the Hubble tension, the mechanism can not resolve it completely.

\subsection{Primordial Black Holes}

Recently a strong interest for primordial black holes (PBHs) as a possible dark matter component of our Universe (see e.g.\ Refs.~\cite{Nesseris:2019fwr, Green:2020jor, Carr:2020xqk, Carr:2020erq}) has developed. 
In particular, in Ref.~\cite{Nesseris:2019fwr} the implications of the Hawking evaporation of light PBHs have been studied, showing that those can affect either $N_{\rm eff}$ or $w_{DE}$, depending on their precise mass, and potentially alleviate the Hubble tension. The authors of Ref.~\cite{Nesseris:2019fwr} find that {\it Planck} 2018 CMB distance priors ({\it Planck} 2018 CMB distance priors + BAO + Pantheon + H(z)) give for this model $H_0=65.37\pm1.92{\rm\,km\,s^{-1}\,Mpc^{-1}}$ ($H_0=67.09\pm1.76{\rm\,km\,s^{-1}\,Mpc^{-1}}$) at 68\% CL, reducing the tension at $3.4\sigma$ ($2.8\sigma$). However, they also speculate that an ultra-light PBH, decaying around the neutrino decoupling period, could raise $H_0=70.49\pm1.34{\rm\,km\,s^{-1}\,Mpc^{-1}}$ at 68\% CL for {\it Planck} 2018 CMB shift, solving the Hubble tension with R20 at $1.4\sigma$.

In Ref.~\cite{Flores:2020drq} a new scenario for the formation of primordial black holes within the dark sector is proposed. This scenario predicts a modification to $N_{\rm eff}$ by:
\begin{equation}
    \Delta N_{\rm eff} = 14\left(\frac{g_1}{g_*(T_D)}\right)^{4/3}\,,
\end{equation}
i.e.\ by about $0.1-0.2$, which could potentially alleviate the Hubble tension. Unfortunately, a full data analysis for this model is missing and therefore it is not possible to fully quantify its effectiveness in solving the $H_0$ tension.

\subsection{Unparticles}

The physics beyond the SM could contain a sector that is conformally invariant in the infra-red region and classically scale-invariant in the ultra-violet limit~\cite{vanderBij:2006pg, Georgi:2007ek}, referred to as the ``unparticle'' and the Banks-Zaks phases~\cite{Banks:1981nn}. Unparticles behave like radiation at high energies, increasing $\Delta N_{\rm eff}$ and therefore reducing the Hubble tension due to their correlation~\cite{Artymowski:2020zwy}. In addition, unparticles may act as a cosmological constant at low energies mimicking the standard $\Lambda$CDM model. A full data analysis for this model is however required to quantitatively assess its ability to resolve the Hubble tension.

\section{Models With Extra Interactions}
\label{InteractSolut}

\begin{figure*}
\centering 
\includegraphics[width=0.75\textwidth, right]{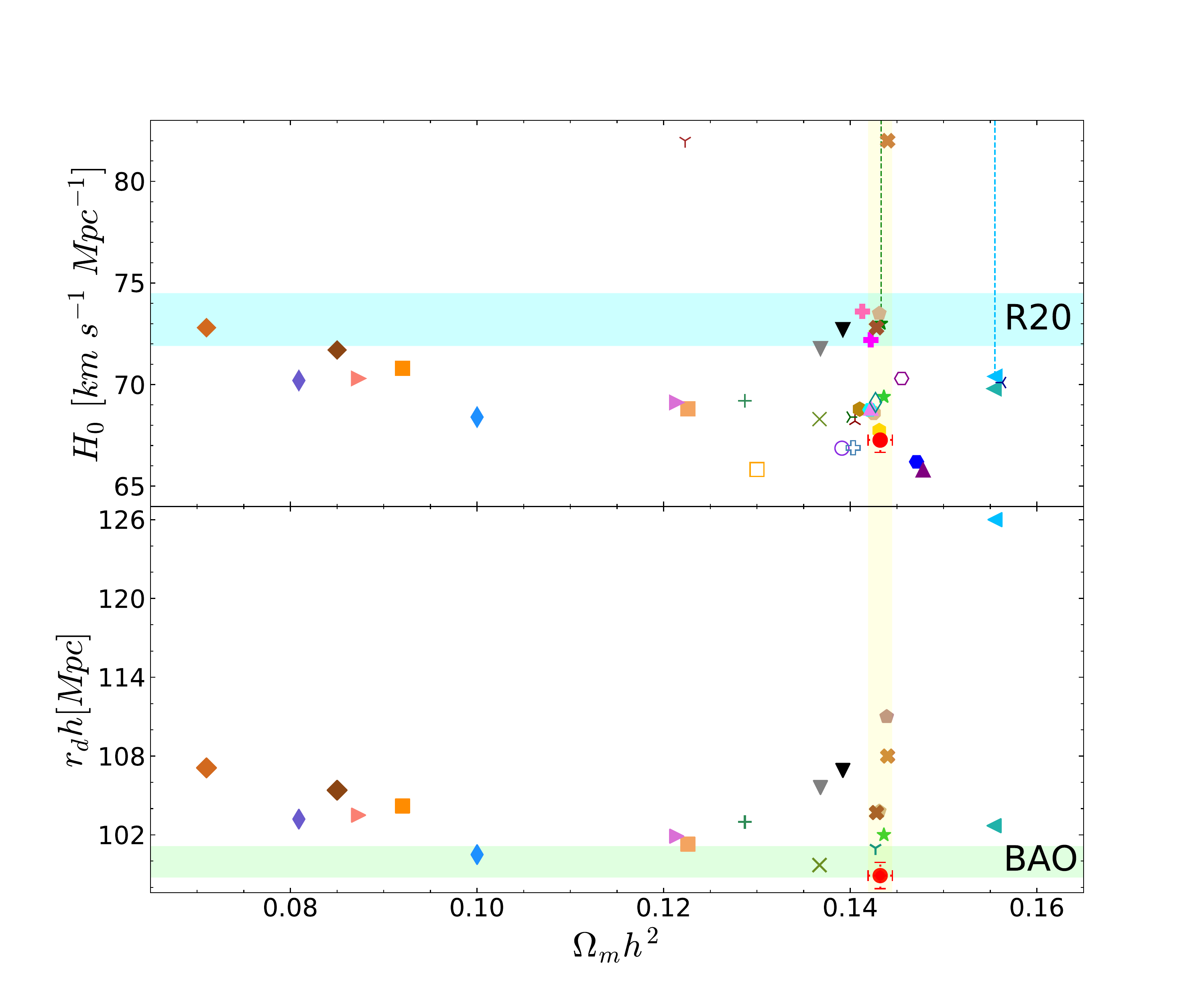}
\includegraphics[width=0.75\textwidth, right]{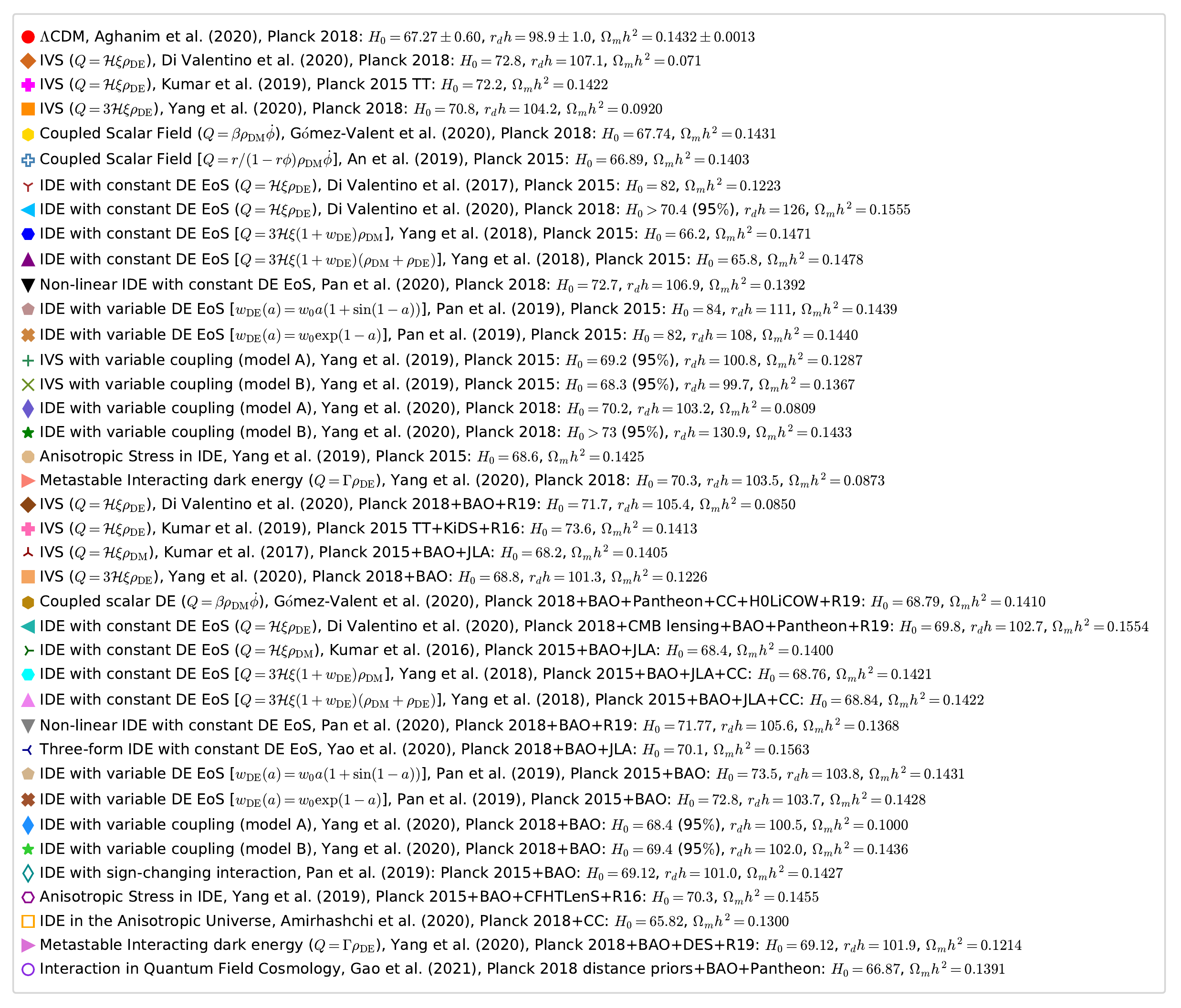}
\caption{Estimated values of the current matter energy density $\Omega_mh^2$, Hubble constant $H_0$ and sound horizon $r_dh$ in terms of various data points for different models discussed throughout the Section~\ref{sec:IDE} of the main Section~\ref{InteractSolut}. The cyan horizontal band corresponds to the $H_0$ value measured by R20~\cite{Riess:2020fzl}, the yellow vertical band to the $\Omega_mh^2$ value estimated by {\it Planck} 2018~\cite{Aghanim:2018eyx} in a $\Lambda$CDM scenario, and the light green horizontal band to the $r_dh$ value measured by BAO data. The points sharing the same symbol refer to the same model in the same paper, and the different colors indicate a different dataset combination. }
\label{fig:chapter8a_H0Om}
\end{figure*}

\begin{figure*}
\includegraphics[width=\textwidth]{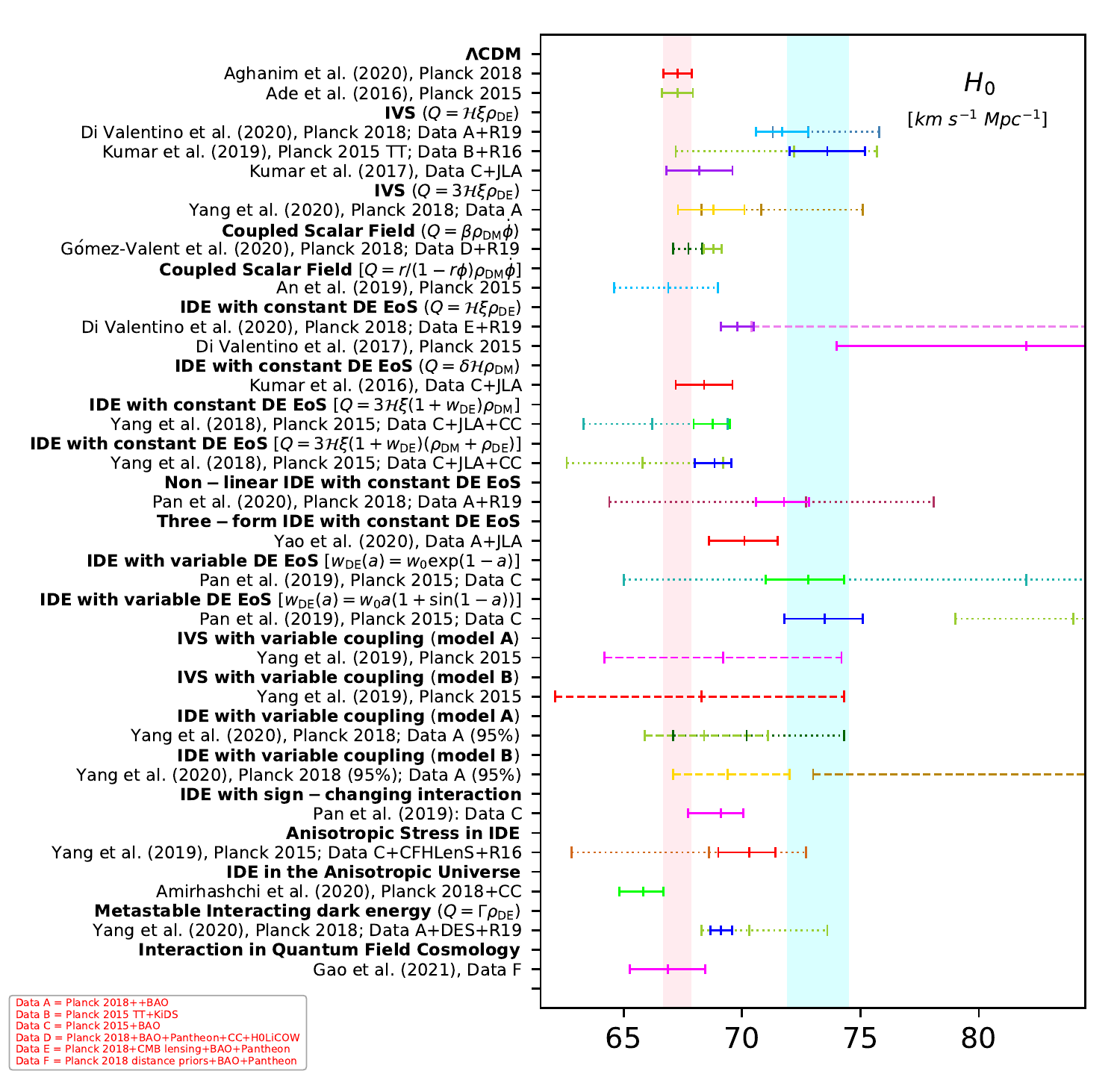}
\caption{Whisker plot with the 68\% (95\% if dashed) marginalized Hubble constant constraints for the models discussed throughout the Section~\ref{sec:IDE} of the main Section~\ref{InteractSolut}. The cyan vertical band corresponds to the $H_0$ value measured by R20~\cite{Riess:2020fzl} and the light pink vertical band corresponds to the $H_0$ value estimated by {\it Planck} 2018~\cite{Aghanim:2018eyx} in a $\Lambda$CDM scenario. For each line, when more than one error bar is shown, the dotted one corresponds to the {\it Planck} only constraint on the Hubble constant, while the solid one to the different dataset combinations reported in the red legend, in order to appreciate the shift due to the additional datasets. }
\label{fig:chapter8a_whisker}
\end{figure*}

\begin{figure*}
\centering
\includegraphics[width=0.85\textwidth, right]{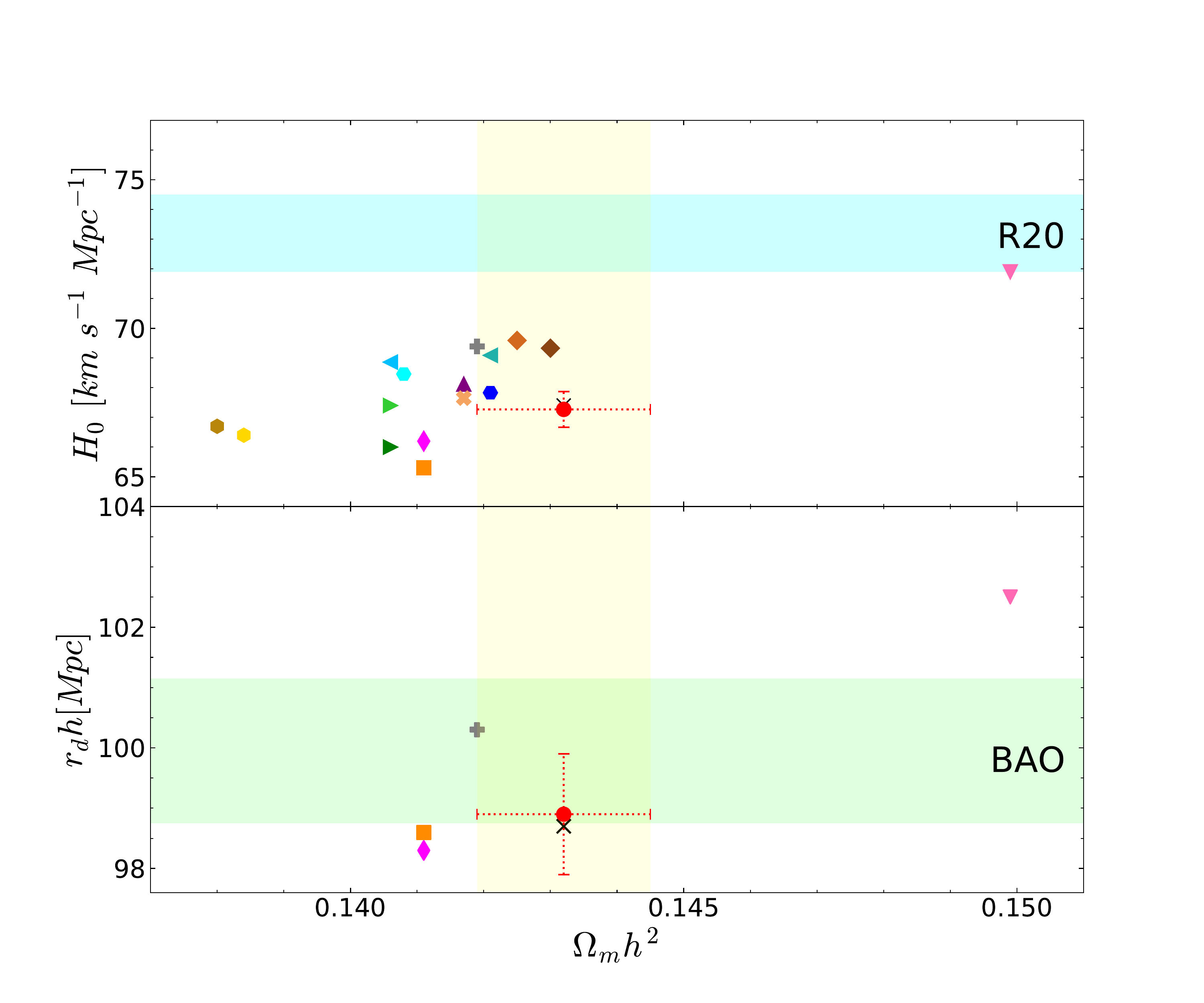}
\includegraphics[width=0.85\textwidth, right]{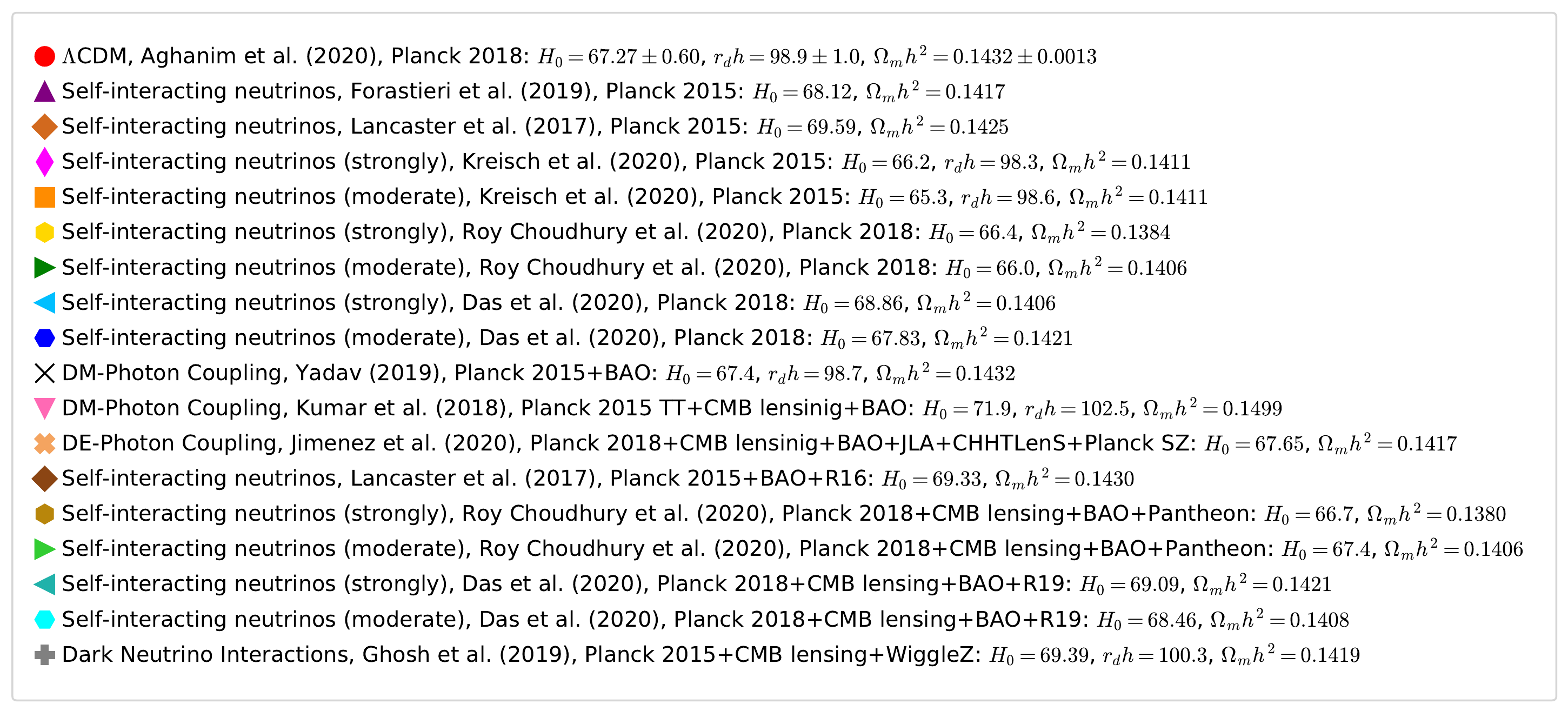}
\caption{Estimated values of the current matter energy density $\Omega_mh^2$, Hubble constant $H_0$ and sound horizon $r_dh$ in terms of various data points for different models discussed throughout the Section~\ref{sec-IDM} of the main Section~\ref{InteractSolut}. The cyan horizontal band corresponds to the $H_0$ value measured by R20~\cite{Riess:2020fzl}, the yellow vertical band to the $\Omega_mh^2$ value estimated by {\it Planck} 2018~\cite{Aghanim:2018eyx} in a $\Lambda$CDM scenario, and the light green horizontal band to the $r_dh$ value measured by BAO data. The points sharing the same symbol refer to the same model in the same paper, and the different colors indicate a different dataset combination.}
\label{fig:chapter8b_H0Om}
\end{figure*}

Cosmological models allowing for a non-gravitational interaction between the components of the Universe have been found to be successful in alleviating the $H_0$ tension. As the number of models in this section is very large, for the clarity in the graphical presentation, we have devoted four figures for this section. Figures~\ref{fig:chapter8a_H0Om} and~\ref{fig:chapter8a_whisker} refer to the models discussed throughout Section~\ref{sec:IDE} of the main Section~\ref{InteractSolut}. Figures ~\ref{fig:chapter8b_H0Om} and Figure~\ref{fig:chapter8a_whisker} cover the models of the remaining sections, i.e.\ Sections~\ref{sec-IDM} to ~\ref{IDnu}.

\subsection{Interacting Dark Energy}
\label{sec:IDE}

Along with the early and late time solutions, a generalized cosmological scenario in which the dark matter and the dark energy interact with each other in a non-gravitational way received massive attention in the literature. These are known as Interacting Dark Energy (IDE) or Coupled Dark Energy (CDE) models~\cite{Amendola:1999er}. The possibility of an interaction was thought to deal with the cosmological constant problem~\cite{Wetterich:1994bg}. Later on, it was argued that an interaction between the dark fluids, namely, dark matter and dark energy, can be used to provide a possible solution to the cosmic coincidence problem~\cite{Huey:2004qv,Cai:2004dk, Pavon:2005yx, Berger:2006db, delCampo:2006vv, delCampo:2008sr, delCampo:2008jx}. Additionally, an interaction in the dark sector can explain the phantom dark energy regime without any scalar field having a negative kinetic term~\cite{Huey:2004qv,Wang:2005jx, Das:2005yj, Sadjadi:2006qb, MohseniSadjadi:2009va, Pan:2014afa}. For those reasons, IDE scenarios have been substantially investigated in the literature, see e.g.\ Refs.~\cite{Barrow:2006hia, Amendola:2006dg, He:2008tn,Valiviita:2008iv, Gavela:2009cy,Koyama:2009gd,Majerotto:2009np,Valiviita:2009nu, Chimento:2009hj,Boehmer:2009tk,He:2010im, Chimento:2011pk,Clemson:2011an,Chimento:2012zz,Chimento:2012aea,Pan:2013rha, Pettorino:2013oxa,Costa:2013sva, Chimento:2013rya,Yang:2014gza, yang:2014vza, Yang:2014hea, Pan:2012ki, Richarte:2014yva,Nunes:2016dlj,Amendola:2016saw, Yang:2016evp, Mukherjee:2016shl,Costa:2016tpb, Erdem:2016hqw, Yang:2017yme, vandeBruck:2017idm, Pan:2017ent, Pan:2016ngu, Sharov:2017iue, Cai:2017yww, Yang:2018xlt, DAmico:2016jbm,Yang:2017zjs, Yang:2017ccc,Bhattacharyya:2018fwb,Yang:2018qec, vonMarttens:2018iav, Yang:2018pej, Paliathanasis:2019hbi, Yang:2019bpr, Asghari:2019qld,Yang:2019vni, Costa:2018aoy, Gonzalez:2018rop, Li:2019loh, Li:2019ajo, vonMarttens:2019ixw, Cheng:2019bkh, Pan:2020zza, Pan:2020mst,DiValentino:2020kpf,Sa:2020fvn,Sa:2021eft,Yang:2021hxg,Zhang:2021yof,Bonilla:2021dql,Johnson:2021wou,Kumar:2021eev} (see also two review articles~\cite{Bolotin:2013jpa, Wang:2016lxa} in this context for a comprehensive reading) with some interesting consequences. 

The IDE models, as examined by several investigators over the last couple of years, can play an effective role to alleviate/solve the Hubble constant tension. 
In this section, we therefore revisit different IDE models which significantly increase the $H_0$ value. The basic framework of this theory is the coupling between dark matter and dark energy characterizing the energy flow between these dark sectors. Due to such a coupling between these dark fluids, the continuity equations can be written as:
\begin{eqnarray}
    \dot{\rho}_{\rm DM}+3{\cal H}\rho_{\rm DM} &=& a\, Q\,;\label{eq:DM_evol} \\
    \dot{\rho}_{\rm DE}+3{\cal H}(1+w_{\rm DE})\rho_{\rm DE} &=& -a\, Q\,, \label{eq:DE_evol}
\end{eqnarray}
where the dot corresponds to the derivative with respect to conformal time $\tau$, $a$ is the scale factor, ${\cal H} \equiv \mathrm{d}\ln a/\mathrm{d}\tau$ is the conformal expansion rate of the Universe, and $\rho_{\rm DM}$, $\rho_{\rm DE}$ are respectively the energy density of dark matter and dark energy. The quantity $Q$ denotes the interaction rate/interaction function/coupling function which characterizes the energy or/and momentum flow between these dark fluids. In the following, we classify the models based on the functional form of $Q$.

\subsubsection{Interacting vacuum energy:}

The simplest interacting scenario is the case where vacuum energy characterized by the equation of state $w_{\rm DE} \equiv p_{\rm DE}/\rho_{\rm DE} = -1$ interacts with dark matter, known as the Interacting Vacuum Scenario (IVS). Consistent observational evidences indicate that IVS can solve the $H_0$ tension~\cite{Kumar:2016zpg, Kumar:2017dnp, Yang:2019uog, DiValentino:2019ffd, Martinelli:2019dau, Kumar:2019wfs, Lucca:2020zjb, Wang:2021kxc} in an exceptional way, even if this result is mostly driven by the existing parameter degeneracies~\cite{DiValentino:2020leo}.

\noindent A possibility is to assume that the rate of the interaction $Q$ between the two dark components is proportional to the dark energy density $\rho_{\rm DE}$ as:
\begin{equation}
    \label{model1}
    Q = \xi \,{\cal H}\, \rho_{\rm DE}\,,
\end{equation}
where $\xi$ is a dimensionless parameter that quantifies the coupling between dark matter and dark energy. An analysis that accounts for {\it Planck} 2015 TT power spectra data ({\it Planck} 2015 TT + R16 + KiDS-450~\cite{Hildebrandt:2016iqg}) results in $H_0=72.2^{+3.5}_{-5.0}{\rm\,km\,s^{-1}\,Mpc^{-1}}$ ($H_0=73.6\pm1.6{\rm\,km\,s^{-1}\,Mpc^{-1}}$) at 68\% CL~\cite{Kumar:2019wfs}, solving the Hubble tension within $1\sigma$.

Repeating the analysis in an extended scenario where the parameters in the neutrino sector are also allowed to vary, and using the {\it Planck} 2015 + JLA + BAO datasets yields the Hubble constant $H_0=68.2\pm1.4{\rm\,km\,s^{-1}\,Mpc^{-1}}$ at 68\% CL~\cite{Kumar:2017dnp}, alleviating the Hubble tension at $2.6\sigma$. For this latter case, results obtained by fitting {\it Planck} observations alone are absent in the literature.

An update to this scenario that considers a flux of energy from dark matter to the dark energy components is presented in Ref.~\cite{DiValentino:2019ffd}, where it is shown that a fit against {\it Planck} 2018 ({\it Planck} 2018 + BAO + R19) data gives $H_0=72.8^{+3.0}_{-1.5}{\rm\,km\,s^{-1}\,Mpc^{-1}}$ ($H_0=71.7\pm1.1{\rm\,km\,s^{-1}\,Mpc^{-1}}$) at 68\% CL, resolving completely the Hubble tension within $1\sigma$. This is in agreement with the findings in Refs.~\cite{Lucca:2020zjb, DiValentino:2020vnx,Kumar:2021eev} and Ref.~\cite{Yang:2019uog}, where instead the interaction term differs by a factor of three, and for which an assessment against {\it Planck} 2018 ({\it Planck} 2018 + BAO) data gives $H_0=70.8^{+4.3}_{-2.5}{\rm\,km\,s^{-1}\,Mpc^{-1}}$ at 68\% CL ($H_0=68.8^{+1.3}_{-1.5}{\rm\,km\,s^{-1}\,Mpc^{-1}}$ at 68\% CL), solving the tension within $1\sigma$ (at $2.4\sigma$). In Refs.~\cite{Yang:2019uog,Kumar:2021eev}, an extended model in which the neutrino sector is also allowed to vary has been considered.

The case in which there is a transition between an interacting and a non-interacting scenario is instead analysed in Ref.~\cite{Martinelli:2019dau}, where it was shown that the Hubble constant resulting from {\it Planck} 2015 data can not solve the tension with R19.

\subsubsection{Coupled Scalar Field:}

We now discuss the Coupled Dark Energy (CDE) scenario, in which dark matter interacts via a dark force mediated by a new scalar field $\phi$, which in turn drives cosmic acceleration~\cite{Amendola:1999er, Amendola:2003wa, Pettorino:2008ez,Boehmer:2015kta,Boehmer:2015sha}. In this model, the coupling in the dark sector is described by the Lagrangian term:
\begin{equation}
    \mathcal{L} = -\frac{1}{2}\partial^\mu\phi\partial_\mu\phi - V(\phi) - m(\phi)\bar\psi\psi + \mathcal{L}_{\rm kin}[\psi]\,,
\end{equation}
in which the mass of DM field $\psi$, $m(\phi)$, is a function of the scalar field $\phi$, $V(\phi)$ is a self-coupling potential, and the last term describes the DM kinetic term. This model has been assessed against WMAP~\cite{Pettorino:2012ts}, {\it Planck} 2013~\cite{Pettorino:2013oxa}, and {\it Planck} 2015~\cite{Ade:2015rim} measurements, by employing a Peebles-Ratra potential~\cite{Peebles:1987ek}:
\begin{equation}
    V(\phi)=V_0\,\phi^{-\alpha}\,,
\end{equation}
where $V_0$ and $\alpha > 0$ are constants. Recently, the model has been reconsidered in light of the {\it Planck} 2018 data ({\it Planck} 2018 + BAO + Pantheon + CC + R19 + H0LiCOW), obtaining the result for the Hubble constant $H_0=67.74^{+0.57}_{-0.66}{\rm\,km\,s^{-1}\,Mpc^{-1}}$ ($H_0=68.79^{+0.35}_{-0.40}{\rm\,km\,s^{-1}\,Mpc^{-1}}$) at 68\% CL~\cite{Gomez-Valent:2020mqn}, a value that is in tension with R20 at $3.9\sigma$ ($3.4\sigma$).

In Ref.~\cite{An:2018vzw} a quintessence model with a Yukawa interaction between dark energy and dark matter has been explored, finding that {\it Planck} 2015 gives $H_0=66.89^{+2.1}_{-2.3}{\rm\,km\,s^{-1}\,Mpc^{-1}}$ at 68\% CL, at $2.5\sigma$ tension with R20.

Recently, in Ref.~\cite{Johnson:2021wou} the authors have explored the scalar field interacting scenario proposed in Ref.~\cite{Johnson:2020gzn}. Using a combination of CC + BAO + High redshift HII galaxy measurements (HIIG) + JLA, the authors find $H_0=69.9^{+0.46}_{-1.02}{\rm\,km\,s^{-1}\,Mpc^{-1}}$ at 68\% CL~\cite{Johnson:2021wou}, alleviating the tension with R20 at $2.4\sigma$.

\subsubsection{IDE with a constant DE equation of state:}

A possible extension to the previous model is an interacting dark energy scenario where the dark energy fluid has a constant equation of state different from the cosmological constant value, $w_{\rm DE} \neq -1$.

\begin{itemize}
\item For the interaction rate of Eq.~\eqref{model1}, i.e.\ $Q = \xi{\cal H}\rho_{\rm DE}$, {\it Planck} 2015 provides $H_0=82^{+10}_{-8}{\rm\,km\,s^{-1}\,Mpc^{-1}}$ at 68\% CL~\cite{DiValentino:2017iww}, in agreement with R20. An update is presented in Ref.~\cite{DiValentino:2019jae} where, for $w_{\rm DE}<-1$, a fit to {\it Planck} 2018 ({\it Planck} 2018 + CMB lensing + BAO + Pantheon + R19) data gives $H_0 > 70.4{\rm\,km\,s^{-1}\,Mpc^{-1}}$ at 95\% CL ($H_0=69.8\pm0.7{\rm\,km\,s^{-1}\,Mpc^{-1}}$ at 68\% CL), in agreement with R20 within $2\sigma$ (at $2.3\sigma$). These results hold when an extended model varying the neutrino sector is considered~\cite{Yang:2020uga}.

\item Ref.~\cite{Wang:2004cp} considered a model in which the rate of the interaction $Q$ is proportional to the DM energy density $\rho_{\rm DM}$ instead of the DE energy density:
\begin{equation}
    \label{model2}
    Q = \delta{\cal H}\rho_{\rm DM}\,,
\end{equation}
where $\rho_{\rm DM} = \rho_{{\rm DM},0}\,a^{-3+\delta}$, $\rho_{{\rm DM},0}$ is the present value of $\rho_{\rm DM}$, and the quantity $\delta > 0$ controls the deviation of the DM scaling from its standard case. A fit to {\it Planck} 2015 + JLA + BAO datasets leads to the Hubble constant $H_0=68.4\pm1.2{\rm\,km\,s^{-1}\,Mpc^{-1}}$ at 68\% CL~\cite{Kumar:2016zpg}, considering a flux of energy from the dark matter sector to the dark energy one and a neutrino mass freely varying. In this case, the Hubble constant tension is at $2.7\sigma$, but results with {\it Planck} data alone are missing.

\item The authors in Ref.~\cite{Yang:2018euj} consider the rate of interaction of the form:
\begin{equation}
    \label{model3}
    Q = 3(1+w_{\rm DE})\,\xi\,{\cal H}\,\rho_{\rm DM}\,,
\end{equation}
which, instead of the coupling parameter $\xi$, a term containing the DE equation of state $w_{\rm DE}$ appears. A fit to {\it Planck} 2015 ({\it Planck} 2015 + BAO + JLA + CC) data gives $H_0=66.2^{+3.2}_{-2.9}{\rm\,km\,s^{-1}\,Mpc^{-1}}$ ($H_0=68.76^{+0.72}_{-0.80}{\rm\,km\,s^{-1}\,Mpc^{-1}}$) at 68\% CL~\cite{Yang:2018euj}, alleviating the Hubble constant tension with R20 at $2\sigma$ ($3\sigma$), shifting considerably the mean value of $H_0$.

\item In Ref.~\cite{Yang:2018euj}, an alternative form for the rate of interaction which depends on the total DM+DE energy density is also considered, as:
\begin{equation}
    \label{model4}
    Q = 3(1+w_{\rm DE})\xi{\cal H}(\rho_{\rm DM}+\rho_{\rm DE})\,.
\end{equation}
An analysis with {\it Planck} 2015 ({\it Planck} 2015 + BAO + JLA + CC) data using this model results in $H_0=65.8^{+3.4}_{-3.2}{\rm\,km\,s^{-1}\,Mpc^{-1}}$ ($H_0=68.84^{+0.70}_{-0.84}{\rm\,km\,s^{-1}\,Mpc^{-1}}$) at 68\% CL~\cite{Yang:2018euj}, alleviating the Hubble constant tension with R20 at $2.1\sigma$ ($2.9\sigma$). The alleviation of the tension for {\it Planck} 2015 alone is mostly due to the large error bars.

\item In Ref.~\cite{Pan:2020bur}, a non-linear interaction rate is considered, of the form:
\begin{equation}
    \label{model5}
    Q = 3\xi\,{\cal H}\,\rho_{\rm DE} \sin\left(\frac{\rho_{\rm DE}}{\rho_{\rm DM}} - 1 \right)\,.
\end{equation}
The sinusoidal function forces the rate $Q$ to change sign according to the relative value of $\rho_{\rm DE}$ and $\rho_{\rm DM}$. A fit to {\it Planck} 2018 ({\it Planck} 2018 + BAO + R19) data leads to the Hubble constant $H_0=72.7^{+5.4}_{-8.3}{\rm\,km\,s^{-1}\,Mpc^{-1}}$ ($H_0=71.77^{+1.05}_{-1.17}{\rm\,km\,s^{-1}\,Mpc^{-1}}$) at 68\% CL~\cite{Pan:2020bur}, a result which is in perfect agreement with R20.

\item Another possibility is the generalized three-form dark energy model proposed in Ref.~\cite{Yao:2020hkw}, that can be regarded as an IDE model with $w_{\rm DE}> - 1$. The analysis of this model against {\it Planck} 2018 + BAO + JLA data gives $H_0=70.1^{+1.4}_{-1.5}{\rm\,km\,s^{-1}\,Mpc^{-1}}$ at 68\% CL~\cite{Yao:2020pji}, alleviating the Hubble constant tension with R20 at $1.6\sigma$.
\end{itemize}

\subsubsection{IDE with variable DE equation of state:}

A further step towards the IDE models in which dark energy has a dynamical equation of state was performed in Refs.~\cite{Yang:2018uae, Pan:2019gop}. In Ref.~\cite{Pan:2019gop}, the authors considered different phenomenological parameterizations for the dark energy equation of state, together with an interaction rate proportional to $\rho_{\rm DE}$. In particular, for the interaction rate $Q = 3 H \xi [ 1+ w_{\rm DE} (a) ] \rho_{\rm DE}$, different variants of $w_{\rm DE} (a)$ as described below were investigated resulting in different estimates of $H_0$:

\begin{itemize}
\item $w_{\rm DE}(a)=w_0\,a\,[1-\log(a)]$: For {\it Planck} 2015 gives $H_0 = 81^{+13}_{-14}{\rm\,km\,s^{-1}\,Mpc^{-1}}$ at 68\% CL and {\it Planck} 2015 + BAO gives $H_0 = 71.0 \pm 1.5{\rm\,km\,s^{-1}\,Mpc^{-1}}$ at 68\% CL. 

\item $w_{\rm DE} (a) = w_0\,a\,\exp(1-a)$: For {\it Planck} 2015 gives $H_0 = 84^{+14}_{-7}{\rm\,km\,s^{-1}\,Mpc^{-1}}$ at 68\% CL and {\it Planck} 2015 + BAO gives $H_0 = 71.7^{+1.5}_{-1.7}{\rm\,km\,s^{-1}\,Mpc^{-1}}$ at 68\% CL. 

\item $w_{\rm DE}(a)=w_0\,a\,[1+\sin(1-a)]$: For {\it Planck} 2015 gives $H_0 = 84^{+12}_{-5}{\rm\,km\,s^{-1}\,Mpc^{-1}}$ at 68\% CL and {\it Planck} 2015 + BAO gives $H_0 = 73.5^{+1.6}_{-1.7}{\rm\,km\,s^{-1}\,Mpc^{-1}}$ at 68\% CL. 

\item $w_{\rm DE}(a)=w_0\,a\,[1+\arcsin(1-a)]$: For {\it Planck} 2015 gives $H_0 = 82^{+14}_{-17}{\rm\,km\,s^{-1}\,Mpc^{-1}}$ at 68\% CL and {\it Planck} 2015 + BAO gives $H_0 = 72.8^{+1.5}_{-1.8}{\rm\,km\,s^{-1}\,Mpc^{-1}}$ at 68\% CL. 
\end{itemize}
Note, that in all the above expressions of $w_{\rm DE} (a)$, $w_0$ denotes the present value of $w_{\rm DE} (a)$. The alleviation of the $H_0$ tension happens at the price of a phantom dark energy equation of state at more than 2-to-3 standard deviations, and for {\it Planck} 2015 data, the $H_0$ tension with R20 is alleviated within $1\sigma$. In the case of the combination {\it Planck} 2015 + BAO, the tension is alleviated within $2\sigma$. The analyses of the same models with {\it Planck} 2018 are pending cases in the literature.

\subsubsection{IVS and IDE with variable coupling:}

In most of the IDE models, the coupling parameter of the interaction model is assumed to be constant. However, the most general case is only realized when a time varying coupling parameter is considered, as there is no theoretical principle that can exclude this possibility. In Ref.~\cite{Yang:2019uzo, Yang:2020tax}, the two following coupling functions were considered:
\begin{eqnarray}
\mbox{Model A:}\quad Q &=& 3 \xi (a) H \rho_{\rm DE}~;\\
\mbox{Model B:}\quad Q &=& 3 \xi (a) H\frac{\rho_{\rm DM}\,\rho_{\rm DE}}{\rho_{\rm DM}+\rho_{\rm DE}}~,
\end{eqnarray}
where $\xi (a)$ is the time dependent and dimensionless coupling parameter having the form 
\begin{equation}
    \xi (a) = \xi_0 + \xi_a \; (1-a)\,.
\end{equation}
The above interaction models together with the variable coupling function 
were investigated for $w_{\rm DE} = -1$ (IVS), in Ref.~\cite{Yang:2019uzo}, and for a constant value of $w_{\rm DE} \neq -1$ in Ref.~\cite{Yang:2020tax} (IDE). 

In Ref.~\cite{Yang:2019uzo}, for Model A (B), {\it Planck} 2015 alone estimates $H_0 = 69.2\pm 5 {\rm\,km\,s^{-1}\,Mpc^{-1}}$ ($68.3^{+6}_{-6.2}{\rm\,km\,s^{-1}\,Mpc^{-1}}$) at 95\% CL. Consequently, due to the large error bars, the $H_0$ tension with R20 is solved at $1.5\sigma$ level for both models. 

In Ref.~\cite{Yang:2020tax}, Model A and Model B are tested for a constant value of $w_{\rm DE} \neq -1$. For Model A, when a quintessence regime $w_{\rm DE}>-1$ is assumed, {\it Planck} 2018 ({\it Planck} 2018 + BAO) gives $H_0 = 70.2^{+4.1}_{-3.1}{\rm\,km\,s^{-1}\,Mpc^{-1}}$ at 68\% CL ($H_0 = 68.4^{+2.7}_{-2.5}{\rm\,km\,s^{-1}\,Mpc^{-1}}$ at 95\% CL). Due to the very large error bars, the $H_0$ tension with R20 is alleviated within $1\sigma$ ($2.6\sigma$). For Model B and $w_{\rm DE}< -1$, a fit to {\it Planck} 2018 ({\it Planck} 2018 + BAO) data provides $ H_0 > 73 {\rm\,km\,s^{-1}\,Mpc^{-1}}$ at 95\% CL ($H_0 = 69.4^{+2.6}_{-2.3}{\rm\,km\,s^{-1}\,Mpc^{-1}}$ at 95\% CL). Notice that the tension is solved within $2\sigma$ (at $2.1\sigma$).

\subsubsection{IDE with sign-changing interaction:}

The energy transfer rate that regulates the conversion between two dark sectors could switch sign during the expansion history. For example, the rate of interaction $Q$ in Eqs.~\eqref{eq:DM_evol}-\eqref{eq:DE_evol} could change the direction in which energy flows~\cite{Sun:2010vz,Wei:2010cs,Forte:2013fua,Guo:2017deu,Arevalo:2019axj,Pan:2019jqh}. In this scenario it is possible to alleviate the Hubble tension~\cite{Pan:2019jqh}. For the rate of interaction $Q = 3 \mathcal{H} \xi (\rho_{\rm DM} - \rho_{\rm DE})$, {\it Planck} 2015 + BAO gives $H_0=69.12^{+0.93}_{-1.39}{\rm\,km\,s^{-1}\,Mpc^{-1}}$ at 68\% CL~\cite{Pan:2019jqh}, reducing the $H_0$ tension at $2.6\sigma$, including the BAO data. One more sign-changing interaction function with two coupling parameters, having a similar feature, can be found in Ref.~\cite{Pan:2019jqh}. Another example for a sign-changing interaction has been described in Ref.~\cite{Pan:2020bur}, see Eq.~\eqref{model5}.

\subsubsection{Anisotropic Stress in IDE:}

An IDE scenario where the anisotropic stress of the large scale inhomogeneities is also considered has been explored in Ref.~\cite{Yang:2018ubt}. In this model, the conservation equations are:
\begin{eqnarray}
    \dot{\rho}_{\rm DM} + 3 {\cal{H}} \rho_{\rm DM} &=& - a Q\,;\\
    \dot{\rho}_{\rm DE} + 3 {\cal{H}}(1+w_{\rm DE})\rho_{\rm DE} &=& a Q\,,
\end{eqnarray}
and the interaction function $Q$ reads:
\begin{equation}
    \label{model6}
    Q = 3(1+w_{\rm DE})\, {\cal{H}}\, \rho_{\rm DE}\,.
\end{equation}
A fit of this model against {\it Planck} 2015 ({\it Planck} 2015 + BAO + R16 + CFHTLenS~\cite{Heymans:2013fya}) data gives $H_0=68.6^{+4.1}_{-5.8}{\rm\,km\,s^{-1}\,Mpc^{-1}}$ ($H_0=70.3^{+1.1}_{-1.3}{\rm\,km\,s^{-1}\,Mpc^{-1}}$) at 68\% CL~\cite{Yang:2018ubt}, solving the tension with R20 at $1.1\sigma$ ($1.7\sigma$). The updated analysis of this scenario using {\it Planck} 2018 measurements is absent from the literature to date.

\subsubsection{Interaction in the Anisotropic Universe:}

The Bianchi cosmological solutions of the Einstein equations break the assumption of an isotropic Universe at its largest scales. In Ref.~\cite{Amirhashchi:2020qep}, a coupling between the dark components within an anisotropic Bianchi type I Universe is considered. A fit of this model to {\it Planck} 2018 + CC data gives $H_0=65.82^{+0.85}_{-0.99}{\rm\,km\,s^{-1}\,Mpc^{-1}}$ at 68\% CL, in disagreement with R20 at $4.6\sigma$.

\subsubsection{Metastable Interacting dark energy:}

A model of metastable interacting DE was studied in Refs.~\cite{Shafieloo:2016bpk, Li:2019san, Yang:2020zuk}. The conservation equations for DE-DM in this scenario follow:
\begin{eqnarray}
    \dot{\rho}_{\rm DE} &=& -\, \Gamma \rho_{\rm DE}\,; \\
    \dot{\rho}_{\rm DM} + 3{\cal H}\rho_{\rm DM} &=& \Gamma \rho_{\rm DE}\,,
\end{eqnarray}
where $\Gamma$ is a constant (for $\Gamma < 0$ DE density increases, while for $\Gamma>0$ DE density decreases). Notice that it is an interacting DE-DM scenario with $Q = \Gamma \rho_{\rm DE}$.

For the combination Pantheon + BAO data the model results in $H_0=71.8^{+4.7}_{-4.6}{\rm \,km\,s^{-1}\,Mpc^{-1}}$ at 68\% CL, showing that the Hubble tension is solved within $1\sigma$~\cite{Li:2019san}. However, in presence of the CMB distance priors from {\it Planck} 2018, the Hubble tension is restored at more than $3\sigma$~\cite{Li:2019san}.

When a full analysis with {\it Planck} 2018 ({\it Planck} 2018 + BAO + DES + R19) is performed with this model, an increase of DE density ($\Gamma < 0$) is supported by the data together with a larger value of the Hubble constant $H_0=70.3^{+3.3}_{-2.0}{\rm \,km\,s^{-1}\,Mpc^{-1}}$ ($H_0=69.12^{+0.46}_{-0.45}{\rm \,km\,s^{-1}\,Mpc^{-1}}$) at 68\% CL~\cite{Yang:2020zuk}, 
solving the tension with R20 within $1\sigma$ (within $2.9\sigma$).

\subsubsection{Quantum Field Cosmology:}

The IDE model proposed in Ref.~\cite{Begue:2017lcw}, that relies on the Einstein-Cartan gravitational theory and considers the Universe in the scale invariant ultra-violet fixed point of the theory (referred to as Quantum Field Cosmology), has been investigated as a possible solution to the Hubble tension. In this model, Newton's gravitational constant $G_N$ and the cosmological constant $\Lambda$ possess a mild dependence on redshift, and vary according to the scaling laws:
\begin{equation}
    \frac{G_N}{G_0} = (1+z)^{-\delta_G}\,, \qquad \hbox{and} \qquad \frac{\Lambda}{\Lambda_0} = (1+z)^{\delta_\Lambda} \,,
\end{equation}
with $\delta_G,\delta_\Lambda \ll 1$, and approaching to an ultraviolet fixed point of $G_0$ and $\Lambda_0$ where the classical Einstein theory is realized.

This picture is extended in Ref.~\cite{Gao:2021xnk} to include the matter and radiation energy densities:
\begin{eqnarray}
    \frac{G_N}{G_0} \rho_{m,r} &=& \rho^0_{m,r} (1+z)^{3(1+w_{m,r})-\delta_G}\,;\\
    \frac{G_N}{G_0} \rho_\Lambda &=& \rho^0_\Lambda (1+z)^{\delta_\Lambda}\,,
\end{eqnarray}
where $\rho^0_{m}$, $\rho^0_{r}$, and $\rho^0_\Lambda$ are the present-day values of $\rho_{m}$, $\rho_{r}$, and $\rho_\Lambda$, and $w_m$, $w_r$ are the equations of state for matter and radiation, respectively. When this model is analysed using the {\it Planck} 2018 distance priors + BAO + Pantheon, the Hubble constant obtained is $H_0=66.87^{+1.57}_{-1.61}{\rm \,km\,s^{-1}\,Mpc^{-1}}$ at 68\% CL~\cite{Gao:2021xnk}, in disagreement with R20 at $3.2\sigma$.

\subsubsection{Interacting Quintom dark energy:}

A quintom model~\cite{Guo:2004fq,Lazkoz:2006pa,Lazkoz:2007mx,Cai:2009zp,Leon:2018lnd}, i.e.\ a dark energy model with two scalar fields where one of them has canonical kinetic energy and the second one a negative kinetic energy term, modified to include an interaction between dark matter and dark energy, can be considered a possible alternative to reconcile the Hubble constant tension, as argued in Ref.~\cite{Panpanich:2019fxq}. 
The addition of this extra component $X$ with negative density, in the Friedmann equation, will leave unaltered the Planck's constraints on the matter and dark energy densities, but will match the requirements for solving the Hubble tension, acting as a phantom field, while the second scalar field will be quintessence. A full data analysis is however missing.

\begin{figure*}
\includegraphics[width=\textwidth]{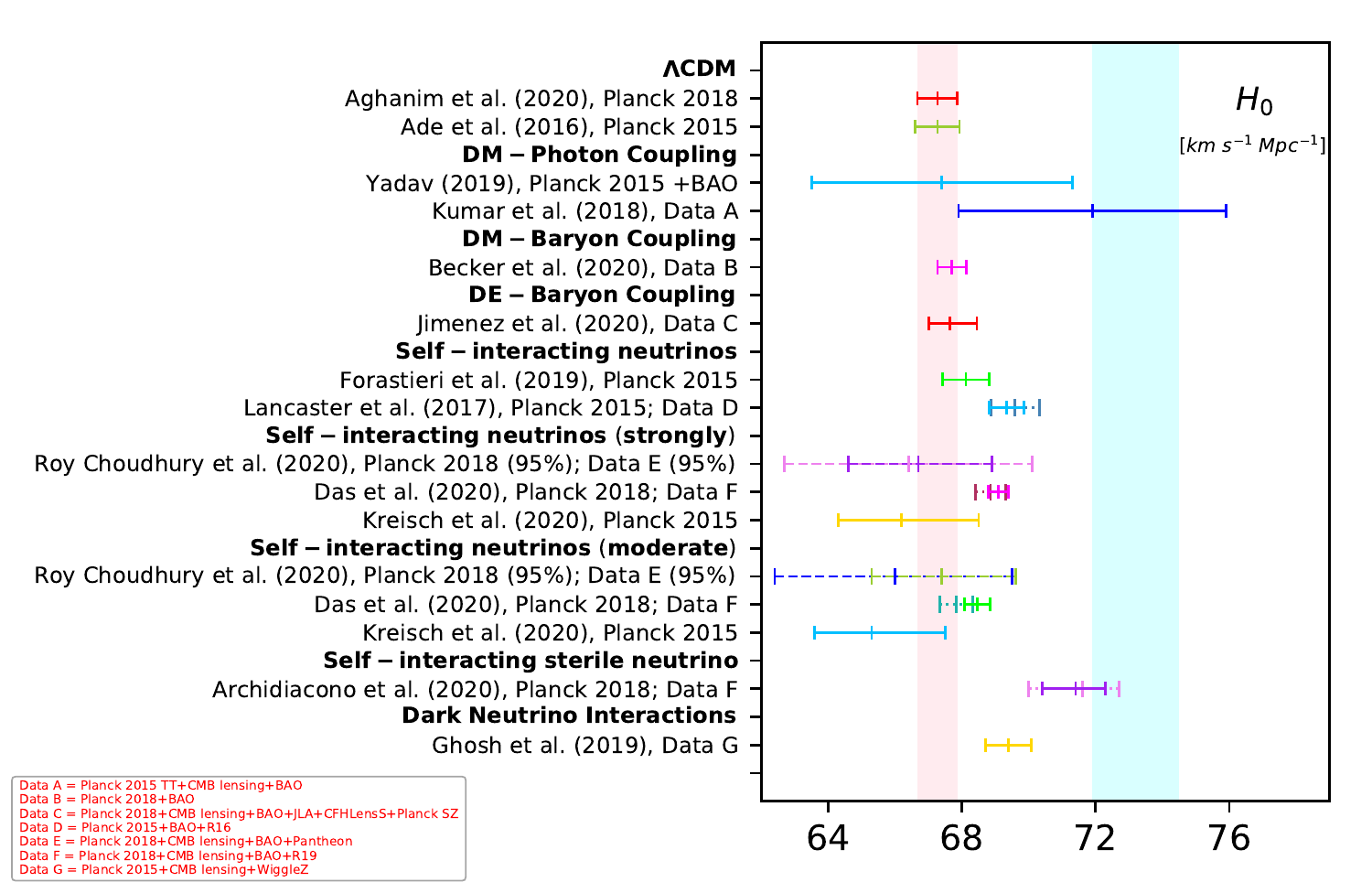}
\caption{Whisker plot with the 68\% (95\% if dashed) marginalized Hubble constant constraints for the models discussed throughout the Sections~\ref{sec-IDM} and~\ref{IDnu} of the main Section~\ref{InteractSolut}. The cyan vertical band corresponds to the $H_0$ value measured by R20~\cite{Riess:2020fzl} and the light pink vertical band corresponds to the $H_0$ value estimated by {\it Planck} 2018~\cite{Aghanim:2018eyx} in a $\Lambda$CDM scenario. For each line, when more than one error bar is shown, the dotted one corresponds to the {\it Planck} only constraint on the Hubble constant, while the solid one to the different dataset combinations reported in the red legend, in order to appreciate the shift due to the additional datasets. } 
\label{fig:chapter8b_whisker}
\end{figure*}


\subsection{Interacting Dark Matter}
\label{sec-IDM}

In the standard $\Lambda$CDM model, dark matter is assumed to be collisionless. Therefore, a possible extension is a dark matter interacting with the other components of the Universe. This process can help in reconciling the Hubble constant tension. We already explored the possibility of self-interacting dark matter, decaying dark matter, and dark matter interacting with neutrinos and DE in the previous sections, therefore we shall restrict ourselves in the following to the remaining cases exclusively.

\subsubsection{DM - Photon Coupling:}

A non-minimal coupling between photons and dark matter~\cite{Wilkinson:2013kia, Boehm:2014vja, Escudero:2018thh, Stadler:2018jin} has also been considered to ameliorate the Hubble tension.
 
The coupling between the dark matter fluid and photons can be described by:
\begin{eqnarray}
\dot\rho_{\rm DM} + 3 {\cal H} \rho_{\rm DM} &=& - Q\,; \\
\dot\rho_{\gamma} + 4 {\cal H} \rho_{\gamma} &=& Q\,,
\end{eqnarray}
where $Q= \Gamma_\gamma {\cal H} \rho_{\rm DM}$. For this scenario, where the neutrino sector is free to vary, {\it Planck} 2015 TT + CMB lensing + BAO gives $H_0=71.9\pm4.0{\rm\,km\,s^{-1}\,Mpc^{-1}}$ at 68\% CL~\cite{Kumar:2018yhh}, solving the $H_0$ tension within $1\sigma$. However, this result has been obtained fitting the CMB temperature power spectrum only. 

An extension of this model has been investigated in Ref.~\cite{Yadav:2019jio}, considering a CPL parameterization for the DE equation of state, obtaining $H_0=67.4\pm3.9{\rm\,km\,s^{-1}\,Mpc^{-1}}$ at 68\% CL for {\it Planck} 2015 + BAO, and alleviating the tension with R20 at \textbf{$1.4\sigma$}.

An updated analysis is instead presented in Ref.~\cite{Becker:2020hzj}, where {\it Planck} 2018 + BAO gives $H_0=67.70\pm0.43{\rm\,km\,s^{-1}\,Mpc^{-1}}$ at 68\% CL, showing a disagreement with R20 at $4\sigma$.

\subsubsection{DM - Baryon Coupling:}

Another possibility explored in the literature to ameliorate the Hubble tension resides in considering DM and baryons interacting~\cite{Barkana:2018lgd, Chen:2002yh, Boehm:2004th, Dvorkin:2013cea, Munoz:2017qpy, Ali-Haimoud:2018dvo, Boddy:2018wzy, Slatyer:2018aqg, Xu:2018efh}. In Ref.~\cite{Becker:2020hzj}, a model in which the DM-baryon interaction modifies the Euler equation that regulates the DM-baryon momentum exchange rate is explored. The analysis against {\it Planck} 2018 + BAO datasets gives $H_0=67.70\pm0.43{\rm\,km\,s^{-1}\,Mpc^{-1}}$ at 68\% CL~\cite{Becker:2020hzj}, showing a disagreement with R20 at $3.9\sigma$.

\subsection{DE - Baryon Coupling}

Contrary to the search for DM, for which realistic particle models motivate the search in direct detection experiments, a laboratory search of DE is conceptually complicated to start with, since the nature of DE is not clear. For example, DE could be due to a theory of gravity beyond GR, or it could be a manifestation of new fields. In the latter case, it is not even clear what the associated mass scale should be; for the case of a light scalar field, for example, we expect a field of a mass of the order of the Hubble constant~\cite{Arvanitaki:2009fg, Visinelli:2018utg}.

Surprisingly, the interaction between dark energy and baryon could proceed through a large Thompson cross section $\sim \mathcal{O}({\rm b})$, with negligible impact on the CMB or structure formation~\cite{Simpson:2010vh, Vagnozzi:2019kvw}. If instead a time-varying cross section is invoked, it is possible to have detectable signatures of an elastic interaction between baryons and DE. In this latter case, an analysis that accounts for the {\it Planck} 2018 + CMB lensing + BAO + JLA + CFHTLensS + {\it Planck} SZ datasets finds $H_0 = 67.65_{-0.64}^{+0.80}{\rm\,km\,s^{-1}\,Mpc^{-1}}$ at 68\% CL~\cite{Jimenez:2020ysu}, and thus in disagreement with R20 at $3.7\sigma$.

\subsection{Interacting neutrinos}
\label{IDnu}

The physics of neutrinos is one of the appealing topics in modern cosmology. The neutrinos may in principle interact with each other or with other cosmic sectors, see for instance Ref.~\cite{Bjaelde:2007ki}. The possibility of an interacting neutrino sector has been explored recently to reconcile the Hubble constant tension. While the possibility of a dark matter sector interacting with neutrinos has been already discussed in Section~\ref{sec:dmnu} in light of a contribution to $\Delta N_{\rm eff}$, here we shall restrict ourselves to previously unexplored models.

\subsubsection{Self-interacting neutrinos:}

A way for increasing the Hubble constant value is considered in Ref.~\cite{Lancaster:2017ksf}. In presence of a ``secret'' self-interacting neutrino mode, {\it Planck} 2015 TT gives $H_0=70.4\pm1.3{\rm\,km\,s^{-1}\,Mpc^{-1}}$ at 68\% CL~\cite{Lancaster:2017ksf}, reducing the Hubble tension at $1.6\sigma$. If the {\it Planck} 2015 high-$\ell$ polarization is included the Hubble estimate becomes $H_0=69.59^{+0.74}_{-0.71}{\rm\,km\,s^{-1}\,Mpc^{-1}}$ at 68\% CL~\cite{Lancaster:2017ksf}, increasing the Hubble tension to $2.3\sigma$ level. For {\it Planck} 2015 + BAO + R16 the Hubble constant is instead $H_0=69.33\pm0.52{\rm\,km\,s^{-1}\,Mpc^{-1}}$ at 68\% CL~\cite{Lancaster:2017ksf}, increasing the Hubble tension to $2.8\sigma$ level.

In Ref.~\cite{Kreisch:2019yzn} instead, it is present a delayed onset of the neutrino free-streaming until the Universe's expansion is very close to the matter-radiation equality epoch, and a neutrino self-interaction in presence of a total neutrino mass different from zero is considered. Therefore, for a strongly interacting neutrino cosmology, {\it Planck} 2015 gives $H_0=66.2^{+2.3}_{-1.9}{\rm\,km\,s^{-1}\,Mpc^{-1}}$ at 68\% CL~\cite{Kreisch:2019yzn}, lowering the Hubble tension down to $2.7\sigma$. For a moderate interacting neutrino scenario, {\it Planck} 2015 gives $H_0=65.3^{+2.2}_{-1.7}{\rm\,km\,s^{-1}\,Mpc^{-1}}$ at 68\% CL~\cite{Kreisch:2019yzn}, showing a disagreement with R20 at $3\sigma$ level.

An update of these results can be found in Ref.~\cite{Choudhury:2020tka}, where for a strongly interacting neutrino cosmology, {\it Planck} 2018 ({\it Planck} 2018 + CMB lensing + BAO + Pantheon) gives $H_0=66.4\pm3.7{\rm\,km\,s^{-1}\,Mpc^{-1}}$ ($H_0=66.7^{+2.2}_{-2.1}{\rm\,km\,s^{-1}\,Mpc^{-1}}$) at 95\% CL~\cite{Choudhury:2020tka}, alleviating at $3.0\sigma$ ($3.8\sigma$) the tension with R20 concerning the Hubble constant, and for a moderate interacting neutrino cosmology, {\it Planck} 2018 ({\it Planck} 2018 + CMB lensing + BAO + Pantheon) gives $H_0=66.0^{+3.5}_{-3.6}{\rm\,km\,s^{-1}\,Mpc^{-1}}$ ($H_0=67.4^{+2.2}_{-2.1}{\rm\,km\,s^{-1}\,Mpc^{-1}}$) at 95\% CL~\cite{Choudhury:2020tka}, reducing the tension to $3.3\sigma$ ($3.4\sigma$). These results show a very good agreement with those derived in Ref.~\cite{Das:2020xke}.

A model where the self-interaction structure is flavor-specific in the three active neutrino framework has been studied in Ref.~\cite{Das:2020xke}. Here, for a scenario with two self-interacting neutrino states, and a strongly interacting neutrino cosmology {\it Planck} 2018 ({\it Planck} 2018 + CMB lensing + BAO + R19) gives $H_0=68.86\pm0.46{\rm\,km\,s^{-1}\,Mpc^{-1}}$ ($H_0=69.09\pm0.31{\rm\,km\,s^{-1}\,Mpc^{-1}}$) at 68\% CL~\cite{Das:2020xke}, in disagreement at $3.1\sigma$ ($3.2\sigma$) with R20, while and for a moderate interacting neutrino cosmology, {\it Planck} 2018 ({\it Planck} 2018 + CMB lensing + BAO + R19) gives $H_0=67.83\pm0.50{\rm\,km\,s^{-1}\,Mpc^{-1}}$ ($H_0=68.46\pm0.38{\rm\,km\,s^{-1}\,Mpc^{-1}}$) at 68\% CL~\cite{Das:2020xke}, in disagreement at $3.8\sigma$ ($3.4\sigma$).

In Ref.~\cite{Lyu:2020lps} electroweak precision observables are taken into account, while in Ref.~\cite{Berbig:2020wve} the effective four-neutrino interaction is supposed to be generated by the exchange of a light mediator.
In Ref.~\cite{Mazumdar:2020ibx} a separate analysis with IceCube data is performed, and this concludes that the strong neutrino self-interactions region preferred by cosmology is disfavoured for both flavour specific and universal cases.
In Ref.~\cite{Forastieri:2019cuf} it is pointed out that neutrino self-interactions induced by a very light or massless mediator can not resolve the Hubble tension below $3.4\sigma$ ($H_0=68.12\pm0.69{\rm\,km\,s^{-1}\,Mpc^{-1}}$ at 68\% CL from {\it Planck} 2015), hence in Ref.~\cite{He:2020zns} self-interacting Dirac neutrinos via a light-dark-photon mediator are explored.

A consequence of a self-interacting neutrino model is, instead, studied in Ref.~\cite{Blinov:2019gcj}, where the experimental constraints on the coupling between the Majoron and the neutrino flavor eigenstates are presented. Following this paper, in Ref.~\cite{Brinckmann:2020bcn} the Majoron coupling is assumed instead to be diagonal to the neutrino mass eigenstates. The authors consider several cases: all neutrino states self-interact plus $N_{\rm eff}$ free to vary; two neutrino species free-stream and one interacts; a variable fraction of neutrinos self-interact with or without $N_{\rm eff}$ free to vary. The conclusions are that all of these cases can not alleviate the Hubble tension better than the case $\Lambda$CDM+$N_{\rm eff}$ alone.

\subsubsection{Self-interacting sterile neutrino model:}

In Ref.~\cite{Archidiacono:2020yey} is considered a cosmological model in which sterile neutrinos are coupled to a new, very light pseudoscalar degree of freedom, firstly introduced in Ref.~\cite{Archidiacono:2014nda} and analysed in Refs.~\cite{Archidiacono:2015oma, Archidiacono:2016kkh}, as a solution of the Hubble tension (see also Ref.~\cite{Beacom:2004yd}). For this pseudoscalar interaction, {\it Planck} 2018 ({\it Planck} 2018 + CMB lensing + BAO + R19) gives $H_0=71.6^{+1.1}_{-1.6}{\rm\,km\,s^{-1}\,Mpc^{-1}}$ ($H_0=71.4^{+0.9}_{-1.0}{\rm\,km\,s^{-1}\,Mpc^{-1}}$) at 68\% CL~\cite{Archidiacono:2020yey}, reducing the Hubble tension within $1\sigma$ (at $1.1\sigma$) level.

\subsubsection{Dark Neutrino Interactions:}

The Dark Neutrino Interactions scenario, introduced in~\cite{Ghosh:2017jdy}, is provided by a component of dark matter that interacts with neutrinos impeding them to free streaming. This produces an enhancement of the Hubble constant, possibly alleviating the tension, without varying $N_{\rm eff}$. In Ref.~\cite{Ghosh:2019tab} the combination of {\it Planck} 2015 + CMB lensing + WiggleZ Dark Energy Survey results in a value of $H_0=69.39^{+0.69}_{-0.68}{\rm\,km\,s^{-1}\,Mpc^{-1}}$ at 68\% CL, ameliorating the Hubble tension down to the $2.5\sigma$ level.

\begin{figure*}
\centering

\includegraphics[width=0.8\textwidth, right]{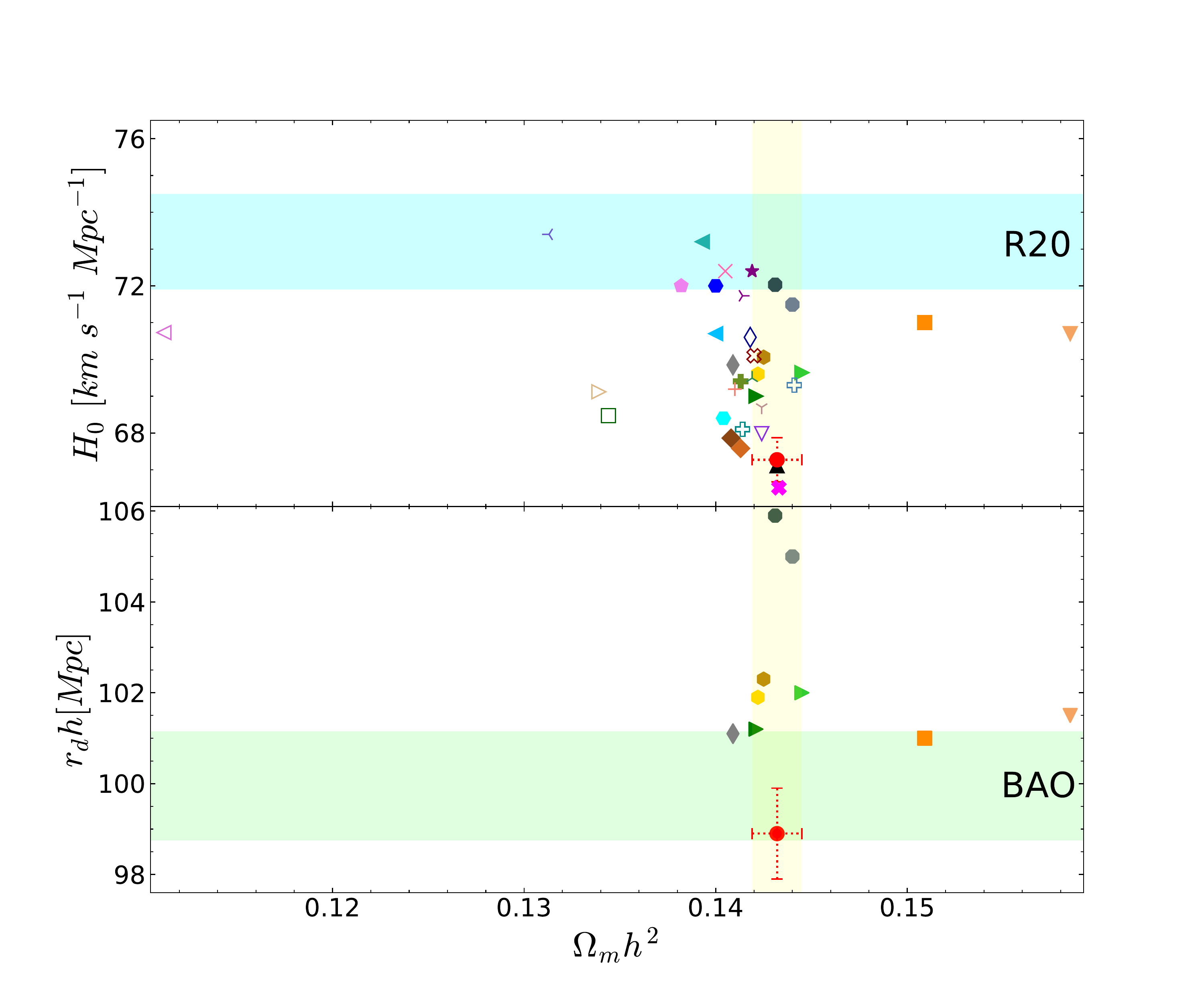}
\includegraphics[width=0.8\textwidth, right]{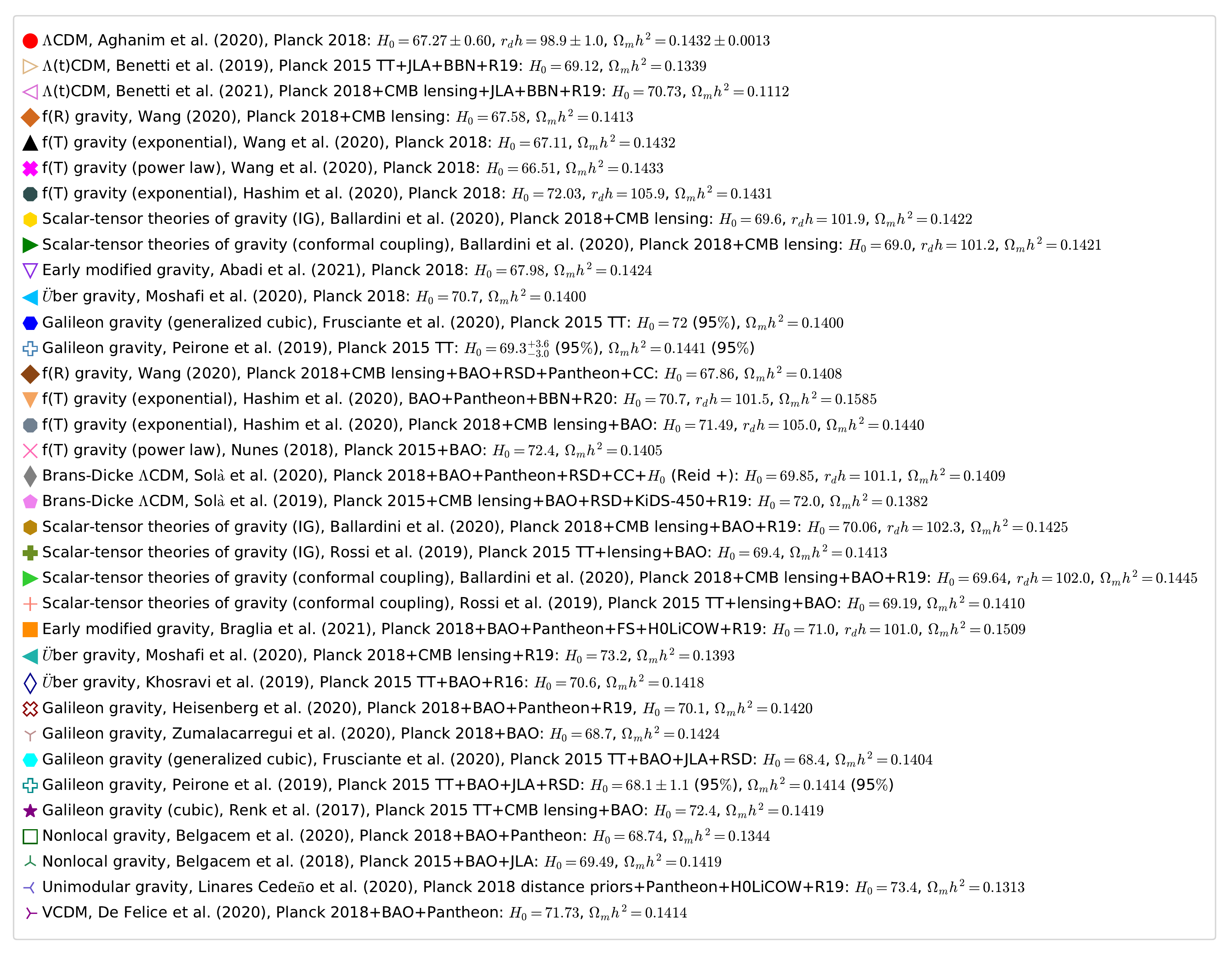}
\caption{Estimated values of the current matter energy density $\Omega_mh^2$, Hubble constant $H_0$ and sound horizon $r_dh$ in terms of various data points for different models discussed throughout the Sections~\ref{Unif} and~\ref{MG}. The cyan horizontal band corresponds to the $H_0$ value measured by R20~\cite{Riess:2020fzl}, the yellow vertical band to the $\Omega_mh^2$ value estimated by {\it Planck} 2018~\cite{Aghanim:2018eyx} in a $\Lambda$CDM scenario, and the light green horizontal band to the $r_dh$ value measured by BAO data. The points sharing the same symbol refer to the same model in the same paper, and the different colors indicate a different dataset combination.}
\label{fig:chapter9_H0Om}
\end{figure*}

\begin{figure*}
\includegraphics[width=\textwidth]{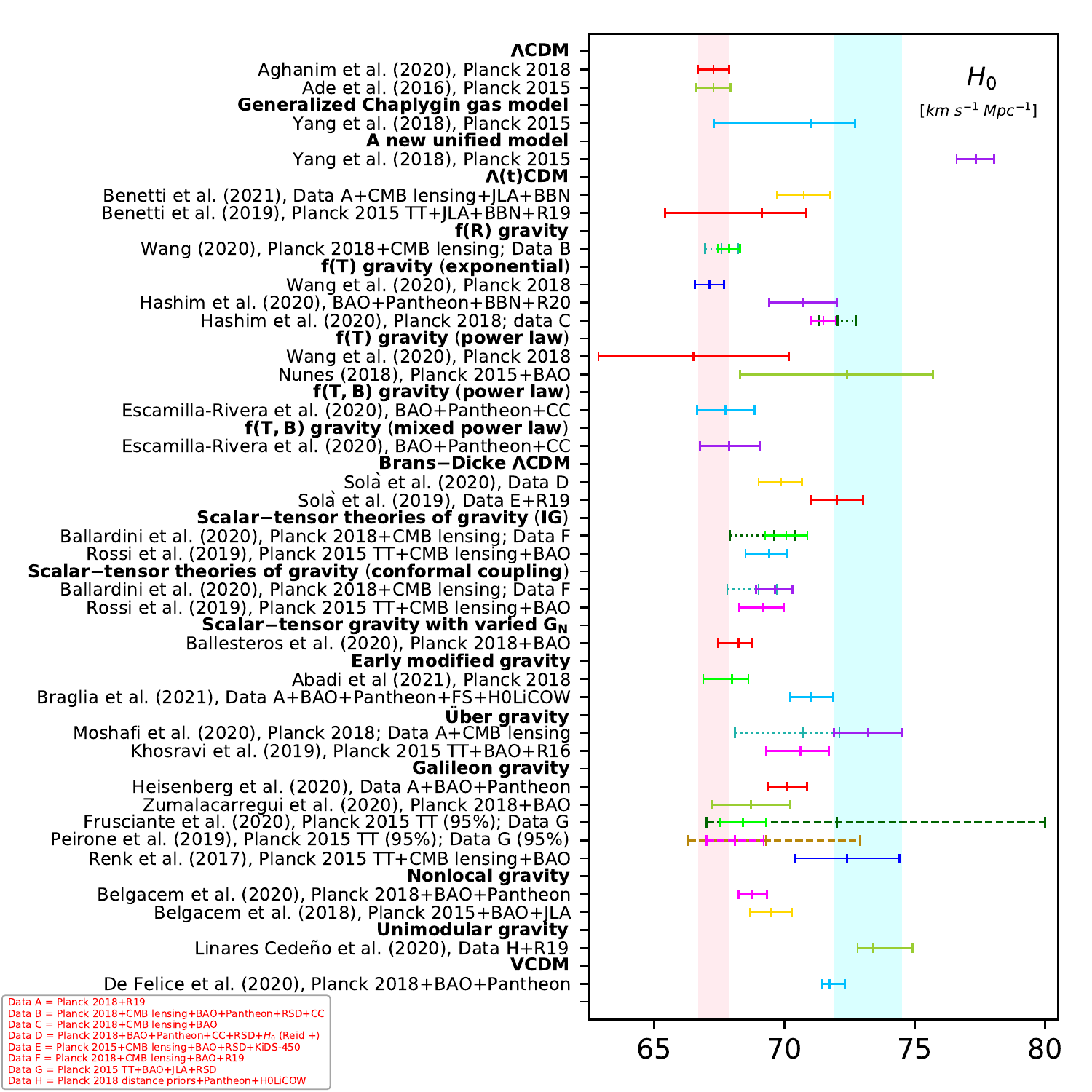}
\caption{Whisker plot with the 68\% (95\% if dashed) marginalized Hubble constant constraints for the models of Sections~\ref{Unif} and~\ref{MG}. The cyan vertical band corresponds to the $H_0$ value measured by R20~\cite{Riess:2020fzl} and the light pink vertical band corresponds to the $H_0$ value estimated by {\it Planck} 2018~\cite{Aghanim:2018eyx} in a $\Lambda$CDM scenario. For each line, when more than one error bar is shown, the dotted one corresponds to the {\it Planck} only constraint on the Hubble constant, while the solid one to the different dataset combinations reported in the red legend, in order to appreciate the shift due to the additional datasets.}
\label{fig:chapter9_whisker}
\end{figure*}

\section{Unified cosmologies}
\label{Unif}

Unified dark fluid models are cosmological scenarios where the dark matter and dark energy behave as a single fluid. This single fluid behaves as dark matter in the early evolution of the Universe and as dark energy at late times. The introduction of unified models in cosmology followed from a work by S. Chaplygin in Ref.~\cite{Chaplygin:1904}, and subsequently, this model, known as Chaplygin model, and its generalizations were extensively investigated by many researchers~\cite{Kamenshchik:2001cp, Bilic:2001cg, Gorini:2002kf, Bento:2002ps, Benaoum:2002zs, Amendola:2003bz, Avelino:2003cf, Multamaki:2003ed, Debnath:2004cd, Gorini:2005nw, Guo:2005qy, Banerjee:2005vy, Banerjee:2006na, delCampo:2008vr, Lu:2008hp, Lu:2008zzb, Lu:2009zzf, Lu:2010zzj, Xu:2010zzb, Liang:2010aa, BouhmadiLopez:2011kw, Xu:2012qx, Xu:2012ca, Lu:2013una, Herrera:2016sov, Marttens:2017njo}. In the following we present how the $H_0$ tension can be reconciled in different unified cosmological models. 

In analogy to earlier sections, we have shown Figures~\ref{fig:chapter9_H0Om} and~\ref{fig:chapter9_whisker}, providing a very comprehensive picture of the unified models along with those from the next Section~\ref{MG}.

\subsection{Generalized Chaplygin gas model}

The generalized Chaplygin gas model (gcg) is characterized by the equation of state: 
\begin{equation}
    \label{eq:gcg}
    p_{\rm gcg} = -\frac{A}{(\rho_{\rm gcg})^{\alpha}}\,,
\end{equation}
where $A$ and $\alpha$ are two real constants, and $p_{\rm gcg}$ and $\rho_{\rm gcg}$ are, respectively, the pressure and energy density of this fluid. For this model, {\it Planck} 2015 estimates $H_0 = 71.0^{+1.7}_{-3.7}{\rm\,km\,s^{-1}\,Mpc^{-1}}$ at 68\% CL~\cite{Yang:2019nhz}, and this solves the tension with R20 at $1\sigma$. The model needs to be updated with the final CMB data from {\it Planck} 2018.

\subsection{A new unified model}

In Ref.~\cite{Hova:2010na}, we find a new type of a unified model based on field theory grounds. In this model, the explicit relation between the pressure $p_u$ and the energy density $\rho_u$ is~\cite{Hova:2010na, Hernandez-Almada:2018osh, Yang:2019jwn}:
\begin{equation}
    \label{model-UM}
    p_u = -\rho_u + \rho_u \, {\rm sinc}\left(\frac{\mu \pi \rho_{u,0}}{\rho_u} \right)\,,
\end{equation}
where ${\rm sinc}(\theta) = \sin \theta /\theta$, $\mu \neq 0$ is a dimensionless quantity, and $\rho_{u,0}$ is the energy density of the unified dark fluid today.

A fit to {\it Planck} 2015 data alone to this model results in $H_0 = 77.33^{+0.71}_{-0.73}$ at 68\% CL~\cite{Yang:2019jwn}, alleviating the tension with R20 at $2.8\sigma$. However, an analysis with the new {\it Planck} 2018 data is absent in the literature.

\subsection{$\Lambda(t)$CDM model}

In Ref.~\cite{Benetti:2019lxu,Benetti:2021div} the $\Lambda(t)$CDM model has been considered to address the Hubble constant tension. The authors of Ref.~\cite{Benetti:2019lxu,Benetti:2021div} analyse a class of interacting models behaving as a generalized Chaplygin gas at the background level, i.e.\ like cold dark matter at early times and a cosmological constant in the asymptotic future. The explicit expression of $\Lambda (t)$ as considered in both the works has a Hubble dependence as: 
\begin{eqnarray}
    \Lambda (t) = \sigma H^{-2\alpha}~,
\end{eqnarray}
where $\alpha >-1$ is the interaction parameter and $\sigma =3 (1- \Omega_m) H_0^{2 (1+\alpha)}$. This is an example of a unified dark sector model where the Hubble rate follows
\begin{equation}\label{eq:E}
\frac{H(z)}{H_0} = \sqrt{\left[ (1-\Omega_m) + \Omega_m (1+z)^{3(1+\alpha)} \right]^{\frac{1}{(1+\alpha)}} + \Omega_r (1+z)^4}\,,
\end{equation}
which recovers the standard $\Lambda$CDM model for $\alpha =0$. 

For the above model of $\Lambda(t)$CDM, {\it Planck} 2015 TT + JLA + BBN + R19 estimates $H_0 = 69.12^{+1.7}_{-3.7}{\rm\,km\,s^{-1}\,Mpc^{-1}}$ at 68\% CL~\cite{Benetti:2019lxu}, and this solves the tension with R20 at $2.4\sigma$, including already a gaussian prior on $H_0$.
An updated analysis is presented in Ref.~\cite{Benetti:2021div}, where {\it Planck} 2018 + CMB lensing + JLA + BBN + R19 gives $H_0 = 70.73\pm1.02{\rm\,km\,s^{-1}\,Mpc^{-1}}$ at 68\% CL~\cite{Benetti:2021div}, solving the Hubble tension at $1.5\sigma$, but always including a Gaussian prior.

\subsection{$\Lambda$-gravity}

In Ref.~\cite{Gurzadyan:2019yir} the possibility that a cosmological constant, describing both the accelerated expansion of the Universe and the dynamics of galaxy groups and clusters, could solve the Hubble tension is taken into account. This theory is called $\Lambda$-gravity and is considered in the modified weak-field limit of GR. In this context it is possible to have a local Hubble constant of a local flow and a global one~\cite{Gurzadyan:2021jrw}, as a consequence of the common nature of dark energy and dark matter, solving naturally the Hubble constant problem.

\section{Modified gravity}
\label{MG}

Alternative gravitational theories including either modified versions of GR or new gravitational theories beyond GR, have been widely studied in the literature for their ability to explain different phases of the Universe, including the late-time cosmic acceleration as well as other aspects~\cite{Capozziello:2002rd, Nojiri:2003ni, Carloni:2004kp, Amarzguioui:2005zq, Capozziello:2005ku, Das:2005bn, Koivisto:2006xf, Li:2008fa, DeFelice:2009rw, Paliathanasis:2011jq, Winther:2011qb, deHaro:2012zt, Chakraborty:2012kj, Chakraborty:2013ywa, Basilakos:2013rua, Paliathanasis:2014iva, Chakraborty:2014joa, Paliathanasis:2015aos, Chakraborty:2015wma, Chakraborty:2015vla, Paliathanasis:2016vsw, Paliathanasis:2016tch, Koivisto:2016jcu, Nunes:2016qyp, Nunes:2016plz, Nunes:2016drj, Paliathanasis:2017efk,Karpathopoulos:2017arc,Paliathanasis:2017htk,Dimakis:2017tvb, Papagiannopoulos:2018mez, Nunes:2018evm, Choudhury:2019zod,Paliathanasis:2019ega} (see the following reviews in this direction~\cite{Capozziello:2007ec, Sotiriou:2008rp, DeFelice:2010aj, Capozziello:2011et, Clifton:2011jh, Nojiri:2017ncd, Cai:2015emx, Quiros:2019ktw}, and the references therein). Throughout this section, we shall discuss how modified gravity may help in alleviating or even solving the $H_0$ tension, obtaining a strong support for these models. The value of $H_0$ from CMB estimates can indeed be shifted towards larges values if the gravity is weaker at intermediate scales.

For example, an EFT approach performing a data-driven reconstruction of gravitational theories and dark energy models on cosmological scales finds that some of the models can alleviate the Hubble tension and are actually preferred against the standard $\Lambda$CDM model~\cite{Raveri:2019mxg}. In particular, this holds for models such as Scalar Horndeski and Full Horndeski theories~\cite{Horndeski:1974wa}.

Modifications of gravity at early times are effective in easing the Hubble tension because of the change induced in the evolution of the gravitational potential fluctuations, which leads to a change in the CMB temperature, polarization, and lensing predictions. Nevertheless, when the background expansion remains unchanged compared to $\Lambda$CDM, the resulting cosmology is still in tension with BAO data~\cite{Lin:2018nxe}. On the other hand, late time modifications induced by modified gravity theories are also beneficial in raising the Hubble constant $H_0$, since they lead to a change in the spectrum of the unlensed CMB temperature fluctuations through the ISW effect and smooth out the CMB acoustic peaks~\cite{Lin:2018nxe}, even if they are in disagreement with lensed CMB data on large scales $\ell \lesssim 400$.

Here, we describe some models of modified gravity in which the Hubble constant tension is alleviated. Figures~\ref{fig:chapter9_H0Om} and~\ref{fig:chapter9_whisker} contain the models of this section together with those from the previous Section~\ref{Unif}.

\subsection{$f(\mathcal{R})$ gravity theory}

Einstein theory of gravitation can be recovered from the principle of least action, once the Einstein-Hilbert action is introduced as
\begin{equation}
    \label{eq:EHaction}
    \mathcal{S} = \frac{1}{2\kappa^2} \int \mathrm{d}^4 x \sqrt{-g} \,\mathcal{R} + \mathcal{S}_m\,,
\end{equation}
where $g$ is the determinant of the metric tensor, $\mathcal{R}$ is the Ricci scalar, $\kappa^2 = 8 \pi G_N$, and $\mathcal{S}_m$ is the action describing any matter fields appearing in the theory.

The simplest generalization of Einstein gravity is the $f(\mathcal{R})$ gravity, in which the Ricci scalar appears in the action in a generic function $f(\mathcal{R})$:
\begin{equation}
    \label{eq:actionfR}
    \mathcal{S} = \frac{1}{2\kappa^2}\int \mathrm{d}^4 x \sqrt{-g}\, f(\mathcal{R}) + \mathcal{S}_m\,.
\end{equation}
The modified gravity theory described in Eq.~\eqref{eq:actionfR} has been widely investigated over the past years, considering various choices of the function $f(\mathcal{R})$. In this context, of particular interest is the Hu-Sawicki $f(\mathcal{R})$ model~\cite{Hu:2007nk},
\begin{equation}
    f(\mathcal{R}) = \mathcal{R} - m^2 \frac{c_1 (\mathcal{R}/m^2)^n}{c_2 (\mathcal{R}/m^2)^n + 1}\,,
\end{equation}
where $c_1, c_2$ are constants, $n > 0$ is an index, and $m^2 = H_0^2 \Omega_m$. In this class of models, an accelerating phase can be achieved without introducing a cosmological constant while satisfying both galactic and solar-system constraints.

Recently, the Hu-Sawicki model has been tested in light of the $H_0$ tension with different conclusions, depending on the cosmological datasets considered. The authors of Ref.~\cite{DAgostino:2020dhv} study the Hu-Sawicki model for $n =1$ using the geometrical data. Using the CC + Pantheon datasets, their best estimate for the Hubble constant is $H_0=69.5\pm 2.0{\rm\,km\,s^{-1}\,Mpc^{-1}}$ at 68\% CL~\cite{DAgostino:2020dhv}, which alleviates the tension with R20 at $1.5\sigma$. In Ref.~\cite{Wang:2020dsc}, the author performed the analyses exploiting {\it Planck} 2018 data in combination with other cosmological probes, leaving the index $n$ free to vary and also considering some specific values of this parameter. For example, for $n=1$ {\it Planck} 2018 + CMB lensing ({\it Planck} 2018 + CMB lensing + RSD + BAO + Pantheon + CC) gives $H_0=67.58\pm 0.64{\rm\,km\,s^{-1}\,Mpc^{-1}}$ ($H_0=67.86\pm 0.42{\rm\,km\,s^{-1}\,Mpc^{-1}}$) at 68\% CL~\cite{Wang:2020dsc}, i.e.\ the $H_0$ tension is not alleviated within this specific $f(\mathcal{R})$ gravity model. Recently, further investigations aimed at testing whether the $H_0$ tension can be solved within the $f(\mathcal{R})$ theory have been performed in Refs.~\cite{Odintsov:2020qzd, Cruz:2020cje}.

\subsection{$f(\mathcal{T})$ gravity theory}

The theory of Einstein-Cartan is an extension of GR that describes gravity in spacetime metrics with a connection that has both torsion and curvature~\cite{Cartan:1923zea}. In the framework of Einstein-Cartan theory, GR is a limit which is formulated based on Levi-Civita connections, for which the spacetime metric is torsion-free and has a possible non-zero curvature. A different limit, in which the spacetime connection has a non-zero torsion tensor $\mathcal{T}{^\lambda}_{\mu\nu}$ and zero curvature (Weitzenb\"ock connection) is teleparallel gravity, see e.g.\ Refs.~\cite{Unzicker:2005in, Maluf:2013gaa}. A torsion scalar $\mathcal{T}$ can be constructed by contractions of the torsion tensor~\cite{Aldrovandi:2013wha}.

Models based on a modification of teleparallel gravity might lead to a successful alternative to inflationary models, resulting in an accelerated expansion rate without introducing an inflaton field~\cite{Ferraro:2006jd, Bengochea:2008gz, Fiorini:2009ux}.\footnote{\, See Ref.~\cite{Awad:2017yod} for other possibilities.} These models are characterized by the inclusion in the action of an arbitrary function $f(\mathcal{T})$ of the torsion scalar:
\begin{equation}
    \mathcal{S} = \frac{1}{2\kappa^2} \int \mathrm{d}^4 x \sqrt{-g}\, f(\mathcal{T}) + \mathcal{S}_m\,.
\end{equation}
An $f(\mathcal{T})$ model can be studied in a EFT framework, in which the action describing perturbations is expanded around a time-dependent background~\cite{ArkaniHamed:2003uy}. In this case, the first Hubble equation is modified as
\begin{equation}
    \label{eq:HubbleT}
    3H^2 = \kappa^2\rho + \frac{1}{2}\left[\mathcal{T} - f(\mathcal{T}) + 2 \mathcal{T} f_{\mathcal{T}}\right]\,,
\end{equation}
where $f_{\mathcal{T}} = \mathrm{d}f/\mathrm{d}\mathcal{T}$.

A simple parameterization for teleparallel gravity is the power-law model~\cite{Li:2018ixg}:
\begin{equation}
    \label{eq:TPGpowerlaw}
    f(\mathcal{T}) = -\mathcal{T} + \alpha \mathcal{T}^b\,,
\end{equation}
where the torsion scalar, in the mostly plus sign convention for the metric signature, is $\mathcal{T} = 6H^2$, and $\alpha$, $b$ are constants. In this model, the GR metric for $\Lambda$CDM is recovered for $b = 0$ and $\alpha = -2\Lambda$. The model described in Eq.~\eqref{eq:TPGpowerlaw} may be able to alleviate the Hubble tension~\cite{Yan:2019gbw}: a fit to {\it Planck} 2015 + BAO data yields $H_0=72.4^{+3.3}_{-4.1}{\rm\,km\,s^{-1}\,Mpc^{-1}}$ at 68\% CL~\cite{Nunes:2018xbm}, in agreement with R20 within $1\sigma$, even in presence of the BAO measurements. An alternative analysis, based on Gaussian processes and $H(z)$ data, is presented in Ref.~\cite{Cai:2019bdh}, where the tension is also efficiently alleviated. In Ref.~\cite{DAgostino:2020dhv}, instead, CC + Pantheon gives $H_0=69.1\pm1.9{\rm\,km\,s^{-1}\,Mpc^{-1}}$ at 68\% CL, in tension at $1.8\sigma$ with R20. An updated analysis is presented in Ref.~\cite{Wang:2020zfv}, where {\it Planck} 2018 for this scenario gives $H_0=66.51\pm 3.65{\rm\,km\,s^{-1}\,Mpc^{-1}}$ at 68\% CL, where the $H_0$ value is shifted towards a lower mean value but with a larger error, alleviating therefore the Hubble tension ($1.7\sigma$).

In order to attain a small variation of the gravitational coupling, Ref.~\cite{Linder:2010py} adopts a $f(\mathcal{T})$ model with an exponential form:
\begin{equation}
    \label{eq:teleparallel1}
    f(\mathcal{T}) = -\mathcal{T} + \frac{1 - \Omega_m}{(1 + p)e^{-p} - 1} \mathcal{T}_0\,\left(1-e^{-p\sqrt{\mathcal{T}/\mathcal{T}_0}}\right)\,,
\end{equation}
where $\mathcal{T}_0 = 6H_0^2$ and $p>0$. The prefactor in Eq.~\eqref{eq:teleparallel1} is obtained by evaluating Eq.~\eqref{eq:HubbleT} at present time. Note, that $\Lambda$CDM is recovered in the limit $p\to +\infty$. For this scenario, a fit to {\it Planck} 2018 data gives $H_0=67.11\pm 0.56{\rm\,km\,s^{-1}\,Mpc^{-1}}$ at 68\% CL, value in disagreement with R20 at the level of $4.4\sigma$~\cite{Wang:2020zfv}.

A different $f(\mathcal{T})$ model with an exponential form has been explored in Ref.~\cite{Bamba:2010wb}:
\begin{equation}
    f(\mathcal{T}) = -\mathcal{T} + \frac{1 - \Omega_m}{(1 + 2 q) e^{-q} - 1} \mathcal{T}_0\, \left(1 - e^{-q\mathcal{T}/\mathcal{T}_0}\right)\,,
\end{equation}
where $q$ is a parameters. For this scenario, a fit to the {\it Planck} 2018 data yields the result $H_0=67.12\pm 0.56{\rm\,km\,s^{-1}\,Mpc^{-1}}$ at 68\% CL~\cite{Wang:2020zfv}, value that shows a $4.4\sigma$ disagreement with R20.

Another $f(\mathcal{T})$ parameterization with an exponential form~\cite{Awad:2017yod}:
\begin{equation}
    f(\mathcal{T}) = -\mathcal{T}\,e^{\beta(\mathcal{T}_0/\mathcal{T})}\,,
\end{equation}
where $\beta$ is found from solving $1 - 2\beta = \Omega_m e^{-\beta}$, has been explored in Ref.~\cite{Hashim:2020sez} in relation with the Hubble tension. A fit to Pantheon + R20 + BBN + BAO gives at the background level $H_0=70.7\pm 1.3{\rm\,km\,s^{-1}\,Mpc^{-1}}$ at 68\% CL~\cite{Hashim:2020sez}, alleviating the Hubble tension at $1.4\sigma$. This estimate, however, already includes a Gaussian prior on the Hubble constant. A full CMB analysis including perturbations has been performed in Ref.~\cite{Hashim:2021pkq}, where {\it Planck} 2018 ({\it Planck} 2018 + CMB lensing + BAO) gives $H_0=72.03\pm 0.70{\rm\,km\,s^{-1}\,Mpc^{-1}}$ ($H_0=71.49\pm 0.47{\rm\,km\,s^{-1}\,Mpc^{-1}}$) at 68\% CL, solving the Hubble tension within $1\sigma$ (at $1.2\sigma$) without the introduction of extra free parameters. 

Finally, Ref.~\cite{Briffa:2020qli} presents constraints on teleparallel gravity and its $f(\mathcal{T})$ extensions using Gaussian processes, and Ref.~\cite{Ren:2021tfi} reconstructs the free function of $f(\mathcal{T})$ gravity in a model-independent manner using different datasets and relieving the Hubble tension.

\subsection{$f(\mathcal{T}, \mathcal{B})$ gravity theory} 

An extension of the $f(\mathcal{T})$ scenario is the $f(\mathcal{T}, \mathcal{B})$ gravity theory where, along with the torsion $\mathcal{T}$, the boundary term $\mathcal{B} = 2\nabla_\mu \mathcal{T}_\nu^{\nu\mu}$ is also included~\cite{Bahamonde:2015zma}. Recently, some specific models of $f(\mathcal{T}, \mathcal{B})$ gravity were examined with the observational data in Ref.~\cite{Escamilla-Rivera:2019ulu}, where the authors argued there that the $H_0$ tension can be weakened in this context. In particular, the authors of Ref.~\cite{Escamilla-Rivera:2019ulu} investigated two different models:
\begin{eqnarray}
    f(\mathcal{T}, \mathcal{B}) &= b_0 \mathcal{B}^k + t_0\mathcal{T}^m & \quad \mbox{(Power-law model)}\,;\\
    f(\mathcal{T}, \mathcal{B}) &= f_0 \mathcal{B}^k \,\mathcal{T}^m & \quad\mbox{(Mixed power-law model)}\,,
\end{eqnarray}where $b_0$, $t_0$, $k$, $m$ and $f_0$ are all arbitrary constants. A fit to BAO + Pantheon + CC datasets gives $H_0 = 67.74 \pm 1.1{\rm\,km\,s^{-1}\,Mpc^{-1}}$ at 68\% CL~\cite{Escamilla-Rivera:2019ulu} for the power-law model, reducing the $H_0$ tension with R20 down to the $3.2\sigma$ level, and $H_0 = 67.86^{+1.2}_{-1.1}{\rm\,km\,s^{-1}\,Mpc^{-1}}$ at 68\% CL~\cite{Escamilla-Rivera:2019ulu} for the mixed power-law model, reducing the tension down to the $3 \sigma$ level. In this context, an analysis with the full CMB data is missing.

\subsection{$f(\mathcal{Q})$ gravity theory}

All models discussed so far assume $\nabla_{\alpha} g_{\mu \nu} = 0$, which is a condition that assures that angles and lengths are preserved under parallel transport. This assumption is dropped in extensions of GR that include non-Riemannian spacetime metrics, introducing a non-zero non-metricity tensor $\mathcal{Q}_{\alpha \mu \nu} = \nabla_{\alpha} g_{\mu \nu}$ (see e.g.\ Refs.~\cite{BeltranJimenez:2017tkd, Jimenez:2019ovq, Mandal:2020buf}). In this framework, the action for a model that includes an arbitrary function $f(\mathcal{Q})$ on a torsion- and curvature-free geometry is:
\begin{equation}
    \mathcal{S} = \frac{1}{2\kappa^2} \int \mathrm{d}^4 x \sqrt{-g}\, f(\mathcal{Q}) + S_m\,,
\end{equation}
where the non-metricity scalar $\mathcal{Q}$ is defined as $\mathcal{Q} = - \mathcal{Q}_{\alpha \mu \nu} P^{\alpha \mu \nu}$, in terms of the non-metricity conjugate:
\begin{equation}
    P^{\alpha}_{\mu\nu} = -\frac{1}{2}\left(\frac{1}{2}\mathcal{Q}^\alpha{}_{\mu\nu} - \mathcal{Q}_{(\mu\nu)}{}^\alpha\right) + \frac{1}{4}\left( \mathcal{Q}^\alpha - \tilde{Q}^\alpha\right) g_{\mu\nu} - \frac{1}{4}\delta^\alpha_{(\mu}\mathcal{Q}_{\nu)}\,.
\end{equation} 
Here, $\mathcal{Q}_\alpha=g^{\mu\nu}\mathcal{Q}_{\alpha\mu\nu}$ and $\tilde{\mathcal{Q}}_\alpha=g^{\mu\nu}\mathcal{Q}_{\mu\alpha\nu}$ are the two independent traces of the non-metricity tensor, and round brackets mean a symmetrisation over the indices.

In Ref.~\cite{Mandal:2020buf}, the $f(\mathcal{Q})$ modified gravity model is tested against the Pantheon sample using a cosmographic approach, in which the parameterization of $f(\mathcal{Q})$ involves an increase in the numbers of derivatives in the theory. More specifically, the authors consider three cosmographic $f(\mathcal{Q})$ models, namely M1, M2, and M3. The estimated values of $H_0$ are higher than R20, since $H_0 = 79.5 \pm 2.5{\rm\,km\,s^{-1}\,Mpc^{-1}}$ at 68\% CL~\cite{Mandal:2020buf} for M1, $H_0 = 79.2 \pm 3.1{\rm\,km\,s^{-1}\,Mpc^{-1}}$ at 68\% CL~\cite{Mandal:2020buf} for M2, and $H_0 = 79.5 \pm 2.6{\rm\,km\,s^{-1}\,Mpc^{-1}}$ at 68\% CL~\cite{Mandal:2020buf} for M3. The tension with R20 is reduced down to the $2.3\sigma$, $1.8\sigma$ and $2.2\sigma$ levels, respectively. However, a full analysis with {\it Planck} CMB data is missing for this theoretical framework.

\subsection{Jordan-Brans-Dicke gravity}

The replacement of Newton's gravitational constant $G_N$ with a coupling that varies with cosmic time, $G_N(t)$, has been proposed for the first time by Brans \& Dicke (BD)~\cite{Brans:1961sx}. In the BD theory, Newton's constant is promoted to a dynamical field that depends on the spacetime coordinates. In more detail, the action describing the BD theory depends on the BD field $\Phi$ as
\begin{equation}
    \label{eq:JBDaction}
    \mathcal{S} = \frac{1}{2\kappa^2}\int \mathrm{d}^4 x \sqrt{-g} \,\left[\Phi\mathcal{R} - \frac{\omega}{\Phi}g^{\mu\nu}\partial_\mu\Phi\partial_\nu\Phi - 2V(\Phi)\right] + \mathcal{S}_m\,,
\end{equation}
where $\omega$ is a new parameter in the theory and $\kappa$ depends on the value of $G_N$ measured today. It can be shown that the GR limit in Eq.~\eqref{eq:EHaction} is recovered for $\omega \to +\infty$.

The Jordan~\cite{Jordan:1959eg}, Brans \& Dicke (JBD) gravity has been extensively studied in the literature (see Refs.~\cite{Chen:1999qh, Banerjee:2000mj, Banerjee:2000gt, Chakraborty:2003ye, Nagata:2003qn, Acquaviva:2004ti, Das:2005yg, Banerjee:2006rp, Banerjee:2007zd, Das:2008iq, Avilez:2013dxa, Li:2013nwa, Tsamparlis:2013aza, Paliathanasis:2014rja, Umilta:2015cta, Paliathanasis:2015arj, Ballardini:2016cvy, Ooba:2016slp, Papagiannopoulos:2016dqw, Leon:2018skk, Paliathanasis:2019luv, Joudaki:2020shz}) and can possibly embed the Running Vacuum Model~\cite{Peracaula:2018dkg, Perez:2018qgw} (see also Section~\ref{sec:RVM}). Models of JBD gravity where the Hubble tension is alleviated have also been discussed, as reviewed below.

\subsubsection{BD-$\Lambda$CDM:}

In Ref.~\cite{Sola:2019jek}, a BD cosmology with an additional cosmological constant term (the BD-$\Lambda$CDM model) is considered in light of easing the Hubble tension.
In this case the action reads
\begin{equation}
    \mathcal{S} = \int \mathrm{d}^4 x \sqrt{-g} \,\left[ \frac{1}{2\kappa^2} \left(\Phi\mathcal{R} - \frac{\omega}{\Phi}g^{\mu\nu}\partial_\mu\Phi\partial_\nu\Phi\right) - \rho_\Lambda \right]  + \mathcal{S}_m\,.
\end{equation}
For this scenario, {\it Planck} 2015 + CMB lensing + BAO + RSD + KiDS-450 + R19 gives $H_0 = 72.0 \pm 1.0{\rm\,km\,s^{-1}\,Mpc^{-1}}$ at 68\% CL, solving the tension within $1\sigma$. Nevertheless, these results include a Gaussian prior for the Hubble constant.

An update with new data has been performed in Ref.~\cite{Sola:2020lba}, where {\it Planck} 2018 + Pantheon + BAO + RSD + CC + $H_0$ from~\cite{Reid:2019tiq} gives $H_0 =69.85^{+0.81}_{-0.85}{\rm\,km\,s^{-1}\,Mpc^{-1}}$ at 68\% CL~\cite{Sola:2020lba}, alleviating the tension down to $2.2\sigma$. However, a prior on the Hubble constant is already included in the analysis.

A result without this prior, and also without CMB polarization measurements, provides instead $H_0 = 68.86 ^{+1.15}_{-1.24}{\rm\,km\,s^{-1}\,Mpc^{-1}}$ at 68\% CL~\cite{Sola:2020lba}, for {\it Planck} 2018 TT + Pantheon + RSD + BAO + CC, reducing the tension down to the $2.6\sigma$ level.

\subsection{Scalar-tensor theories of gravity:}

The JBD theory can be reformulated to include the equivalent formulation of induced gravity (IG) in a scalar-tensor model~\cite{Zee:1979hy, Cooper:1982du}:
\begin{equation}\label{action-scalar-tensor}
    \mathcal{S} = \frac{1}{2\kappa^2}\int \mathrm{d}^4 x \sqrt{-g}\, \left[F(\sigma)\mathcal{R} -g^{\mu\nu}\partial_\mu\sigma\partial_\nu\sigma - 2V(\sigma) - 2\Lambda \right] + \mathcal{S}_m\,,
\end{equation}
where $\sigma$ is the scalar field in units of $M_{\rm Pl}$ which is responsible for generating Newton's gravitational constant $G_N$ through the spontaneous breaking of scale invariance and moving in a potential $V(\sigma)$, while $F(\sigma) = N_{\rm Pl}^2 + \xi \sigma^2$ where $N_{\rm Pl}$ is a parameter and $\xi>0$ is the coupling to the Ricci scalar. The conformal coupling case is $\xi=-1/6$ and $N_{\rm Pl} = 0$ ($\xi=0$ and $N_{\rm Pl} = 1$) for IG (GR).

In Ref.~\cite{Rossi:2019lgt}, an extended JBD is considered to alleviate the Hubble tension, assuming an effectively massless scalar field $\sigma$ with a potential $V \propto F^2$. A fit to {\it Planck} 2015 TT + CMB lensing + BAO data gives $H_0 = 69.4^{ +0.7}_{-0.9}{\rm\,km\,s^{-1}\,Mpc^{-1}}$ at 68\% CL~\cite{Rossi:2019lgt}, reducing the tension with R20 at $2.5\sigma$. An update of this model is performed in Ref.~\cite{Ballardini:2020iws}, where {\it Planck} 2018 + CMB lensing ({\it Planck} 2018 + CMB lensing + BAO + R19) gives $H_0 = 69.6^{+0.8}_{-1.7}{\rm\,km\,s^{-1}\,Mpc^{-1}}$ ($H_0 = 70.06\pm0.81{\rm\,km\,s^{-1}\,Mpc^{-1}}$) at 68\% CL~\cite{Ballardini:2020iws}, alleviating the Hubble tension at $2.4\sigma$ ($2.1\sigma$).

For the conformal coupling model, a fit to {\it Planck} 2015 TT + CMB lensing + BAO data gives $H_0 = 69.19 ^{+0.77}_{-0.93}{\rm\,km\,s^{-1}\,Mpc^{-1}}$ at 68\% CL~\cite{Rossi:2019lgt}, in disagreement with R20 at $2.7\sigma$. An update of this model is performed in Ref.~\cite{Ballardini:2020iws}, where {\it Planck} 2018 + CMB lensing ({\it Planck} 2018 + CMB lensing + BAO + R19) gives $H_0 = 69.0 ^{+0.7}_{-1.2}{\rm\,km\,s^{-1}\,Mpc^{-1}}$ ($H_0 = 69.64^{+0.65}_{-0.73}{\rm\,km\,s^{-1}\,Mpc^{-1}}$) at 68\% CL~\cite{Ballardini:2020iws}, with the Hubble tension still in disagreement at $2.8\sigma$ ($2.4\sigma$).

An extension of the previous scenario is studied in Ref.~\cite{Braglia:2020iik}, where the non-minimal coupling of the scalar field to the Ricci scalar is:
\begin{equation}
    F(\sigma) = 1 + \xi \sigma^n\,.
\end{equation}
Unfortunately, all the cases considered in the context of this model are in disagreement with R20 at more than $3\sigma$.

A similar scenario, where a variation of the Newton's gravitational constant $G_N$ between the early and the late Universe is accounted for, in the context of a scalar field model which is non-minimally and quadratically coupled to gravity, is considered in Ref.~\cite{Ballesteros:2020sik}. The $H_0$ value for this model using {\it Planck} 2018 + BAO is estimated to be $H_0 = 68.24^{+0.5}_{-0.79}{\rm\,km\,s^{-1}\,Mpc^{-1}}$ at 68\% CL~\cite{Ballesteros:2020sik} and the disagreement with R20 is at the level of $3.5\sigma$.

\subsubsection{Early modified gravity:}

The scenario described by the action in Eq.~\eqref{action-scalar-tensor} with $F(\sigma) = 1 + \xi \sigma^2$ and with the potential $V(\sigma) =\lambda \sigma^4/4$ ($\lambda$ is a free parameter) has been recently studied in Ref.~\cite{Braglia:2020auw}, where it has been named ``Early Modified Gravity model''.\footnote{\, Note, that a different class of models with the same name ``Early Modified Gravity'' exists in the literature~\cite{Brax:2013fda, Pettorino:2014bka, Lima:2016npg}.}

For this model, the combination of the {\it Planck} 2018 + BAO + FS + Pantheon + R19 + H0LiCOW datasets gives $H_0 = 71.00^{+0.87}_{-0.79}{\rm\,km\,s^{-1}\,Mpc^{-1}}$ at 68\% CL~\cite{Braglia:2020auw}, reducing the Hubble tension at $1.4\sigma$ with R20. The analysis already includes a Gaussian prior on $H_0$.

This result is in agreement with the same analysis performed independently in Ref.~\cite{Abadi:2020hbr} and called Conformally Coupled Modified Gravity. For {\it Planck} 2018 this model gives $H_0 = 67.98^{+0.63}_{-1.1}{\rm\,km\,s^{-1}\,Mpc^{-1}}$ at 68\% CL~\cite{Abadi:2020hbr}, reducing the Hubble tension at $3.7\sigma$.

\subsubsection{Screened Fifth Forces:}

The reduction of the fifth force strength that occurs in regions of strong gravitational field (known as {\it screening}) is a fairly generic property of scalar-tensor gravity theories, see e.g.\ Refs.~\cite{Burrage:2017qrf, Sakstein:2018fwz}. Due to this behaviour, the distance ladder inferred from Cepheid measurements could be altered if a screened fifth force is present~\cite{Jain:2012tn}.

In Ref.~\cite{Desmond:2019ygn} the assumption that the physics of Cepheid stars is identical across the galaxies used to build the cosmic distance ladder is questioned. The authors consider different models in which a screened fifth force is realized and show how altering the Cepheid calibration of supernova distances leads to a possible reduction of the disagreement in the Hubble constant measurements. In addition, in Ref.~\cite{Desmond:2020wep} it is shown that a fifth force is also effective for the TRGB calibration of the distance ladder, lowering the inferred $H_0$ value.

\subsection{\"{U}ber-gravity}

The \"{u}ber gravity model is a fixed point in the space of the gravity models obtained from varying the Ricci scalar~\cite{Khosravi:2016kfb}. This model mimics the Einstein-Hilbert theory in the high-curvature regime, while in the low-curvature regime it predicts a sharp transition at a model-dependent Ricci scale $R_0$. The cosmological model embedded in this theory, the \"{u}$\Lambda$CDM model, is characterised by a density-dependent transition between $\Lambda$CDM and a phase in which the Ricci scalar is constant~\cite{Khosravi:2017aqq}.

This scenario has been proposed to alleviate the Hubble tension in Ref.~\cite{Khosravi:2017hfi}. The combined analysis of {\it Planck} 2015 TT + R16 + BAO for this model estimates $H_0=70.6^{+1.1}_{-1.3}{\rm\,km\,s^{-1}\,Mpc^{-1}}$ at 68\% CL~\cite{Khosravi:2017hfi}, in agreement with R20 at $1.5\sigma$. However, this result relies exclusively on {\it Planck} temperature data at high multipoles, on a Gaussian prior on $H_0$, as measured by R16, and on BAO measurements. In Ref.~\cite{Khosravi:2017hfi} it is shown that for {\it Planck} 2015 alone the constraints are largely relaxed, and therefore a possible agreement with R20 would be possible within one standard deviation. An updated analysis for this scenario has been performed in Ref.~\cite{Moshafi:2020rkq}, and {\it Planck} 2018 ({\it Planck} 2018 + CMB lensing + R19) measurements provide a value of $H_0 = 70.7^{+1.4}_{-2.6}{\rm\,km\,s^{-1}\,Mpc^{-1}}$ at ($H_0 = 73.2\pm1.3{\rm\,km\,s^{-1}\,Mpc^{-1}}$) at 68\% CL~\cite{Moshafi:2020rkq}, in agreement with R20 within $1.3\sigma$ ($1\sigma$).

\subsection{Galileon gravity}

The Covariant Galileon model is a theory of modified gravity in which the accelerated expansion rate of the Universe is driven by a scalar field $\varphi$, whose Lagrangian is invariant under the Galilean shift symmetry by a constant vector $b_\mu$, $\partial_\mu\varphi \to \partial_\mu\varphi + b_\mu$~\cite{Nicolis:2008in, Deffayet:2009wt}. One aspect of this model is that the background component of the Galileon field $\phi$ is described by a ``tracker'' evolution, $H\mathrm{d}\phi/\mathrm{d}t \propto \xi$, where $\xi$ is a constant~\cite{DeFelice:2010pv}. Once the Galileon field has reached the tracker solution, its energy density contributes appreciably to the total energy budget,  reaching the present value
\begin{equation}
    \Omega_{\phi 0} = \frac{c_2}{6}\xi^2 - 2c_3\xi^3 + c_4\frac{15}{2}\xi^4 + c_5\frac{7}{3}\xi^5\,.
\end{equation}
Depending on the highest exponent for $\xi$, we refer either to the cubic ($c_4 = c_5 = 0$), quartic $(c_5 = 0)$, or quintic Galileon model.

In Ref.~\cite{Renk:2017rzu} the Galileon gravity scenario has been proposed to solve the Hubble tension. When the total neutrino mass is allowed to vary in addition to the standard parameters, the combination {\it Planck} 2015 TT + CMB lensing + BAO leads to the values $H_0=71.6\pm 2.1{\rm\,km\,s^{-1}\,Mpc^{-1}}$, $H_0=72.4\pm 2.0{\rm\,km\,s^{-1}\,Mpc^{-1}}$ and $H_0=72.3\pm 2.1{\rm\,km\,s^{-1}\,Mpc^{-1}}$ for the several possible Galileon scenarios (cubic, quartic and quintic, respectively), all with 95\% CL errors. The Hubble tension is therefore reduced in these cases within $1\sigma$, even if the BAO observations are also included. Updated results with {\it Planck} 2018 are not yet available.

A similar scenario has been analysed in Ref.~\cite{Frusciante:2019puu}, where the authors studied a generalized cubic covariant Galileon scenario. {\it Planck} 2015 TT ({\it Planck} 2015 TT + BAO + RSD + JLA) gives in this case $H_0=72^{+8}_{-5}{\rm\,km\,s^{-1}\,Mpc^{-1}}$ ($H_0=68.4\pm0.9{\rm\,km\,s^{-1}\,Mpc^{-1}}$) at 95\% CL~\cite{Frusciante:2019puu}, solving the Hubble tension within $1\sigma$ ($3.4\sigma$). Once the {\it Planck} high-$\ell$ polarization is included, the Hubble tension is however restored.

In Ref.~\cite{Zumalacarregui:2020cjh}, instead, it has been argued that the problem can be overcome in the Enhanced Early Gravity model, i.e.\ an exponentially coupled cubic Galileon scenario, where {\it Planck} 2018 + BAO gives $H_0=68.7\pm1.5{\rm\,km\,s^{-1}\,Mpc^{-1}}$ at 68\% CL, relaxing the Hubble tension down to $2.3\sigma$ level. 

Within the subclass of Generalized Proca interactions~\cite{Heisenberg:2014rta,Allys:2015sht,Jimenez:2016isa}, the authors of Ref.~\cite{Heisenberg:2020xak} focus on the cubic Galileon scenario, based on a vector field for the solution of the Hubble constant tension, mainly due to the phantom-like behaviour of dark energy. Using a combination of {\it Planck} 2018 + BAO + Pantheon + R19, they find $H_0=70.1\pm0.76{\rm\,km\,s^{-1}\,Mpc^{-1}}$ at 68\% CL~\cite{Heisenberg:2020xak}, alleviating the Hubble tension at $2.4\sigma$, but including a Gaussian prior on $H_0$. The same dataset combination without R19 restores the tension above $3\sigma$.

Finally, in Ref.~\cite{Peirone:2019aua} a Galileon ghost condensate model has been studied to alleviate the Hubble constant tension. For this scenario {\it Planck} 2015 TT ({\it Planck} 2015 TT + BAO + RSD + JLA) gives $H_0=69.3^{+3.6}_{-3.0}{\rm\,km\,s^{-1}\,Mpc^{-1}}$ ($H_0=68.1\pm1.1{\rm\,km\,s^{-1}\,Mpc^{-1}}$) at 95\% CL~\cite{Peirone:2019aua}, solving the Hubble tension within $1.8\sigma$ ($3.6\sigma$).

\subsection{Nonlocal gravity}

The introduction of quantum gravity effects in the Einstein-Hilbert action leads to the presence of non-local effects that typically signal the presence of quantum properties corresponding to the local fundamental action of gravity, including the effect of quantum fluctuations~\cite{Birrell:1982ix, Mukhanov:2007zz}. Among the possible nonlocal gravity models there is the RR scenario, in which the Einstein-Hilbert action in Eq.~\eqref{eq:EHaction} is modified as~\cite{Maggiore:2014sia}
\begin{equation}
    \label{eq:RRaction}
    \mathcal{S} = \frac{1}{2\kappa^2} \int \mathrm{d}^4 x \sqrt{-g} \,\left[\mathcal{R} - \frac{m^2}{6}\mathcal{R}\frac{1}{\Box^2}\mathcal{R}\right] + \mathcal{S}_m\,,
\end{equation}
where $m$ is a new mass parameter. In in this scenario, the nonlocal term acts as an effective dark energy with a phantom equation of state.

The RR model is relevant for solving the Hubble tension~\cite{Belgacem:2017cqo}. For this particular scenario, the combination of {\it Planck} 2015 + BAO + JLA results in $H_0=69.49^{+0.79}_{-0.80}{\rm\,km\,s^{-1}\,Mpc^{-1}}$ at 68\% CL~\cite{Belgacem:2017cqo}, reducing the Hubble constant tension down to $2.5\sigma$. 

Updated results for nonlocal gravity models have been performed in Ref.~\cite{Belgacem:2020pdz}. While the RR model does not satisfy the Lunar Laser Ranging constraints, the RT model~\cite{Maggiore:2013mea} works better and for {\it Planck} 2018 + Pantheon + BAO the minimal case gives $H_0=68.74^{+0.59}_{-0.51}{\rm\,km\,s^{-1}\,Mpc^{-1}}$ at 68\% CL~\cite{Belgacem:2020pdz}, reducing the tension with R20 at $3.2\sigma$.

\subsection{\, Unimodular gravity}

Another gravitational theory having close resemblance to the Einstein gravity is the unimodular gravity. The unimodular gravity is obtained by adding the unimodular condition~\cite{Anderson:1971pn} to the Einstein-Gravity action through the Lagrange multiplier $\lambda$~\cite{LinaresCedeno:2020uxx}.

The possibility that this gravitational theory could help in resolving of the $H_0$ tension has been recently discussed in Ref.~\cite{LinaresCedeno:2020uxx}. The $H_0$ tension can be alleviated by allowing for a non-gravitational interaction between the dark matter and the dark energy fluids within this gravitational context. Considering four different interaction rates between the dark matter and dark energy, namely, Sudden Transfer Model (Model 1), Anomalous Decay of the Matter Density (Model 2), Barotropic Model (Model 3) and Continuous Spontaneous Localization (Model 4), the best estimations of the Hubble constant for the combined dataset including {\it Planck} 2018 CMB distance priors + Pantheon + R19 + H0LiCOW, are, respectively, $H_0 = 73.4^{+1.5}_{-0.6}{\rm\,km\,s^{-1}\,Mpc^{-1}}$ at 68\% CL~\cite{LinaresCedeno:2020uxx} (Model 1), $H_0 = 73.2^{+1.4}_{-0.9}{\rm\,km\,s^{-1}\,Mpc^{-1}}$ at 68\% CL~\cite{LinaresCedeno:2020uxx} (Model 2), $H_0 = 70\pm1{\rm\,km\,s^{-1}\,Mpc^{-1}}$ at 68\% CL~\cite{LinaresCedeno:2020uxx} (Model 3) and $H_0 = 72\pm1{\rm\,km\,s^{-1}\,Mpc^{-1}}$ at 68\% CL~\cite{LinaresCedeno:2020uxx} (Model 4). For Model 1, Model 2 and Model 4, the $H_0$ values are in agreement with R20 within $1\sigma$, while for Model 3 the $H_0$ tension with R20 is reduced to $2 \sigma$. However, when only CMB data are considered, the value of $H_0$ is unconstrained, and therefore the Gaussian priors on the Hubble constant are essential in the analysis to constrain it. A complete data analysis to {\it Planck} 2018 observations is missing.

\subsection{\, Scale $-$ dependent scenario of gravity}

A cosmological model with a scale $-$ dependent scenario of gravity has been proposed in Ref.~\cite{Alvarez:2020xmk} to potentially alleviate the Hubble tension. Unfortunately, the data analysis is missing.

\subsection{\, VCDM}

A cosmological theory where the cosmological constant term $\Lambda$ of the standard $\Lambda$CDM scenario is replaced by a free function $V (\phi)$, without introducing any extra physical degrees of freedom, was proposed in Ref.~\cite{DeFelice:2020eju}. The `$V$' of $V$CDM therefore stands for the free function $V (\phi)$. The authors of Ref.~\cite{DeFelice:2020cpt} studied a specific model in this context finding that the $H_0$ tension can be alleviated. {\it Planck} 2018 + BAO + Pantheon gives $H_0 = 71.73^{+0.58}_{-0.29}{\rm\,km\,s^{-1}\,Mpc^{-1}}$ at 95\% CL~\cite{DeFelice:2020cpt}, alleviating the tension with R20 at $1.1\sigma$.

\section{Inflationary models}
\label{inflat}

Inflation~\cite{Brout:1977ix, Guth:1980zm, Sato:1980yn} is a period of accelerated expansion that is believed to take place at the very early stages in the history of the Universe. It was first proposed to explain the homogeneity, isotropy, and flatness observed in the CMB, as well as the lack of relic monopoles~\cite{Guth:1982ec, Starobinsky:1982ee}.

A period of inflation can be achieved when the expansion rate of the Universe is driven by the energy density of a rolling scalar field $\phi$, the inflaton~\cite{Linde:1981mu, Albrecht:1982wi}. In this framework, the quantum fluctuations of the inflaton field seed the density perturbations that are observed in the CMB, and later develop into the large scale structures observed~\cite{Colless:2001gk, Blanton:2017qot}. Within a specific model of inflation, it is possible to characterize various observables, such as the scalar spectral index $n_s$ and its running $\mathrm{d}n_s /\mathrm{d} \log k$, the tensor-to-scalar ratio $r$, the spectral index of tensor perturbations $n_T$, and the non-Gaussianity parameter $f_{\rm NL}$. To date, the most stringent constraints on the theory of inflation come from the observations of the CMB by the {\it Planck} satellite, which include the features of the power spectrum~\cite{Akrami:2018odb} and the bispectrum~\cite{Akrami:2019izv} of temperature anisotropies.

With such a successful beginning, the theory of inflation got a wide attention in the cosmological community and consequently this theory was intensively investigated over the years, see e.g\ Refs.~\cite{Barrow:1986jd,Burd:1988ss,Barrow:1988xh,Barrow:1988xi,Gottlober:1990um,Barrow:1993hn,Liddle:1994dx,Barrow:1994nt,Barrow:1995fj,Barrow:1995xb,Linde:1995rv,Linde:1998iw,Kofman:2002cj,Linde:2003hc,Albrecht:2004ke,Nunes:2005ra,Bousso:2006ge,Barrow:2006dh,Watson:2006px,Rocher:2006fh,Weinberg:2008hq,Turner:2008zza,Koivisto:2008xf,Baumann:2008bn,Koivisto:2009sd,Ringstrom:2009zz,Pal:2009sd,Weinberg:2009wa,Kaneda:2010qv,Yamauchi:2011qq, Mithani:2013ed,Tsujikawa:2013ila,Ellis:2013iea,Wan:2014fra,Creminelli:2014fca,Freese:2014nla,Basilakos:2015sza,Barrow:2016qkh,Barrow:2016wiy,Paliathanasis:2017apr} (see also Refs.~\cite{Linde:2000kn, Brandenberger:2016uzh} and references therein).\footnote{\, In this context, we refer to a very interesting class of models known as ``quintessential inflation''~\cite{Peebles:1998qn, Peloso:1999dm,Kaganovich:2000fc, Yahiro:2001uh,Sami:2004xk, Rosenfeld:2005mt,Geng:2015fla, deHaro:2012dn, deHaro:2016hpl, deHaro:2016hsh,deHaro:2016ftq, deHaro:2016cdm,deHaro:2017nui, Geng:2017mic,AresteSalo:2017lkv, Haro:2018zdb,Haro:2019gsv,deHaro:2019oki, Haro:2019umj,Dimopoulos:2019gpz, Haro:2019peq} that try to connect two distant accelerating phases of the Universe $-$ inflation and quintessence.} In this section, we shall point out some recent works where modifications of the early Universe physics, either through a suitable choice of the inflationary potential or by modifying the primordial power spectrum, allow the Hubble constant tension to be alleviated.

Since the data points (referring to the number of models constraining $\Omega_m h^2$, $H_0$ and $r_d h$) in Sections~\ref{inflat}--\ref{others} are very small in number, we have combined the Sections~\ref{inflat}--\ref{others} into two Figures~\ref{fig:chapter11-14_H0Om} and~\ref{fig:chapter11-14_whisker}.

\subsection{Exponential inflation}

In the single-field inflation model, there exists a degeneracy between the spectral index of the primordial scalar power spectrum, $n_s$, and $N_{\rm eff}$, see e.g.\ Ref.~\cite{Bowen:2001in}. Therefore, in principle, it is possible to build a model in which the interplay between the inflationary mechanism and the presence of additional dark radiation may alleviate the Hubble tension, due to the strong correlation between $\Delta N_{\rm eff}$ and $H_0$. In the following, we shall discuss the inflationary models that embed additional dark radiation.

The authors of Ref.~\cite{Tram:2016rcw} re-examine various inflationary models in light of the presence of additional dark radiation. They study the large-field inflation scenario with a potential $V(\phi) \propto \phi^2$~\cite{Linde:1983gd}, the natural inflation model~\cite{Freese:1990rb, Adams:1992bn}, the Starobinsky model~\cite{Starobinsky:1980te}, and the power-law inflation paradigm (PLI)~\cite{Lucchin:1984yf}, in which the potential is given in terms of an amplitude $M$ and an index $\alpha>0$ by:\footnote{\, PLI earns its name from the fact that the exact solution for the scale factor in the model is $a(t) \propto t^{2/\alpha^2}$.}
\begin{equation}
    V(\phi) = M^4 \exp\left(-\alpha\frac{\phi}{M_{\rm Pl}}\right)\,.
\end{equation}
In the $\Lambda$CDM, the PLI model is excluded since it predicts a value of the tensor-to-scalar ratio $r = 8(1-n_s)$ that lies above the limit from current observations~\cite{Planck:2013jfk}. However, when including an extra component of dark radiation, the relatively large value of $n_s$ predicted within PLI, turns into an opportunity to address the Hubble tension since, using {\it Planck} 2015 TT + CMB lensing datasets, the Hubble constant results in $H_0=73.6 \pm 0.95{\rm\,km\,s^{-1}\,Mpc^{-1}}$ at 68\% CL~\cite{Tram:2016rcw}, in agreement with R20 within $1\sigma$ (see also Ref.~\cite{Guo:2017qjt} for similar results).
For the very same scenario, the addition of the CMB polarization data and a lower value for the optical depth $\tau$, as preferred by the new {\it Planck} 2018 power spectra, restores the Hubble constant tension above 3 standard deviations~\cite{DiValentino:2016ucb}. The same results are confirmed even when an origin of the Universe from the quantum landscape multiverse is considered (see Ref.~\cite{DiValentino:2016ziq}).

\subsection{Reconstructed Primordial Power Spectrum}

A possibility for solving the Hubble tension is to change the primordial power spectrum (PPS). 
In Ref.~\cite{Liu:2019dxr} it has been shown that band-limited features in the PPS can not resolve the Hubble tension.

In Ref.~\cite{Hazra:2018opk}, instead, the shape of the PPS is reconstructed by implementing a Modified Richardson-Lucy algorithm (MRL), and assuming the fitting of the {\it Planck} 2015 TT data, $H_0$ from R18 and $S_8$ from the cosmic shear data. This reconstructed model allows the data to be perfectly in agreement with the measured R20 Hubble constant. The reconstructed form of the PPS will have a suppression of power at large scales and sharp fluctuations at wave numbers larger than $0.02{\rm\,Mpc^{-1}}$.

A generalization is performed in Ref.~\cite{Keeley:2020rmo}, where a class of PPS, that continuously deforms between the best-fit power-law and the MRL-reconstructed PPS, is parameterized. This interpolation is called ``deformation model'', and the Hubble constant is degenerate with the new degree of freedom in the PPS. Using {\it Planck} 2018 TT, the Hubble constant is $H_0=70.2\pm 1.2{\rm\,km\,s^{-1}\,Mpc^{-1}}$ at 68\% CL~\cite{Keeley:2020rmo}, solving the tension with R20 at $1.7\sigma$. However, it is unclear whether this result holds once polarization data are included in the analysis.

\subsection{Lorentzian Quintessential Inflation}

In Ref.~\cite{AresteSalo:2021lmp} the authors show that the quintessential inflation, coming from the Lorentzian distribution introduced in Refs.~\cite{Benisty:2020xqm,Benisty:2020qta}, agrees with the recent observations and is in agreement with R20. In particular they find for {\it Planck} 2018 CMB distance priors + BAO + Pantheon + CC + R19, $H_0=71.75\pm 0.89{\rm\,km\,s^{-1}\,Mpc^{-1}}$ at 68\% CL~\cite{AresteSalo:2021lmp}.

\subsection{Harrison-Zel'dovich spectrum}

Because of the existing degeneracy between $n_s$, $N_{\rm eff}$, and $H_0$, in Ref.~\cite{DiValentino:2018zjj} the authors pointed out that in light of the Hubble tension a Harrison-Zel'dovich~\cite{Harrison:1969fb, Peebles:1970ag, Zeldovich:1972zz} primordial power spectrum $n_s=1$ is not ruled out by the data.

\section{Modified recombination history}
\label{RecombH}

Early recombination scenarios can also be a possible route to obtain a higher values of the Hubble constant and thus alleviate the $H_0$ tension. We refer to Figures~\ref{fig:chapter11-14_H0Om} and~\ref{fig:chapter11-14_whisker} for an overall idea about the models in this section.

In Ref.~\cite{Chiang:2018xpn} a general phenomenological model that modifies the timing and width of the recombination processes has been considered. {\it Planck} 2015 ({\it Planck} 2015 + BAO) gives $H_0=67.17^{+2.04}_{-2.17}{\rm\,km\,s^{-1}\,Mpc^{-1}}$ ($H_0=68.17^{+1.18}_{-1.14}{\rm\,km\,s^{-1}\,Mpc^{-1}}$) at 68\% CL~\cite{Chiang:2018xpn}, alleviating the tension at $2.5 \sigma$ ($2.8\sigma$). In Ref.~\cite{Liu:2019awo} it is possible to find a different approach.

\subsection{Effective Electron Rest Mass}
\label{sec:effectiveelectronrestmass}

A modified effective electron rest mass $m_e$ during the cosmological recombination era~\cite{Hart:2017ndk} could provide a mechanism to reduce the Hubble constant tension. In Ref.~\cite{Hart:2019dxi} it has been shown that {\it Planck} 2018 + BAO gives $H_0=69.1\pm1.2{\rm\,km\,s^{-1}\,Mpc^{-1}}$ at 68\% CL~\cite{Hart:2019dxi} for a varying $m_e$, lowering the $H_0$ tension down to the $2.3\sigma$ level. The concordance model results in a larger electron rest mass $m_e = (1.0078 \pm 0.0067) m_{e,0}$ at 68\% CL~\cite{Hart:2019dxi}.

\subsection{Time Varying Electron Mass}

In Ref.~\cite{Sekiguchi:2020teg}, an explicit model showing how the recombination history of the Universe can be modified has been proposed, in which a time varying electron mass $m_{e}$ plays a key role. Specifically, a time varying electron mass can shift the recombination epoch $z_*$ and the drag epoch $z_d$ from the baseline model without affecting the CMB power spectra. Thus, considering the varying electron mass within the $\Omega_k\Lambda$CDM model, the best estimated value of the Hubble constant for {\it Planck} 2018 + BAO + Pantheon is $H_0 = 72.3^{+2.7}_{-2.8}{\rm\,km\,s^{-1}\,Mpc^{-1}}$ at 68\% CL~\cite{Sekiguchi:2020teg}. This reconciles the tension with R20 at less than $1\sigma$.

\subsection{Axi-Higgs model}

The authors of Ref.~\cite{Fung:2021wbz} present a simple model in which a light axion is coupled to the Higgs field. 
In this scenario, the Higgs vacuum expectation value in the early Universe is larger than its measured value, thus modifying the electron rest mass and possibly alleviating the Hubble tension (see Section~\ref{sec:effectiveelectronrestmass}). The largest estimate presented in Ref.~\cite{Fung:2021wbz} for the Hubble constant is achieved with an analysis a posteriori of the results obtained in~\cite{Hart:2019dxi} for {\it Planck} 2018 + BAO, with $m_e$ free to vary. Assuming a model with non-linear BBN, the analysis gives $H_0 = 69.24\pm0.68{\rm\,km\,s^{-1}\,Mpc^{-1}}$ at 68\% CL~\cite{Fung:2021wbz}, alleviating the tension with R20 at $2.6\sigma$. A complete full CMB data analysis is however missing.

\subsection{Primordial magnetic fields}

Additional small-scale, mildly non-linear inhomogeneities in the baryon density changing the recombination history could be a possible route to alleviate the $H_0$ tension~\cite{Jedamzik:2020krr}. These might be caused by the evolution of primordial magnetic fields (PMF) prior to recombination.

Using the model proposed in Ref.~\cite{Jedamzik:2011cu} and analysed in Ref.~\cite{Jedamzik:2018itu}, the combination of {\it Planck} 2018 + CMB lensing + R19 + H0LiCOW + MCP gives $H_0 = 71.03\pm0.74{\rm\,km\,s^{-1}\,Mpc^{-1}}$ at 68\% CL~\cite{Jedamzik:2020krr}, reducing the Hubble tension at $1.4\sigma$. However, Gaussian priors on the Hubble constant are already included in this analysis, inducing a possible bias in the result.

Another possibility is to have a weaker impact on recombination, with only a tiny fraction of the total volume in high density regions. {\it Planck} + CMB lensing + R19 + H0LiCOW + MCP results in a value of $H_0 = 69.81\pm0.62{\rm\,km\,s^{-1}\,Mpc^{-1}}$ at 68\% CL~\cite{Jedamzik:2020krr}, alleviating the Hubble tension down to the $2.4\sigma$ level. Nevertheless, the Gaussian priors are also already included in this analysis, as in the previous case, which may bias the result.


\begin{figure*}
\centering

\includegraphics[width=0.85\textwidth, right]{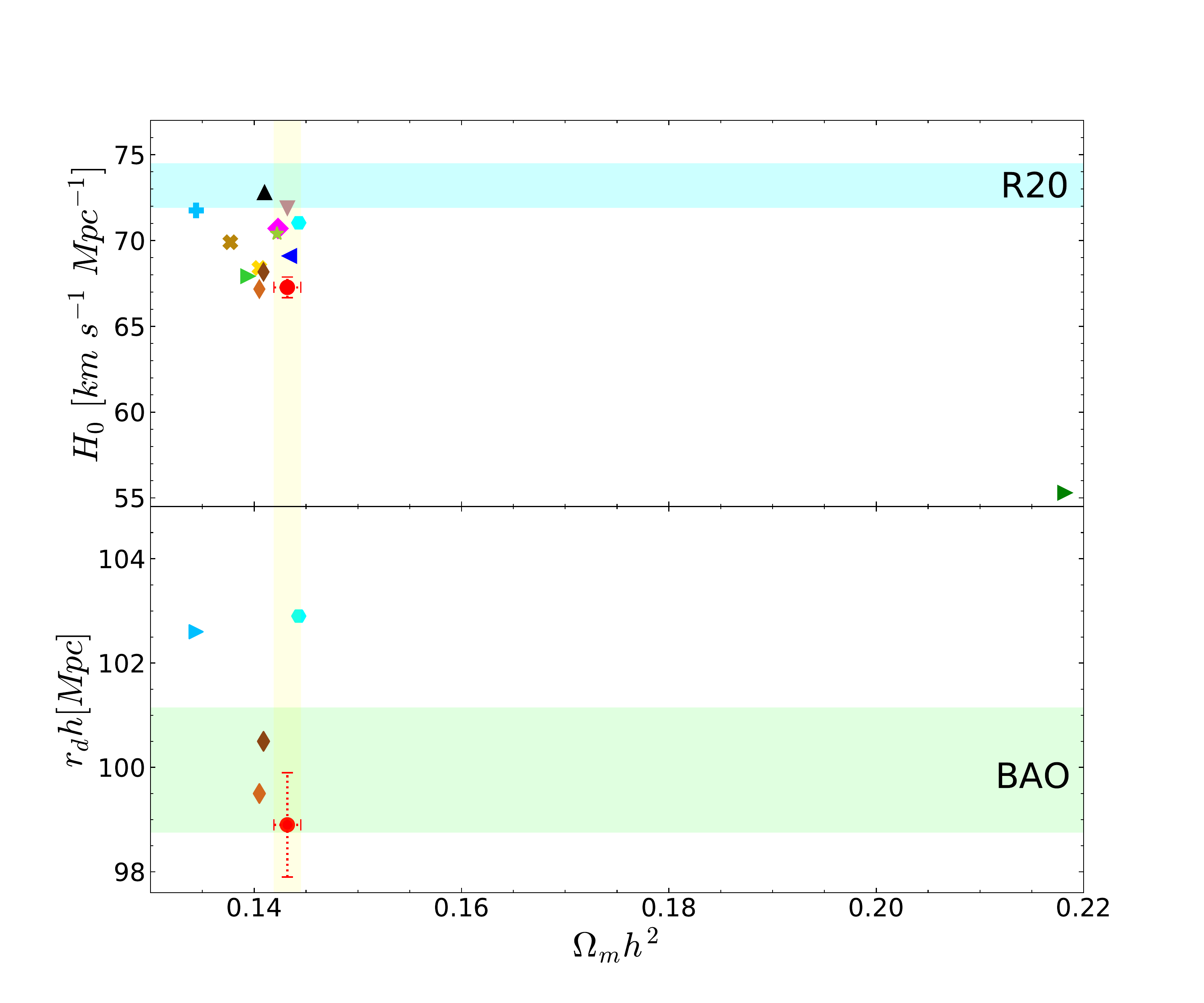}
\includegraphics[width=0.85\textwidth, right]{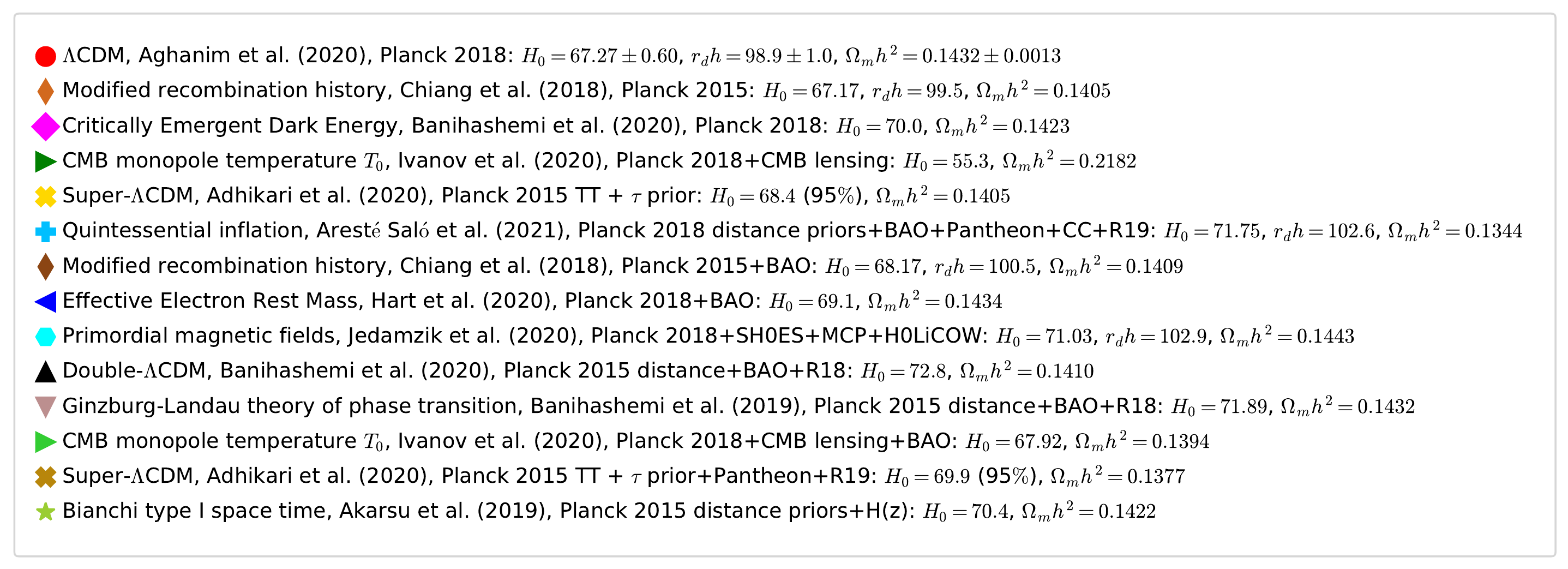}
\caption{Estimated values of the current matter energy density $\Omega_mh^2$, Hubble constant $H_0$ and sound horizon $r_dh$ in terms of various data points for different models discussed throughout Sections~\ref{inflat}-\ref{others}. The cyan horizontal band corresponds to the $H_0$ value measured by R20~\cite{Riess:2020fzl}, the yellow vertical band to the $\Omega_mh^2$ value estimated by {\it Planck} 2018~\cite{Aghanim:2018eyx} in a $\Lambda$CDM scenario, and the light green horizontal band to the $r_dh$ value measured by BAO data. The points sharing the same symbol refer to the same model in the same paper, and the different colors indicate a different dataset combination.}
\label{fig:chapter11-14_H0Om}
\end{figure*}

\begin{figure*}
\includegraphics[width=\textwidth]{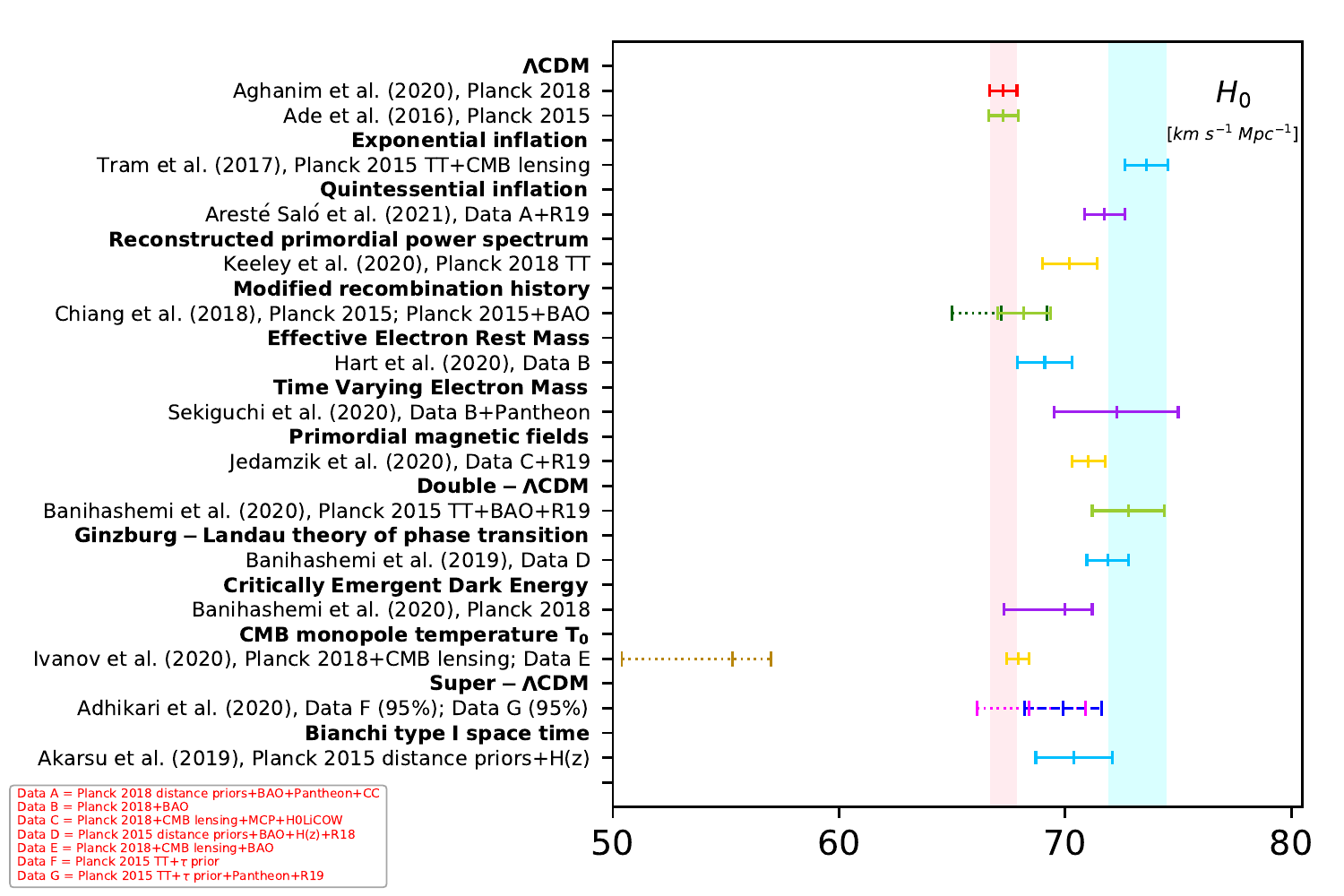}
\caption{Whisker plot with the 68\% (95\% if dashed) marginalized Hubble constant constraints for the models of Sections~\ref{inflat}-\ref{others}. The cyan vertical band corresponds to the $H_0$ value measured by R20~\cite{Riess:2020fzl} and the light pink vertical band corresponds to the $H_0$ value estimated by {\it Planck} 2018~\cite{Aghanim:2018eyx} in a $\Lambda$CDM scenario. For each line, when more than one error bar is shown, the dotted one corresponds to the {\it Planck} only constraint on the Hubble constant, while the solid one to the different dataset combinations reported in the red legend, in order to appreciate the shift due to the additional datasets.}
\label{fig:chapter11-14_whisker}
\end{figure*}

\section{Physics of the critical Phenomena}
\label{CritPhen}

Since the physics operating at late time seems to be different from the physics of early time, yet another interesting possibility could be a phase transition in the dark sector. The critical phenomena studied extensively the idea of a phase transition, in which local interactions of a many-body system produce a global phase transition, if a free parameter of the model is lowered beyond a critical point. 

We refer to Figures~\ref{fig:chapter11-14_H0Om} and~\ref{fig:chapter11-14_whisker} summarizing the performance of the models discussed in this section in light of the Hubble constant tension.

\subsection{Double-\texorpdfstring{$\Lambda$}{Lambda}CDM ($\metalambda$CDM)}

The $Double-\Lambda$ Cold Dark Matter ($\metalambda$CDM) scenario is inspired by the Ising model, a classic model of critical phenomena describing the phase transition from para-magnet to ferro-magnet at Curie temperature. This cosmological scenario assumes a cosmological constant with two values before a transition redshift and with a single value afterwards.
In Ref.~\cite{Banihashemi:2018oxo} it has been shown that, with this phase transition in the dark sector, the Hubble constant tension can be solved. 
Considering a $\chi^2$ analysis, and the combination of {\it Planck} 2015 TT + BAO + R19, the Hubble constant is $H_0 = 72.8 \pm 1.6 {\rm\,km\,s^{-1}\,Mpc^{-1}}$ at 68\% CL~\cite{Banihashemi:2018oxo}. The $H_0$ tension with R20 is therefore solved within $1\sigma$, but including already a Gaussian prior on $H_0$.

\subsection{Ginzburg-Landau theory of phase transition}

In the Ginzburg-Landau theory of dark energy a phase transition happens, causing a spontaneous symmetry breaking, in the Landau approximation. Considering a $\chi^2$ analysis, and the combination of {\it Planck} 2015 CMB distance priors on the angular size of horizon at decoupling and $\Omega_mh^2$ + BAO + R18 + quasars $H(z)$ data, the Hubble constant is $H_0=71.89 \pm 0.93{\rm\,km\,s^{-1}\,Mpc^{-1}}$ at 68\% CL~\cite{Banihashemi:2018has}. While the $H_0$ tension with R20 is therefore solved within $1\sigma$, this result relies on a non full CMB data analysis and includes already a Gaussian prior on $H_0$.

\subsection{Critically Emergent Dark Energy}

Based on the physics of the critical phenomena, a dark energy model named as Critically Emergent Dark Energy was recently proposed in Ref.~\cite{Banihashemi:2020wtb}. The evolution of the dark energy in this model takes the form:
\begin{equation}
    \Omega_{\rm DE} (z) = (1- \Omega_m - \Omega_r)\; \sqrt{\frac{z_c -z}{z_c}},
\end{equation}
where $z_c$ is the transition redshift from which dark energy starts to emerge, and the corresponding DE equation of state is
\begin{equation}
    w_{\rm DE}(z)=-1-\frac{1+z}{6(z_c-z)}\,.
\end{equation}
For this model, a fit to {\it Planck} 2018 data results in a value of $H_0=70.0^{+1.2}_{-2.7}{\rm\,km\,s^{-1}\,Mpc^{-1}}$ at 68\% CL~\cite{Banihashemi:2020wtb}, which solves the tension with R20 within $1.8\sigma$.

\section{Alternative proposals}
\label{others}

In this section we include a number of models that can not be catalogued in any of the sections detailed before, i.e.\ other than dark energy, dark radiation, interacting models or modified gravity scenarios. In Figures~\ref{fig:chapter11-14_H0Om} and~\ref{fig:chapter11-14_whisker} we have shown the viability of the models in light of the Hubble constant tension.

\subsection{Local Inhomogeneity}

Inhomogeneities in the density distribution could lead to a modification of the expansion rate over some finite region of spacetime; a domain average density in the locally observed region would lead to a modification of the estimate for the local Hubble parameter from the value inferred from using the global background energy density, see Section~\ref{sec:InhomogeneousAndAnisotropicSolutions} for the relevant literature.

In Ref.~\cite{Kasai:2019yqn} the possibility that the Hubble tension could be solved within the general relativistic framework of perturbation theory in an inhomogeneous Universe is investigated. The authors find that the crucial point is the first-order effect due to inhomogeneities at linear order in perturbation theory.

\subsection{Bianchi type I spacetime}

In Ref.~\cite{Akarsu:2019pwn} a simple anisotropic correction to the standard $\Lambda$CDM model by replacing the spatially flat FLRW metric with the Bianchi type-I metric has been investigated. Adopting a compilation of 36 $H(z)$ measurements from CC, BAO signal in both galaxy and Ly-$\alpha$ forest distributions, the authors estimate $H_0=70.4\pm1.7{\rm\,km\,s^{-1}\,Mpc^{-1}}$ at 68\% CL~\cite{Akarsu:2019pwn} in the anisotropic $\Lambda$CDM scenario, in combination with the {\it Planck} 2015 distance prior, which is in agreement with both the CMB and the R20 values within $2\sigma$. A full analysis considering CMB data is still missing, but the scenario is highly promising.

\subsection{Scaling Solutions}
\label{sec:scaling_solutions}

The inclusion of the backreaction from inhomogeneities in the cosmological expansion within GR has been proposed in Refs.~\cite{Buchert:1999er, Buchert:2001sa, Buchert:2011sx, Buchert:2019mvq}, see also Section~\ref{sec:InhomogeneousAndAnisotropicSolutions}. This scheme is generally referred to in the literature as the ``Buchert equations''.

In Ref.~\cite{Heinesen:2020sre} a class of `scaling solutions' satisfying this scalar averaging scheme has been proposed to solve the Hubble tension, while at the same time being in agreement with a slightly positively curved Universe as measured by Planck~\cite{Aghanim:2018eyx, DiValentino:2019qzk, Handley:2019tkm}. In fact, in this generic average model, there is a dynamical curvature, i.e.\ curvature and structure in the matter distribution are dynamically coupled within GR and without the necessity of introducing a Dark Energy component. A full data analysis is however missing for this interesting possibility.

In Ref.~\cite{Bolejko:2017fos}, a GR fluid simulation of the LSS that includes a nonlinear evolution of structures leads to a negative emerging spatial curvature and to a value of the Hubble constant $H_0 = 72.5 \pm 2.1{\rm \,km\,s^{-1}\,Mpc^{-1}}$. Instead, neglecting the inhomogeneities, the same simulation finds a lower value $H_0 = 68.1 \pm 2.0{\rm \,km\,s^{-1}\,Mpc^{-1}}$, in agreement with the findings in Ref.~\cite{Heinesen:2020sre}. Instead, an independent analysis following a fully inhomogeneous, anisotropic relativistic simulation finds that these effects alone are not sufficient to reconcile the discrepancy represented by the Hubble tension~\cite{Macpherson:2018akp}.

\subsection{CMB monopole temperature \texorpdfstring{$T_0$}{T0}}

In Ref.~\cite{Ivanov:2020mfr} the possibility of varying the CMB monopole temperature $T_0$, typically fixed when considering the CMB data, is explored to solve the Hubble tension. This is in fact fixed because of the extremely good precision of measurements of $T_0$ from the Cosmic Background Explorer (COBE) Far Infrared Absolute Spectrophotometer (FIRAS) data, molecular lines, and balloon-borne experiments, but should in principle be considered as an extra free cosmological parameter to be varied in the Bayesian analysis~\cite{Yoo:2019dyl}. Using {\it Planck} 2018 + CMB lensing ({\it Planck} 2018 + CMB lensing + BAO), Ref.~\cite{Ivanov:2020mfr} finds $H_0=55.3^{+1.7}_{-4.9}{\rm\,km\,s^{-1}\,Mpc^{-1}}$ ($H_0=67.92^{+0.49}_{-0.51}{\rm\,km\,s^{-1}\,Mpc^{-1}}$) at 68\% CL, increasing the tension with R20 at several standard deviations ($3.8\sigma$) when varying $T_0$. Therefore, even if a strong anti-correlation is present between $T_0$ and $H_0$, this is not enough for solving the Hubble tension because the CMB data prefer the wrong direction.

\subsubsection{Open and Hotter Universe:}

Another interesting possibility to alleviate the Hubble constant tension may arise from considering a non-zero spatial curvature together with a free CMB temperature. In Ref.~\cite{Bose:2020cjb}, the authors consider such a scenario in light of {\it Planck} 2015, BAO and R19 data. The results show that both {\it Planck} 2015 and BAO prefer an open and hotter Universe with significantly higher expansion rate and the estimated values of $H_0$ are in agreement with its local measurements from R20. 

The possibility of a hotter Universe has been explored in Ref.~\cite{Bengaly:2020vly}, where the currently available temperature-redshift $T(z)$ measurements have been analysed. The authors find a good agreement with the FIRAS measurement and a discrepancy above $1.9\sigma$ from the $T_0$ value needed to solve the Hubble tension.

\subsection{Super-$\Lambda$CDM}

In Ref.~\cite{Adhikari:2019fvb} it has been assumed that a non-Gaussian covariance, due to possibly non-Gaussian primordial fluctuations, can be extracted from a four-point correlation function. This non-Gaussian covariance can be modeled through two additional degrees of freedom describing the trispectrum in the theoretical CMB angular power spectrum, and the resulting model has been named as Super-$\Lambda$CDM. The combination of {\it Planck} 2015 TT + $\tau$ prior ({\it Planck} 2015 TT + $\tau$ prior + R19 + Pantheon) gives $H_0 = 68.4^{+2.5}_{-2.3}{\rm\,km\,s^{-1}\,Mpc^{-1}}$ ($H_0 = 69.9\pm1.7{\rm\,km\,s^{-1}\,Mpc^{-1}}$) at 95\% CL~\cite{Adhikari:2019fvb}, reducing the Hubble tension at $2.7\sigma$ ($2.1\sigma$). It should be checked if this result holds after the inclusion in the fit of the {\it Planck} 2018 polarization data.

\subsection{Heisenberg uncertainty}

In Ref.~\cite{Capozziello:2020nyq} it has been studied how the Heisenberg principle can affect the reliability of cosmological measurements. The authors ascribe the Hubble tension as the effect due to the indetermination associated to the comparison of kinematical versus dynamical measurements. They conclude that the uncertainty on a possible photon mass not accounted for, can be the reason for the $H_0$ disagreement.

\subsection{Diffusion}

Another possible way to alleviate the Hubble tension, considering an effective energy flux from the matter sector into dark energy has been proposed in Ref.~\cite{Perez:2020cwa}. This scenario results naturally from a combination of unimodular gravity and an energy diffusion process. While the two simple models proposed in this study (one of them assuming a quick transfer of energy from the matter density to the cosmological constant sector, and a second one in which a diffusion process decreases anomalously the matter density) may be able to solve the Hubble tension. A complete data analysis is absent in the literature.

\subsection{Casimir Cosmology}

In Ref.~\cite{Leonhardt:2020qam} the extrapolation of physics of the Quantum Vacuum~\cite{Leonhardt:2020fdi}, a theory well-tested in atomic, molecular and optical physics, has been proposed to solve the Hubble tension. In this model the vacuum energy is time-dependent because of the Casimir forces, and therefore $\Lambda$ varies with the cosmic expansion, allowing a larger value for $H_0$.

\subsection{Surface forces}

In Ref.~\cite{Ortiz:2020noa}, the author argues that the inclusion of the surface forces of the homogeneous and isotropic Universe from the Euler Cauchy stress principle can explain the present accelerating expansion of the Universe without any dark energy fluid. The model was constrained using a joint analysis of Hubble parameter measurements and the Pantheon sample. The resulting value of the Hubble constant is enhanced ($H_0= 74.63^{+3.2}_{-2.7}{\rm\,km\,s^{-1}\,Mpc^{-1}}$ at 68\% CL~\cite{Ortiz:2020noa}) solving perfectly the tension with R20. A full analysis involving {\it Planck} 2018 data is missing.

\subsection{\ Milne Model}

In Ref.~\cite{Vishwakarma:2020paa}, the author considers alleviating the Hubble tension within the Milne model~\cite{1935rgws.book.....M}. However, the model needs to be fitted with the observational data in order to be more conclusive in this direction. 

\subsection{\ Running Hubble Tension}

In Ref.~\cite{Krishnan:2020vaf} a running of $H_0$ as a function of redshift is proposed to alleviate the Hubble tension, although a full data analysis is missing. A similar idea has been explored in Ref.~\cite{Dainotti:2021pqg} to analyse the Pantheon SNIa data.

\subsection{\ Rapid Transition in the effective Gravitational Constant}

In Ref.~\cite{Marra:2021fvf} a rapid transition in the value of the relative effective gravitational constant is proposed to explain the lower luminosity of local SNIa and solve the $H_0$ crisis. In particular, the authors assume that there is a transition of this luminosity at $z_t=0.01$, with a 10\% higher luminosity at $z>0.01$, due to a gravitational transition, and they argue that this is a defined testable assumption which would fully resolve both the $H_0$ and the $S_8$ tensions, in addition to provide an equally good fit to BAO, SNIa and Planck. A full data analysis is however missing.

\subsection{\ Causal Horizons}

In Ref.~\cite{Fosalba:2020gls} it has been argued that CMB maps show ``causal horizons'' where cosmological parameters within each horizon can differ significantly, because those regions of the Universe have never been in causal contact. Within these causal horizons (see also Ref.~\cite{Gaztanaga:2020ksy}) $H_0$ takes values which differ up to 20\%, and therefore, if similar ``causal horizons'' are present in the local Universe, variations between the local and high-$z$ measures of the Hubble constant are expected.

\subsection{\ Milgromian Dynamics}

In Ref.~\cite{Haslbauer:2020xaa} a Milgromian dynamics has been proposed as a possibility to solve the Hubble tension. Assuming a cosmological MOND model extended with the presence of sterile neutrinos with mass $11{\rm \,eV/c^2}$, it has been shown that the Keenan-Barger-Cowie (KBC) void~\cite{Keenan:2013mfa} can arise, despite being highly unexpected within the $\Lambda$CDM framework, and can naturally resolve the Hubble tension. 

\subsection{\ Charged Dark Matter}

In Ref.~\cite{Jimenez:2020bgw} a model of Charged Dark Matter has been explored to solve the Hubble tension. In this scenario, the Dark Matter is charged under a dark non-linear electromagnetic force which features a screening of the K-mouflage type~\cite{BeltranJimenez:2021imo}. The idea is that the expansion of different shells is modified by the presence of the electric repulsion, and therefore the $H_0$ value measured locally (inner shells) can be larger for the expansion rate due to the electric interaction with respect to the outer shells (see also Ref.~\cite{BeltranJimenez:2020csl}). A full data analysis is however missing for this proposal.

\section{Summary and Conclusions}
\label{concl}

The $\Lambda$CDM cosmological model, a simple and elegant framework, has been found to provide a very good fit to almost all of the observational probes available until present. Despite its great success, the model is based on the assumption of three basic ingredients (CDM, a cosmological constant, and inflation) whose underlying physics are largely unknown.

The significant discrepancy in the Hubble constant measurements by early and local observations has raised a giant question mark over the $\Lambda$CDM scenario. Along this review, we have focused in this timely and top-priority problem from a number of different perspectives. 

The estimated value of $H_0$ from early time data by the {\it Planck} satellite within the $\Lambda$CDM paradigm~\cite{Aghanim:2018eyx} is significantly differing (at $4.2\sigma$) from the measured values of $H_0$ in model-independent approaches, e.g.\ using the latest local distance ladders by SH0ES collaboration~\cite{Riess:2020fzl}. This has been confirmed by other astronomical missions as well (see Section~\ref{sec:exp} and references therein) leading to a serious and desperate crisis in cosmology. Understanding this large discrepancy in the different observational techniques of the Hubble constant is one of the most serious issues in modern cosmology. Over the last few years, the scientific community has taken a very active role in deciphering this problem. A very large number of possible solutions that could lead to a statistically convincing agreement between the early and late time values of $H_0$ have been investigated. We have classified the proposed models and theories in the following categories: Early Dark Energy (Section~\ref{earlyDE}), Late Dark Energy (Section~\ref{lateDE}), dark energy models with six degrees of freedom and their extensions (Section~\ref{DE6Dof}), models with extra relativistic degrees of freedom (Section~\ref{DR}), models with extra interactions (Section~\ref{InteractSolut}), Unified cosmologies (Section~\ref{Unif}), Modified gravity (Section~\ref{MG}), Inflationary models (Section~\ref{inflat}), Modified recombination history (Section~\ref{RecombH}), Physics of the critical Phenomena (Section~\ref{CritPhen}), and Alternative proposals (Section~\ref{others}). 

The cosmological models arising from each category have been found to resolve the $H_0$ tension with a significance ranging from the $1\sigma $ to the $4\sigma$ level. Based on this, one could first try to categorize these cosmological solutions as {\it excellent}, {\it good}, or {\it moderate}, depending on their ability to solve the $H_0$ tension within $1\sigma$, $2\sigma$, and $3\sigma$, respectively, considering {\it Planck} data alone (see Table~\ref{listPlanck}). Rather than being a quantitative model comparison method, this a priori simple and qualitative taxonomy provides nevertheless a very practical and sharp criteria to classify the large number of the proposed solutions. In fact, these a priori successful cosmological models are often not in agreement with additional cosmological probes, such as Baryon Acoustic Oscillations (BAO) or Pantheon data. Moreover, the Hubble constant tension is alleviated due to an increase in the error bars of $H_0$, rather than by an increase in the Hubble constant itself.

Clearly, this classification could appear extremely basic since it is just based on how well the proposed mechanism solves the tension while ignoring either the physics behind the model or the agreement with other cosmological observables such as, for example, BAO, as well as the effect of the correlation between the datasets, that could cause a fake solution.\footnote{\, See for example Ref.~\cite{DiValentino:2020leo}.} Still, it seems that models based on modifications of the dark energy sector (either dynamical or interacting dark energy) are somewhat more efficient in solving the tension than models based on early dark energy or neutrino-dark matter interactions.

For this reason, we can again now categorize these cosmological solutions as {\it excellent}, {\it good}, or {\it promising}, depending on their ability to solve the $H_0$ tension within $1\sigma$, $2\sigma$, and $3\sigma$, respectively, considering {\it Planck} in combination with external data (mainly BAO, Pantheon, and R19), see Table~\ref{listPlanck+}. In this case, we are accounting for the overall ability of the model to agree with all the available cosmological data. Even if the datasets combinations are not the same for each model, however, this can give a good overview of the most promising proposals, with the details of the datasets combinations used in the text and Figures.

We see that, while no specific proposal makes a strong case for being highly likely or far better than all others, solutions to the Hubble puzzle present in both the Tables~\ref{listPlanck} and~\ref{listPlanck+}, i.e. involving EDE models, DE in extended parameters space, Dynamical DE, Metastable DE, PEDE, VM and its extension, IDE, self-interacting neutrinos, Galileon Gravity, $f(\cal T)$ gravity, \"{U}ber-Gravity, Decaying DM, or Interacting Dark Radiation scenarios, can provide clear improvements to the fit of the cosmological data and thus offer the best options until a better solution comes along. Obviously, this is a priori classification method and a quantitative model comparison should be performed to make this statement more robust. 

Note, that the list of potential cosmological models is quite large and therefore the phenomenology to explore is extremely rich. With the increased sensitivity in the experimental data and the precise measurements of the Hubble constant from various astronomical missions, it seems to us that the journey through the Hubble constant has just began. The measurements of the Hubble constant by the SH0ES collaboration in 2016~\cite{Riess:2016jrr} ($H_0= 73.24 \pm 1.74{\rm\,km\,s^{-1}\,Mpc^{-1}}$ at 68\% CL), 2018~\cite{Riess:2018uxu} ($H_0 = 73.48 \pm 1.66{\rm\,km\,s^{-1}\,Mpc^{-1}}$ at 68\% CL), 2019~\cite{Riess:2019cxk} ($H_0 = 74.03 \pm 1.42{\rm\,km\,s^{-1}\,Mpc^{-1}}$ at 68\% CL) and 2020~\cite{Riess:2020fzl} ($H_0= 73.2 \pm 1.3{\rm\,km\,s^{-1}\,Mpc^{-1}}$ at 68\% CL), have led to a striking tension, and consequently, to the strong need for an alternative physical scenario beyond $\Lambda$CDM.

With this manuscript we aimed to the ambitious goal of presenting the most complete and up-to-date review of the proposed theoretical solutions to the Hubble tension. While we let the reader judge whether we have achieved our goal, we think to have clearly demonstrated how alternative cosmologies, beyond the canonical $\Lambda$CDM paradigm, could play a crucial role in alleviating or solving this problem. The overwhelming effort in the field to find a new cosmological concordance scenario that could accommodate current tensions between complementary datasets that probe vastly different scales and times, strongly suggests that we are now facing a critical phase. While upcoming astronomical observations will shed light on this issue, a synergy of both new theoretical scenarios and improved experimental measurements will be mandatory to solve the Hubble constant puzzle.


\section*{Acknowledgments}

EDV acknowledges the support of the Addison-Wheeler Fellowship awarded by the Institute of Advanced Study at Durham University. OM is supported by the Spanish grants FPA2017-85985-P, PROMETEO/2019/083 and by the European ITN project HIDDeN (H2020-MSCA-ITN-2019//860881-HIDDeN). SP acknowledges the Mathematical Research Impact-Centric Support Scheme [File No.\ MTR/2018/000940] of the Science and Engineering Research Board (SERB), Govt.\ of India. LV acknowledges support from the European Union's Horizon 2020 research and innovation programme under the Marie Sk{\l}odowska-Curie grant agreement No.\ 754496 (H2020-MSCA-COFUND-2016 FELLINI). WY is supported by the National Natural Science Foundation of China under Grants No. 11705079 and No. 11647153, and Liaoning Revitalization Talents Program under Grant no. XLYC1907098. 
AM thanks TASP, iniziativa specifica INFN, for support. DFM thanks the Research Council of Norway for their support. Computations were performed using resources provided by UNINETT Sigma2 -- the National Infrastructure for High Performance Computing and Data Storage in Norway.

\appendix

\newpage


\section{List of conventions and acronyms used}
\begin{table*}[ht]
\footnotesize
\begin{center}
\begin{tabular}{|l|l|}
\hline
\hspace{0.5cm} Greek small letters $\mu, \nu,$... & \hspace{0.5cm} Spacetime coordinates indices\\
\hspace{0.5cm} Latin small letters $i,j,k$... & \hspace{0.5cm} Space coordinates indices\\
\hspace{0.5cm} $g_{\mu\nu}$ & \hspace{0.5cm} Metric tensor \\
\hspace{0.5cm} $\nabla_{\mu}$ & \hspace{0.5cm} Covariant derivative \\
\hspace{0.5cm} $\left(-, +, +, +\right)$ & \hspace{0.5cm} Metric signature \\
\hspace{0.5cm} $\Gamma^{\mu}_{\phantom{\mu} \nu \rho}$ & \hspace{0.5cm} Levi-Civita connection\\
\hspace{0.5cm} $\mathcal{R}^{\mu}_{\phantom{\mu} \nu \alpha \beta}$ & \hspace{0.5cm} Riemann curvature tensor\\
\hspace{0.5cm} $\mathcal{R}_{\mu \nu} = \mathcal{R}^{\alpha}_{\phantom{\alpha} \mu \alpha \nu}$ & \hspace{0.5cm} Ricci tensor\\
\hspace{0.5cm} $\mathcal{R}=\mathcal{R}^{\alpha}_{\phantom{\alpha} \alpha}$ & \hspace{0.5cm} Ricci scalar\\
\hspace{0.5cm} $G_{\mu \nu} = \mathcal{R}_{\mu \nu} - \frac{1}{2} g_{\mu \nu} \mathcal{R}$ & \hspace{0.5cm} Einstein tensor\\
\hspace{0.5cm} $T^{\mu \nu}$ & \hspace{0.5cm} Energy-momentum tensor \\
\hspace{0.5cm} $a(t)$ & \hspace{0.5cm} Scale factor as a function of cosmic time $t$ \\
\hspace{0.5cm} $H(t) \equiv \frac{1}{a}\frac{\mathrm{d}a}{\mathrm{d}t}$ & \hspace{0.5cm} Hubble expansion rate at cosmic time $t$ \\
\hspace{0.5cm} $\tau = \int \frac{\mathrm{d}t}{a(t)}$ & \hspace{0.5cm} Conformal time\\
\hspace{0.5cm} $\dot{v} \equiv \frac{\mathrm{d}v}{\mathrm{d}\tau}$ & \hspace{0.5cm} Conformal time derivative of $v$\\
\hspace{0.5cm} ${\cal H}(\tau) \equiv \frac{1}{a}\frac{\mathrm{d}a}{\mathrm{d}\tau}$ & \hspace{0.5cm} Conformal Hubble expansion rate\\
\hspace{0.5cm} $\rho_m$, $\rho_{\rm DM}$, $\rho_b$ & \hspace{0.5cm} Energy density of matter, dark matter, baryons\\
\hspace{0.5cm} $\rho_r$, $\rho_\nu$ & \hspace{0.5cm} Energy density of radiation and neutrinos\\
\hspace{0.5cm} $\rho_{\rm DE}$, $p_{\rm DE}$ & \hspace{0.5cm} Energy density and pressure of dark energy\\
\hspace{0.5cm} $w_0$ & \hspace{0.5cm} Equation of state with a constant value\\
\hspace{0.5cm} $w_{\rm DE} \equiv p_{\rm DE}/\rho_{\rm DE}$ & \hspace{0.5cm} Equation of state for dark energy ($z$-dependent)\\
\hspace{0.5cm} $M_{\rm Pl} \equiv 1/\sqrt{8\pi G_N}$ & \hspace{0.5cm} Reduced Planck mass\\
\hspace{0.5cm} $\kappa \equiv \sqrt{8\pi G_N}$ & \hspace{0.5cm} Gravitational constant\\
\hspace{0.5cm} $\rho_{{\rm crit},0} \equiv 3H_0^2M_{\rm Pl}^2$ & \hspace{0.5cm} Present critical energy density\\
\hspace{0.5cm} $\mathcal{T}^{\mu}{}_{\nu\rho}$; $\mathcal{T}$ & \hspace{0.5cm} Torsion tensor; torsion scalar\\ 
\hspace{0.5cm} $\mathcal{Q}_{\alpha\mu\nu} \equiv \nabla_{\alpha}g_{\mu\nu}$; $\mathcal{Q}$ & \hspace{0.5cm} Non-metricity tensor; non-metricity scalar\\ 
\hspace{0.5cm} $N_{\rm eff}$; $N_{\rm eff}^{\rm SM} = 3.046$ & \hspace{0.5cm} Effective number of neutrino species; SM value of $N_{\rm eff}$ used here~\cite{Mangano:2005cc,deSalas:2016ztq,Akita:2020szl}\\
\hspace{0.5cm} $r_{\rm s}^*$ & \hspace{0.5cm} Comoving sound horizon at CMB last scattering\\
\hspace{0.5cm} $r_d$ & \hspace{0.5cm} Comoving sound horizon at the end of baryon-drag epoch\\
\hline
\hspace{0.5cm} SM     & \hspace{0.5cm} Standard Model\\
\hspace{0.5cm} (C)DM  & \hspace{0.5cm} (Cold) dark matter\\
\hspace{0.5cm} DE     & \hspace{0.5cm} Dark energy\\
\hspace{0.5cm} CMB    & \hspace{0.5cm} Cosmic Microwave Background\\
\hspace{0.5cm} WMAP   & \hspace{0.5cm} Wilkinson Microwave Anisotropy Probe\\
\hspace{0.5cm} DES    & \hspace{0.5cm} Dark Energy Survey\\
\hspace{0.5cm} SDSS   & \hspace{0.5cm} Sloan Digital Sky Survey\\
\hspace{0.5cm} BAO    & \hspace{0.5cm} Baryon acoustic oscillations\\
\hspace{0.5cm} BOSS   & \hspace{0.5cm} Baryon Oscillation Spectroscopic Survey\\
\hspace{0.5cm} ACTPol & \hspace{0.5cm} Atacama Cosmology Telescope Polarimeter\\
\hspace{0.5cm} SPTPol & \hspace{0.5cm} South Pole Telescope Polarimeter\\
\hspace{0.5cm} {\it Planck} 2015/2018 TT & \hspace{0.5cm} {\it Planck} 2015/2018 temperature power spectrum at high-$\ell$\\
\hspace{0.5cm} {\it Planck} 2015/2018 & \hspace{0.5cm} {\it Planck} 2015/2018 temperature and polarization power spectra at high-$\ell$\\
\hspace{0.5cm} BBN    & \hspace{0.5cm} Big bang nucleosynthesis\\
\hspace{0.5cm} HST    & \hspace{0.5cm} Hubble Space Telescope\\
\hspace{0.5cm} LMC    & \hspace{0.5cm} Large Magellanic Cloud\\
\hline
\end{tabular}
\end{center}
\caption{List of conventions and acronyms used in the review.}
\label{tabnotation}
\end{table*}

\newpage
\section{Successful Models in light of the Hubble constant tension}

\begin{table*}[ht]
\scriptsize
\begin{center}
\begin{tabular}{|l|l|l|}
\hline
tension $\leq 1\sigma$ {\it ``Excellent models''} & tension $\leq 2\sigma$ {\it ``Good models''}  & tension $\leq 3\sigma$ {\it ``Promising models''}  \\
\hline
Dark energy in extended parameter spaces~\cite{DiValentino:2019dzu} & Early Dark Energy~\cite{Murgia:2020ryi} & Early Dark Energy~\cite{Hill:2020osr}\\
Dynamical Dark Energy~\cite{Yang:2018qmz} & Phantom Dark Energy~\cite{Aghanim:2018eyx} & Decaying Warm DM~\cite{Blinov:2020uvz}\\
Metastable Dark Energy~\cite{Yang:2020zuk} & Dynamical Dark Energy~\cite{Aghanim:2018eyx,Yang:2021flj,Yang:2018qmz} & Neutrino-DM Interaction~\cite{DiValentino:2017oaw}\\
PEDE~\cite{Pan:2019hac,Yang:2021egn} & GEDE~\cite{Li:2020ybr} & Interacting dark radiation~\cite{Blinov:2020hmc}\\
Elaborated Vacuum Metamorphosis~\cite{DiValentino:2017rcr,DiValentino:2020kha,DiValentino:2021zxy} & Vacuum Metamorphosis~\cite{DiValentino:2021zxy} & Self-Interacting Neutrinos~\cite{Lancaster:2017ksf,Kreisch:2019yzn}\\
IDE~\cite{Kumar:2019wfs, DiValentino:2019ffd, Yang:2019uog, DiValentino:2017iww,Pan:2020bur,Pan:2019gop, Yang:2019uzo, Yang:2020tax, Yang:2020zuk} & IDE~\cite{DiValentino:2019jae,Yang:2018euj,Pan:2019gop,Yang:2020tax,Yang:2018ubt,Yang:2020zuk} & IDE~\cite{Yang:2018euj}\\
Self-interacting sterile neutrinos~\cite{Archidiacono:2020yey} & Critically Emergent Dark Energy~\cite{Banihashemi:2020wtb}  & Unified Cosmologies~\cite{Yang:2019jwn}\\
Generalized Chaplygin gas model~\cite{Yang:2019nhz} & $f(\mathcal{T})$ gravity~\cite{Wang:2020zfv} & Scalar-tensor gravity~\cite{Ballardini:2020iws}\\
Galileon gravity~\cite{Frusciante:2019puu,Peirone:2019aua} & \"{U}ber-gravity\cite{Moshafi:2020rkq} & Modified recombination~\cite{Chiang:2018xpn}\\
Power Law Inflation~\cite{Tram:2016rcw} & Reconstructed PPS~\cite{Keeley:2020rmo} & Super $\Lambda$CDM~\cite{Adhikari:2019fvb}\\
$f(\mathcal{T})$~\cite{Hashim:2021pkq}&&Coupled Dark Energy~\cite{An:2018vzw}\\
\hline
\end{tabular}
\end{center}
\caption{Models solving the $H_0$ tension with R20 within the $1\sigma$, $2\sigma$ and $3\sigma$ confidence levels considering the {\it Planck} dataset only.}
\label{listPlanck}
\end{table*}

\begin{table*}[ht]
\scriptsize
\begin{center}
\begin{tabular}{|l|l|l|}
\hline
tension $\leq 1\sigma$   {\it ``Excellent models''}   &   tension $\leq 2\sigma$   {\it ``Good models''} & tension $\leq 3\sigma$   {\it ``Promising models''}  \\
\hline
Early Dark Energy~\cite{Smith:2019ihp,Chudaykin:2020acu,Ye:2020btb,Murgia:2020ryi} & Early Dark Energy~\cite{Poulin:2018cxd,Hill:2020osr,Chudaykin:2020igl,Braglia:2020bym} & DE in extended parameter spaces~\cite{DiValentino:2019dzu}\\
Exponential Acoustic Dark Energy~\cite{Yin:2020dwl} & Rock `n' Roll~\cite{Agrawal:2019lmo} & Dynamical Dark Energy~\cite{Yang:2021flj,Yang:2018qmz}\\
Phantom Crossing~\cite{DiValentino:2020naf} & New Early Dark Energy~\cite{Niedermann:2020dwg} & Holographic Dark Energy~\cite{Guo:2018ans}\\
Late Dark Energy Transition~\cite{Benevento:2020fev} & Acoustic Dark Energy~\cite{Lin:2019qug} & Swampland Conjectures~\cite{Agrawal:2019dlm}\\
Metastable Dark Energy~\cite{Yang:2020zuk} & Dynamical Dark Energy~\cite{Yang:2018qmz} & MEDE~\cite{Benaoum:2020qsi}\\
PEDE~\cite{Yang:2021egn} & Running vacuum model~\cite{Sola:2017znb} & Coupled DM - Dark radiation~\cite{Becker:2020hzj}\\
Vacuum Metamorphosis~\cite{DiValentino:2021zxy} & Bulk viscous models~\cite{daSilva:2020mvk,Yang:2019qza} & Decaying Ultralight Scalar~\cite{Gonzalez:2020fdy}\\
Elaborated Vacuum Metamorphosis~\cite{DiValentino:2020kha,DiValentino:2021zxy} & Holographic Dark Energy~\cite{Guo:2018ans} & BD-$\Lambda$CDM~\cite{Sola:2020lba}\\
Sterile Neutrinos~\cite{Carneiro:2018xwq} & Phantom Braneworld DE~\cite{Alam:2016wpf} & Metastable Dark Energy~\cite{Yang:2020zuk}\\
Decaying Dark Matter~\cite{Vattis:2019efj} & PEDE~\cite{Li:2019yem,Pan:2019hac} & Self-Interacting Neutrinos~\cite{Lancaster:2017ksf}\\
Neutrino-Majoron Interactions~\cite{Escudero:2019gvw} & Elaborated Vacuum Metamorphosis~\cite{DiValentino:2020kha} & Dark Neutrino Interactions~\cite{Ghosh:2019tab}\\
IDE~\cite{Kumar:2019wfs,DiValentino:2019ffd,Pan:2020bur,Pan:2019gop} & IDE~\cite{Yao:2020pji,Yang:2018ubt}& IDE~\cite{Kumar:2016zpg, Kumar:2017dnp,Yang:2019uog,DiValentino:2019jae,Yang:2018euj,Yang:2020tax,Pan:2019jqh}\\
DM - Photon Coupling~\cite{Kumar:2018yhh} & Interacting Dark Radiation~\cite{Blinov:2020hmc} & Scalar-tensor gravity~\cite{Rossi:2019lgt,Ballardini:2020iws}\\
$f(\mathcal{T})$ gravity theory~\cite{Nunes:2018xbm} & Decaying Dark Matter~\cite{Pandey:2019plg,Blinov:2020uvz}  & Galileon gravity~\cite{Zumalacarregui:2020cjh,Heisenberg:2020xak}\\ 
BD-$\Lambda$CDM~\cite{Sola:2019jek} & DM - Photon Coupling~\cite{Yadav:2019jio} & Nonlocal gravity~\cite{Belgacem:2017cqo}\\
\"{U}ber-Gravity\cite{Moshafi:2020rkq} & Self-interacting sterile neutrinos~\cite{Archidiacono:2020yey} & Modified recombination~\cite{Chiang:2018xpn}\\
Galileon Gravity~\cite{Renk:2017rzu} & $f(\mathcal{T})$ gravity theory~\cite{Hashim:2020sez} & Effective Electron Rest Mass~\cite{Hart:2019dxi}\\
Unimodular Gravity~\cite{LinaresCedeno:2020uxx} & \"{U}ber-Gravity~\cite{Khosravi:2017hfi} & Super $\Lambda$CDM~\cite{Adhikari:2019fvb}\\
Time Varying Electron Mass~\cite{Sekiguchi:2020teg} & VCDM~\cite{DeFelice:2020cpt} & Axi-Higgs~\cite{Fung:2021wbz}\\
$\metalambda$CDM~\cite{Banihashemi:2018oxo} & Primordial magnetic fields~\cite{Jedamzik:2020krr} & Self-Interacting Dark Matter~\cite{Hryczuk:2020jhi}\\
Ginzburg-Landau theory~\cite{Banihashemi:2018has} & Early modified gravity~\cite{Braglia:2020auw} & Primordial Black Holes~\cite{Nesseris:2019fwr}\\
Lorentzian Quintessential Inflation~\cite{AresteSalo:2021lmp} & Bianchi type I spacetime~\cite{Akarsu:2019pwn} & \\
Holographic Dark Energy~\cite{Dai:2020rfo}& $f(\mathcal{T})$~\cite{Hashim:2021pkq}&\\
\hline
\hline
\end{tabular}
\end{center}
\caption{Models solving the $H_0$ tension with R20 within $1\sigma$, $2\sigma$ and $3\sigma$ considering {\it Planck} in combination with additional cosmological probes. Details of the combined datasets are discussed in the main text.}
\label{listPlanck+}
\end{table*}
\begin{figure*}
\includegraphics[width=0.9\textwidth, right]{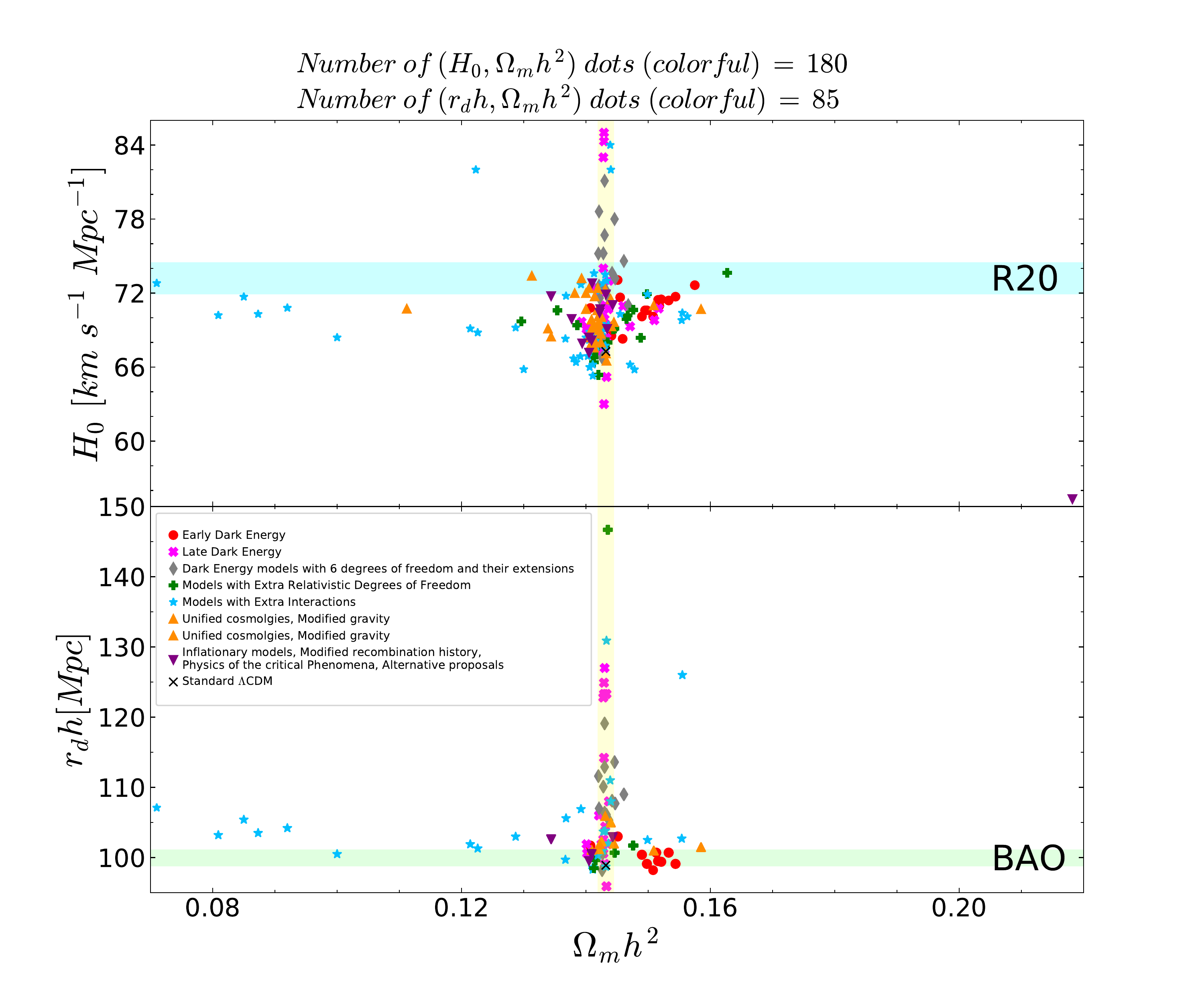}
\caption{In this plot we have an estimate of the density of the available cosmological models proposed to solve or alleviate the Hubble constant tension over the past couple of years. We have therefore accumulated the values of $\Omega_mh^2$, $H_0$ and $r_dh$ from various earlier figures (i.e.\ Figures \ref{fig:chapter4_H0Om},~\ref{fig:chapter5_H0Om},~\ref{fig:chapter6_H0Om},~\ref{fig:chapter7_H0Om},~\ref{fig:chapter8a_H0Om},~\ref{fig:chapter8b_H0Om},~\ref{fig:chapter9_H0Om},~\ref{fig:chapter11-14_H0Om}) into a single plot for a better understanding on the entire theme.  The cyan horizontal band corresponds to the $H_0$ value measured by R20~\cite{Riess:2020fzl}, the yellow vertical band to the $\Omega_mh^2$ value estimated by {\it Planck} 2018~\cite{Aghanim:2018eyx} for the base-$\Lambda$CDM model, and the light green horizontal band to the $r_dh$ value measured by BAO data (see the Planck Legacy Archive \href{https://pla.esac.esa.int}{https://pla.esac.esa.int}). }
\label{fig:all}
\end{figure*}

\clearpage
\bibliographystyle{iopart-num}
\bibliography{H0}

\printindex
\end{document}